\documentclass[preprint]{elsarticle}
\pdfoutput = 1

\usepackage{floatrow}
\usepackage{subfig}

\usepackage[export]{adjustbox}  
\usepackage{booktabs, tabularx} 

\usepackage[utf8]{inputenc}
\usepackage[T1]{fontenc}
\usepackage{lipsum}
\usepackage{xargs} 
\usepackage[pdftex,dvipsnames,table]{xcolor} 
\usepackage{parskip} 
\usepackage{booktabs,arydshln}
\usepackage{algpseudocode} 
\usepackage{algorithm} 
\usepackage{soul} 
\usepackage{outlines}

\soulregister\cite7 
\soulregister\ref7
\soulregister\textsubscript7
\soulregister\textsuperscript7

\usepackage[section]{placeins} 
\makeatletter
\AtBeginDocument{%
 \expandafter\renewcommand\expandafter\subsection\expandafter{%
  \expandafter\@fb@secFB\subsection
 }%
}
\makeatother

\usepackage{amssymb}
\usepackage{mathtools}
\usepackage{mycommands}
\usepackage{lipsum}
\usepackage{boldline}
\usepackage{multirow}
\usepackage{booktabs,siunitx}
\usepackage{nicematrix}
\usepackage{caption}
\usepackage[shortcuts]{extdash}
\usepackage{nameref}
\usepackage{geometry}
\usepackage{appendix}
\usepackage{pdfpages}

\usepackage[colorinlistoftodos,prependcaption,textsize=tiny]{todonotes}

\newwrite\figuresusedout
\immediate\openout\figuresusedout=\jobname.figs%
 
\NewDocumentCommand{\FigureUsedOut}{m}{%
 \immediate\write\figuresusedout{#1}
}

\NewDocumentCommand{\IncludeGraphics}{ O{} m }{%
 \includegraphics[#1]{#2}\FigureUsedOut{#2}%
}

\usepackage{fancyhdr}
\journal{ArXiv}
\begin{document}

\begin{frontmatter}

\title{Advancements in Constitutive Model Calibration: Leveraging the Power of Full-Field DIC Measurements and In-Situ Load Path Selection for Reliable Parameter Inference}
\author[1]{D. E. Ricciardi\corref{cor1}}
\ead{dericci@sandia.gov}

\author[1]{D. T. Seidl}
\author[1]{B. T. Lester}
\author[1]{A. R. Jones}
\author[1]{E. M. C. Jones}

\cortext[cor1]{Corresponding author}
\affiliation[1]{organization={Sandia National Laboratories},
  addressline={PO Box 5800},
  postcode={87185},
  city={Albuquerque, NM},
  country={USA}}

\begin{abstract}

Accurate material characterization and model calibration are essential for computationally-supported high-consequence engineering decisions. Historically, characterization and calibration methods have (1) use simplified test specimen geometries and global data, (2) cannot guarantee that sufficient characterization data is collected for a specific model of interest, (3) use deterministic methods that provide best-fit parameter values with no uncertainty quantification, and (4) are sequential, inflexible, and time-consuming.

This work brings together several recent advancements into an improved workflow called Interlaced Characterization and Calibration (ICC) that advances the state-of-the-art in constitutive model calibration. The ICC paradigm (1) employs tools to efficiently use full-field data to calibrate high-fidelity material models, (2) aligns the data needed with the data collected by adopting an optimal experimental design protocol, (3) quantifies parameter uncertainty through Bayesian inference, and (4) incorporates these advancements into a quasi real-time feedback loop. The ICC framework is demonstrated here on the calibration of a material model using simulated full-field data for an aluminum cruciform specimen being deformed bi-axially. The cruciform is actively driven through the myopically preferred load path using Bayesian optimal experimental design, which selects load steps that yield the maximum expected information gain (EIG). Principal component analysis (PCA) is performed on the model predictions of full-field displacements, and fast surrogate models are built to approximate the input-output relationships of the expensive finite element model. The tools developed and demonstrated here show that high-fidelity constitutive models can be efficiently and reliably calibrated with quantified uncertainty, thus supporting credible decision-making and potentially increasing the agility of solid mechanics modeling by enabling utilization of computational simulations at earlier stages of the design cycle.\blfootnote{SAND2024-15320O}

\end{abstract}


\end{frontmatter}

\includegraphics[width=\textwidth]{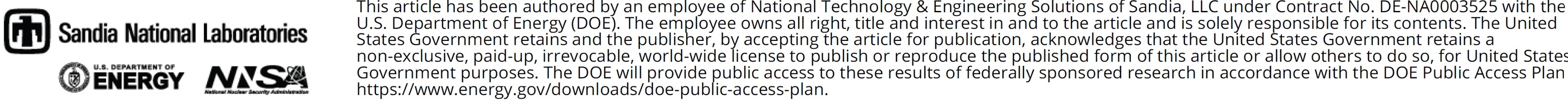}

\floatsetup[figure]{style=plain,subcapbesideposition=top,font=large}
\section{Introduction}\label{sec: introduction}

Computational simulation is relied upon to understand complex engineering problems and provides critical information for decision-making. In finite element analysis (FEA) of solid mechanics, constitutive relationships describe a material's response to external stimuli. These constitutive equations typically contain material-dependent parameters that are determined through material characterization and model calibration. The calibration and subsequent validation of these equations and parameters are crucial for accurate simulations and structural analyses, necessitating an emphasis on efficiency and accuracy in the calibration process to establish confidence in the engineering predictions. 

Historically, material characterization and model calibration have frequently been performed sequentially and independently of each other, often requiring multiple experimental campaigns to gather adequate data for calibration \cite{karlson2016sandia}. This process typically involves assuming phenomenological features based on subject expertise, designing and executing characterization experiments based on those assumptions and calibrating a single material model deterministically. The accuracy of the calibration is not evaluated until the end of the process, and if the results are inadequate, the entire workflow may need to be repeated.

In addition to the procedure, the type of data collected has contributed to the historical challenge of material characterization and model calibration. Traditionally, tests which target specific stress states (i.e.\ uniaxial tension, notched tension, top hat shear, etc.) have been used for calibration. These tests provide limited information per specimen, necessitating large test matrices which include various tests, sample geometries and cut directions in order to reveal complex material behaviors \cite{corona2021anisotropic}. Further, such simplified states are often not reflective of real-world loading conditions. As a result of the described characterization and calibration workflow in tandem with traditional testing procedures, the model calibration process is elongated and expensive, thus delaying the support of computational simulation for design and engineering decisions.

Recent advancements have contributed to significant improvements in constitutive model calibration to attempt to address these shortcomings. The use of digital image correlation (DIC) to collect rich full-field data through camera-based measurements has led to a paradigm shift in material characterization. Also known by the community as Material Testing 2.0 (MT2.0) \cite{pierron2021towards}, full-field DIC data collected from complex testing configurations (specimen shape plus loading) has greatly improved the information content that can be garnered from a single experiment for model calibration \cite{jones2018parameter, jones2019investigation, paranjape2021probabilistic, seidl2022calibration, hamel2023calibrating, ilg2019constitutive, moussawi2013constitutive, bertin2016optimization, rossi2018general, bertin2016integrated, aquino2019design}. Moreover, researchers have developed several different methods---including Chebychev and Zernicke polynomials \cite{Wang2012,Berke2016,Silva2019,Sebastian2012,Wang2009,Wang2011} and various types of wavelets \cite{17salloum,Salloum:2018,Salloum:2020}---for decomposing the field data into a spectral domain in order to reduce the dimensionality of the data while retaining its rich information content. 

The availability of full-field measurements has led to various methods for identification of mechanical properties and parameters from this data \cite{avril2008overview}. Updating techniques, such as Finite Element Model Updating (FEMU) and Constitutive Equation Gap Method (CEGM), compute displacement and stress fields using finite element analysis and optimize a cost function to align computed results with experimental data. In contrast, non-updating methods like the Virtual Fields Method (VFM) and Equilibrium Gap Method (EGM) derive stress fields from measured displacements based on a postulated constitutive model, using equilibrium equations to solve for the desired parameters. Finally data-driven approaches, such as demonstrated in \cite{leygue2018data}, identify the strain-stress relationship of non-linear elastic materials without relying on a specific constitutive equation. These methods construct a database of strain-stress pairs from non-homogeneous strain fields, such as those obtained through DIC, to capture the material's mechanical response.

Even with the improved characterization and parameter identification methods, selection of the proper experimental configuration, or \emph{design}, which greatly influences the suitability of the data for model calibration, remains a challenging task. Thus, another area of improvement for model calibration is the development of procedures for optimal experimental design within the context of solid mechanics. In general, there are two different approaches which have been adopted: 1) optimization of the geometry of the specimen \cite{thoby2022robustness, barroqueiro2020design, souto2016numerical, bertin2016optimization}, or 2) optimization of the load path of deformation \cite{villarreal2023design, ricciardi2024bayesian, fayad2025identification}. Both approaches use various techniques to construct an experiment which will yield the most useful data for model calibration and, consequently, improve constitutive model calibrations and predictions. 

While the advancements made in material characterization and experimental design are significant, they have been, until this point, developed independently of each other. The contribution of this work is to drive forward the state-of-the-art in model calibration by bringing together these recent advancements into one improved workflow by 1) developing tools to efficiently use full-field data to calibrate high-fidelity material models and 2) aligning the data needed with the data collected by adopting the optimal experimental design protocol introduced in Ricciardi et al. \cite{ricciardi2024bayesian}. The new workflow, which has been coined \emph{Interlaced Characterization and Calibration} (ICC), actively controls the load path of a complex specimen \emph{in situ}. The ICC framework was validated in \cite{ricciardi2024bayesian} for a simplified exemplar problem of deforming a material point and calibrating with stress-strain data. This work extends the ICC framework from a material point to a complete structural problem of a cruciform specimen tested in a planar biaxial load frame. The ICC framework is demonstrated synthetically with the simulated deformation of the cruciform specimen in order to calibrate a plasticity model for an aluminum alloy. The development of the tools required for this demonstration are a critical stepping stone in moving towards utilization of the ICC framework with real-time data collection and model calibration.

The remainder of this paper is organized as follows: First, the components of the ICC framework as applied to load path selection of a cruciform specimen in a biaxial load frame are presented (Sec.~\ref{sec: framework}). Next, a discussion of the experimental setup, constitutive model, statistical methods and dimension reduction of displacement field data, as well as all other details of the framework are described in detail (Sec.~\ref{sec: methods}). The ICC framework is demonstrated on an exemplar problem to calibrate an elasto-plastic model using synthetically generated full-field DIC and global load data (Sec.~\ref{sec: hosford example}). Finally, various ICC algorithmic decisions and considerations for future work are discussed (Sec.~\ref{sec: discussion}). In summary, this work brings together several recent advancements in model calibration into one enhanced calibration framework to improve both the efficiency and results of constitutive model calibration.

\section{Interlaced Characterization and Calibration (ICC) Framework}\label{sec: framework}

The ICC framework integrates material characterization and model calibration to enhance efficiency, reduce cost and accelerate the application of computational simulations in design. By prioritizing the collection of the most relevant data, ICC reduces uncertainties in material model parameters and increases confidence in model predictions to support informed decision-making.

In preparation for using the ICC framework with real-time data collection and model calibration, a simulated experiment of a cruciform specimen being deformed biaxially was used for tool development. Instead of simulating a traditional cruciform test, which applies equal loading on both arms and yields homogenous stress data in the central gauge region, the movement on each axis was isolated in order to introduce complex stress fields which are representative of real-world loading conditions. Thus, the experimental configuration in tandem with the complex geometry of the cruciform provides rich information on hardening and anisotropy.

\begin{figure}[ht!]
\centering
\includegraphics[width=140mm]{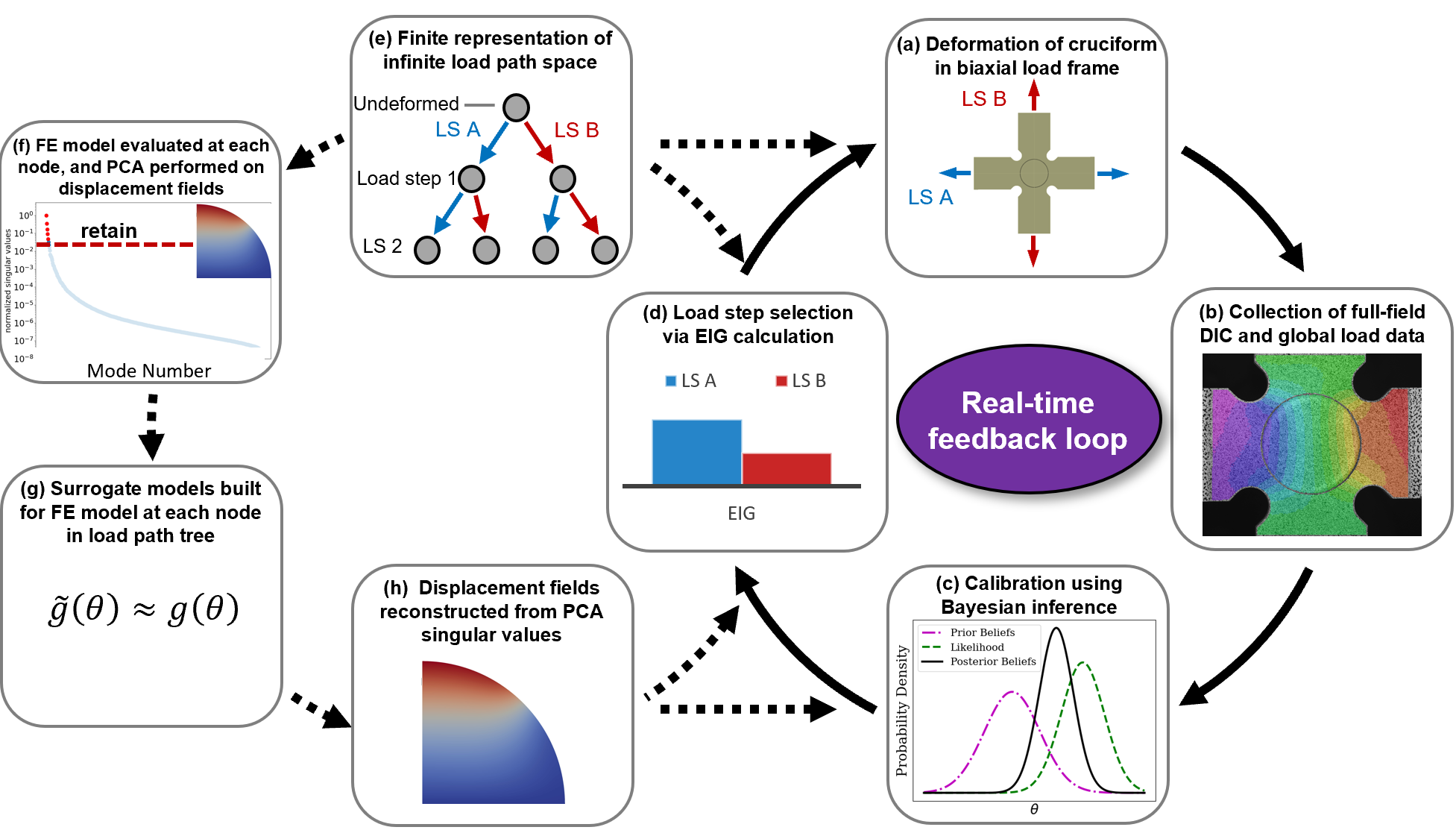}
\caption{The ICC framework is shown for a cruciform specimen being deformed in an actively controlled biaxial load frame. Solid arrows point in the direction the ICC steps proceed, and dashed arrows stem from steps that happen offline prior to entering the ICC feedback loop and point towards steps where they are utilized. An initial displacement is imposed on the cruciform along a single axis, either the X (horizontal) or Y(vertical) axis corresponding to load step (LS) A or B, respectively, while keeping the other axis fixed (a). Full-field DIC and global load data are collected (b). Given the collected data and any prior knowledge on the model parameters, Bayesian inference is used to update the knowledge about the parameters with quantified uncertainty (c). The next load step is chosen as the one which has the greater EIG (d). Prior to entering the feedback loop, the infinite load path space is reduced to a finite representation through a load path tree (e). The FE model is evaluated at each node for various parameter combinations, and PCA is performed on the displacement fields (f). Surrogate models are constructed in the spectral domain to approximate the expensive FE simulation at each node in the load path tree (g), which are the locations where data may be collected. The displacement fields are reconstructed from the surrogate predictions of the PCA singular values before being used in the feedback loop (h), along with the predicted load.}
\label{fig: ICC framework}
\end{figure}

The steps of the framework are illustrated in Fig.~\ref{fig: ICC framework}. The solid arrows point to each successive step in the ICC framework, where data collection and model calibration occur in an iterative, quasi-real-time feedback loop. The dashed arrows stem from tasks that occur offline prior to entering the feedback loop and point to the steps where they are utilized. Starting with plot (a), the experiment begins with an initial load step---deforming the cruciform specimen along one of the two axes, which correspond to load step (LS) A or B, while holding the other axis fixed (displacement = 0). DIC field data and global load data are collected in (b). In (c), model calibration is performed via Bayesian inference, which yields a distribution that describes the parameter uncertainty. The expected information gain (EIG) is then calculated for the load step options (applying an increment of displacement along either of the two axes), and the load step that results in the greater EIG is determined to be the next load step in (d). The process then returns to (a), where the selected load step is taken. The feedback loop continues for some pre-determined number of cycles (equivalently, load steps) and yields a calibrated model with quantified uncertainty given the load path that is chosen. 

Several tasks occur offline prior to entering the feedback loop. In order to control the experiment, the infinite load path space is reduced to a finite load path tree, as seen in (e), which enables load step selection in (d) and aids in the construction of surrogate models in (g). The FE model is evaluated at each node in the load path tree for various parameter combinations, and PCA is performed on the displacement fields (f).  Finally, in order to efficiently perform steps (c) and (d) for the expensive finite element (FE) model, a surrogate replacement of the model is used that approximates the input-output relationship of the expensive model at each node in the load path tree and is computationally cheap to evaluate: The surrogate model output consists of global load values and PCA singular values for the dimensionally-reduced displacement fields. The displacement fields are reconstructed from the surrogate predictions of the PCA singular values in (h) before being used in the feedback loop. Each of the components which comprise the ICC framework in this setting is discussed in detail in Sec.~\ref{sec: methods}.
    
In reality, the load path space from which the myopically preferred load path is being selected is infinite when considering strain rate, step size, direction, etc. In order to enable the selection of the load path in real-time, the decision space must be moved from infinite to finite dimensions. This is done by reducing the infinite-dimensional load path space of material deformation to a discrete space through the introduction of a load path tree.

\begin{figure}[!ht]
\begin{center}
  \includegraphics[width=75mm]{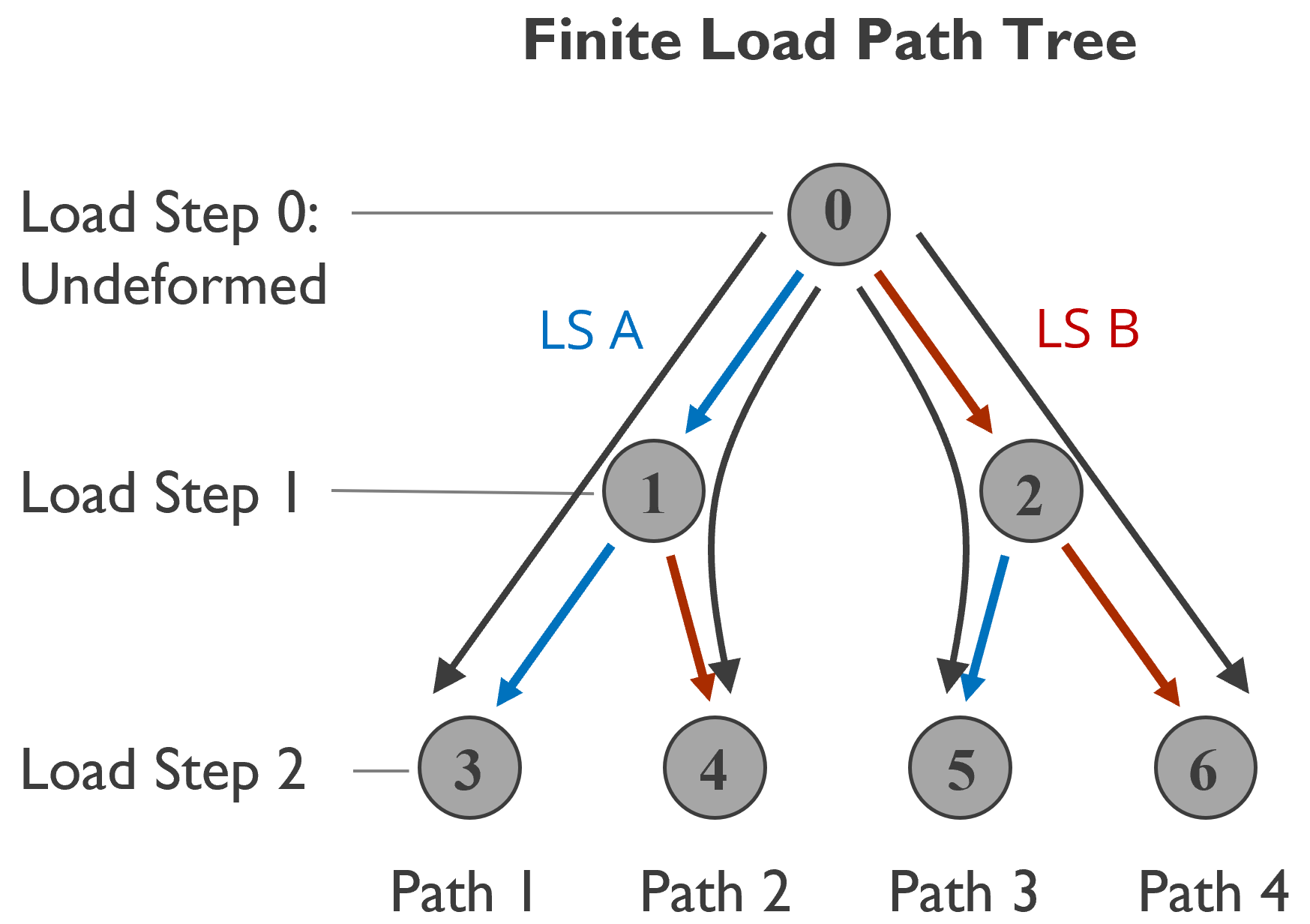}
  \caption{The load path tree reduces the infinite-dimensional load path space to finite dimensions. Here, a binary tree is shown which stands in for the deformation of a cruciform specimen in a biaxial load frame. The top node represents the specimen in its undeformed state. Each parent node has two children corresponding to applying an increment of displacement along either the X (horizontal) axis---load step (LS) A---or the Y (vertical) axis---LS B---while holding the alternate axis fixed. In the example tree, four possible load paths (1-4) are shown, each consisting of two load steps.}
\label{fig: loadpath_tree}
\end{center}
\end{figure}

The same graph structure that was introduced in \cite{ricciardi2024bayesian} is adopted here. The graph structure is depicted in Fig.~\ref{fig: loadpath_tree}, which is a binary tree that stands in for the biaxial loading of the cruciform specimen. The BOED workflow begins at the root node with the cruciform in its undeformed state. Each parent node has two children nodes that correspond to applying a fixed displacement increment along either the horizontal axis or the vertical axis of the cruciform while holding the alternate axis fixed. The ICC framework is not restricted to this tree structure, but it was chosen for this demonstration as a simple starting point. Other more complex, perhaps more appropriate, options are possible and are discussed in Sec.~\ref{sec: discussion}. In this work, the two load step options are generically referred to as A and B, which correspond to applying an increment of displacement along the X (horizontal) or Y (vertical) axis, respectively.

In summary, the ICC framework utilizes a combination of complex specimen geometry, discovery of the preferred load path via BOED and full-field DIC calibration data, leading to fewer experiments and increased information content for model calibration compared to traditional approaches. As a result, well-calibrated constitutive models with quantified parameter uncertainty, a natural output of BOED, supports decision-making at earlier stages in the design cycle. 

\section{Methods}\label{sec: methods}

In this section, the various methods and techniques utilized to perform material characterization and model calibration in the ICC framework are discussed. The model of the physical system is discussed in Sec.~\ref{sec: constitutive model}, with details of the model form in Sec.~\ref{sec: material model form}, the cruciform and finite element analysis in Sec.~\ref{sec: cruciform description} and the synthetic data generation in Sec.~\ref{sec: data generation}. A description of dimension reduction of the displacement field data through PCA is found in Sec.~\ref{sec: dimension reduction}, and details of the surrogate model construction are found in Sec.~\ref{sec: surrogate model}. Bayesian techniques for inference and experimental design are discussed in Sec.~\ref{sec: bayesian inference} and Sec.~\ref{sec: boed}, respectively.

\subsection{Model of the Physical System}\label{sec: constitutive model}

The selection of the proper model to represent the phenomena of interest is an important consideration and is crucial for producing credible predictions of material behavior. The process is simplified in this synthetic example as the same model being used to generate the data is also the one being calibrated. Thus, model-form error is removed from consideration, and in the absence of noise, the model fully captures the data generation process.

\subsubsection{Material Model Form}\label{sec: material model form}

\indent For the current study, the calibration of an elastoplastic constitutive model is being explored. The (synthetic) experiments considered in this work are conducted at close to quasistatic rates and nominally room-temperature. As such, a rate and temperature-independent model is leveraged. Further, while non-proportional loadings are being probed, reverse and/or cyclic paths are not being pursued. Thus, plastic hardening is taken to be isotropic, and considerations of multiple and/or alternative hardening forms are left to future work. 

\indent The final consideration in the selection of the model is the yield function. In the current study, multiaxial loadings are being leveraged enabling the consideration of non-quadratic and/or anisotropic yield functions. Multiple yield functions in those classes have been used to probe the response of various aluminum alloys and corresponding forms~\cite{KandK2008,KandK2008_PartII,Kohar2017,corona2021anisotropic,jones2021anisotropic}. Each model has a different degree of complexity and corresponding number of calibration parameters. As this effort represents the first attempt at the novel ICC calibration approach, the relatively simple Hosford yield function~\cite{hosford} is considered to enable testing of the proposed scheme. While isotropic, the non-quadratic Hosford surface needs only a single calibration coefficient (the exponent, $a$) to describe different strengths in shear versus tension. The single parameter enables keeping the dimensionality of the calibration optimization problem appropriate while introducing a term sensitive to the yield surface shape. Further, the Hosford model has been shown to reasonably reproduce various responses in aluminum alloys (e.g. ~\cite{KandK2008,Kohar2017}).

\indent Combining these various considerations, the constitutive response of interest is taken to be,

\begin{equation}\label{eq: constitutive_response}
  \sigma_{ij}=\mathbb{C}_{ijkl}\left(\varepsilon_{ij}-\varepsilon_{ij}^{\text{p}}\right),
\end{equation}

\noindent in which $\sigma_{ij}$, $\mathbb{C}_{ijkl}$, $\varepsilon_{ij}$ and $\varepsilon_{ij}^{\text{p}}$ are the Cauchy stress, elastic stiffness (assumed isotropic), total strain and plastic strain, respectively. The yield function, $g$, is then written,

  \begin{equation}\label{eq: yield function}
    g = \phi(\sigma_{ij}) - \bar{\sigma}(\kappa),
  \end{equation}

\noindent with $\kappa$ being the isotropic hardening variable (found via the consistency condition), $\phi$ the effective stress describing the shape of the yield function and $\bar{\sigma}$ the flow stress capturing the size of the yield surface. The effective stress is taken to be that of Hosford and given as,

  \begin{equation}\label{eq: hosford effective stress}
    \phi(\sigma_{ij}) = {\frac{1}{2}[|\sigma_1 - \sigma_2|^a + |\sigma_2 - \sigma_3|^a + |\sigma_3 - \sigma_1|^a]}^{\frac{1}{a}},
  \end{equation}

\noindent where $\sigma_1,~\sigma_2$ and $\sigma_3$ are the (ordered) principle stresses and $a$ is the exponent. As is commonly done with aluminum materials, a Voce-type saturation expression is used for the flow stress such that,

  \begin{equation}\label{eq: voce hardening}
    \bar{\sigma}(\kappa) = \sigma_y + A\left(1 - \exp(-n\kappa)\right).
  \end{equation}

\noindent In Eqn.~\eqref{eq: voce hardening}, $\sigma_y$, $A$ and $n$ are the initial yield stress, hardening modulus and hardening coefficient to be found via calibration. An associative flow rule is also utilized such that,

\begin{equation}\label{eq: associative_flow_rule}
  \dot{\varepsilon}_{ij}^{\text{p}}=\dot{\kappa}\frac{\partial\phi}{\partial\sigma_{ij}}.
\end{equation}

\indent The described constitutive model has been implemented as part of the Library of Advanced Materials for Engineering (LAM\'{E})~\cite{LAME} in the Sierra/SolidMechanics finite element code~\cite{UserGuide}. It is implemented via a traditional implicit, closest point projection return mapping 
algorithm~\cite{simo2006computational,Lester2017,Lester2021}.   

\indent The model parameters being inferred from the data are the initial yield stress $\sigma_y$ and isotropic hardening variables $A$ and $n$ from Eqn.~\eqref{eq: voce hardening} and the exponent $a$ from Eqn.~\eqref{eq: hosford effective stress}. These parameters are collectively referred to as the \emph{unknown} parameters such that $\pars := [\sigma_y,\, A,\, n,\,a]^T$. The elastic properties -- Young's modulus, $E$, and Poisson ratio, $\nu$ -- are taken to be known as they are well established.

\subsubsection{Cruciform Specimen}\label{sec: cruciform description}

The key design parameters for the cruciform geometry used in the ICC framework were (1) a multi-axial/ complex state of stress, (2) multiple load-steps before failure and (3) no yielding in the arms (i.e.\ failure would occur in the central gauge area instead of the arms).  Many cruciform geometries exist in the literature \cite{creuziger2017,muller1996,kuwabara2007,abu2009,tiernan2014,xiao2019,iadicola2014,demmerle1993,mamros2022,makinde1992}, but these geometries are not always in strict alignment with the above criteria. The cruciform for this work was designed according to the criteria presented in Marmos et al. \cite{mamros2022}, which capitalizes on using re-entrant radii and a thinned gauge area. The geometry is shown notionally in Fig.~\ref{fig: cruciform schematic}, with complete dimensions included in Fig.~\ref{fig:cruciform_geometry} in the Supplementary Materials. Further details and considerations of the specimen design can be found in \cite{jones2024interlaced}.

Exploiting symmetry about the three principal planes enables consideration of an eighth symmetry model as depicted in Fig.~\ref{fig: cruciform schematic} to preserve computational resources. Corresponding symmetry displacement boundary conditions were used on these planes to enforce this condition.  To achieve the desired experimental loadings, a constant and uniform displacement was applied at the distal nodes of each arm along the coordinate direction of either the X (horizontal) or Y (vertical) axis---corresponding to taking load step A or load step B in Fig.~\ref{fig: loadpath_tree}.  Surfaces without displacement-specified boundary conditions were taken to be traction-free. The mesh was comprised of linear hexahedral elements with a selective-deviatoric integration rule. The elements in the center region were considerably smaller than those in the arms---Fig.~\ref{fig: cutout quarter symmetry mesh}. 

The deformation of the specimen was driven by displacement boundary conditions applied at a constant rate, and the displacement increment was 0.25~mm per load step applied to each arm. For example, load path 1 in Fig.~\ref{fig: loadpath_tree} consists of two sequential A load steps---denoted as load path AA---in which the end of each horizontal arm of the cruciform is displaced by 0.25~mm outward at each load step while the ends of the vertical arms are held in fixed in displacement control. Given the symmetry boundary conditions, the displacement of each load step equates to 0.5~mm total extension along a given axis. 

\floatsetup[figure]{style=plain,subcapbesideposition=top,font=small}
\begin{figure}%
  \centering
  \sidesubfloat[]{{\includegraphics[width=55mm]{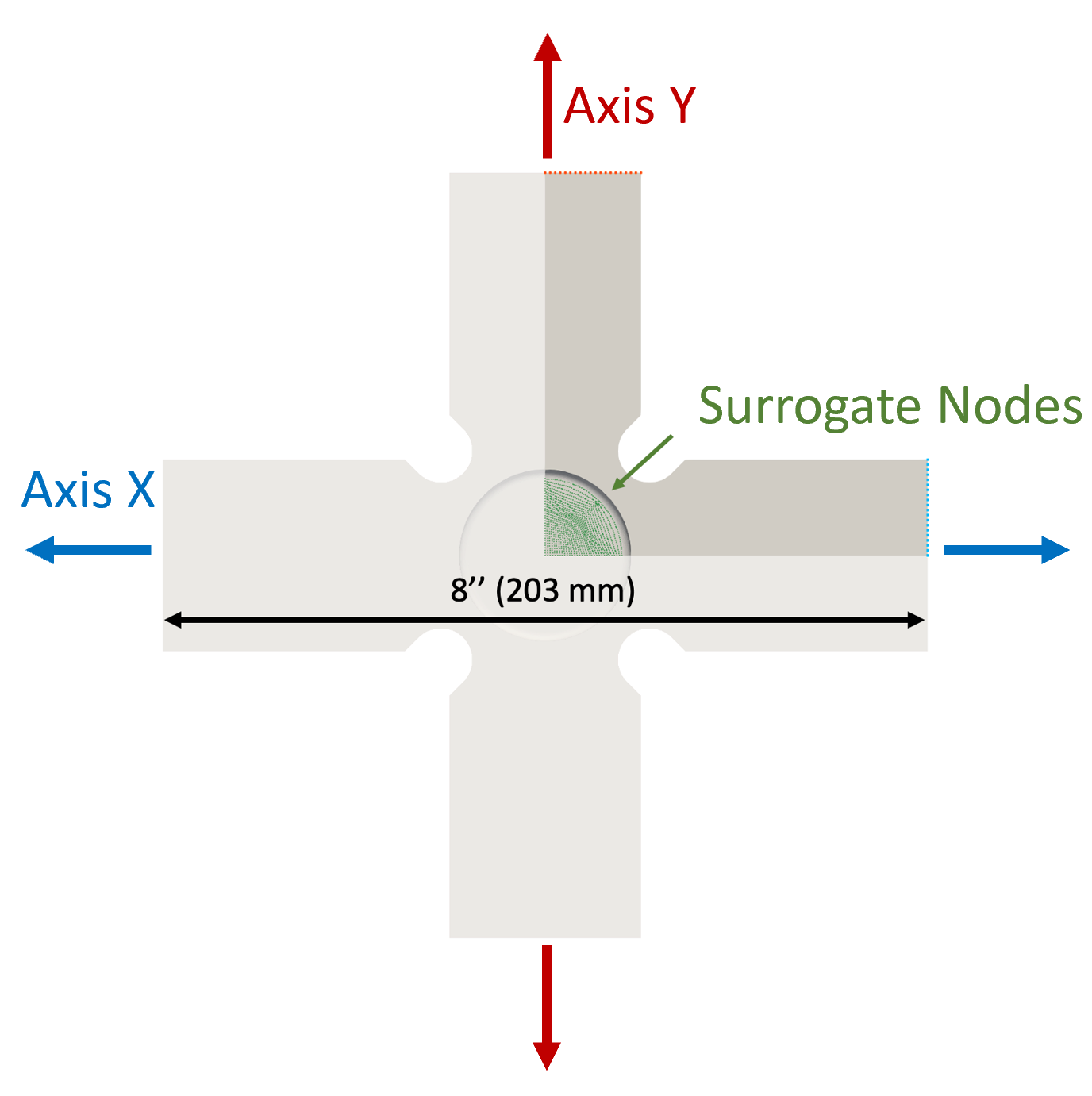}\label{fig: annotated cutout cruciform} }}%
  \sidesubfloat[]{{\includegraphics[width=50mm]{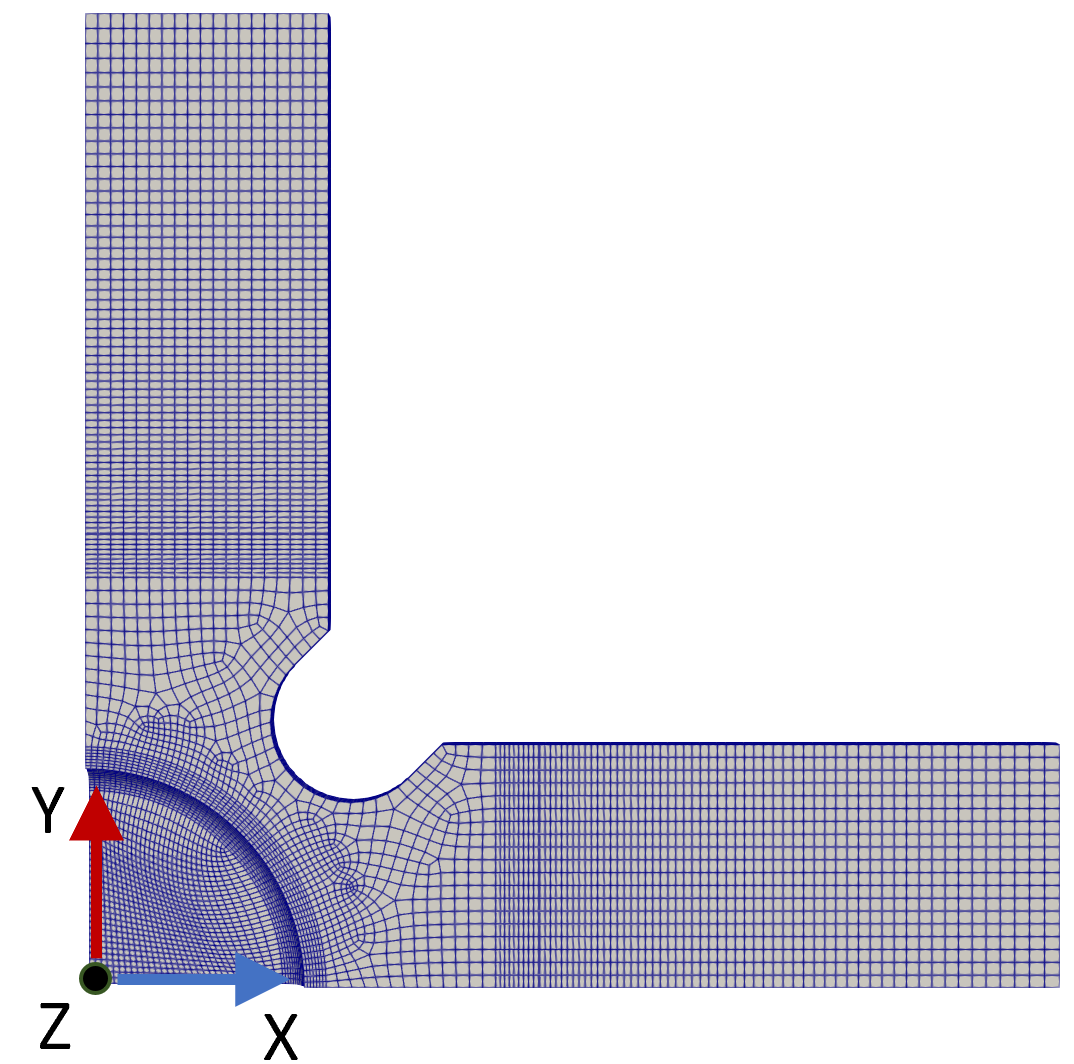}\label{fig: cutout quarter symmetry mesh} }}\\
  \caption{ Cruciform geometry and mesh. The full geometry of the cruciform specimen is shown in (a) with the opaque region denoting the eighth-symmetry domain considered for computation. Displacement boundary conditions for the X axis and Y axis were prescribed at the nodes at the end of each arm. The displacement surrogates predict values at nodes in the circular region (green). The mesh of the computational domain used for the simulation is shown in (b) depicting the heterogeneous element size distribution.}.%
  \label{fig: cruciform schematic}%
\end{figure}

An implicit, quasistatic FEA was performed using Sierra/SM \cite{UserGuide} with the given geometry and material model form (Sec.~\ref{sec: material model form}). An adaptive time-stepping algorithm was employed for the backward Euler time integrator used with the global equilibrium solution. The quantities of interest (QoIs) extracted from the FEA, which can in principle be computed from experimental measurements, include nodal displacements and resultant forces along each axis of the cruciform. The QoIs are returned for both the X and Y directional components. Thus the input-output relationship of the physical model is written as $g(\pars) = [\text{disp}_i, \text{load}_i], \, i \in [X, Y]$.

\subsubsection{Synthetic Data Generation}\label{sec: data generation}

To generate synthetic experimental data of the cruciform specimen, the FEA was run with a set of model parameters which are referred to as the true parameters, $\pars^{true}$, and are reported in Table~\ref{table: cruciform parameters}. The values in $\pars^{true}$ were loosely based on a traditional calibration \cite{lu2021solid,DFSAND} of the same material under consideration here so that this synthetic demonstration would closely mimic the expected behavior of a physical experiment. 

\begin{table}[ht!]
\centering
\begin{tabular}{lll}
\hline
\textbf{Parameter} & \textbf{Symbol} & \textbf{Value} \\ 
\hlineB{4}
Young's modulus & $E$ & 9900 ksi (68.3 GPa) \\
Poisson's ratio & $\nu$  & 0.33 \\
Initial yield stress & $\sigma_{y}$ & 42.51 ksi (293.1 MPa) \\
Hardening modulus & $A$ & 13.63 ksi (94.0 MPa) \\
Hardening exponent & $n$ & 14.35 \\
Hosford exponent & $a$ & 11.19 \\
\hline
\end{tabular}
\caption{Material model parameter values used for synthetic data generation, $\pars^{true}$.}\label{table: cruciform parameters}
\end{table}

The number of load steps and the displacement increment per load step were chosen to keep the maximum total strain imposed on the cruciform within the plastic regime of deformation for an aluminum alloy, away from failure \cite{lu2021solid}. The synthetic experiment was carried out through 5 load steps with each load step imposing a displacement increment of 0.25~mm per arm. Thus a load path of 5 identical load steps, i.e. pulling on the same axis for the duration of the experiment, would result in a total displacement of 1.25~mm along that axis (and a total displacement of 2.50~mm when considering symmetry) and 0~mm of displacement along the alternate axis. 

\begin{figure}%
    \centering
    \sidesubfloat[]{{\includegraphics[width=75mm]{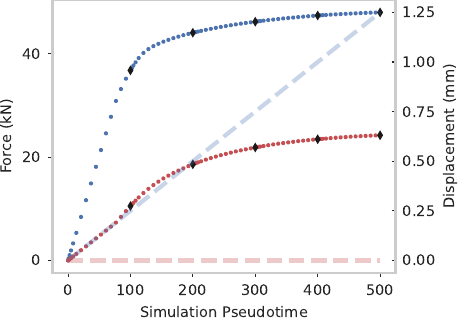}\label{fig: AAAAA bcs} }}%
    \sidesubfloat[]{{\includegraphics[width=75mm]{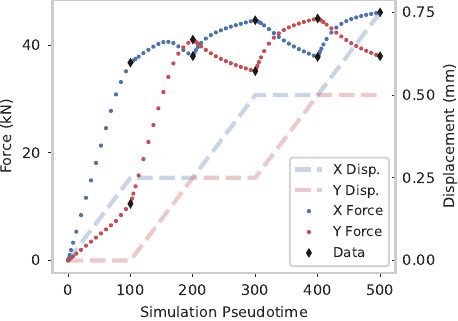}\label{fig: ABABA bcs} }}\\
   \caption{The left plot shows the resulting global forces from following load path AAAAA and the right plot from following load path ABABA using parameter values from Table~\ref{table: cruciform parameters}. Locations of the measured data points are shown with black diamonds, which correspond to the nodes in the load path tree. The right axes of the plots show the associated displacement for each case.}%
    \label{fig: cruciform bcs}%
\end{figure}

To aid in the description of the simulated cruciform experiment, the displacements and resulting forces from two contrasting load paths are shown in Fig.~\ref{fig: cruciform bcs}. The left plot shows results from a load path pulling on the X axis for 5 consecutive load steps (AAAAA), and the right plot shows an alternating load path (ABABA). The resulting forces for each directional component are plotted vs. simulation pseudotime using parameter values from Table~\ref{table: cruciform parameters}, and applied displacement boundary conditions are plotted on the right axis. The 5 diamonds correspond to nodes in the load path tree, which is where measurements were collected for calibration. 

Two data types were extracted from the FEA to be used as the synthetic experimental data: these were 1) the nodal displacements (which simulate full-field DIC displacement measurements) collected from 1,022 nodes in the circular region of the cruciform geometry, indicated by the green nodes in Fig.~\ref{fig: cruciform schematic}, and 2) the resultant forces along each axis of the cruciform. The QoIs were stored at the end of each load step, which coincide with each node of the load path tree (Fig.~\ref{fig: loadpath_tree}, but extended to 5 load steps). Independent and identically distributed (i.i.d.) Gaussian noise was added to the synthetic data to mimic what is seen experimentally. For the forces, a noise variance of 582.11~$\text{N}^2$ was used, and for the nodal displacements, a noise variance of 4~{\textmu}m\textsuperscript{2} was used. In the remainder of this paper, \emph{observed data} refers to data that was synthetically generated. An example of the synthetic displacement data for the AAAAA and ABABA load paths can be found in Figs.~\ref{fig: AAAAA data x}-\ref{fig: ABABA data y} in the Supplementary Material.

\floatsetup[figure]{style=plain,subcapbesideposition=top,font=small}

\subsection{Dimension Reduction of Displacement Field Data}\label{sec: dimension reduction}

The displacement field data provides rich information content for calibration, and when calibrating the expensive FE model, fast surrogate replacements are used to predict the model output---load and displacement---efficiently (Sec.~\ref{sec: surrogate model}). The high-dimensional nature of the displacement field data challenges the construction of the surrogate models, so dimension reduction is used in order to build surrogates for the expensive FEA.

There are various methods to choose from when considering dimension reduction of full-field data. Methods based on Chebychev and Zernicke polynomials have been developed \cite{Wang2012,Berke2016,Silva2019,Sebastian2012,Wang2009,Wang2011}; however, these are often sensitive to missing data and difficult to perform on irregular geometries. There are also methods based on various types of wavelets \cite{17salloum,Salloum:2018,Salloum:2020}, including the hierarchical polynomial wavelet decomposition (HPWD) method \cite{Kury2022HierarchalPW}. HPWD combines features from Alpert wavelets along with QR factorization to create an orthogonal basis for data representation. Thus, HPWD works directly on the domain, or point cloud, with no dependence on the field measurements (e.g.\ full-field DIC displacement data), and inherently, without consideration of the material model. Furthermore, HPWD does not need collocated points. HPWD can operate between different point clouds associated with the same geometry, for instance, nodes from an FE mesh from a simulation and a point cloud from experimental DIC data; it does not require data to be remapped from one point cloud to the other. Finally, there are singular value decomposition (SVD)-based techniques, such as principal component analysis (PCA) \cite{greenacre2022principal} or proper orthogonal decomposition (POD) \cite{chatterjee2000introduction}, which require training data from the FEA simulation to obtain a transformation basis. HPWD and PCA were the two methods considered for this work as both handle missing data and irregular geometries well and are capable of capturing both large and small scale patterns in the data. After an analysis of the two methods for the specific case of dimension reduction for full-field data from a cruciform specimen, PCA was chosen as it yielded a smaller reconstruction error with fewer retained components (or modes). The remainder of this section will present the high-level procedure for this dimension reduction technique.

PCA is a method for dimension reduction which performs a linear transformation on a data set $\mathcal{A}\in \mathbb{R}^{u\times v}$ in order to project it onto a new lower-dimensional orthogonal basis, a.k.a.\ principal components, such that the maximum amount of variance in the data is preserved. This dimension reduction is accomplished through a decomposition of the original matrix $\mathcal{A}$ into three new matrices via SVD \cite{klema1980singular} such that $\mathcal{A} = USV^T$. The maximum number of principal components that can be extracted from the data matrix is determined by the rank $r$ of matrix $\mathcal{A}$; given a full-rank matrix, $ r \coloneqq \text{rank}(\mathcal{A}) = \min\{u, v\}$.

\begin{figure}[hb!]
  \centering
  \includegraphics[width=140mm]{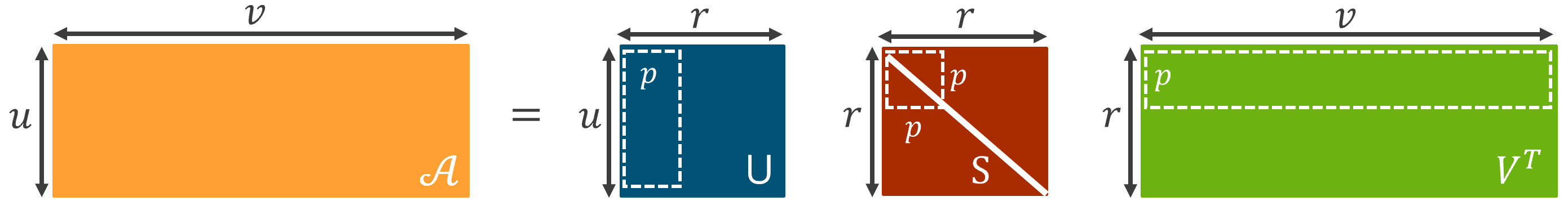}
  \caption{Schematic representation of singular value decomposition (SVD) of a non-square matrix $\mathcal{A} \in \mathbb{R}^{u \times v}$ with $u < v$. SVD factorizes the matrix into three new matrices $\mathcal{A} = USV^T$, with $U \in \mathbb{R}^{u \times r}$, $S \in \mathbb{R}^{r \times r}$ and $V^T \in \mathbb{R}^{r \times v}$, where $r$ is the rank of $\mathcal{A}$. The columns and rows of $U$ and $V^T$ contain the left and right singular vectors of $\mathcal{A}$, respectively. The singular values are contained in $S$. By retaining a number of principal components $p < r$, matrix $\mathcal{A}$ can be expressed in a reduced-dimensional space.}
  \label{fig: svd}
\end{figure}

Fig.~\ref{fig: svd} shows a schematic representation of SVD for a rectangular matrix $\mathcal{A}$, where $u < v$. The matrices $U\in\mathbb{R}^{u\times r}$ and $V^T \in \mathbb{R}^{r \times v}$ are orthonormal bases with columns and rows, respectively, consisting of the left and right singular vectors of $\mathcal{A}$, and $S \in \mathbb{R}^{r \times r}$ is a diagonal matrix of singular values. The directions of the principal components are explained by the rows of $V^T$ and their magnitudes by the diagonal entries of $S$. The principal components are ordered according to the amount of variance they explain in the original data set from greatest to least. Thus, their relative importance is revealed, and a subset of components can be used to explain a given amount of variance in the original data set. By retaining a number $p < r$ of principal components, the data may be represented in a reduced-dimensional space with very little information loss. Further details and nuances of PCA are left to \cite{greenacre2022principal}.

The schematic representation of SVD in Fig.~\ref{fig: svd} is reflective of the context in which PCA was exercised in this work. Training data for the PCA decomposition was generated by first obtaining samples of the unknown model parameters. For this task, Halton samples \cite{halton1960efficiency}, which consist of a deterministic sequence of prime numbers to produce a space-filling design, were used. 400 training samples were generated within the bounds specified for the unknown parameters in Table~\ref{Tab: parameter bounds}, $\pars^{hal} \in \mathbb{R}^{u \times D}$, where $D$ is the dimensionality of the unknown parameter vector $\pars$ ($D=4$ in this case) and $u=400$---all other parameters were held fixed at the values recorded in Table~\ref{table: cruciform parameters}. Each parameter sample was then used as input to the FE model, $g(\pars^{hal})$, to produce corresponding output, which was the nodal displacements and resultant forces for both the $X$ and $Y$ directional components. However, PCA was performed only for the high-dimensional displacement field, not the resultant forces. 

\begin{table}[ht!]
\centering
\begin{tabular}{llll}
\hline
\textbf{Parameter} & \textbf{Symbol} & \textbf{Bounds} \\ 
\hlineB{4}
Initial yield stress & $\sigma_y$ & $[32.0, 50.0]$ ksi \\
Hardening modulus & $A$ & $[1.0, 20.0]$ ksi \\
Hardening exponent & $n$ & $[0.5, 20.0]$\\
Hosford exponent & $a$ & $[4.0, 16.0]$\\
\hline
\end{tabular}
\caption{Bounds for the unknown parameters.}\label{Tab: parameter bounds}
\end{table}

\begin{figure}
  \centering
  \includegraphics[width=140mm]{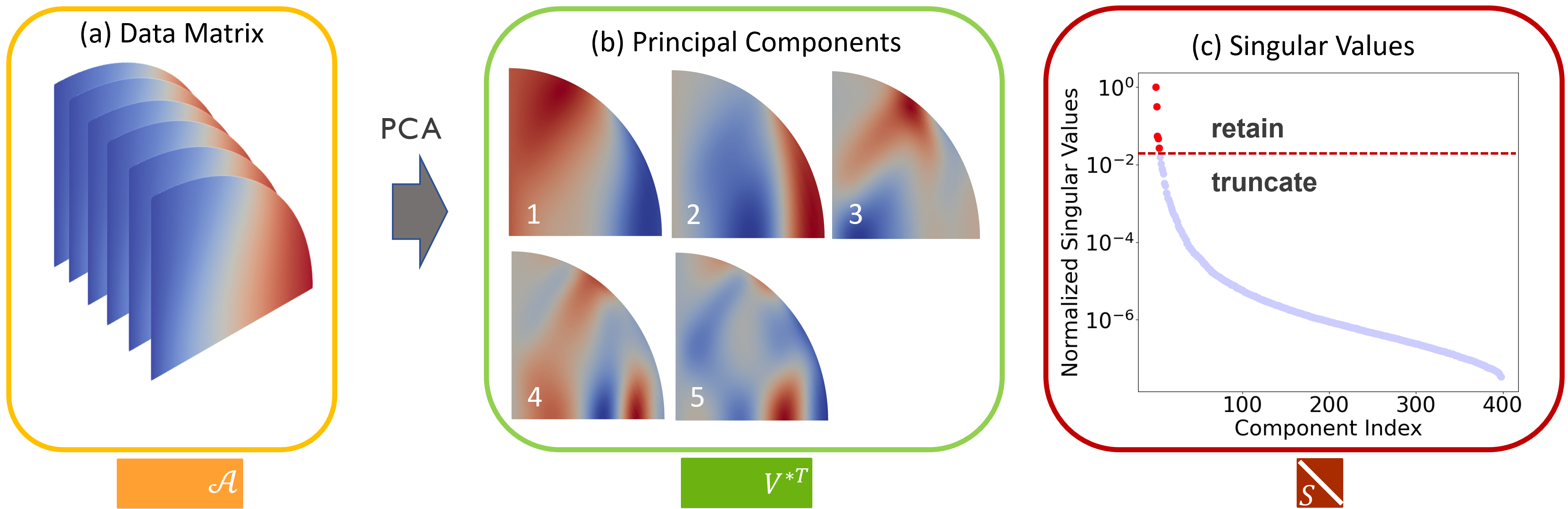}
  \caption{PCA uses SVD to factor a data matrix into three new matrices $\mathcal{A}=USV^T$ by which dimension-reduction proceeds; this approach was used to reduce the full-field displacement measurements from the FE model. The first step was to generate $u$ parameter samples, $\pars^{hal} \in \mathbb{R}^{u \times D}$, and evaluate the FE model at the sampled parameters, $g(\pars^{hal})$. The output was collected into the data matrix, $\mathcal{A}$, in (a). The first five principal components in $V^T$ were retained to produce $V^{*T}$ and are shown in (b). The corresponding singular values of the retained components in $S$ are shown with red dots in (c), and the singular values of the components that were truncated are shown with blue dots. The images are shown for the decomposition of the X directional component after a single load step along Axis A (Node 1 in Fig.~\ref{fig: loadpath_tree}).}
  \label{fig: cruciform pca}
\end{figure}

The generated displacement data comprised matrix $\mathcal{A}$, which was a data set of $u = 400$ nodal displacement files which contained $v = 1,022$ values each (green nodes in Fig.~\ref{fig: cruciform schematic}). The important first step was to make $\mathcal{A}$ zero-mean by subtracting the mean of the data set such that, $\mathcal{A}' = \mathcal{A}-\mathcal{A}^{\text{mean}}, \quad \mathcal{A}^{\text{mean}} \in \mathbb{R}^{1 \times v}$. Fig.~\ref{fig: cruciform pca} shows the PCA analysis for the $X$ directional component of displacement after a single load step applied along the X (horizontal) axis (which corresponds to Node 1 in Fig.~\ref{fig: loadpath_tree}). Note, separate PCA analyses were performed for each directional component of displacement ($X$ and $Y$) and for each node in the load path tree. A small portion of the training data set is shown in Fig.~\ref{fig: cruciform pca}(a), and the retained principal components, $V^{*T}\subseteq V^T$, $V^{*T} \in \mathbb{R}^{p\times v}$, and singular values of $S$ from the SVD are shown in Fig.~\ref{fig: cruciform pca}(b) and (c), respectively. The singular values decrease in magnitude quickly, indicating a waning level of information content in the corresponding principal components. A number of components $p=5$ was retained, thus, reducing dimensionality of the data set from $400 \times 1,022$ to $400 \times 5$ after performing the linear transformation $\mathcal{A}^{\text{pca}} = \mathcal{A}'V^{*} \in \mathbb{R}^{u \times p}$. The interested reader may refer to Figs.~\ref{fig: x pca modes} and \ref{fig: y pca modes} in the Supplementary Material to view the retained principal components at Node 1 in the load path tree.

\begin{figure}
  \centering
  \includegraphics[width=150mm]{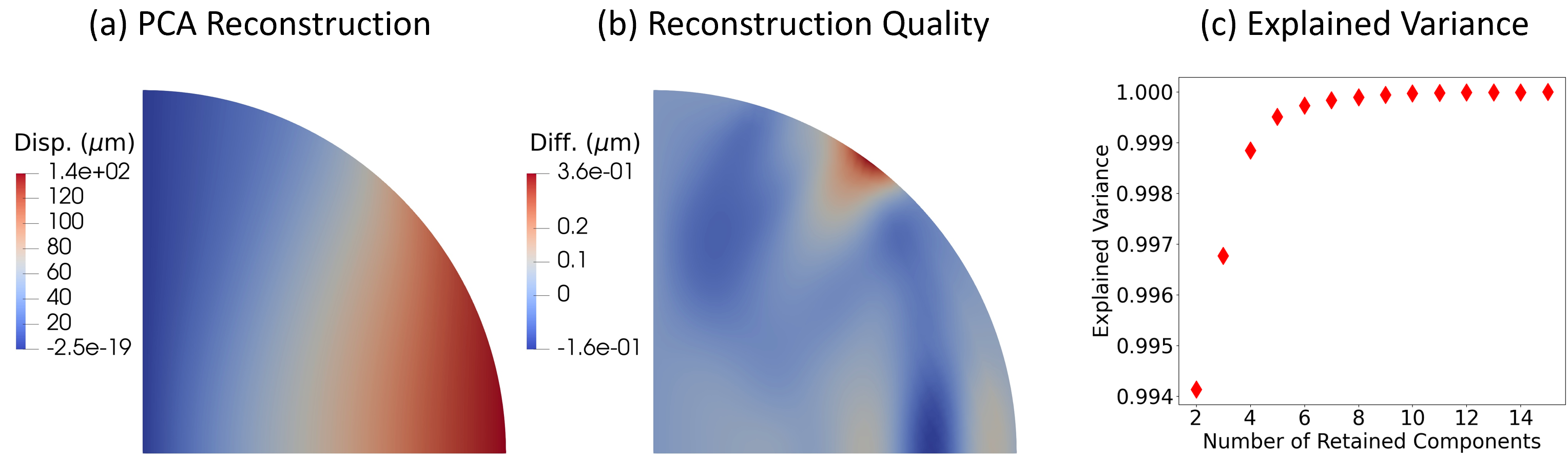}
  \caption{The quality of the reduced-dimensional representation of the displacement field through PCA can be visualized by (a) reconstructing the displacement field with the $p=5$ retained principal components, $\mathcal{A}^{\text{recon}}$ and (b) calculating the point-wise difference between the real field and reconstructed field, $\mathcal{A} - \mathcal{A}^{\text{recon}}$. The low error seen here is expected since the 5 principal components explain (or capture) over 99.9\% of the variance in the original data set (c). Thus, very little information was lost with the truncation of the remaining principal components. The example reconstruction is shown for the displacement field of the X directional component after a single load step along Axis A (Node 1 in Fig.~\ref{fig: loadpath_tree}). Note, the displacement control was prescribed at the end of the cruciform arms, thus resulting in a slightly lower maximum displacement in the region of interest.}
  \label{fig: pca quality}
\end{figure}

The original data set can be reconstructed from the retained components with $\mathcal{A}^{\text{recon}} = \mathcal{A}^{\text{pca}}V^{*T} + \mathcal{A}^{\text{mean}} \approx \mathcal{A}$. Fig.~\ref{fig: pca quality}(a) shows an example reconstruction of one of the data sets which comprised $\mathcal{A}$. The point-wise difference between the original and reconstructed data is shown in Fig.~\ref{fig: pca quality}(b), which is very small relative to the magnitude of the measurements. Fig.~\ref{fig: pca quality}(c) shows that with only 5 retained principal components, over $99.9\%$ of the information content in the original data set was preserved. Thus, retaining just the first few components successfully describes the majority of the variation in the data set and results in minimal information loss. Furthermore, there is a negligible benefit in retaining any additional components.

This work focuses on calibration in the spatial domain, which requires reconstruction of the displacement fields from the surrogate predictions of the PCA singular values as previously described. However, calibration may also be performed in the spectral domain, in which case the surrogate predictions of the PCA singular values would be used directly in the calibration along with dimensionally reduced experimental data. The steps needed to transform the measured DIC data are briefly discussed here. The new orthogonal basis which is obtained can be used to transform any new DIC data which is on the same grid as that used for training. In an experimental example, the DIC data would need to pre-processed before applying the PCA transformation in order to obtain measured values on an identical grid as that used for the PCA training. If the same model were being used for the data generation process and model calibration---as in this case, no pre-processing of the data would be necessary. Thus, as experimental data is collected (or simulated), the PCA transformation can be used to express the high-dimensional data $\mathcal{A}^{\text{new}} \in \mathbb{R}^{1 \times v}$ in a low-dimensional space, $\mathcal{A}^{\text{new}}V^{*} \in \mathbb{R}^{1 \times p}$. Further considerations of the two approaches for calibration---field vs spectral domain---can be found in Sec.~\ref{sec: discussion}.

Dimension-reduction of the high-dimensional displacement fields with minimal information loss is a crucial component of being able to use the highly-informative DIC data for model calibration. With the use of PCA, the construction of the surrogate models for the displacement field, discussed subsequently in Sec.~\ref{sec: surrogate model}, was made viable, enabling calibration with rich displacement fields.

\subsection{Surrogate Model}\label{sec: surrogate model}
Surrogate models have proven to be invaluable in model calibration within applied mechanics and engineering, particularly for mitigating the computational cost of high-fidelity simulations \cite{hoffer2021mesh, kudela2022recent}. By providing efficient approximations of complex constitutive models, surrogate-based approaches facilitate efficient model calibration and Expected Information Gain (EIG) calculations in this work. Their ability to accelerate computations without sacrificing accuracy ensures that these tasks can be performed reliably within a significantly reduced time-frame, making them an essential tool in the ICC framework.
 
The combined Bayesian calibration via the Laplace approximation (Sec.~\ref{sec: bayesian inference}) and EIG calculation (Sec.~\ref{sec: boed}) requires a minimum of $10^4$ model evaluations.\footnote{This number may vary based on the EIG sampling strategy and other selected settings.} Each run of the FEA conservatively takes 30 minutes; clearly, this computational demand is not feasible for the quasi-real-time setting the ICC framework is intended to operate within. Therefore, in order to cost-effectively perform calibration of the constitutive model within the ICC framework, the model evaluation time must be reduced significantly, which is accomplished through the construction of a computationally inexpensive surrogate model replacement of the expensive FEA. 

\begin{figure}[hb!]
  \centering
  \includegraphics[width=100mm]{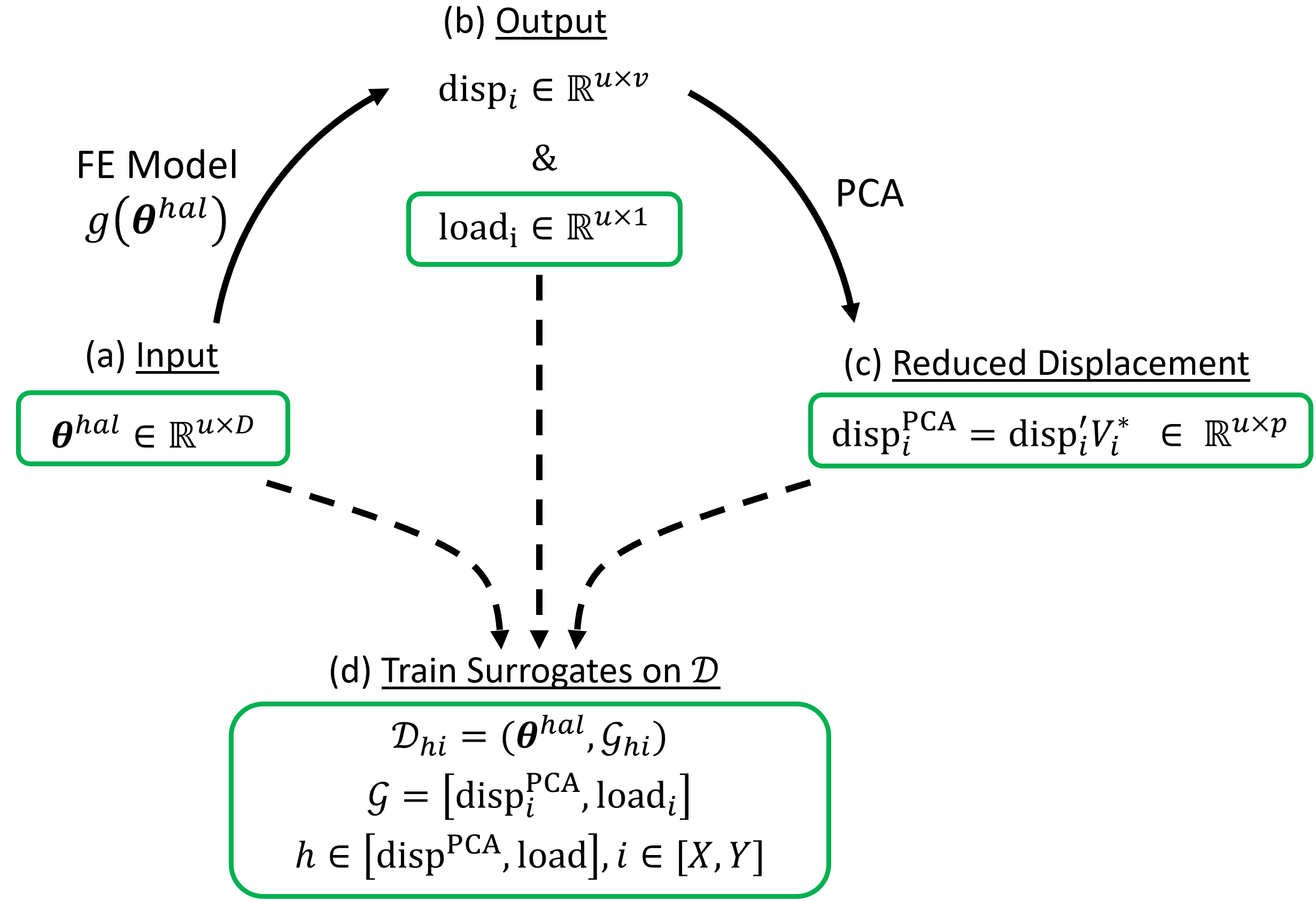}
  \caption{A workflow showing how the surrogate models were built. The solid arrows show the steps taken to prepare the surrogate training data, and the dashed arrows connect the surrogate training data (enclosed with green boxes) to the training step. The first step was to (a) generate $u$ parameter samples of $D$ dimensions ($D=4$) with a space-filling Halton sequence to serve as input to the FE model, $g(\pars^{hal})$. The FE model yielded nodal displacements and load data for the horizontal and vertical directional components, $i \in [X, Y]$ (b). The displacement data was reduced via PCA in (c) to yield a reduced-dimensional representation of displacement through singular values of the $p$ retained principal components. The input parameters, loads and nodal displacement singular values created the input-output pairs in $\mathcal{D}$ used to train the surrogate models in (d). A separate surrogate model was trained for each QoI, $h \in [\text{disp}^{\text{PCA}}, \text{load}]$ and directional component $i \in [X, Y]$ at each node in the load path tree in Fig.~\ref{fig: loadpath_tree}. The surrogate predictions of the PCA singular values were used to reconstruct the displacement fields before performing inference and calculating the EIG according to the procedure described in Sec.~\ref{sec: dimension reduction}.
  }
  \label{fig: surrogate workflow}
\end{figure}

Surrogate models were built to approximately map constitutive model parameters to the FE model output $\tilde{g}(\pars) \approx g(\pars)$ at each node in the load path tree (besides the root node which represents the undeformed state). Fig.~\ref{fig: surrogate workflow} shows the workflow for building a surrogate replacement for the FEA at a single node. The solid arrows connect the steps taken to prepare the data for surrogate training, and the dashed arrows connect the surrogate training data, enclosed with green boxes, to the training step. A Halton sequence was used to generate 400 sets of model inputs $\pars^{hal}$--Fig.~\ref{fig: surrogate workflow}(a)--which then produced 400 corresponding outputs from the FEA for the given node---Fig.~\ref{fig: surrogate workflow}(b). The outputs are the full-field displacements and load values for the $X$ and $Y$ directional components, $g(\pars^{hal}) = [\text{disp}_{i}, \text{load}_{i}], \, i\in [X, Y]$. The displacement values were then dimensionally reduced via PCA to obtain the corresponding singular values of the retained principal components---Fig.~\ref{fig: surrogate workflow}(c)---via the process described in Sec.~\ref{sec: dimension reduction}. The resulting collection of input-output pairs constitute the training data for the surrogates, $\mathcal{D}_{hi} = (\pars^{hal}, \mathcal{G}_{hi})$, where $\mathcal{G}=[\text{disp}_{i}^{\text{PCA}}, \text{load}_{i}], \, i\in [X, Y]$ and $h \in [\text{disp}^{\text{PCA}}, \text{load}]$---Fig.~\ref{fig: surrogate workflow}(d). The subscript $hi$ is to indicate that a separate surrogate model was constructed for each QoI and directional component. 

Each trained surrogate model then takes the model parameters $\pars$ as input and returns either an approximation to the displacement singular values or load for one of the directional components. Thus, in order to approximate the full FE model output at a given node in the load path tree, a collection of surrogates is needed, $\tilde{g}(\pars)_{hi}, h \in [\text{disp}^{\text{PCA}}, \text{load}], i \in [X, Y]$. The displacement field is then reconstructed from the singular values via $\text{disp}_i^{\text{recon}} = \text{disp}^{\text{PCA}}_iV^{*T}_i + \text{disp}^{\text{mean}}_i$ before performing inference and calculating the EIG. Note, whereas the experimental design parameters (i.e.\ load path in this case) along with the model parameters are traditionally input to surrogate models in OED applications, here, the design parameters are omitted since a separate surrogate model is built for each node in the load path tree. Therefore, obtaining a surrogate output for a specific design, or load path, corresponds to selecting a combination of surrogate models corresponding to the nodes in that load path. For example, a surrogate approximation for Path 1 in Fig.~\ref{fig: loadpath_tree} is obtained by collecting the output from the surrogate models built for Nodes 1 and 3.

For the given load path tree, the total number of surrogates needed is $|\Xi|^\mathcal{T} \times 2(p + 1)$. In the adopted notation, $\Xi$ is the set containing the design options at each decision point (i.e.\ $\Xi = \{\text{A}, \text{B}\}$), and $|\Xi|$ is the cardinality of the set---equal to 2 in this example. $\mathcal{T}$ is the number of load steps in the load path tree ($\mathcal{T}=2$ in the abbreviated load path tree illustrated in Fig.~\ref{fig: loadpath_tree} and $\mathcal{T}=5$ for the full load path tree employed here), and $p$ is the number of retained principal components ($p=5$ in this example). Multiplication by 2 is to account for both directional components (X and Y), and the addition of 1 accounts for the global load values. By building surrogates for each node under the prescribed boundary conditions described in Sec.~\ref{sec: cruciform description} (i.e.\ 0.25~mm displacement increments), the history dependence which is inherent in elastoplastic models was accounted for, consistent with the approach taken in \cite{ricciardi2024bayesian}. In reality, experimental boundary conditions may be more complex than those assumed here; this point is further discussed in Sec.~\ref{sec: discussion}. The surrogate model used in this work is a Gaussian process (GP) \cite{williams2006gaussian} with an anisotropic squared exponential kernel.

The surrogates generally exhibited low error across all QoIs and load steps when evaluated against data reserved for testing. The absolute error for the predicted load was less than 0.003\% of the true load for all load steps and directional components, and the absolute error for the predicted displacement field (which includes both the PCA + surrogate error) was less than 0.1\% of the test fields. Parity plots of the predicted load, PCA singular values and displacements can be found in \ref{parity.app} in Figs.~\ref{fig: pca parity plots}~\&~\ref{fig: load and disp parity plots}. Thus, with the aid of a fast surrogate approximation of the expensive high-fidelity model, parameter inference and load step selection were able to be performed in a rapid manner. The cycle time for a single load step was reduced from months to just several minutes, thus significantly alleviating the ICC computational burden.

\subsection{Bayesian Techniques} \label{sec: Bayesian techiques}
A brief discussion of the Bayesian paradigm for inference and experimental design is contained in this section. The reader is referred to \cite{ricciardi2024bayesian} for further discussion on the theory and tools found in this work. 

\subsubsection{Bayesian Inference}\label{sec: bayesian inference}

In each cycle of the ICC feedback loop, a single load step is applied to the specimen, and the collected data is used to perform model calibration. In lieu of a deterministic approach for model calibration, Bayesian inference is used, which provides an elegant way to quantify parameter uncertainty. In this work, inference was performed with both load and displacement field measurements, although it could have been performed using the dimensionally-reduced displacement measurements in the PCA space as well; see Sec.~\ref{sec: discussion}.

The first step is to model any knowledge regarding the unknown parameters of the forward model, $\pars \in \Pars \subset \mathbb{R}^{D}$, that is available \textit{a priori} to observing data. In the adopted notation, the parameters $\pars$ belongs to subspace $\Pars$ of dimension $D$. Data $\data \in \Data \subset \mathbb{R}^{N_{obs}}$ is then collected which belongs to the data space $\Data$ with dimensionality $N_{obs}$, which is the number of measurements. The \emph{likelihood function} is a probability model that describes the model fitness to the data (or the likelihood of observing the data) for a given parameter value, $f(\data \given \pars)$. Once this data become available, the prior model is conditioned upon it and updated to the posterior distribution, $\pi(\pars \given \data)$, via Bayes' rule,
  
\begin{equation}\label{eq: bayes rule}
  \pi\left(\pars \given \data\right) = \frac{f\left(\data \given \pars\right)\pi\left(\pars\right)}{\int_{\Pars}f\left(\data \given \pars\right)\pi\left(\pars\right)d\pars} \propto f\left(\data \given \pars\right)\pi\left(\pars\right).
\end{equation}

Thus, both the data and prior beliefs are taken into account when calculating the posterior distribution, which is, generally speaking, not available in closed-form due to the dimensionality of the parameter space. For this reason, simulation techniques such as Markov Chain Monte Carlo (MCMC) \cite{tierney1994markov,gamerman2006markov,smith1993bayesian,andrieu2003introduction,liu2001monte} are widely used to obtain an approximation of the posterior distribution through one or more of a variety of sampling algorithms. In the ICC framework, the posterior distribution must be approximated after each load step, with the total number of load steps predetermined to be 5 in this case. As the goal is to eventually use this method in real-time decision making with experiments, less than an hour is allotted to posterior estimation within each cycle of the ICC framework in order to complete the experiment (including setup time) in one day.\footnote{This time allotment would reduce with additional load step options or a larger number of total load steps.} Therefore, in this application of real-time decision-making, the expense of MCMC would be challenging to overcome. Thus, within the ICC framework, a Laplace approximation to the posterior distribution is adopted due to its relative speed and ease to compute. 

Laplace's approximation, based on a Gaussian assumption, provides an analytical expression for the posterior distribution. Define $J(\pars) \coloneqq \log \pi(\pars \given \data)$, and let $\hat{\pars}$ be the parameter vector which maximizes $J(\pars)$, $\hat{\pars} = \argmax_{\pars \in \Pars} J(\pars)$---also known as the \emph{maximum a posteriori} (MAP) estimate. The covariance of the approximated posterior is determined by the inverse of the Hessian, $\bs{H}$, of $J(\hat{\pars})$, $\Sigma = \bs{H}_{J}(\hat{\pars})^{-1}$. In this work, the Hessian is calculated through central finite difference. Thus, the Laplace approximation to the posterior is a Gaussian distribution characterized by the MAP parameter estimate and the local curvature at the MAP estimate, $\tilde{\pi}^{L}(\pars \given \data) = \mathcal{N}_{D}(\hat{\pars}, \Sigma)$. This approach for approximating the posterior is very efficient and makes quasi-real-time step selection and model calibration a viable task. An MCMC simulation was performed and included in \ref{contours.app} to confirm that the posterior is well-approximated by a Gaussian distribution in this example.

\subsubsection{Bayesian Optimal Experimental Design}\label{sec: boed}

The ICC framework actively drives the experiment through the load path that will yield informative calibration data by leveraging Bayesian optimal experimental design (BOED). Let $\des$ denote the load path of the specimen from start to finish and $\dest_t$ denote each load step from $t = 1, \dots, \mathcal{T}$, where $\mathcal{T}$ is the total number of load steps in the load path ($\mathcal{T}=5$ in this example). Also, let $\data_t$ be the data which is collected at load step $t$.

Within each cycle of the ICC framework, Bayesian inference is performed (Fig.~\ref{fig: ICC framework}(c)), which produces a posterior probability distribution that quantifies the parameter uncertainty conditioned on the data collected thus far. For example, after collecting data at load step $t$, the posterior is obtained via $\pi(\pars \given \data_{1:t}, \des_{1:t}) \propto f(\data_{1:t} \given \pars, \des_{1:t})\pi(\pars)$. Next, load step $\dest_{t+1}$ must be selected (Fig.~\ref{fig: ICC framework}(d)). While this selection has traditionally been accomplished through subject matter expertise, BOED accomplishes this task in the ICC framework by taking into account previously collected data and the current state of knowledge about the unknown parameters to determine which load step, from a set of candidate load steps $\Xi$, will be the most informative.

Specifically, in this work, the decision is made through the calculation of the expected information gain (EIG), which is derived from the Kullback-Leibler (KL) divergence. The KL divergence provides a measure of the statistical distance between two probability distributions; in this setting, the two distributions of interest are the prior $\pi(\pars)$, which contains all the information up to this point (details following), and the posterior $\pi(\pars \given \data, \xi)$ after performing candidate load step $\xi$. The candidate load steps are the branches in the load path tree, which here are applying either a displacement increment along the X (horizontal) axis (load step A) or the Y (vertical) axis (load step B) of the cruciform specimen, $\xi \in \Xi = \{\text{A}, \text{B}\}$. The preferred load step is then the one that has the greater expected KL divergence, which means it is expected to yield a posterior distribution that has the greater statistical distance from the prior and therefore contains more information about the parameters. Since the KL divergence is calculated before the experiment is performed and the data collected, in practice, the expectation of the KL divergence over the marginal distribution of all possible outcomes is used to yield the EIG of candidate load step $\xi$,

\begin{equation}\label{eq: EIG}
  \mbox{EIG}(\xi) = \mathbb{E}_{\data \given \xi} [\infdiv{\pi\left(\pars \given \data, \xi\right)}{\pi\left(\pars\right)}],
\end{equation}

where $\infdiv{\cdot}{\cdot}$ denotes the KL divergence between the two given distributions.

After some manipulation, the final form of the EIG is,

\begin{equation} \label{eq: EIG final}
   \mbox{EIG}(\xi) = \int_{Y} \int_{\Theta} f(\data \given \pars, \xi)\pi(\pars) \big[ \log f(\data \given \pars, \xi) - \log f(\data \given \xi) \big] d\pars d\data,
\end{equation}

where $f(\data \given \xi)$ is called the \emph{evidence}.\footnote{The interested reader is referred to \cite{ricciardi2024bayesian} for a more complete derivation of the EIG from the KL divergence.} Since the EIG is intractable for high dimensions, a double-nested Monte Carlo estimator is used to approximate the EIG,

\begin{equation}\label{eq: DNMC EIG}
  \begin{aligned}
  \widehat{EIG}_{MC}(\xi) = \frac{1}{N} \sum_{i=1}^{N} \log \frac{f\left(\data_{i} \given \pars_{i,0},\xi \right)}{\frac{1}{M} \sum_{j=1}^{M} f\left(\data_{i} \given \pars_{i,j},\xi\right)}, \quad \pars_{i,j} \sim \pi(\pars), \quad \data_{i} \sim f(\data \given \pars_{i,0}, \xi).
  \end{aligned}
\end{equation}

The approximation in Eqn.~\eqref{eq: DNMC EIG} uses samples from the likelihood and prior to evaluate the inner and outer sum of the Monte Carlo estimator. The quality of the estimate is largely determined by $N$ and $M$, the number of samples used to evaluate the outer and inner sums, which control the estimator variance and bias, respectively \cite{ryan2003estimating}. In order avoid numerical underflow, the log-sum-exp of the log likelihoods is used to stabilize the inner sum \cite{stan_user_guide}.

The next load step is then chosen as the candidate load step which maximizes the approximated EIG in Eqn.~\eqref{eq: DNMC EIG},

\begin{equation}\label{eq: optimal design}
 \xi_{t+1} = \argmax_{\xi}\widehat{EIG}_{MC}(\xi).
\end{equation}

In order to incorporate previously collected data into the EIG calculation, at each decision point, the prior in Eqn.~\eqref{eq: DNMC EIG} is defined as the posterior from the previous step. For example, when choosing load step $t+1$, $\pi(\pars) \coloneqq \pi(\pars \given \data_{1:t}, \dest_{1:t})$, resulting in parameter samples being drawn from the calculated posterior distribution conditioned on data from load steps $1,\dots,t$: $\pars_{i,j} \sim \pi(\pars \given \data_{1:t}, \des_{1:t})$. To start the EIG process and determine the first load step from the undeformed state, either (a) the EIG may be calculated for the two candidate load steps by sampling from the prior, or (b) the first load step may be pre-determined, in which case there is no EIG calculation for the first load step. In this work, the second option was pursued. Finally, in this work, a myopic (a.k.a. \emph{greedy}) BOED algorithm is used, meaning a local or short-sighted decision is made for each load step, as opposed to a global decision which considers the entire load path. This helps preserve computational resources as the EIG can be very costly to compute, and the cost would increase exponentially if calculating for the globally preferred path. Searching for the globally preferred full load path may be made possible with the use of a sequential OED approach, as was done in \cite{shen2023bayesian}; however, exploration of this approach is reserved for future endeavors as sequential OED would be particularly challenging for this application due to the non-Markovian nature of materials.

In this work, $M=1\times 10^{3}$ and $N=1\times10^4$, and the initial load step was chosen to be $\xi_{1} = \text{A}$. Pre-determining the first load step preserved computational resources as this reduced the number of surrogate models that needed to be built: they were only built for the portion of the tree which followed Node 1 in Fig.~\ref{fig: loadpath_tree}. Additionally, as was observed in \cite{ricciardi2024bayesian}, if there is no previously collected data (i.e.\ prior to the first load step), the EIG is determined solely by the prior modeling choice and physical model. So in cases such as this where the prior is only weakly informative, the EIG is unlikely to show strong preference between the two load steps. After the initial load step, the EIG was calculated to determine all subsequent load steps in the load path.
\section{Synthetic Cruciform Exemplar}\label{sec: hosford example}

In this section, the exemplar problem is presented and discussed. The statistical model and other algorithmic settings within the ICC framework are discussed (Sec.~\ref{sec: exemplar set up}) and results of the load path selection and calibration are presented (Sec.~\ref{sec: exemplar results}).

\subsection{Problem Setup}\label{sec: exemplar set up}

The exemplar problem explored in this work is the calibration of a constitutive model for a cruciform specimen made from an aluminum alloy being deformed in a biaxial load frame. The model form under consideration is the elastoplastic constitutive model described in Sec.~\ref{sec: constitutive model} with unknown parameter vector $\pars := [\sigma_y,\, A,\, n,\,a]^T$. These are the initial yield stress $\sigma_y$ and the isotropic hardening variables $A$ and $n$ from Eqn.~\eqref{eq: voce hardening} and the exponent $a$ from Eqn.~\eqref{eq: hosford effective stress}.

The parameters were constrained to take on values within the bounds specified in Table~\ref{Tab: parameter bounds}, which are the same as those use for surrogate training. The prior model for each parameter was chosen to be a truncated normal probability distribution (TN) which enforced these bounds during the calibration,

\begin{equation}\label{eq: exemplar prior settings}
    \pi(\pars_{d}) = TN\left(\pars_{d}; \mu_{(\pars_{d})}, \delta^{2}_{(\pars_{d})}, lb_{(\pars_{d})}, ub_{(\pars_{d})}\right), \quad d = 1, \ldots, D.
\end{equation}

The distribution means $\mu$ were chosen to yield an expected value within the bounds, and the variances $\delta^{2}$ were chosen to be moderately diffuse. The support of each distribution, defined by $lb$ (the lower bound) and $ub$ (the upper bound), was chosen to be the same as that used for the surrogate training. All distribution parameters are detailed in Table~\ref{Tab: exemplar priors} for each parameter. The truncated normal prior distributions were only mildly informative, showing just a slight preference for the mean of the bounding region; a comparative analysis using a uniform prior yielded results with no noticeable difference.

\begin{table}[ht!]
\centering
\begin{tabular}{lllll}
\hline
\textbf{Parameter}  & $\boldsymbol{\mu_{\theta}}$ & $\boldsymbol{\delta^{2}_{\theta}}$ & $\boldsymbol{lb}$ & $\boldsymbol{ub}$ \\ 
\hlineB{4}
$\sigma_y$ (ksi)   & 40.0    & 225.0   & 32.0 & 50.0  \\
$A$ (ksi)    & 10.0    & 225.0   & 1.0 & 20.0 \\
$n$    & 10.0    & 225.0   & 0.5 & 20.0  \\
$a$    & 10.0    & 225.0   & 4.0 & 16.0 \\
\hline
\end{tabular}
\caption{Means $\mu_{\theta}$, variances $\delta^{2}_{\theta}$ and bounds $lb$ and $ub$ for the truncated normal prior model on each parameter, Eqn.~\eqref{eq: exemplar prior settings}, used in the exemplar problem. Values for parameters $\sigma_{y}$ and $A$ have units of ksi.}\label{Tab: exemplar priors}
\end{table}

Simulated experimental data was generated by running the expensive FEA under assumed true parameter values $\pars^{true}$ (detailed in Table~\ref{table: cruciform parameters}) with added i.i.d. Gaussian error. This synthetic data is used here for model calibration to demonstrate the ICC framework under controlled conditions in preparation for later using it for real, physical experiments. The data consisted of the nodal displacements and global load values for both the $X$ and $Y$ components, $\data = [\text{disp}_{i}, \text{load}_{i}], \, i \in [X, Y]$. The FEA model was evaluated and measurements collected at the end of each load step, corresponding to the nodes in the load path tree.

The data generation process was modeled under the assumption that the observed data is equal to the surrogate model output with added Gaussian error $e$,

\begin{equation}\label{eq: data model}
    \data_{hi} = \tilde{g}(\pars)^{r}_{hi} + e_{hi}, \qquad e_{hi} \sim \mathcal{N}_{N_{hi}}\left(0,\Psi_{hi}\right), \, h \in [\text{disp}, \text{load}], \, i \in [X, Y],
\end{equation} 

where $\data_{hi}$ and $\tilde{g}(\pars)^{r}_{hi}$ are the data the surrogate replacement of the forward model for each QoI and directional component evaluated at $\pars$. The $r$ superscript for the surrogate output indicates that the full-field displacements were reconstructed from the predicted surrogate model PCA singular values via the process described in Sec.~\ref{sec: dimension reduction}. $N_{hi}$ is the number of measured observations for each QoI and directional component: $N_{\text{disp}} = 1,022$ and $N_{\text{load}} = 1$---the same for both $X$ and $Y$. The equivalent likelihood function for the expression in Eqn.~\eqref{eq: data model} is a multivariate normal distribution centered at the surrogate replacement of the forward model with a covariance matrix $\Psi_{hi}$ of size $N_{hi} \times N_{hi}$,

\begin{equation}\label{eq: likelihood}
    f\left(\data_{hi} \given \pars, \des \right) = \mathcal{N}_{N_{hi}}\left(\data_{hi}; \tilde{g}(\pars)^{r}_{hi}, \Psi_{hi}\right) \qquad h \in [\text{disp}, \text{load}], \, i \in [X, Y].
\end{equation}

The observation noise variance is assumed to be independent and identically distributed (i.i.d.)\ such that $\Psi_{hi} = \psi^{2}_{hi}\mathbb{I}_{N_{hi}}$, where $\mathbb{I}_{N_{hi}}$ is an $N_{hi} \times N_{hi}$ identity matrix and $\psi^{2}_{hi}$ is the observation noise variance---assumed to be known and fixed at $\psi^{2}_{\text{disp}} = 4${\textmu}m\textsuperscript{2} and $\psi^{2}_{\text{load}} = 582.11\,\mbox{N}^{2}$ for both the $X$ and $Y$ directional components. The value chosen for the load measurement noise equates to a standard deviation of approximately 0.05\% of the highest expected force at the final load step predicted from simulations with a nominal set of parameters. Likewise, the value chosen for the displacement measurement noise mimics the expected noise of DIC displacement measurements, based on an assumed DIC noise of 0.05~px (typical of a well-controlled DIC measurement in a laboratory environment), a field-of-view of 100~mm and a 5~MP camera with resolution of $2048 \times 2448$~px\textsuperscript{2}. The likelihood is written as a product over the 4 data sets contained in $\data$ (load and displacement for the $X$ and $Y$ directional components) under the assumption of independence,

\begin{equation}
    f\left(\data \given \pars, \des\right) = \prod_{h}\prod_{i} f\left(\data_{hi} \given \pars, \des\right), \quad h \in [\text{disp}, \text{load}], \quad i \in [X, Y].
\end{equation}

The ICC algorithm was carried out to a total of $\mathcal{T}=5$ load steps, with each load step applying a displacement increment of 0.25~mm along the specified axis. The first load step was pre-determined to be $\xi_{1} = \text{A}$. After the initial load step, the EIG was calculated to determine all subsequent load steps in the load path---$[\xi_{2}, \xi_{3}, \xi_{4}, \xi_{5}]$.

\subsection{Results}\label{sec: exemplar results}

Model calibration was first performed with all possible load paths with an initial load step along the X axis for a total of 16 calibrations with pre-determined load paths. For ease of reading, A is used to denote a load step along the X axis (horizontal direction) and B for a load step along the Y axis (vertical direction). Table~\ref{Tab: static results} reports the expected values, $\mathbb{E}_{\pars \given \data, \des}$, marginal variances, $\mathbb{V}_{\pars \given \data, \des}$ and 95\% credible intervals (CI) for parameter vector $\pars = [\sigma_y, A, n, a]$ for all 16 calibrations, with $\sigma_y$ and $A$ being reported in units of ksi. The marginal variance is the variance (or uncertainty) of one variable when looked at in isolation, without consideration of the others, and the 95\%~CI is the interval in which the parameter has a 95\% probability to fall within. Since a Laplace approximation to the posterior distribution was used, the marginal posterior distributions are symmetric, and the 95\%~CIs are reported with a single value, $\zeta$, where the expected value plus and minus $\zeta$ defines the interval: $95\% \, \text{CI} = \mathbb{E}_{\pars \given \data, \des} \, \pm \zeta$. Additionally, the generalized variance, which is a valuable measure of overall variance in a multivariate distribution, defined as the determinant of the posterior covariance $\det(\Sigma)$, is included. It captures the joint variability of all variables, providing a single scalar value that reflects the spread and orientation of the data. A larger generalized variance indicates greater dispersion among the variables, while a smaller value suggests closer clustering around the mean.

The resultant expected values exhibited less sensitivity to the load path than the uncertainty about their values. Parameter expected values were very similar for each load path. Specifically, $\mathbb{E}_{\sigma_y \given \data, \des}$ for the initial yield stress was in agreement to the tenth for all cases, and the expected value of all other parameters were in close agreement to the tenth. The marginal variances for $\sigma_{y}$ and $a$ were similar for all cases; however, there were orders of magnitude difference in the marginal variances reported for $A$ and $n$ among the different load paths. There was likewise orders of magnitude difference in the reported generalized variance among the different load paths. A similar trend was seen in the 95\%~CI; the interval was similar for parameters $\sigma_y$ and $a$ for all cases, but showed more deviation case to case for parameters $A$ and $n$. Thus parameters $A$ and $n$ were the most sensitive to the applied load path and resulting calibration data. In summary, the pre-determined AAAAA load path resulted in the least amount of parameter uncertainty (highlighted in green), expressed through the various reported uncertainty metrics, thus demonstrating the highest confidence in the parameter values; conversely, the ABABA resulted in the greatest parameter uncertainty (highlighted in red).

\begin{table}[ht!]
\centering
\resizebox{\textwidth}{!}{\begin{tabular}{l l l@{} r@{\hspace{1.0\tabcolsep}} r@{\hspace{1.0\tabcolsep}} r@{\hspace{1.0\tabcolsep}} r@{} l l l}
\hline
\multirow{2}{*}{\textbf{Load path}} & \multirow{2}{*}{$\boldsymbol{\mathbb{E}_{\pars \given \data, \des}}$} & \multicolumn{6}{l}{$\boldsymbol{\mathbb{V}_{\pars \given \data, \des}}$} & \multirow{2}{*}{$\zeta$} & \textbf{Gen. Var.}\\ 
& & \multicolumn{6}{l}{$\boldsymbol{(1\times 10^{-4})}$} &  & $\boldsymbol{(1\times10^{-14})}$\\
\hlineB{4}
\cellcolor{green!30} AAAAA & [42.50, 13.65, 14.29, 11.19] & [&1.6,& 2.5,& 15.8,& 5.3&] & [0.02, 0.03, 0.08, 0.04] & 0.23\\
AAAAB & [42.51, 13.56, 14.45, 11.17] & [&1.9,& 17.4,& 84.0,& 5.0&] & [0.03, 0.08, 0.18, 0.04] & 1.8\\
AAABA & [42.50, 13.46, 14.66, 11.17] & [&1.9,& 21.8,& 113.7,& 5.2&] & [0.03, 0.09, 0.21, 0.04] & 3.6\\
AAABB & [42.50, 13.63, 14.32, 11.16] & [&1.8,& 49.4,& 145.3,& 6.3&] & [0.03, 0.14, 0.24, 0.05] & 10.1\\
AABAA & [42.51, 13.54, 14.52, 11.19] & [&1.8,& 15.8,& 89.6,& 6.4&] & [0.03, 0.08, 0.19, 0.05] & 4.6\\
AABAB & [42.50, 13.53, 14.50, 11.19] & [&1.8,& 97.5,& 305.9,& 6.1&] & [0.03, 0.19, 0.34, 0.05] & 35.9\\
AABBA & [42.52, 13.59, 14.46, 11.23] & [&2.9,& 61.5,& 230.8,& 8.1&] & [0.03, 0.15, 0.30, 0.06] & 51.4\\
AABBB & [42.50, 13.61, 14.56, 11.23] & [&2.5,& 13.7,& 96.9,& 7.5&] & [0.03, 0.07, 0.19, 0.05] & 6.9\\
ABAAA & [42.50, 13.59, 14.39, 11.17] & [&1.4,& 11.7,& 68.7,& 10.8&] & [0.02, 0.07, 0.16, 0.06] & 5.2\\
ABAAB & [42.50, 13.54, 14.50, 11.19] & [&2.3,& 145.2,& 524.9,& 8.2&] & [0.03, 0.24, 0.45, 0.06] & 61.5\\
\cellcolor{red!30} ABABA & [42.53, 13.60, 14.31, 11.17] & [&2.2,& 193.6,& 580.2,& 9.6&] & [0.03, 0.27, 0.47, 0.06] & 249.6\\
ABABB & [42.51, 13.60, 14.44, 11.24] & [&1.9,& 28.8,& 130.7,& 13.0&] & [0.03, 0.11, 0.22, 0.07] & 48.9\\
ABBAA & [42.51, 13.52, 14.52, 11.17] & [&1.4,& 31.6,& 97.7,& 10.7&] & [0.02, 0.11, 0.19, 0.06] & 28.7\\
ABBAB & [42.51, 13.71, 14.20, 11.21] & [&1.6,& 158.1,& 442.7,& 7.1&] & [0.03, 0.25, 0.40, 0.05] & 87.7\\
ABBBA & [42.54, 13.71, 14.12, 11.20] & [&1.8,& 121.1,& 376.8,& 7.5&] & [0.03, 0.22, 0.38, 0.05] & 29.5\\
ABBBB & [42.50, 13.62, 14.36, 11.20] & [&1.3,& 8.0,& 47.5,& 10.0&] & [0.02, 0.06, 0.14, 0.06] & 2.50\\
\hline
\end{tabular}}
\vspace{1em} 
\centering
$\pars^{true} = [\sigma_y, A, n, a]^{true} = [42.51, 13.63, 14.35, 11.19]$
\caption{Posterior summaries for 16 pre-determined load paths, each starting within an initial A load step. The parameter expected values, $\mathbb{E}_{\pars \given \data, \des}$, marginal variances, $\mathbb{V}_{\pars \given \data, \des}$ and 95\%~CI are reported for parameter vector $\pars = [\sigma_{y}, A, n, a]$ ($\sigma_y$ and $A$ reported in units of ksi) along with the generalized variance, which is the determinant of the posterior covariance. The 95\%~CI is represented as $\mathbb{E}_{\pars \given \data, \des} \, \pm \zeta$, where $\zeta$ is the reported value. For visual simplicity, horizontal load steps are indicated with an A and vertical load steps with a B. The best (AAAAA) and worst (ABABA) case scenarios as determined by parameter uncertainty are highlighted green and red, respectively. The material model parameters used for data generation were $\pars^{true} = [42.51, 13.63, 14.35, 11.19]$.}\label{Tab: static results}
\end{table}

\begin{figure}%
    \centering
    \includegraphics[width=75mm,valign=c]{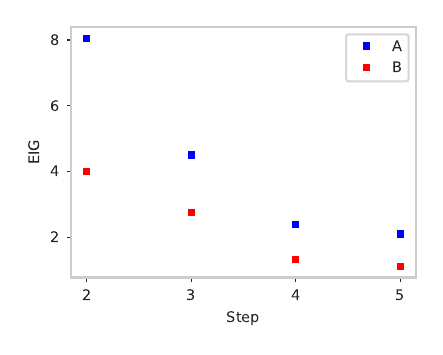}%
    \caption{The EIG was calculated for each candidate load step (A or B) at each decision point. Since the first load step was pre-determined to be A (pulling on the horizontal axis of the cruciform), the EIG was only calculated for steps 2-5. Load step A had the greater EIG at each load step, thus resulting in a total load path of AAAAA.}\label{fig: EIG}%
\end{figure}

In practice, not all 16 possible load paths would be evaluated experimentally, as the goal of this work is to identify the one preferred path based on its expected information content. Therefore, the load path was then chosen adaptively, as would be done in reality with the ICC framework, by calculating the EIG of the candidate load steps in $\Xi$ at each decision point as the cruciform was guided through the load path tree. Figure~\ref{fig: EIG} illustrates the EIG for each candidate load step at each decision point. As the load steps progress, the EIG values steadily decrease, indicating that additional data from either load step has a waning influence on the calibration. In this study, an A load step was selected each time, resulting in a total load path of AAAAA. This choice was confirmed by the pre-determined load paths in Table~\ref{Tab: static results}, showing that AAAAA had the least collective parameter uncertainty given the data. While it was not guaranteed, this myopic decision-making led to the globally preferred full load path within the current tree structure. The selection of an AAAAA load path over alternatives, like ABABA, is not necessarily intuitive \emph{a priori}. This outcome demonstrates that the BOED algorithm effectively identified the load path that yielded more informative data compared to other paths which might have been chosen based on subject-matter expertise.

\begin{figure}%
    \centering
    \includegraphics[width=75mm]{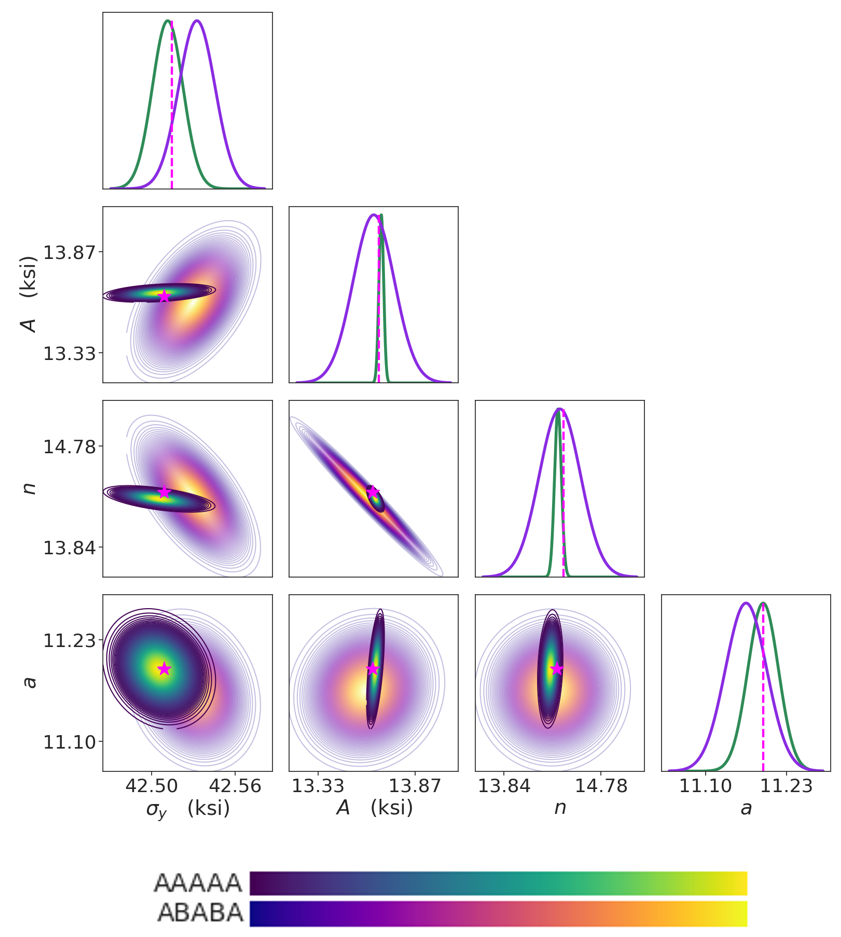}\label{fig: contour comparison}\\
   \caption{The posterior contours for load paths AAAAA and ABABA. Marginal posterior densities are plotted on the diagonals and joint-marginal densities on the off-diagonals. Marginal densities for the AAAAA calibration are plotted with a green line and for the ABABA calibration with a purple line. $\pars^{true}$ is indicated with a dashed pink line and star.}%
    \label{fig: posterior contours}%
\end{figure}

Fig.~\ref{fig: posterior contours} shows a side-by-side comparison of the posterior contours from the AAAAA and ABABA calibrations so that a visual comparison can be made between the cases with least and greatest parameter uncertainty. Contours from each calibration are shown separately in Fig.~\ref{fig: all posterior contours} in \ref{contours.app}. The joint marginal posterior contours shown on the off-diagonal plots show the posterior distribution of two parameters simultaneously, which provide insights into their likely values based on the observed data and prior information. Each contour line represents a level of constant posterior density, with innermost contours indicating regions of higher probability and outer contours indicating lower probability for those parameter combinations. The contour shapes provide insight into the relationship between the parameters. Elliptical or elongated contours suggest a correlation between the two parameters, which is indicative of a trade-off relationship between the given parameters such that changing the parameters along the direction of correlation results a model output that is relatively stable (for a diffuse prior); circular contours imply independence. The spread of these distributions reflects uncertainty; wider distributions indicate greater variability and less confidence in the parameter estimates, whereas narrower distributions, with concentration around specific values, imply more precise estimates. The marginal posterior distribution of each parameter can be inferred by projecting the contours of the joint distributions onto the axes, providing a clearer understanding of individual parameter uncertainty. The marginal distributions are shown on the diagonals plots along with $\pars^{true}$ (pink dashed line). $\pars^{true}$ is also included with the joint marginal contours, indicated by a pink star. 

The two load paths resulted in posterior distributions that exhibited different levels of joint and marginal parameter uncertainty as seen by the shape and spread of the contours, as well as different levels of parameter correlation as seen by the distinct amount of contour elongation (specific values of parameter correlation are reported in \ref{contours.app}). While the expected values align closely (Table~\ref{Tab: static results}), the marginal and joint posterior plots reveal subtle differences in the means when considering the small uncertainties in the parameters. This visual comparison elucidates the distinct outcomes of both calibrations; importantly, load path AAAAA resulted in lower parameter uncertainty compared to load path ABABA.

\begin{figure}[ht!]
\centering
\includegraphics[width=110mm]{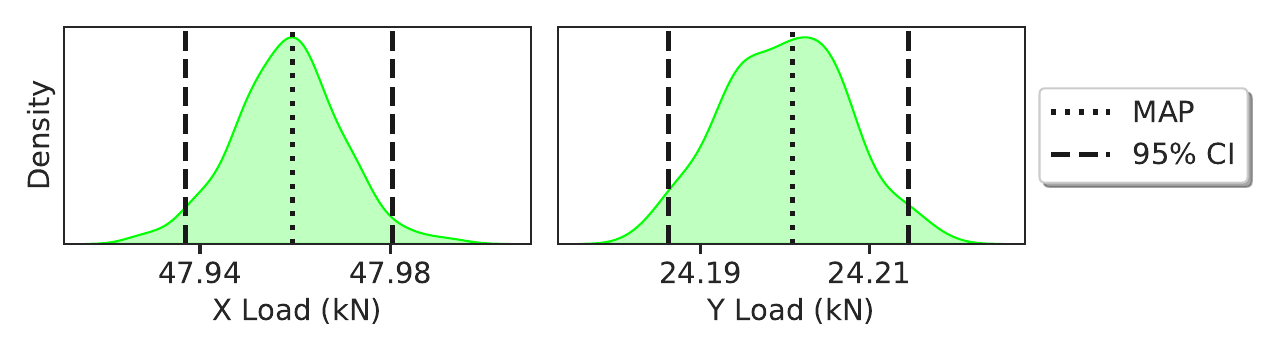}
\caption{Posterior probability over the load for the AAAAA calibration for both the $X$ (left column) and $Y$ (right column) directional components. The load at the \textit{maximum a posteriori} (MAP) parameter estimate is shown with a dotted black line, and the 95\% credible interval (CI) of the distribution over the load is shown with dashed black lines, indicating the interval in which the load values have a 95\% probability of residing within.}\label{fig: AAAAA posterior probabilty over load}
\end{figure}

The posterior probability is visualised over the surrogate model output by drawing 200 parameter samples from the posterior distribution $\pars^{draw} \sim \pi(\pars \given \data, \des)$ and evaluating the surrogate model at each sample $\tilde{g}(\pars^{draw})$. The corresponding posterior probability of the load values at load step 5 is shown in Fig.~\ref{fig: AAAAA posterior probabilty over load} for the AAAAA calibration. The model output at the MAP parameter estimate is shown with a dotted black line, and the 95\% CI of the distribution over the load is shown with dashed black lines, indicating the interval in which the load values have a 95\% probability of residing within.

\begin{figure}%
    \centering
    \sidesubfloat[]{\includegraphics[width=45mm]{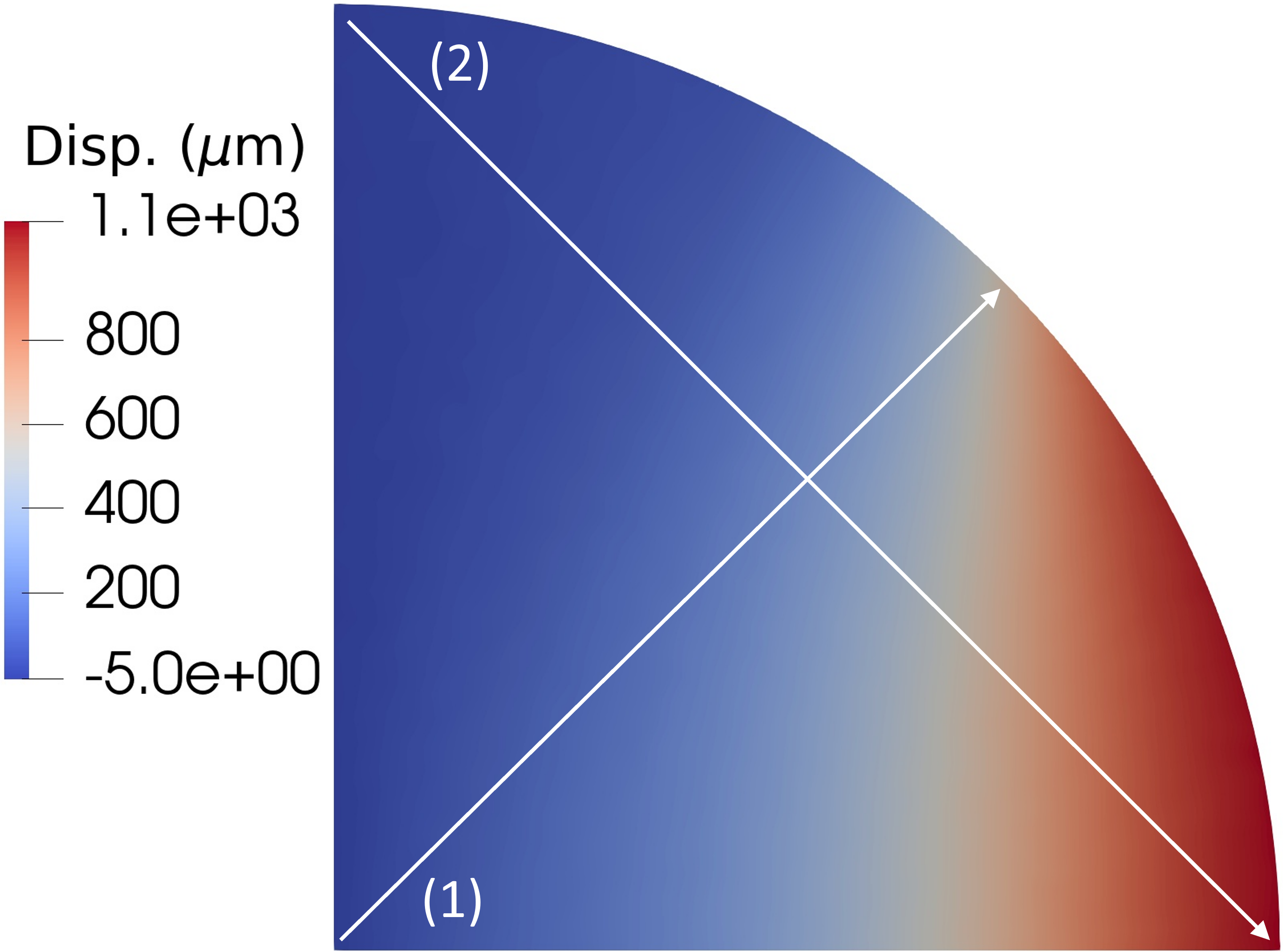}\label{fig: AAAAA ux cal data step 4}} %
    \sidesubfloat[]{\includegraphics[width=45mm]{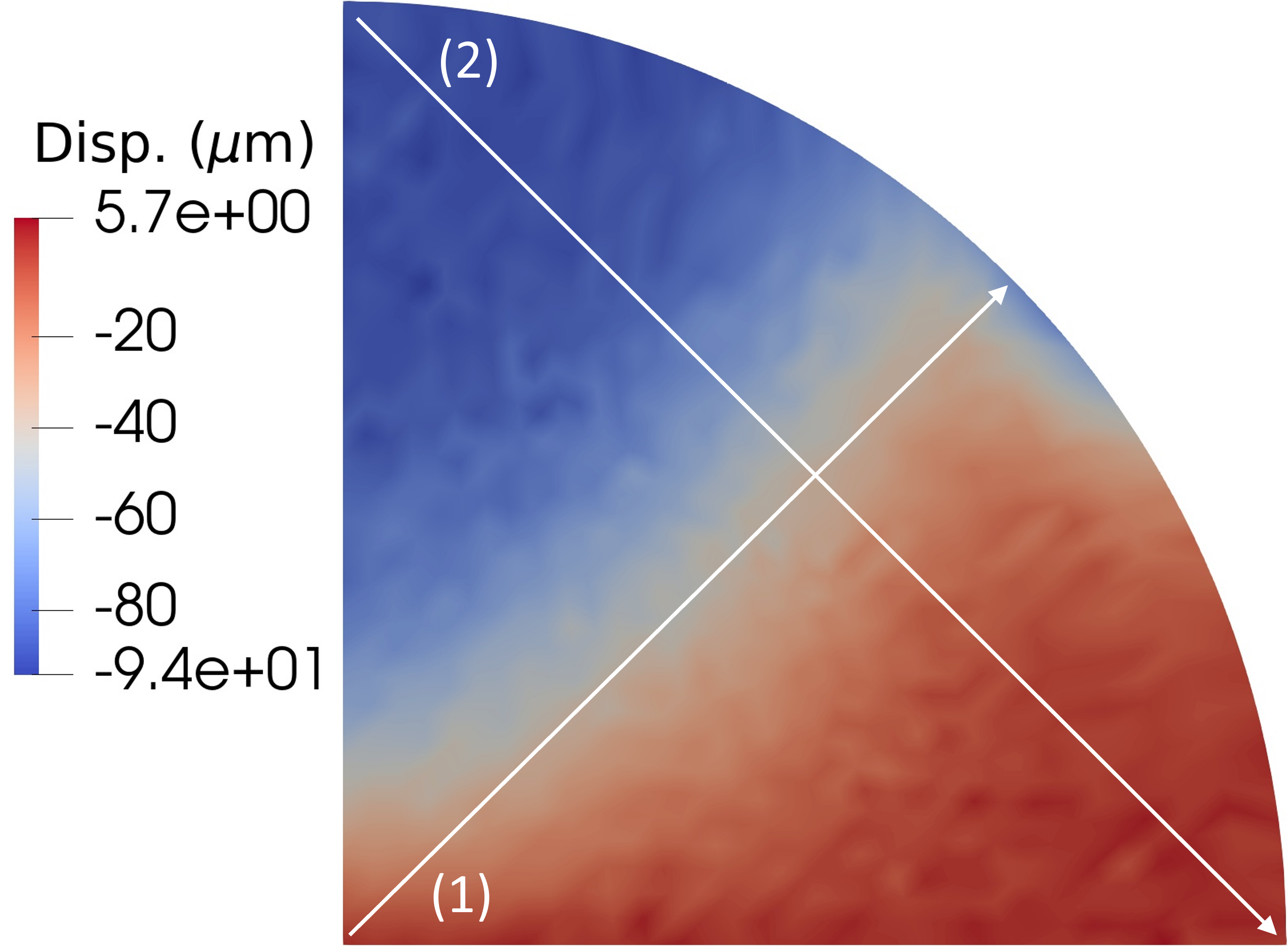}\label{fig: AAAAA uy cal data step 4}} %
    \caption{The observed displacement field generated with $\pars^{true}$ and added noise for the (a) X directional component and (b) Y directional component of the AAAAA load path at the final load step reported in microns. Lines (1) and (2) indicate the space over which the posterior probability of the displacement field is plotted in Fig.~\ref{fig: posterior over displacement AAAAA}.}%
    \label{fig: AAAAA step 4 cal data with lines}%
\end{figure}

\begin{figure}%
    \centering
    \sidesubfloat[]{\includegraphics[width=110mm]{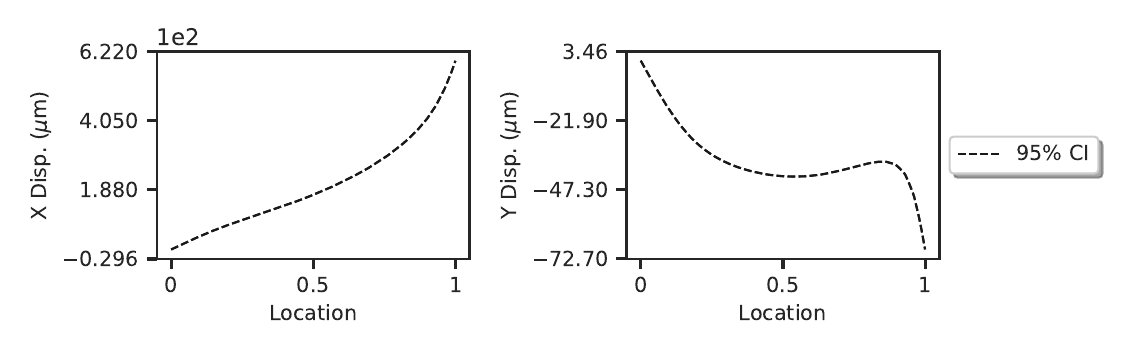}}\label{fig: AAAAA diag} \\ %
    \sidesubfloat[]{\includegraphics[width=110mm]{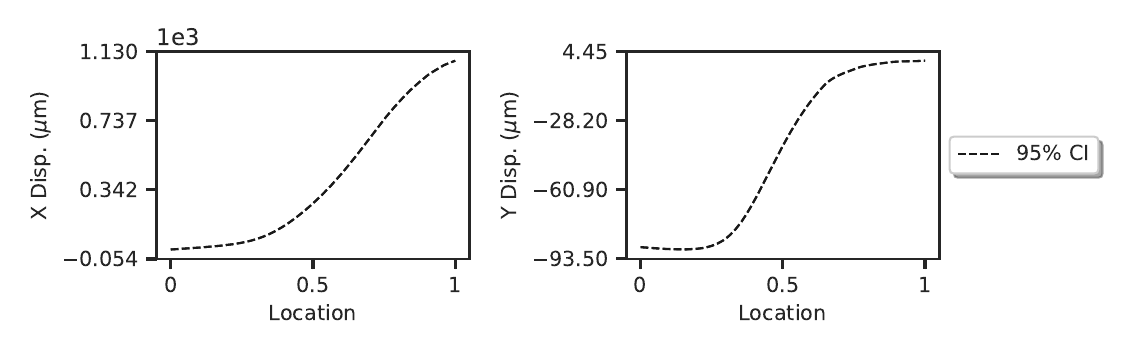}}\label{fig: AAAAA cross}%
    \caption{Posterior probability over the nodal displacements for the AAAAA calibration across the two lines indicated in Fig.~\ref{fig: AAAAA step 4 cal data with lines}. The $X$ (left column) and $Y$ (right column) directional components are shown along Line 1 (top row) and Line 2 (bottom row). The 95\% CI of the distribution over the displacement is shown with dashed black lines. Given the relative scale of the displacements and the low uncertainty, the 95\% CI interval appears as a single line. The x-axis is the normalized location across each line scan.}%
    \label{fig: posterior over displacement AAAAA}%
\end{figure}

The singular values obtained from $\tilde{g}(\pars^{draw})$ were then used along with the 5 retained principal components to reconstruct the displacement field, $\text{disp}_i^{\text{recon}} = \text{disp}^{\text{PCA}}_iV^{*T}_i + \text{disp}^{\text{mean}}_i, \, i \in [X, Y]$, so that there were 200 reconstructed instances of the displacement field. Fig.~\ref{fig: AAAAA step 4 cal data with lines} shows the observed displacement field generated with $\pars^{true}$ at the final load step of the AAAAA load path along with two line scans, across which the 95\%~CI was computed from the 200 reconstructed fields and plotted in Fig.~\ref{fig: posterior over displacement AAAAA}. Given the relative scale of the displacements and the low uncertainty, the 95\% CI interval appears as a single line. Overall, the high confidence in the parameter values translates to a high confidence in the resulting model output. The reader may refer to the Supplementary Material to view the posterior density over all QoIs and directional components for both the AAAAA and ABABA calibrations---global load values are shown in Fig.~\ref{fig: posterior probabilty over load} and the displacements in Figs.~\ref{fig: SUPP AAAAA step 4 cal data with lines}-\ref{fig: posterior over displacement ABABA}. Similar results were observed for the ABABA calibration: The posterior probability over the parameters translated to high confidence in the model output and recovered displacement field.

\subsection{Model Validation}\label{sec: validation}

Once the calibration is complete, the results are checked with external validation by using the model to make predictions on data not seen in calibration and comparing them to observed data. 
If the model obtained through the calibration fits the data well, then data generated under the model should look similar to new observed data. The new data could be taken from a new load path or specimen geometry, or it could simply be data from the same load path and geometry that could have been observed but was not (i.e.\ if the same experiment were to be performed again tomorrow).

Data for a new load path was predicted by drawing 200 samples from the joint posterior distribution $\pi(\data^{pred}, \pars \given \data)$. This is known as the posterior predictive distribution, which is the distribution of the predicted data if the calibrated model is true,

\begin{equation}\label{eq: posterior_predictive}
  \pi(\data^{pred} \given \data) = \int \pi(\data^{pred} \given \pars)\pi(\pars \given \data)d\pars.
\end{equation}

Although any of the 16 possible load paths starting with and A load step could have been chosen, an ABBBA load path was used for validation here. By comparing observed data to the posterior predictive distribution, the fit of the posterior distribution can be evaluated.

\begin{figure}[ht!]
\centering
\includegraphics[width=70mm]{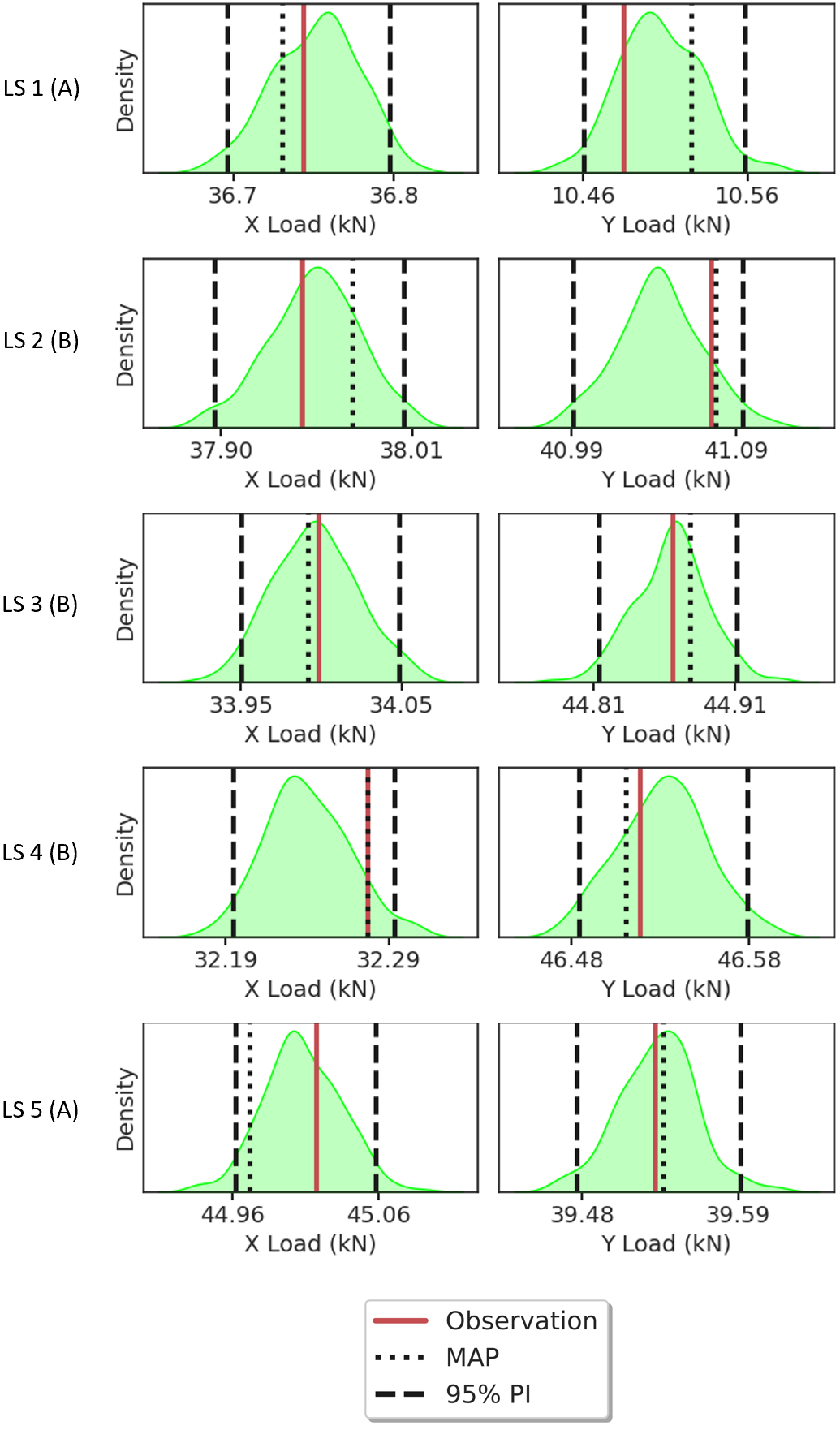}
\caption{The AAAAA posterior predictive distribution for the new ABBBA load path is shown over the load for the $X$ (left column) and $Y$ (right column) directional components and at each load step. The observed data is shown with a solid red line, the load at the \textit{maximum a posteriori} (MAP) parameter estimate is shown with a dotted black line, and the 95\% prediction interval (PI) is shown with dashed black lines.}\label{fig: AAAAA posterior predictive load for ABBBA}
\end{figure}

Figure~\ref{fig: AAAAA posterior predictive load for ABBBA} presents the posterior predictive distribution over the load for the new ABBBA load path, derived from the AAAAA-calibrated model. The distribution, calculated from 200 samples, is organized by load step (rows) and directional component (columns). The figure also includes the observed data (solid red line), which was generated by running the model with the true parameters $\pars^{true}$ for the ABBBA load path, incorporating noise to simulate real-world conditions. Additionally, the predicted load at the MAP parameter estimate is represented by the dotted black line, while the point-wise 95\%~prediction intervals (PI) are indicated by the dashed black lines. These intervals illustrate the uncertainty surrounding future observations based on the model's predictions. Notably, for each load step and for both directional components, the observed data falls comfortably within the 95\%~PI, indicating a high probability under the assumed model. This alignment between the observed data and the predictive intervals provides compelling evidence that the model is well-calibrated, reinforcing its reliability for future predictions. Furthermore, the ability of the model to capture the underlying distribution of the load suggests that it effectively accounts for the inherent variability in the data.

\begin{figure}[ht!]
\centering
\includegraphics[width=\linewidth]{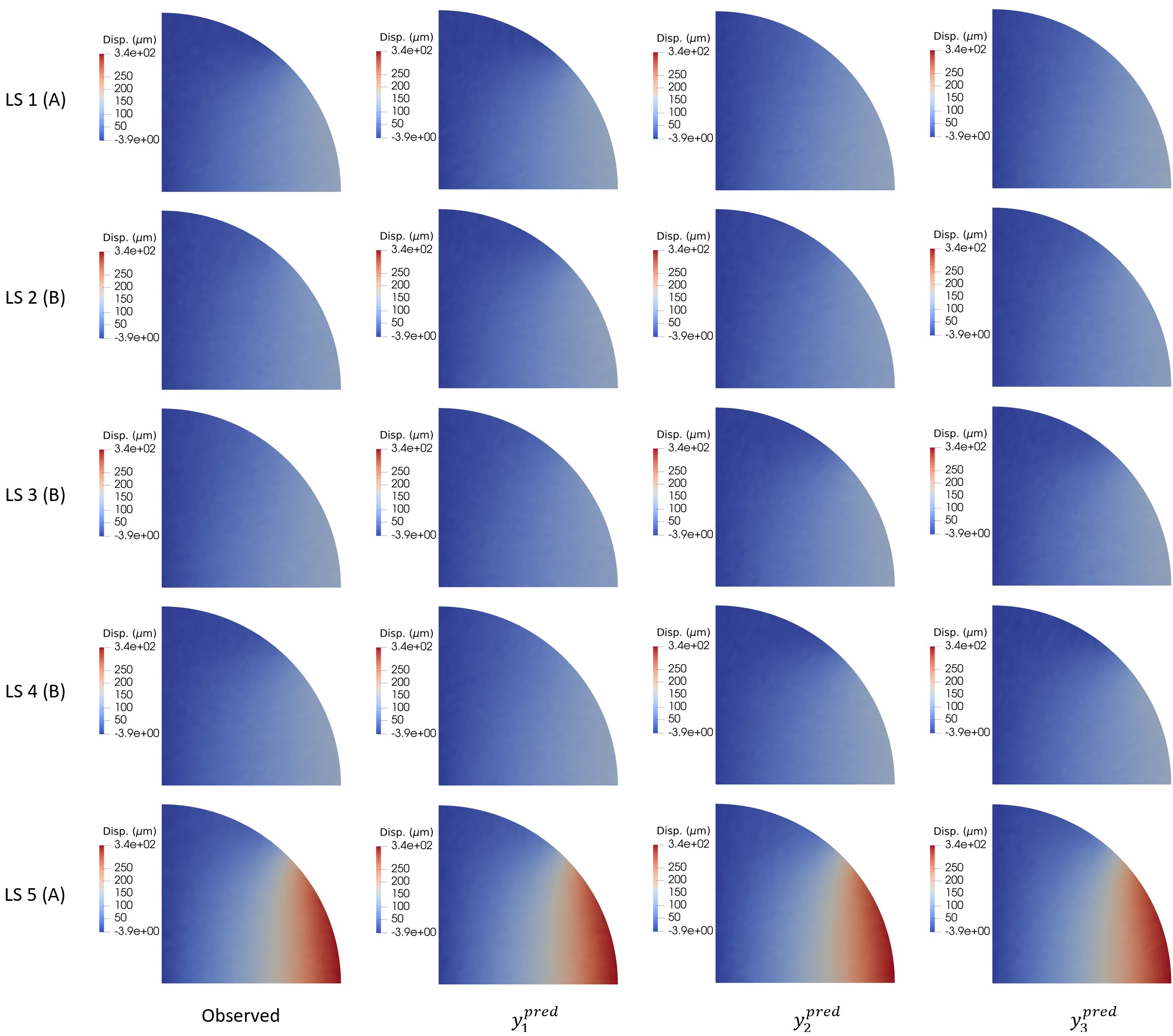}
\caption{Samples from the AAAAA posterior predictive distribution are shown for the $X$ directional component of the displacement field for the new ABBBA load path. The left column shows the observed data for each of the 5 load steps (LS) and the three right columns show three different predictive samples.}
\label{fig: AAAAA ABBBA X field validation}
\end{figure}

Although it is challenging to visualize the posterior predictive distribution over the entire displacement field, two complementary visualizations are utilized here to gain insight on the predicted uncertainty. Of the 200 samples from the posterior predictive distribution, 3 were chosen to reconstruct corresponding displacement fields and are visualized in Fig.~\ref{fig: AAAAA ABBBA X field validation} for the $X$ directional component. The data collected at each load step in the ABBBA load path is shown along with 3 samples from the predictive distribution. This allows for a comparison to be made between the displacement field that is observed and that which is expected to be observed under the posterior distribution that was obtained. The multiple predictive samples serve to establish repeatability, or that the patterns seen in the displacement field did not merely happen by chance. There is no visible systematic discrepancy between the observed and predicted data, and it is reasonable to suggest that the observed data came from the same distribution as the predictions. The error between the validation data and the predictive fields is shown in Fig.~\ref{fig: AAAAA ABBBA X field validation diff} in \ref{appen:disp posterior predictive}. This figure provides additional evidence of the repeatability of the results and confirms the absence of systematic discrepancies. Similar results were seen for the Y directional component in Figs.~\ref{fig: SUPP AAAAA ABBBA Y field validation}~and~\ref{fig: SUPP AAAAA ABBBA Y field validation diff} in the Supplementary Material.

\begin{figure}%
    \centering
    \sidesubfloat[]{\includegraphics[width=55mm]{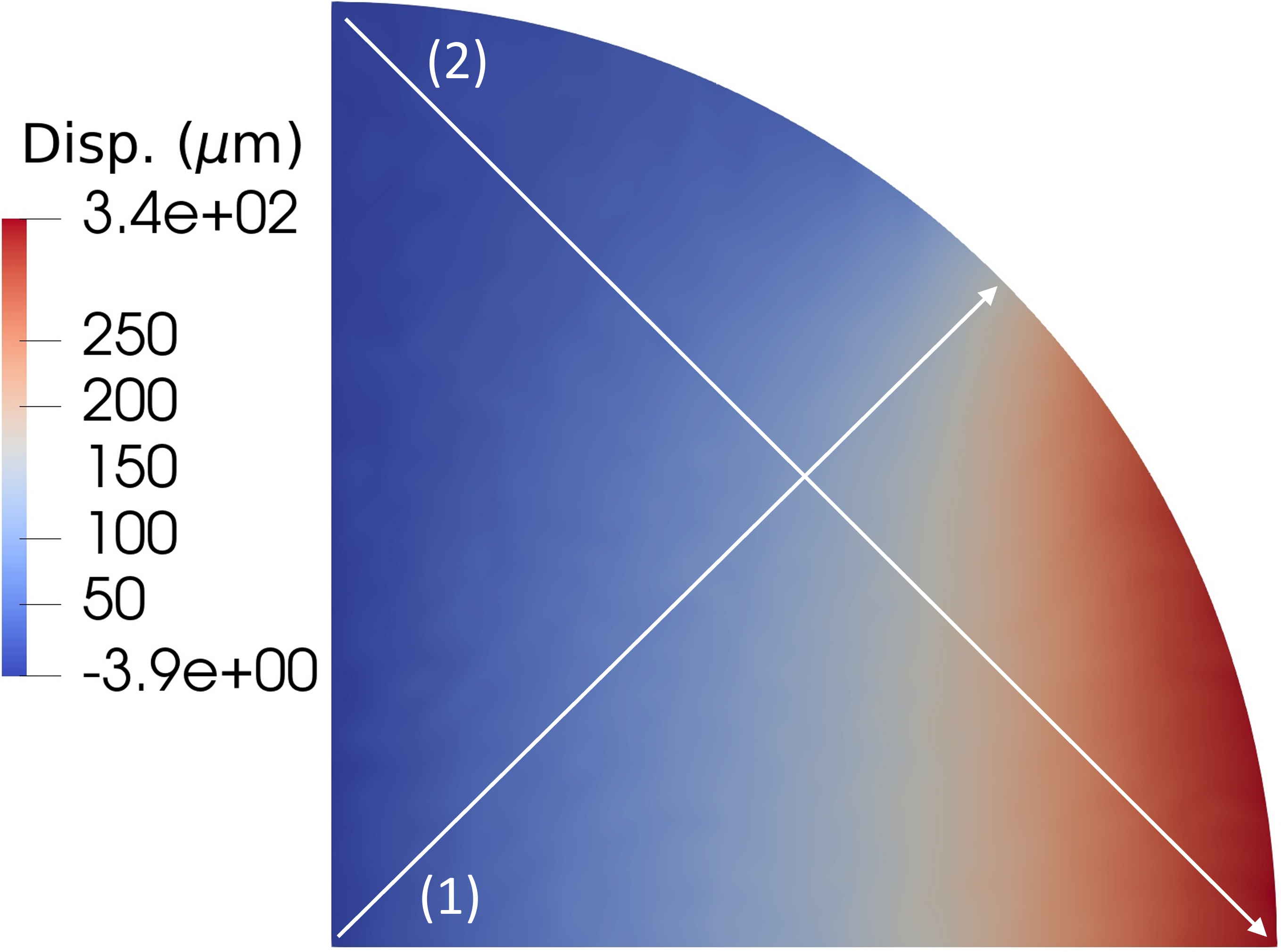}\label{fig: ABBBA ux val data step 4}} %
    \sidesubfloat[]{\includegraphics[width=55mm]{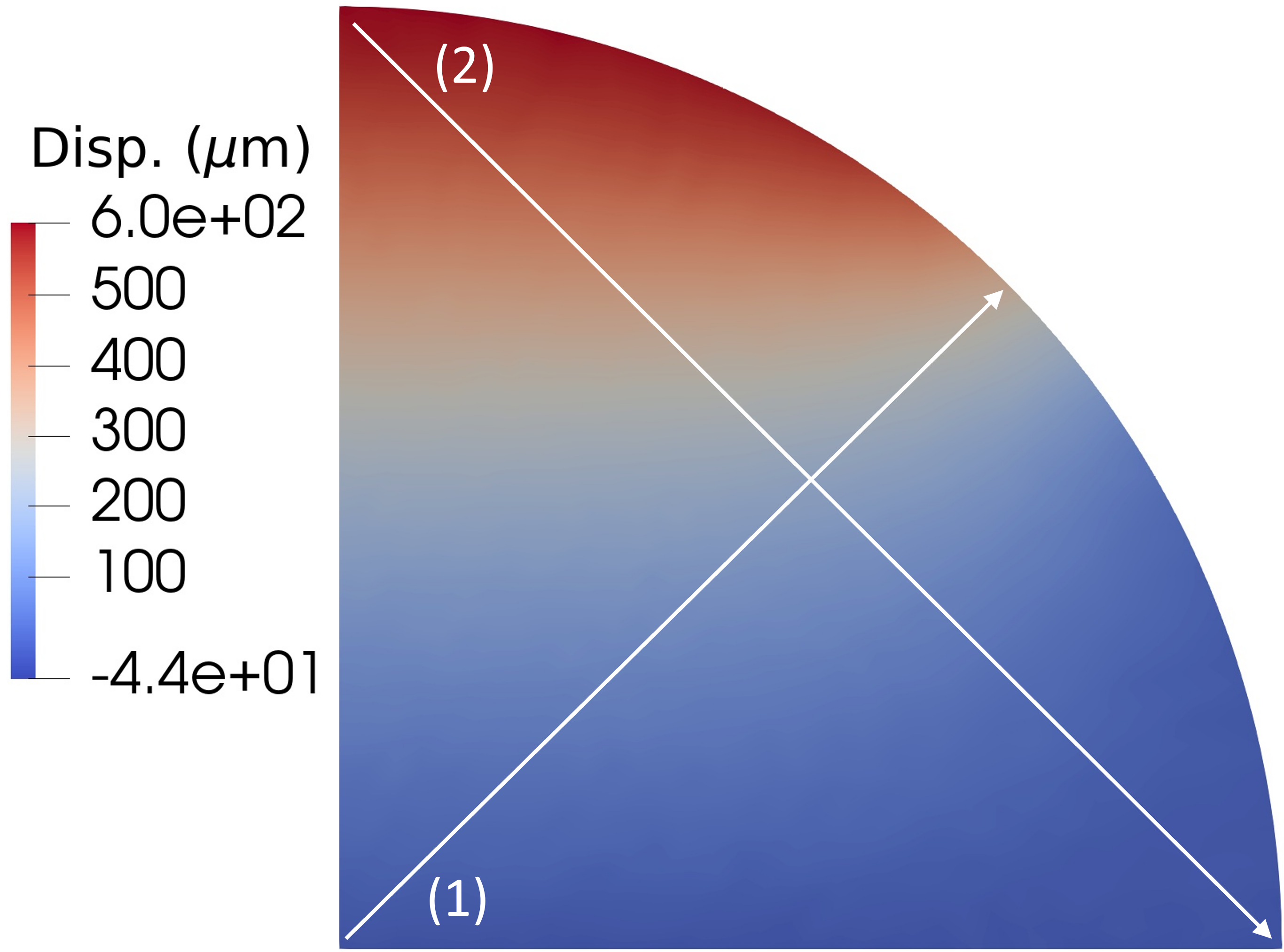}\label{fig: ABBBA uy val data step 4}} %
    \caption{The observed displacement field for the (a) X and (b) Y directional components of the ABBBA load path at the final load step reported in microns. Lines 1 marks the space over which the posterior predictive distribution is plotted in Fig.~\ref{fig: AAAAA prediction of ABBBA field diag line}---the posterior predictive is shown over Line 2 in Fig.~\ref{fig: SUPP AAAAA prediction of ABBBA field cross line} in the Supplementary Material.}\label{fig: ABBBA step 4 val data with lines}%
\end{figure}

\begin{figure}%
    \centering
    \includegraphics[width=165mm]{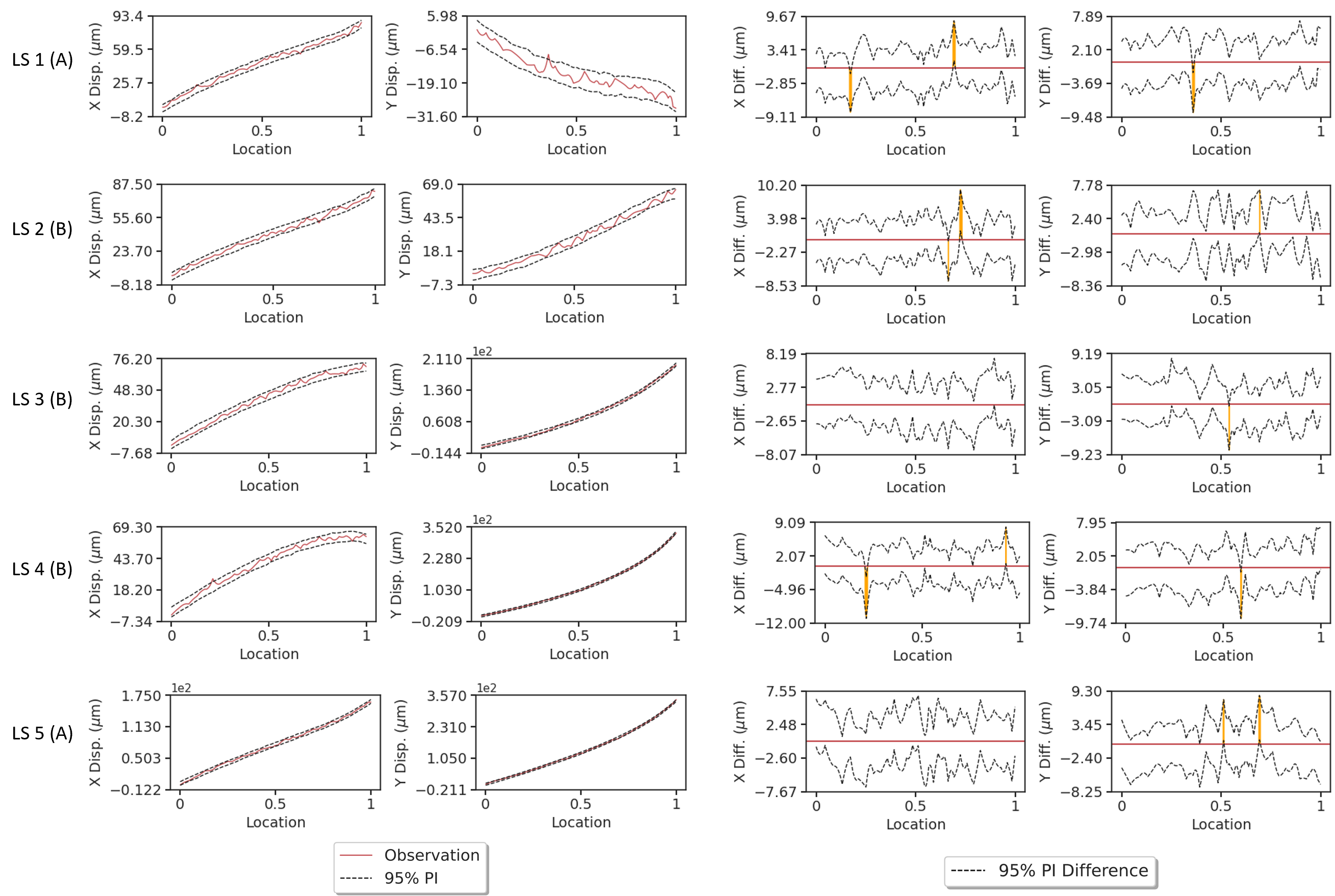}\label{fig: AAAAA post pred of ABBBA along line 1 (diag)} 
    \caption{The AAAAA posterior predictive distribution is plotted over the Line 1 displacements (Fig.~\ref{fig: ABBBA step 4 val data with lines}) for load path ABBBA. The two left columns show the 95\% prediction interval (PI) of the distribution as well as the observations for the $X$ and $Y$ directional components for all 5 load steps (LS). The two right columns show the difference between the 95\%~PI and the observations. Regions where the observations fall outside the 95\%~PI are colored orange, and 0 is marked with a red line. The x-axis is the normalized location along Line 1.}%
    \label{fig: AAAAA prediction of ABBBA field diag line}%
\end{figure}

The 95\%~PI of the distribution was calculated over the field point-wise from the 200 samples. Fig.~\ref{fig: ABBBA step 4 val data with lines} shows a full-field contour plot from which the 95\% PI over Line 1 was extracted and plotted in Fig.~\ref{fig: AAAAA prediction of ABBBA field diag line} for all 5 load steps to show the predictive uncertainty. Plots in the left column show the 95\% PI (dashed lines) along with the observed data (red line). Since distinguishing between the lines is difficult in some plots, the plots in the right column show the difference between the 95\% PI and the observed data. The solid line represents the observations (centered at 0 difference), and the difference between the observations and the PIs are shown with dashed lines. Everywhere the observations fall outside the 95\% PI coincides with locations where the upper bound difference is negative or the lower bound difference is positive; these regions are colored orange. Between 0\% and 2.3\% of the points along Line 1 have observations that fall outside the 95\% PI for all load steps and both directional components, less than the expected amount of 5\%. Similar observations were made for the predictive distribution over Line 2 in Fig.~\ref{fig: SUPP AAAAA prediction of ABBBA field cross line} in the Supplementary Material.

Similar plots were made for the posterior predictive distribution from the ABABA calibration for the new ABBBA load path to serve as a visual comparison of the two cases and may be found in the Supplementary Material---Fig.~\ref{fig: SUPP posterior predictive load for ABBBA} shows the posterior predictive density over the global load values and Figs.~\ref{fig: SUPP AAAAA ABBBA X field validation}-\ref{fig: ABABA prediction of ABBBA field cross line} show the same over the displacement field for both the AAAAA--- and ABABA---calibrated models. Additionally, a quantitative comparison is made between the predictive accuracy of the AAAAA vs. ABABA calibrations in \ref{log_predictive_density.app} through the log predictive density (or log likelihood). The results show that the AAAAA load path generally yielded greater predictive probability than the ABABA load path.

In summary, the model validation showed that the model obtained through the calibration fit the data well. The posterior predictive distribution was visualized over the load at each load step as well as over the displacement field. There were no systematic discrepancies seen between the observed data and the predicted data. Additionally, the observed data looked plausible under the posterior predictive distribution; generally speaking, the observed data had high probability under the assumed model and fell within the 95\%~PI. Thus, the well-calibrated and validated model with uncertainty metrics can be used for decision-making with a quantified level of confidence. Importantly, the ICC framework achieved this efficiently and with greater credibility than would have been attained through alternative load paths. 
\section{Discussion}\label{sec: discussion}

\paragraph{Algorithmic Decisions: Assumptions and Simplifications}
In the construction of the ICC framework, certain assumptions and simplifications were made to ease the burden of algorithm development. Some of these constraints may be relaxed or adjusted in future applications; however, others are necessary in order to make ICC feasible with real-time data collection and model calibration, which is the future application space of the ICC framework. These assumptions are bulleted and discussed below along with some suggestions of relaxing these constraints in the future where possible.

\begin{itemize}
    \descitem{Posterior Approximation} The posterior distribution was obtained with a Laplace approximation that assumes the posterior is Gaussian. Another method for approximating the posterior distribution which does not make such a strict assumption is Markov chain Monte Carlo (MCMC), which is widely used but very expensive. While the posterior is not guaranteed to be Gaussian, the Laplace approximation is obtained in less than a minute, while MCMC could easily take hours or more, conservatively, even with a fast surrogate model. Since the posterior is calculated after each load step, it is quickly realized that completing an entire experiment consisting of 5 or more load steps would not be feasible with such a computational demand. The Laplace approximation serves to quickly provide an estimate of parameter uncertainty which is then used in the EIG calculation to decide the next load step. It is possible that MCMC could result in a different adaptively selected load path if the posterior distribution is non-Gaussian; however, this possibility is secondary to efficient and practical decision-making. At the end of the data collection, MCMC could optionally be performed offline for a final calibration if desired. Other methods with reduced time requirements for posterior approximation compared to MCMC include variational inference \cite{zhang2018advances, fox2012tutorial} and sequential Bayesian methods \cite{naesseth2019elements, bernardo2011sequential}, which may be considered in the future.
    
    In order to support the use of the Laplace approximation in this work, a validation of the posterior approximation was obtained by performing an MCMC simulation with the collected data, and the results are included in \ref{contours.app}. Results showed that the posterior distribution was well-approximated with a Gaussian distribution in this case. The posterior obtained via MCMC took approximately 1.5 hours to obtain; however, this time requirement could increase significantly in future applications if the posterior is difficult to sample from. In comparison, posterior approximation took less than a minute with the Laplace approximation.

    \descitem{Measurement Error} The measurement noise variance for both the nodal displacements as well as the global load values was assumed to be known and i.i.d. In reality, noise from a load cell is a percent of the reading instead of constant, and displacement field data has a noise structure that is spatially correlated \cite{fayad2025identification}. The statistical model would be more flexible and provide more reliable uncertainty metrics in the experimental application if the noise variance were also inferred from the data. The incorporation of unknown hetereoscedastic noise into the BOED framework has been left to future work, and the correlated noise of the field data can be addressed by performing inference in the spectral domain, where PCA singular values offer a decorrelated representation as addressed in the \emph{Inference in Field vs. PCA Space} discussion point.
    
    \descitem{Binary Load Path Tree} A cruciform specimen in a planar biaxial load frame was selected for this synthetic experimental configuration. The corresponding load path tree had two branches per node, corresponding to tensile displacement in either the horizontal or vertical arms of the cruciform, with a fixed number of load steps and a fixed displacement increment for the load path tree. This was a simple starting point; however, ICC is not restricted to this tree structure. Other, more complex, and perhaps even more appropriate, branch options may include both positive and negative displacements, combinations of displacements applied in both directions simultaneously, a variable displacement increment, etc. Additionally, other deformation modes---such as tension-torsion, or even a combination of planar biaxial and torsional loading given the capabilities of the load frame---could be considered.
    
    \descitem{Symmetry of the Cruciform Specimen} Eighth-symmetry was used in constructing the FE model to reduce the size and computational cost. While this is a reasonable idealization for the synthetic demonstration, laboratory results often deviate from such approximations.  For instance, recent investigations on complex specimens~\cite{jones2021anisotropic} have demonstrated that utilization of ideal in place of actual boundary conditions can impact predicted results.  Therefore, when used in practice, such symmetries may not be exploitable.
    
    \descitem{Likelihood Contributions from Multiple QoIs} In engineering applications, it is common to combine multiple QoIs for calibration to capture various aspects of material behavior. For example, this work utilized both full-field displacements and global load values to calibrate a material model. While the calibration could have been performed using only the load data or only the displacement data, using both data sources and combining the unique information they contain greatly improves the calibration. Table~\ref{Tab: qoi posterior comparison} shows the posterior summaries using solely load data, solely displacement or both sources combined. The expected values indicate that all options effectively infer parameter values from the available data. However, when both sources are used together, parameter uncertainty is reduced. Figure~\ref{fig: posterior contours comparing qois} visually compares the different cases, clearly demonstrating the advantages of incorporating both data sources in the calibration process.
    
    The posterior uncertainty significantly decreases when adding displacement data to the load-only calibration (a). Conversely, transitioning from the displacement-only calibration to the combined calibration with load data (b) results in a much smaller reduction in posterior uncertainty. This difference highlights the varying information content of each QoI: the displacement data which has over 1,000 measurements provides rich information, while the load data which provides a single measurement contributes less. While balancing contributions from multiple QoIs may be necessary in some cases---for example, to avoid over-fitting---both data sources in this example, as shown in Fig.~\ref{fig: posterior contours comparing qois} and supported by posterior predictive checks in Sec.~\ref{sec: validation}, meaningfully contributed to the inference.   

    \begin{table}[ht!]
    \centering
    \resizebox{.9\textwidth}{!}{\begin{tabular}{l l l l@{} r@{\hspace{1.0\tabcolsep}} r@{\hspace{1.0\tabcolsep}} r@{\hspace{1.0\tabcolsep}} r@{} l l}
    \hline
    \multirow{2}{*}{\textbf{Load path}} & \multirow{2}{*}{\textbf{Data Space}} & \multirow{2}{*}{$\boldsymbol{\mathbb{E}_{\pars \given \data, \des}}$} & \multicolumn{6}{l}{$\boldsymbol{\mathbb{V}_{\pars \given \data, \des}}$} & \textbf{Gen. Var.}\\ 
    & & & \multicolumn{6}{l}{$\boldsymbol{(1\times 10^{-4})}$} &  $\boldsymbol{(1\times10^{-14})}$ \\
    \hlineB{4}
    \multirow{2}{*}{AAAAA} & Both & [42.50, 13.65, 14.29, 11.19] & [&1.6,& 2.5,& 15.8,& 5.3&]  & 0.23 \\
      & Disp. & [42.49, 13.64, 14.28, 11.17] & [&9.4,& 3.2,& 19.0,& 6.9&]  & 2.06\\
      & Load & [42.53, 13.70, 14.18, 11.23] & [&30.5,& 275.8,& 1922.4,& 32.4&]  & $3.06\times10^{4}$ \\
    \hline
    \end{tabular}}
    \vspace{1em} 
    \centering
    $\pars^{true} = [42.51, 13.63, 14.35, 11.19]$
    \caption{Posterior summaries obtained for load path AAAAA using solely load data, solely displacement data, or both data sources combined.}\label{Tab: qoi posterior comparison}
    \end{table}
    
    \floatsetup[figure]{style=plain,subcapbesideposition=top,font=small}
    \begin{figure}[ht!]%
        \centering
        \sidesubfloat[]{{\includegraphics[width=65mm]{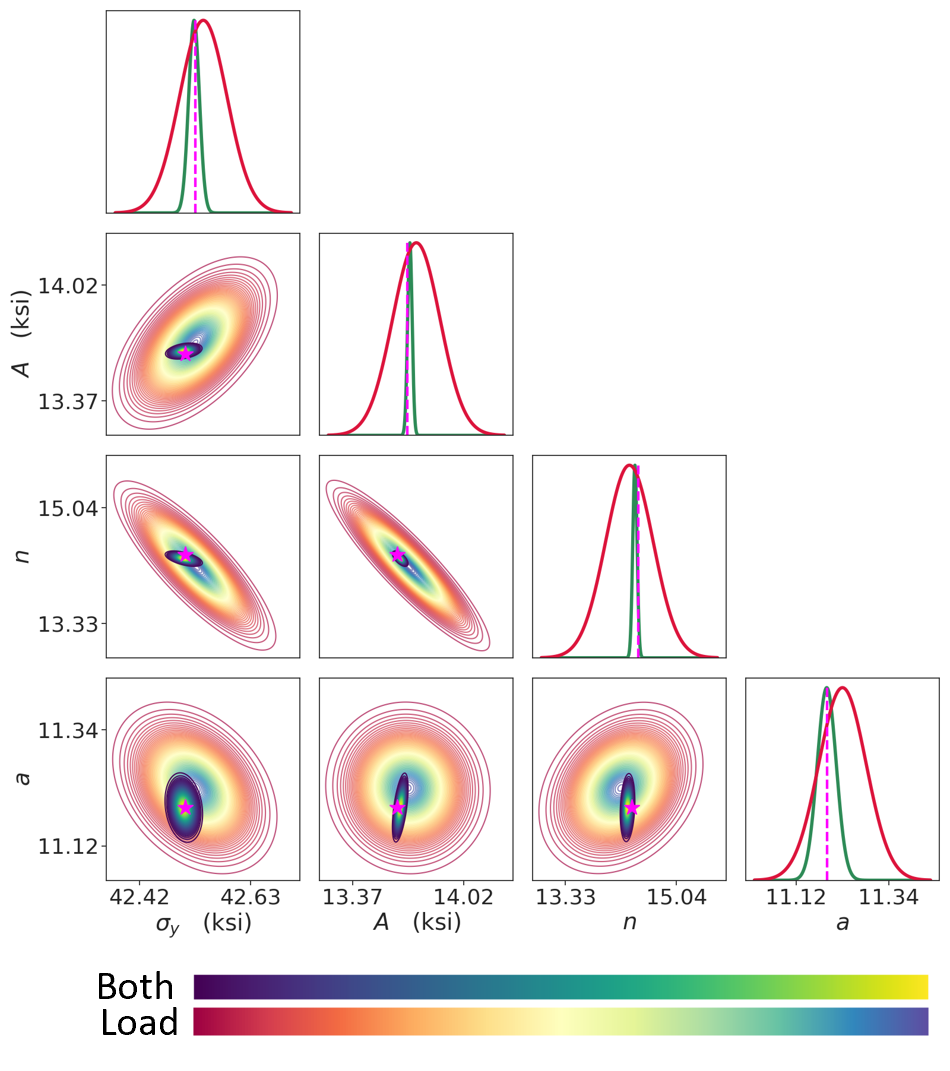}\label{fig: contour both vs load} }}%
        \sidesubfloat[]{{\includegraphics[width=65mm]{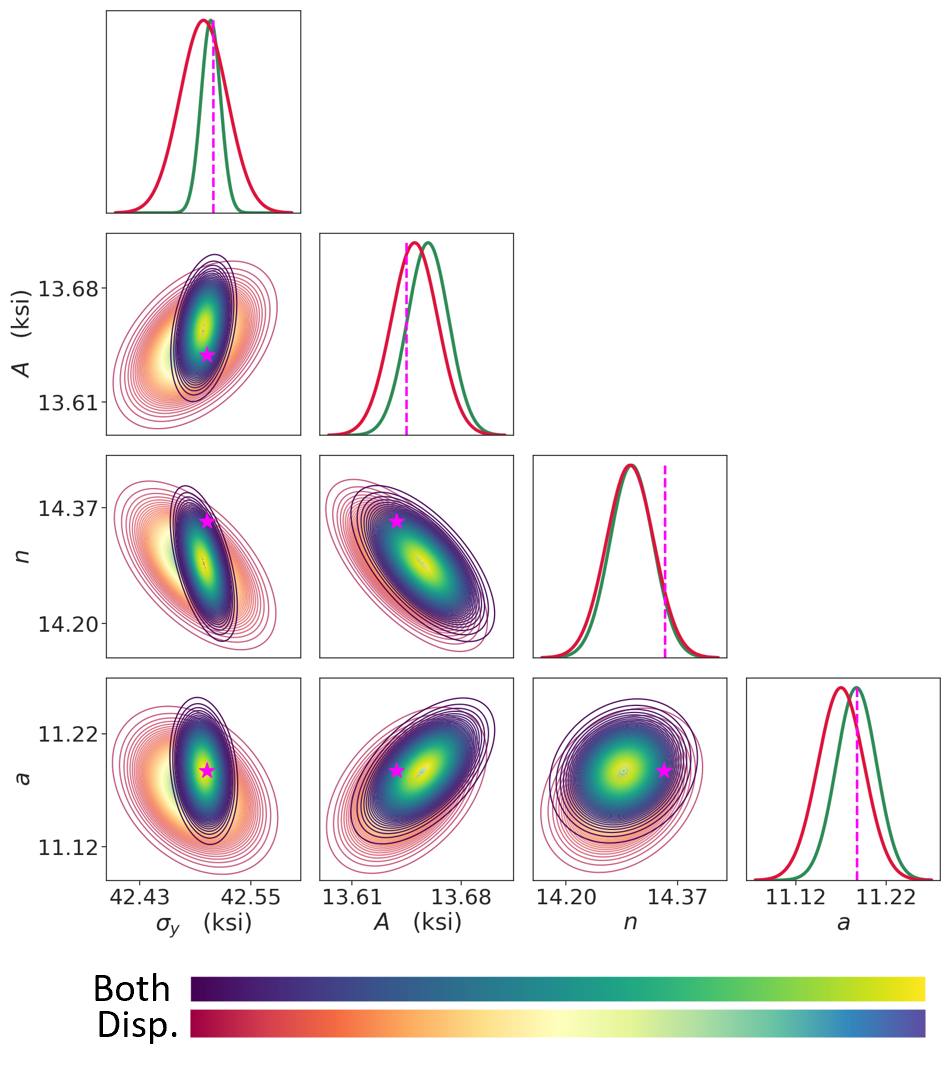}\label{fig: contour both vs disp} }}\\
       \caption{Posterior contour comparisons for a) calibrating with both full-field data and load data vs. solely load data, and b) calibrating with both full-field data and load data vs. solely displacement data for load path AAAAA. Marginal posterior densities are plotted on the diagonals and joint-marginal densities on the off-diagonals. Marginal densities for the calibration performed with both data sources are plotted with a green line and with a single data source with an orange line. $\pars^{true}$ is indicated with a dashed pink line and star. The contours shown for the combined calibration are the same as those shown in Fig.~\ref{fig: posterior contours}.}%
        \label{fig: posterior contours comparing qois}%
    \end{figure}
    
    \descitem{Myopic Decision-Making} The load step selection was performed with a \emph{greedy} (a.k.a.\ myopic) decision-making protocol, where each load step was chosen based on the immediate expected gain in information---looking only one step ahead. The alternative is to make a global decision by calculating the EIG for the entire load path; however, this increases the cost of calculating the EIG exponentially. For example, at load step $t$, instead of calculating the EIG for 2 candidate load steps in a mypocially-driven decision (load step A vs. load step B), the EIG would need to be calculated for the $(\mathcal{T}-t)^2$ possible load paths which follow from all possible future load steps in a global decision. So, while the myopic algorithm is not guaranteed to result in the globally preferred load path, the resulting time savings make ICC possible in quasi-real-time and may yet result in the globally preferred decision, as in this example. As previously mentioned, sequential OED approaches \cite{shen2023bayesian} may allow for the selection of the globally preferred full load path with a manageable computational demand. However, due to the non-Markovian nature of materials, such an approach would be challenging in this application and exploration of its use is reserved for future endeavors.
    
    \descitem{Surrogate replacement of the FE model} A fast Gaussian process (GP) surrogate replacement of the FE model was used for calibration and load step selection. A nice feature of GPs is that uncertainty surrounding the surrogate model can be incorporated into the inference problem. In this case, a simplified approach was taken, and the mean of the surrogate model was used without taking into account uncertainty of the surrogate model, but future work may be expanded to include the uncertainty. In general, the mean absolute error between the surrogate output and the model (evaluated at locations in the parameter space reserved for testing) was less than 0.003\% of the load and the less than 0.1\% of the displacements (which included any error from the PCA transformation as well). Thus, the error due to the surrogate model approximation was significantly less than the expected measurement noise for both QoIs. Additionally, because the surrogate model is trained on a grid of parameter values, the inference is restricted to this training space with truncated normal prior distributions. The parameter bounds used for training were chosen to be very conservative. However, if the support of the target posterior distribution falls outside the parameter bounds, it is likely the case that the chosen model-form is not appropriate for the specific application and should be revisited.

    \descitem{Inference in Field vs. PCA Space} When calibrating a model using full-field data, the choice between performing inference in PCA space versus the original field space presents both advantages and disadvantages. Calibrating in field space allows for the direct incorporation of experimental data, enhancing the accuracy of parameter estimates. The simplified model used in this work assumed the noise was i.i.d. While this is an appropriate assumption for synthetically generated data, when moving to a live demonstration of ICC with experimental data, correlated or structured noise in the field data \cite{fayad2025identification} may lead to underestimated uncertainty or biased parameter estimates. In such a case, a more complex model which accounts for any such noise characteristics may be necessary. Conversely, PCA space offers the benefits of dimension reduction, decorrelation of the noise structure and simplification of the calibration process, significantly speeding up tasks like the EIG approximation. For instance, in this synthetic example, calculating the EIG with field data took approximately 40 minutes, while using PCA singular values reduced this time to just 30 seconds. However, since the PCA modes are derived from model evaluations devoid of noise, they may not fully capture the noise structure of the experimental data, potentially contributing to model form error and parameter bias. To fully leverage the benefits of PCA-based inference while mitigating potential bias, it may be necessary to incorporate the noise structure of the experimental data directly into the PCA basis or to model the residual noise explicitly in the spectral domain.
    
    Calibrations in both the field space and the PCA space were performed in this work, with comparisons of the posterior summaries and contours shown in Table~\ref{Tab: field vs pca posterior comparison} and Figure~\ref{fig:field_pca_contour_comparison}. The results are very similar for this example; however, the posterior contours and marginal variances indicate that a higher posterior probability is placed on the true values with the field data, resulting in a less-biased parameter expected value.  As a result, the main body of the work discusses inference in the field space, allowing a direct incorporation of experimental data and reducing parameter bias. 
    
    While the time to calculate the EIG in the field space (40~min at each step) was not prohibitive for this synthetic demonstration, it would be during a live demonstration. A comparison of computing the EIG in the PCA space (while maintaining the calibration in the field space) vs. the field space in Table~\ref{Tab: field vs pca EIG comparison} reveals that the two methods yield identical approximations---to the 11th digit. Thus, when moving to an experimental demonstration, the preferred method is to perform the EIG approximation in the PCA space. Future decisions of performing inference in the field or PCA domain must reflect a careful consideration of the trade-offs between accuracy, computational efficiency, and the need to capture the inherent complexities of the experimental data.
    
    \begin{table}[ht!]
    \centering
    \resizebox{.7\textwidth}{!}{\begin{tabular}{l l l l@{} r@{\hspace{1.0\tabcolsep}} r@{\hspace{1.0\tabcolsep}} r@{\hspace{1.0\tabcolsep}} r@{} l l}
    \hline
    \multirow{2}{*}{\textbf{Load path}} & \multirow{2}{*}{\textbf{Data Space}} & \multirow{2}{*}{$\boldsymbol{\mathbb{E}_{\pars \given \data, \des}}$} & \multicolumn{6}{l}{$\boldsymbol{\mathbb{V}_{\pars \given \data, \des}}$} & \textbf{Gen. Var.}\\ 
    & & & \multicolumn{6}{l}{$\boldsymbol{(1\times 10^{-4})}$} &  $\boldsymbol{(1\times10^{-15})}$ \\
    \hlineB{4}
    \multirow{2}{*}{AAAAA} & Full-field & [42.50, 13.65, 14.29, 11.19] & [&1.6,& 2.5,& 15.8,& 5.3&]  & 2.34\\
      & PCA & [42.50, 13.67, 14.28, 11.21] & [&1.6,& 2.5,& 15.9,& 5.3&]  & 2.37\\
    \hline
    \end{tabular}}
    \vspace{1em} 
    \centering
    $\pars^{true} = [42.51, 13.63, 14.35, 11.19]$
    \caption{Posterior summaries obtained for load path AAAAA using the load plus displacement data in the field space vs. the reduced PCA space (i.e.\ PCA singular values).}\label{Tab: field vs pca posterior comparison}
    \end{table}
     
    \begin{figure}
        \centering
        \includegraphics[width=0.4\linewidth]{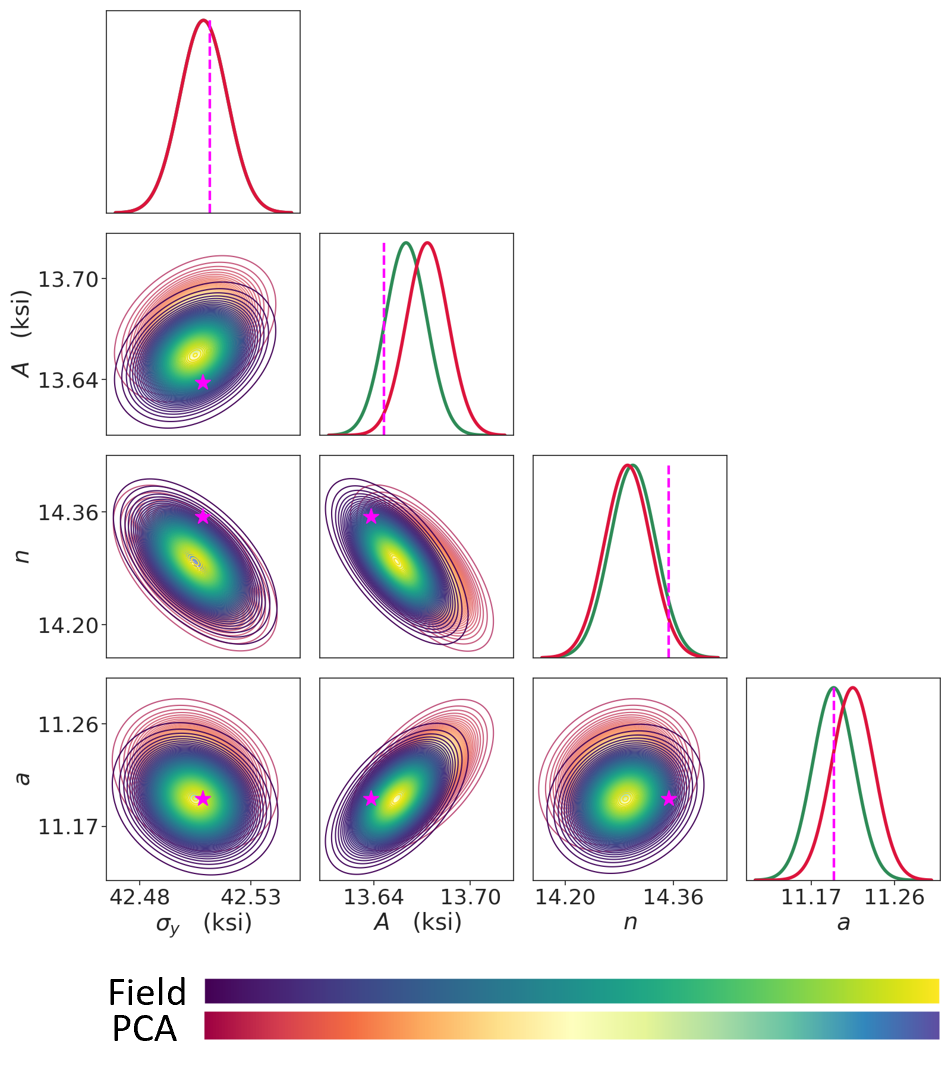}
        \caption{The posterior contours for load path AAAAA using full-field displacement data and PCA singular values. Marginal posterior densities are plotted on the diagonals and joint-marginal densities on the off-diagonals. Marginal densities for the full-field calibration are plotted with a green line and for the PCA calibration with an orange line. $\pars^{true}$ is indicated with a dashed pink line and star. The contours shown for the calibration in the field space are the same as those shown in Fig.~\ref{fig: posterior contours}.}
        \label{fig:field_pca_contour_comparison}
    \end{figure}
    
    \begin{table}[ht!]
    \centering
    \resizebox{.5\textwidth}{!}{\begin{tabular}{l l l l}
    \hline
        \multirow{2}{*}{\textbf{Load path}} & \multirow{2}{*}{\textbf{Load Step}} & \textbf{Field EIG} & \textbf{PCA EIG} \\ 
        & & [A, B] & [A, B] \\
    \hlineB{4}
    \multirow{5}{*}{AAAAA} & 1 &  N/A & N/A \\
    & 2 & [8.578, 4.144] & [8.578, 4.144] \\
    & 3 & [4.388, 2.643] & [4.388, 2.643]\\
    & 4 & [2.423, 1.344] & [2.423, 1.344] \\
    & 5 & [2.139, 1.125] & [2.139, 1.125] \\ 
    \hline
    \end{tabular}}
    \centering
    \caption{A comparison of the EIG calculated in the field space vs. the PCA space for both load step options (A and B) and for each load step except the first; the first load step was pre-determined to be A. The values are identical to the 11th digit, with the PCA-based calculations taking only a fraction of the time as the field-based calculations.}\label{Tab: field vs pca EIG comparison}
    \end{table}
    
\end{itemize}
    
\paragraph{Ending Criteria}
The ending criteria for the ICC feedback loop was to exit after a pre-determined number of load steps with the aim to keep the material sample in the plastic regime of deformation before failure. However, there are other ending criteria that could be employed. For one, the ending criteria could be the difference in the EIG between two consecutive load steps---if the EIG does not change by some threshold amount, then there is very little information expected to be gained from collecting additional data. Another approach could be to determine a threshold level of allowable uncertainty for a given application. These ending criteria are all different ways to define what is ``good enough'' in model calibration, which is highly application-dependent. Whatever is decided, the value of performing ICC remains the same, which is to efficiently perform model calibration for high-fidelity models with credibility that is both improved from traditional approaches and quantified.

\paragraph{Calibration of Multiple Models}
In this phase of algorithm development, the ICC framework was constructed and implemented for the calibration of a single material model. However, the intent is to eventually use this framework for the calibration of multiple models as there is generally a level of uncertainty regarding the best model form for a given application. The Library of Advanced Materials for Engineering (LAM\'{E})~\cite{LAME}, part of the Sierra/SolidMechanics~\cite{UserGuide} finite element software, offers a modular framework for model development from which various model forms of differing degrees of fidelity may be explored. With the consideration of multiple models, there is the additional factor of determining how the step selection will proceed---this is a topic of future work. 

\paragraph{Preferred Load Path Insights}

In this study, the AAAAA load path was identified as the preferred load path for generating the most informative data for calibration, leading to a greater degree of confidence in the parameter estimates. The preference of the AAAAA load path over all others is influenced by a number of factors, including the specimen geometry, model form, relative impact of material phenomenologies and the structure of the load path tree. Even with the same experimental set-up, the resultant preferred path may change depending on the material response. For instance, the preference of the AAAAA load path for a cruciform specimen and the calibration of an isotropic model contrasts with the results presented in Ricciardi et al. \cite{ricciardi2024bayesian}, which indicated that an alternating ABABA load path was more advantageous for a material point undergoing similar deformation and the calibration of an anisotropic model, assuming the same load path tree in both examples. This discrepancy underscores the complexity of the issue and suggests that while optimized experimental strategies implicitly assume that richer information content leads to better calibration, understanding the reason behind the superiority of one load path over another is non-trivial. Such activities, while important, are left to future studies.

\paragraph{Experimental Considerations}
This work presented a demonstration of the ICC framework in a tightly-controlled synthetic environment. The synthetic demonstration was constructed to mimic a planned experiment as closely as possible, which is to deform an aluminum cruciform specimen in a planar biaxial load frame and perform calibration with full-field DIC data. A first demonstration of the ICC framework has been performed experimentally, and a primary challenge moving to the real-world laboratory conditions was found to be machine control. The surrogate models were constructed under assumed boundary conditions, and reliable use of these models requires that the realized experimental boundary conditions match the ones used for surrogate training as closely as possible. With some adjustments to the ICC framework \cite{jones2024interlaced}, the strict boundary conditions were closely met and ICC  was completed successfully. However, future demonstrations may require that surrogates be trained over a range of boundary conditions in order to relax the requirements on machine control. Additionally, in this synthetic demonstration, the calibration and load step selection were performed in the absence of model-form error since the same model was used to both generate the data and perform the calibration. In a real-world application, no model can perfectly capture the the true behavior of the material. If there is an observed systematic discrepancy between the data and the model, model-form error will need to be taken into account in order to obtain credible uncertainty estimates from the calibration.

\paragraph{Calibration Efficiency}
In this synthetic demonstration, each load step required approximately 40 minutes in total, with less than a minute needed to approximate the posterior and around 40 minutes to estimate the EIG for both load step options. Thus, the entire synthetic demonstration of the ICC framework took less than 3.5 hours for 5 total load steps. Transitioning towards implementing the ICC framework with real-time data collection and model calibration, it is anticipated that material characterization and model calibration can be completed within a single day. Considering the setup time for both the experiments and the surrogates, it is projected that a fully developed ICC framework could accomplish the entire characterization and calibration process (excluding the time required for specimen machining) in just 1 to 2 weeks. In contrast, the traditional approach typically allocates one week for each experimental setup, which could extend the timeline to 6 to 7 weeks for a complete test matrix. Additionally, an extra 2 to 3 weeks is often needed for data reduction, FEA meshing and setup and calibration, resulting in an overall duration of 8 to 10 weeks. Therefore, the ICC paradigm has the potential to achieve a remarkable 5 to 10 times increase in efficiency compared to traditional methodologies.
\section{Conclusions}\label{sec: conclusion}

This work brings together several recent advancements made in material characterization and model calibration into an improved workflow---Interlaced Characterization and Calibration (ICC)---that improves not only the process of obtaining a reliable material model calibration, but also the quality of the calibration itself. The ICC framework achieves these improvements by 1) replacing uniaxial test specimens with a cruciform specimen that reveals complex stress states, 2) replacing global load-displacement curves with full-field displacement data, 3) replacing deterministic calibrations and best-fit parameter estimates with Bayesian inference, which produces a distribution quantifying parameter uncertainty, 4) adaptively selecting the load steps---within the confines of a load path tree---as the cruciform is deformed through Bayesian optimal experimental design (BOED) and 5) performing steps 1-4 in a quasi-real-time feedback loop. The result is that informative experimental data is collected in a highly-efficient manner to yield a credible material model that can be used for early decision-making.

A synthetic demonstration of the ICC framework was presented which considered the deformation of a cruciform specimen in a biaxial load frame to calibrate an elastoplastic constitutive model. In the exemplar problem, calibration was performed with a surrogate replacement of the expensive FE model, and full-field displacement data and global load data were synthetically generated for calibration. The displacement field was dimensionally reduced via principal component  analysis (PCA) to aid in the construction of surrogate models; the surrogate predictions of the PCA singular values were then used to reconstruct the displacement field to perform inference and calculated the EIG in the field space. A non-alternating AAAAA load path (pulling on the same axis of the cruciform for all five load steps) was preferred by the algorithm and resulted in a calibration that had the least parameter uncertainty compared to all other load paths---a non-intuitive result. This outcome reinforced the utility of BOED algorithms to identify load paths that collect more informative data than load paths selected \textit{a priori} by subject matter experts. 

The tools developed for this synthetic demonstration make a significant step in moving towards a demonstration of the ICC framework in the lab with real-time data collection. By using this approach, high-fidelity material models can be calibrated with a great increase in efficiency over traditional methods. Additionally, the proposed approach improves the quality of the calibrations with guided data collection and quantified uncertainty. These results improve the state-of-the-art in standard calibration procedures.

\newpage
\section{Acknowledgements}\label{sec: acknowledgements}
This work was supported by the Laboratory Directed Research and Development program at Sandia
National Laboratories, a multimission laboratory managed and operated by National Technology
\& Engineering Solutions of Sandia, LLC, a wholly owned subsidiary of Honeywell International
Inc., for the U.S. Department of Energy’s National Nuclear Security Administration under
contract DE-NA0003525.

This paper describes objective technical results and analysis. Any subjective views or opinions
that might be expressed in the paper do not necessarily represent the views of the U.S. Department
of Energy or the United States Government.
\section{CRediT Authorship Contribution Statement}\label{sec: credit}

\textbf{D.E. Ricciardi}: Data curation, Formal analysis, Investigation, Methodology, Software, Validation, Visualization, Writing – original draft, Writing – review \& editing. \textbf{D.T. Seidl}: Data curation, Methodology, Resources, Supervision, Validation, Writing – review \& editing. \textbf{B.T. Lester}: Resources, Supervision, Writing – review \& editing. \textbf{A.R. Jones}: Resources, Writing – review \& editing. \textbf{E.M.C. Jones}: Conceptualization, Funding acquisition, Project administration, Supervision, Writing – review \& editing.

\biboptions{sort&compress}
\bibliographystyle{unsrt}
\bibliography{refs}

\appendix
\section{Surrogate Parity Plots}\label{parity.app}
\setcounter{figure}{0}  
\setcounter{table}{0}  

\begin{figure}%
    \centering
    \sidesubfloat[]{\includegraphics[width=50mm]{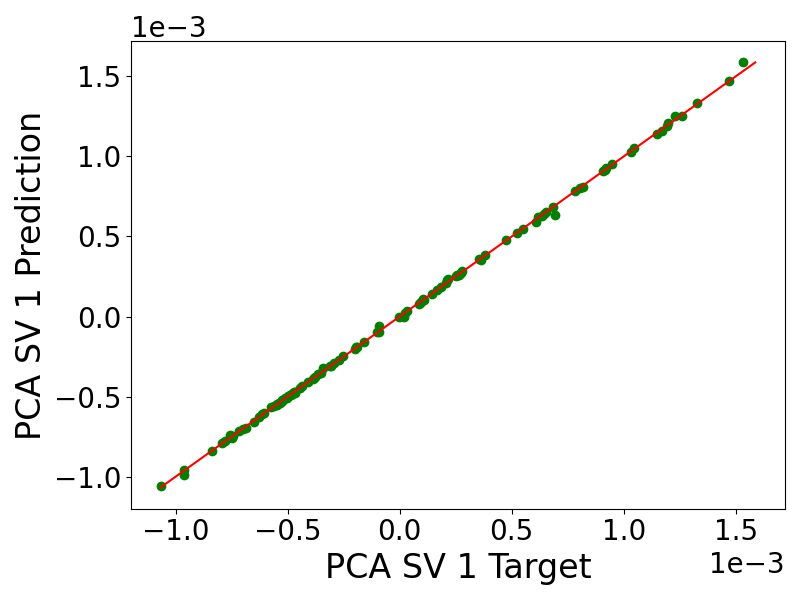}\label{fig: parity x mode 0} }
    \sidesubfloat[]{\includegraphics[width=50mm]{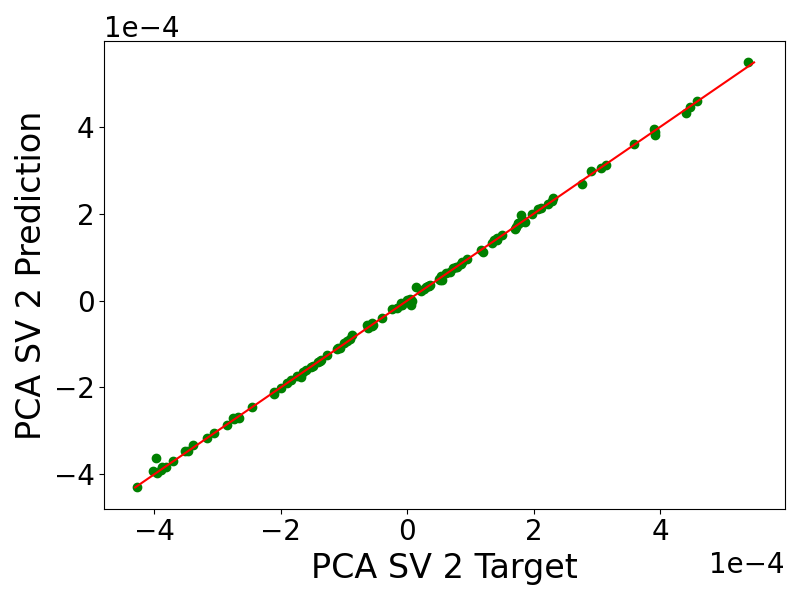}\label{fig: parity x mode 1} }%
    \sidesubfloat[]{\includegraphics[width=50mm]{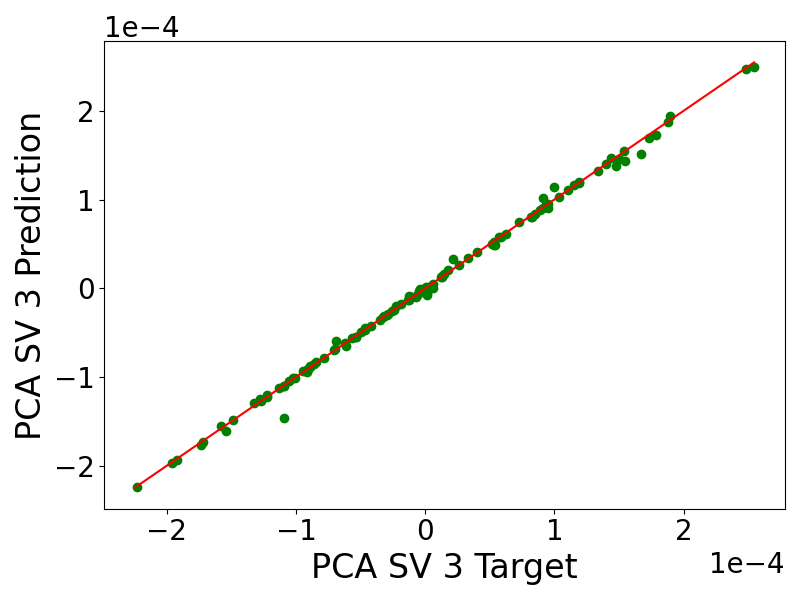}\label{fig: parity x mode 2} } \\ %
    \sidesubfloat[]{\includegraphics[width=50mm]{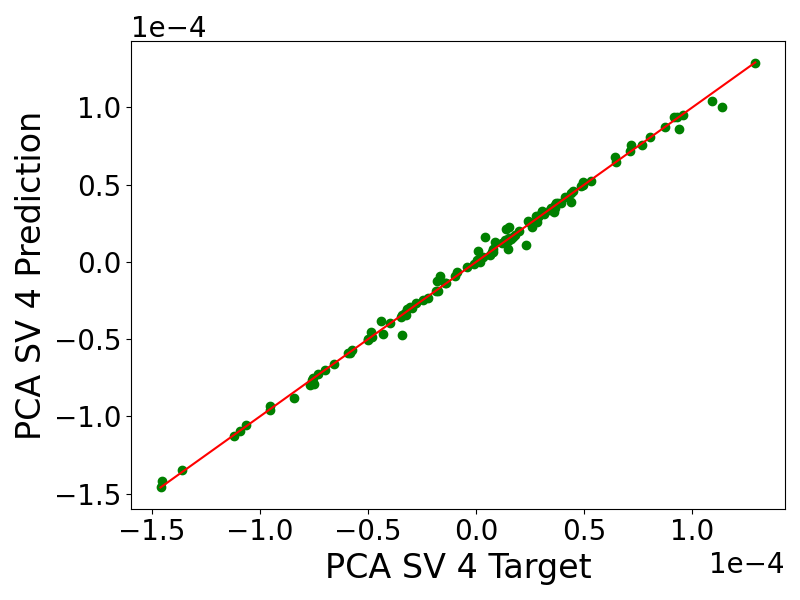}\label{fig: parity x mode 3} }%
    \sidesubfloat[]{\includegraphics[width=50mm]{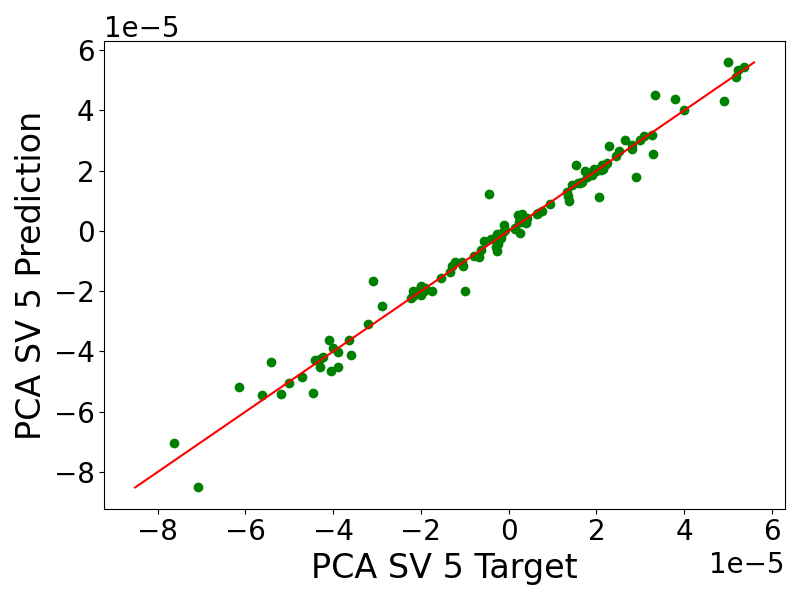}\label{fig: parity x mode 4} }%
    \caption{Parity plots for the predicted PCA singular values (SV) from the surrogate model at the final load step of an ABABA load path for the $X$ directional component. The observed data is plotted on the x-axis and the surrogate prediction is plotted on the y-axis. The red 45-degree line is the line of perfect agreement.}%
    \label{fig: pca parity plots}%
\end{figure}

\begin{figure}%
    \centering
    \sidesubfloat[]{\includegraphics[width=55mm]{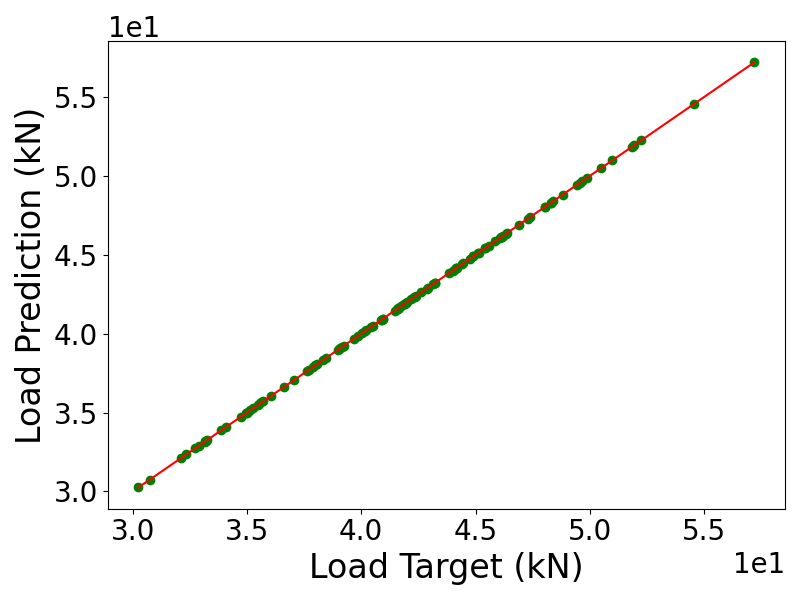}\label{fig: parity x load}
    }
    \sidesubfloat[]{\includegraphics[width=55mm]{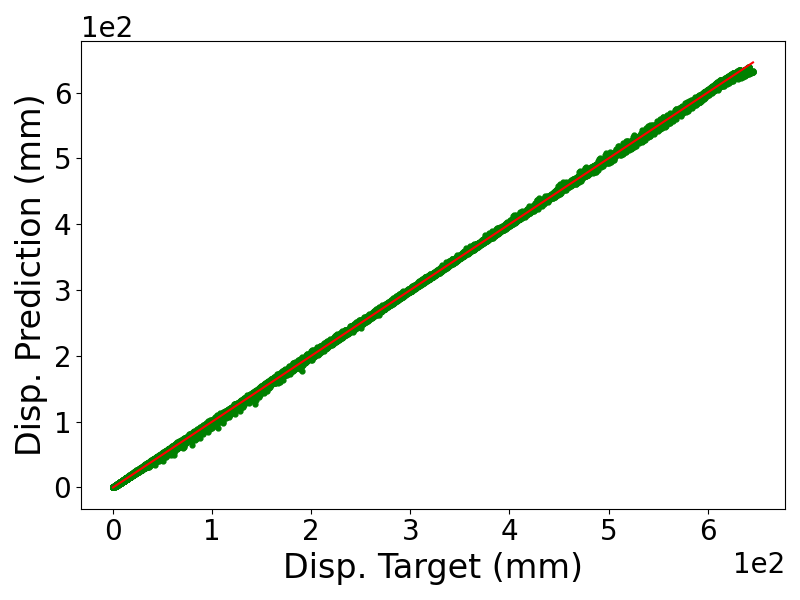}\label{fig: parity x disp}
    }
    \caption{Parity plots for the surrogate-predicted global load values (a) and reconstructed displacement fields (b) at the final load step of an ABABA load path for the $X$ directional component. The observed data is plotted on the x-axis and the surrogate prediction is plotted on the y-axis. The red 45-degree line is the line of perfect agreement.}%
    \label{fig: load and disp parity plots}%
\end{figure}

Parity plots are a graphical tool used here to assess the performance of the surrogate models by comparing predicted values against observed values. Here, there observed data is FE model output that was reserved for testing. In these plots, the x-axis represents the observed values, while the y-axis displays the predicted values. Ideally, if the model's predictions are accurate, the points will lie along a 45-degree line, known as the line of perfect agreement. Deviations from this line indicate discrepancies between the predicted and observed values. The parity plots shown in this section assess the performance of the surrogate models at the final load step of an ABABA load path for the $X$ directional component. Fig.~\ref{fig: pca parity plots} shows parity plots for the displacement PCA singular values (SV), and Fig.~\ref{fig: load and disp parity plots} contains parity plots for the global load values, which are the two quantities the surrogate models predict. Fig.~\ref{fig: load and disp parity plots} also contains a parity plot for the displacement field, which was obtained by reconstructing field data from the surrogate prediction of the PCA singular values; any error between the true values and predicted values in this plot is due to the surrogate error as well as error originating from the PCA reconstruction. In all cases, there is very good alignment between the predicted values and true values, and there is not evidence of systematic discrepancy between the two.

\section{Posterior Contours}\label{contours.app}
\setcounter{figure}{0}  
\setcounter{table}{0} 

\begin{figure}[ht!]%
    \centering
    \sidesubfloat[]{{\includegraphics[width=75mm]{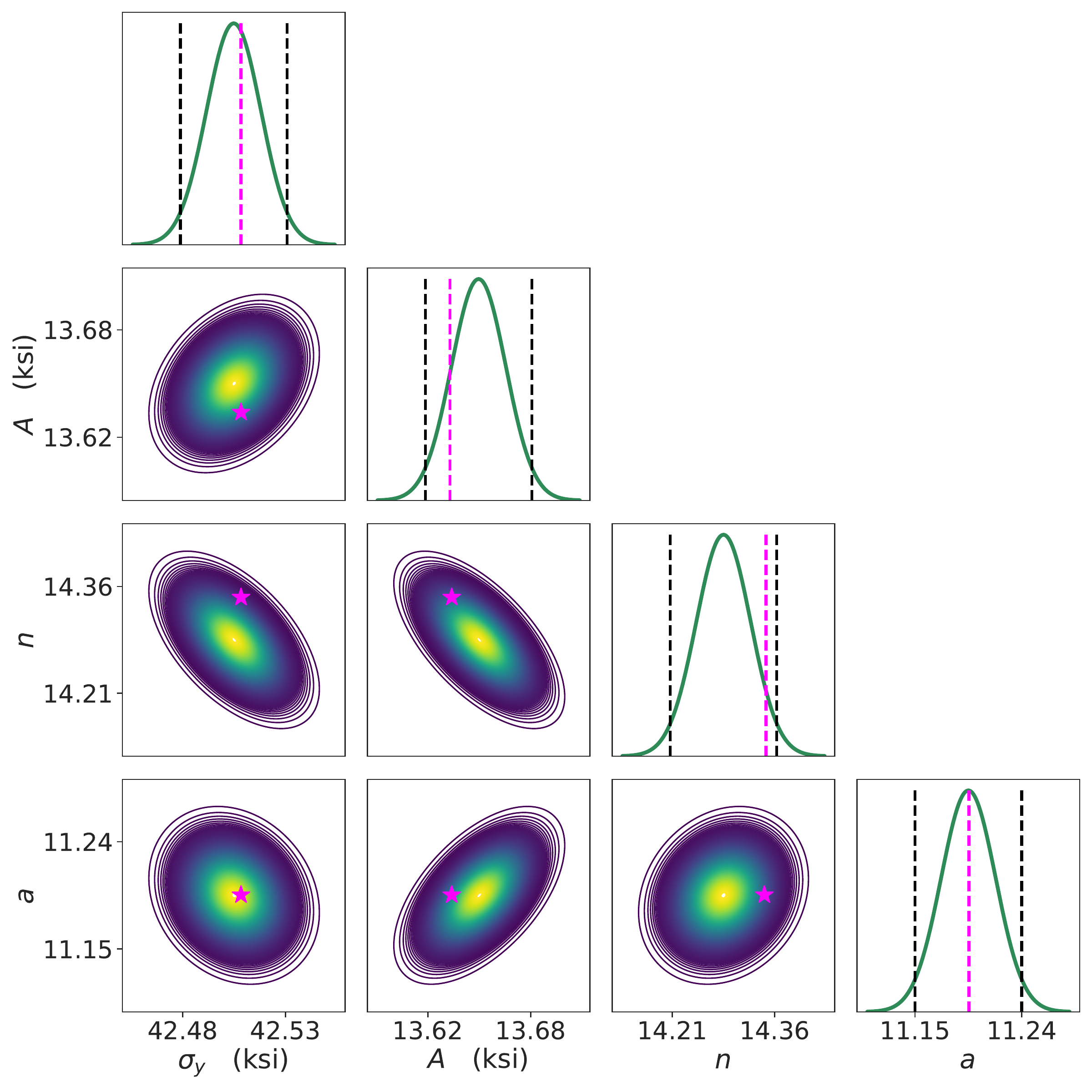}\label{fig: contour AAAAA} }}%
    \sidesubfloat[]{{\includegraphics[width=75mm]{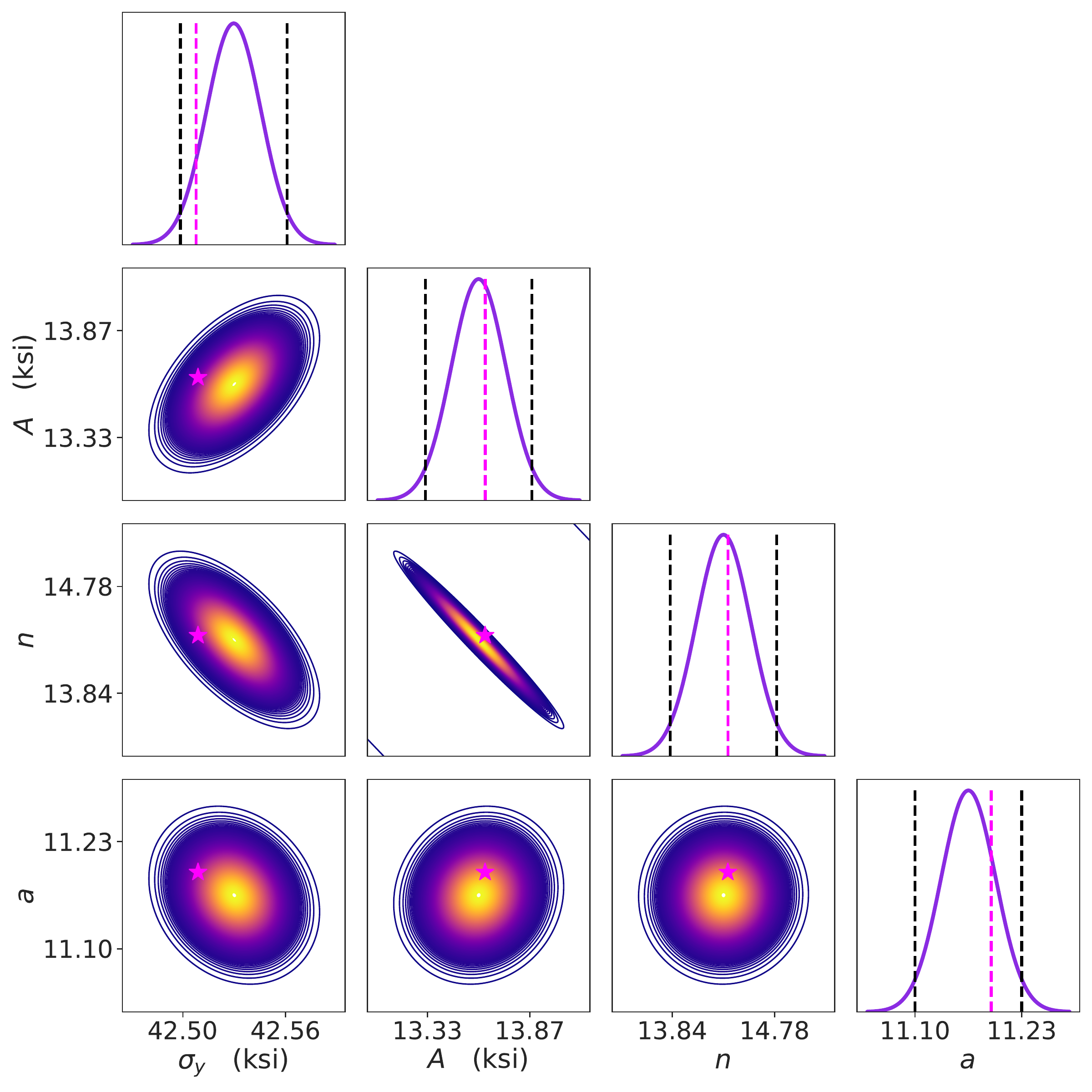}\label{fig: contour ABABA} }}\\
    \sidesubfloat[]{{\includegraphics[width=75mm]{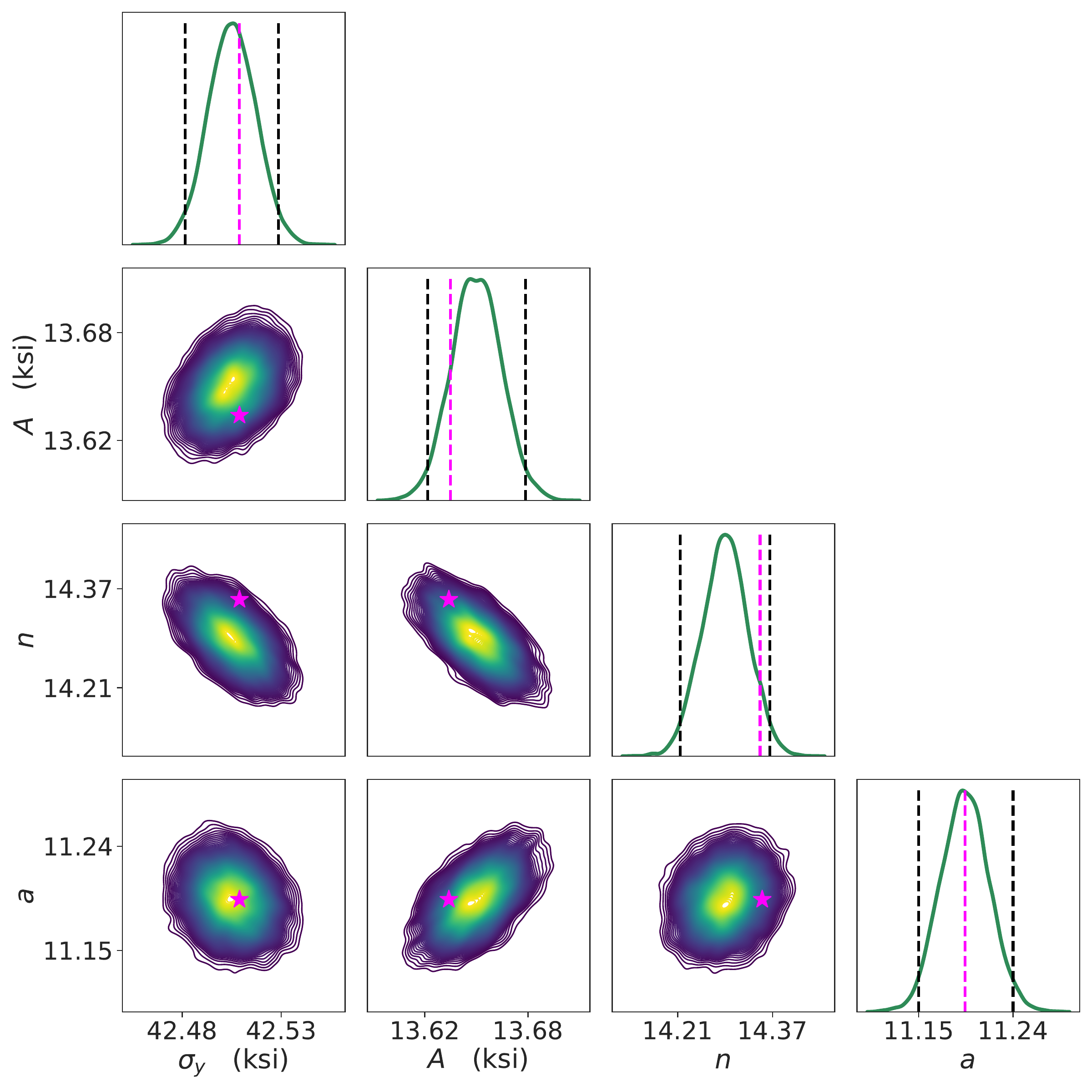}\label{fig: MCMC contour AAAAA} }}%
    \sidesubfloat[]{{\includegraphics[width=75mm]{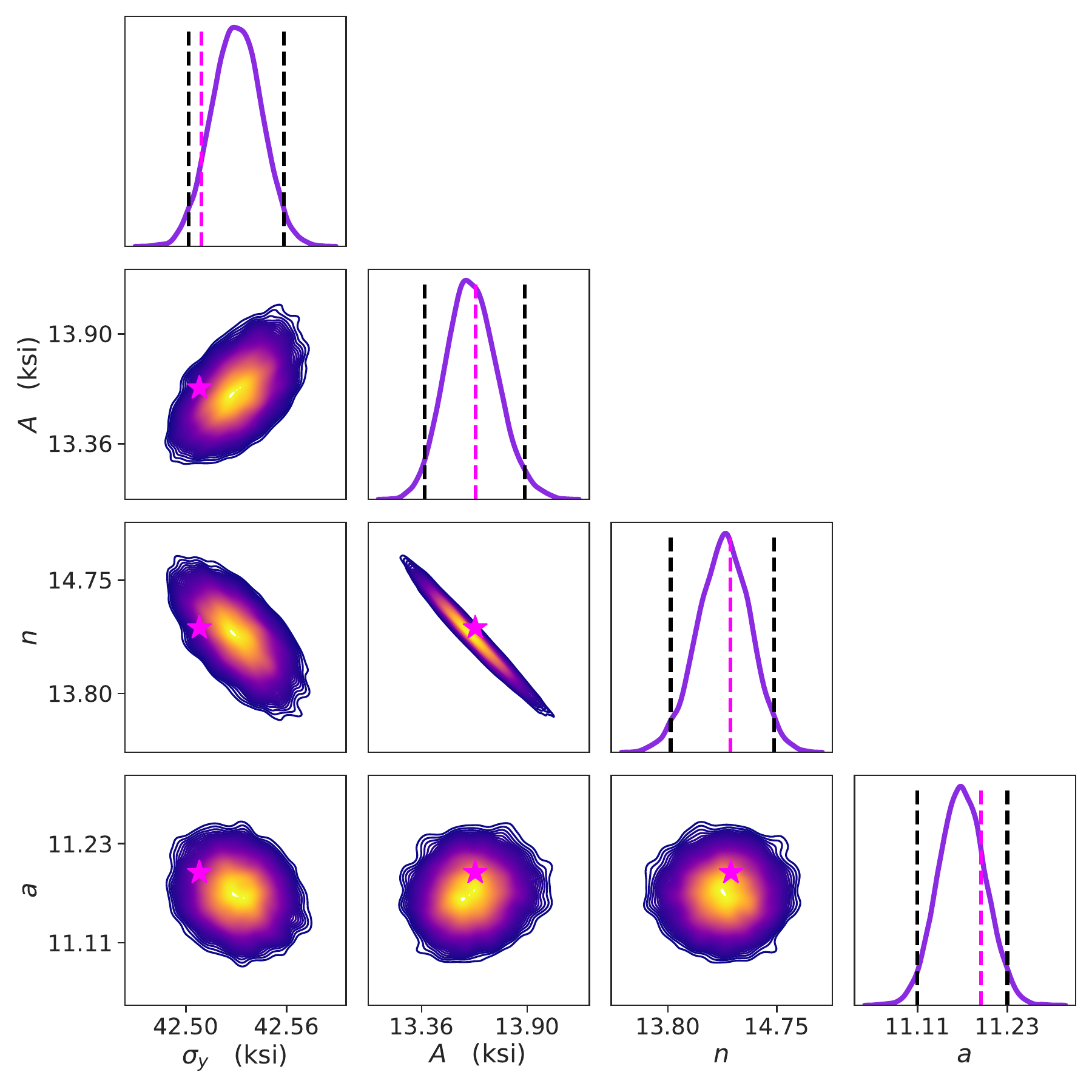}\label{fig: MCMC contour ABABA} }}\\
   \caption{The posterior contour from the (a) AAAAA and (b) ABABA calibrations obtained via the Laplace approximation are shown on the top row. The bottom row shows posterior contours obtained via MCMC as a validation for (c) AAAAA and (d) ABABA. Marginal posterior densities are plotted on the diagonals and joint-marginal densities on the off-diagonals. $\pars^{true}$ is indicated with a dashed pink line and star and the 95\%~CI with dashed black lines.}%
    \label{fig: all posterior contours}%
\end{figure}

Fig.~\ref{fig: all posterior contours} shows the posterior contours for the AAAAA and ABABA calibrations obtained via the Laplace approximation (top row) as well as through MCMC simulation (bottom row) as a validation. MCMC was performed with an adaptive Metropolis-Hastings sampling algorithm; 120,000 samples were obtained with 20,000 samples discarded as burn-in. The chain was thinned by a factor of 10 to produce independent samples. The marginal distributions are shown on the diagonals plots along with the 95\% credible interval (CI) (dashed black lines) and $\pars^{true}$ (pink dashed line). The off-diagonals show the joint marginal posterior distributions along with $\pars^{true}$ indicated by a pink star. Additional quantitative validation is contained in Tables~\ref{Tab: mcmc-laplace posterior comparison}~-~\ref{Tab: MCMC posterior correlation}. Table~\ref{Tab: mcmc-laplace posterior comparison}, reports posterior summaries obtained via the Laplace approximation as well as MCMC. Tables~\ref{Tab: Laplace posterior correlation} and \ref{Tab: MCMC posterior correlation} report posterior correlations obtained via the Laplace approximation and MCMC, respectively, for both the AAAAA and ABABA calibrations. The posterior correlation describes the trade-off relationship between different parameters. Values closer to 1 (perfect correlation) indicate a stronger relationship between parameters than those closer to 0 (independence). The reported values correspond to the posterior contours shown in Fig.~\ref{fig: all posterior contours}, where elongated contours suggest a correlation between the given parameters, while circular contours suggest little to no parameter correlation.

\floatsetup[figure]{style=plain,subcapbesideposition=top,font=small}

\begin{table}[ht!]
\centering
\resizebox{0.65\textwidth}{!}{\begin{tabular}{l l l l@{} r@{\hspace{1.0\tabcolsep}} r@{\hspace{1.0\tabcolsep}} r@{\hspace{1.0\tabcolsep}} r@{} l}
\hline
\multirow{2}{*}{\textbf{Load path}} & \multirow{2}{*}{Method} & \multirow{2}{*}{$\boldsymbol{\mathbb{E}_{\pars \given \data, \des}}$} & \multicolumn{6}{l}{$\boldsymbol{\mathbb{V}_{\pars \given \data, \des}}$}\\ 
& & & \multicolumn{6}{l}{$\boldsymbol{(1\times 10^{-4})}$}\\
\hlineB{4}
\multirow{2}{*}{AAAAA} & Laplace & [42.50, 13.65, 14.29, 11.19] & [&1.6,& 2.5,& 15.8,& 5.3&]  \\
  & MCMC & [42.50, 13.65, 14.29, 11.19] & [&1.6,& 2.5,& 15.9,& 5.2&]  \\
\multirow{2}{*}{ABABA} & Laplace & [42.53, 13.60, 14.31, 11.17] & [&2.2,& 193.6,& 580.2,& 9.6&] \\
  & MCMC & [42.53, 13.62, 14.29, 11.17] & [&2.2,& 194.7,& 574.0,& 9.6&]\\
\hline
\end{tabular}}
\caption{Posterior summaries obtained via a Laplace approximation and MCMC simulation using data collected from the respective load paths.}\label{Tab: mcmc-laplace posterior comparison}
\end{table}

\begin{table}[ht!]
\centering
\begin{tabular}{lllll}
\hline
 & $\sigma_y$ & $A$ & $n$ & $a$ \\ 
\hlineB{4}
$\sigma_y$ & 1. &&& \\
$A$ & 0.33 & 1. && \\
$n$ & -0.55 & -0.64 & 1. & \\
$a$ & -0.19 & 0.56 & 0.19 & 1. \\
\hline
\end{tabular}
\qquad
\begin{tabular}{lllll}
\hline
 & $\sigma_y$ & $A$ & $n$ & $a$ \\ 
\hlineB{4}
$\sigma_y$ & 1. &&& \\
$A$ & 0.52 & 1. && \\
$n$ & -0.63 & -0.98 & 1. & \\
$a$ & -0.19 & 0.10 & 0.01 & 1. \\
\hline
\end{tabular}
\caption{Element-wise posterior correlation for load path AAAAA (left) and ABABA (right) obtained via the Laplace approximation.}\label{Tab: Laplace posterior correlation}
\end{table}

\begin{table}[ht!]
\centering
\begin{tabular}{lllll}
\hline
 & $\sigma_y$ & $A$ & $n$ & $a$ \\ 
\hlineB{4}
$\sigma_y$ & 1. &&& \\
$A$ & 0.35 & 1. && \\
$n$ & -0.56 & -0.64 & 1. & \\
$a$ & -0.18 & 0.56 & 0.20 & 1. \\
\hline
\end{tabular}
\qquad
\begin{tabular}{lllll}
\hline
 & $\sigma_y$ & $A$ & $n$ & $a$ \\ 
\hlineB{4}
$\sigma_y$ & 1. &&& \\
$A$ & 0.52 & 1. && \\
$n$ & -0.63 & -0.98 & 1. & \\
$a$ & -0.18 & 0.10 & 0.01 & 1. \\
\hline
\end{tabular}
\caption{Element-wise posterior correlation for load path AAAAA (left) and ABABA (right) obtained with MCMC.}\label{Tab: MCMC posterior correlation}
\end{table}

\subsection{Displacement Posterior Predictive Fields}\label{appen:disp posterior predictive}

Figure \ref{fig: AAAAA ABBBA X field validation} in Sec.~\ref{sec: validation} displays draws from the posterior predictive distribution for the displacement field to establish that the observed data is reasonable under the calibrated model. Here, complementary information is shown, which is the difference between the observed field and the predictive fields. From this, the error at each load step and for each predictive field can be observed. There are no observed systematic discrepancies and the error is similar among the multiple draws.

\begin{figure}
    \centering
    \includegraphics[width=\linewidth]{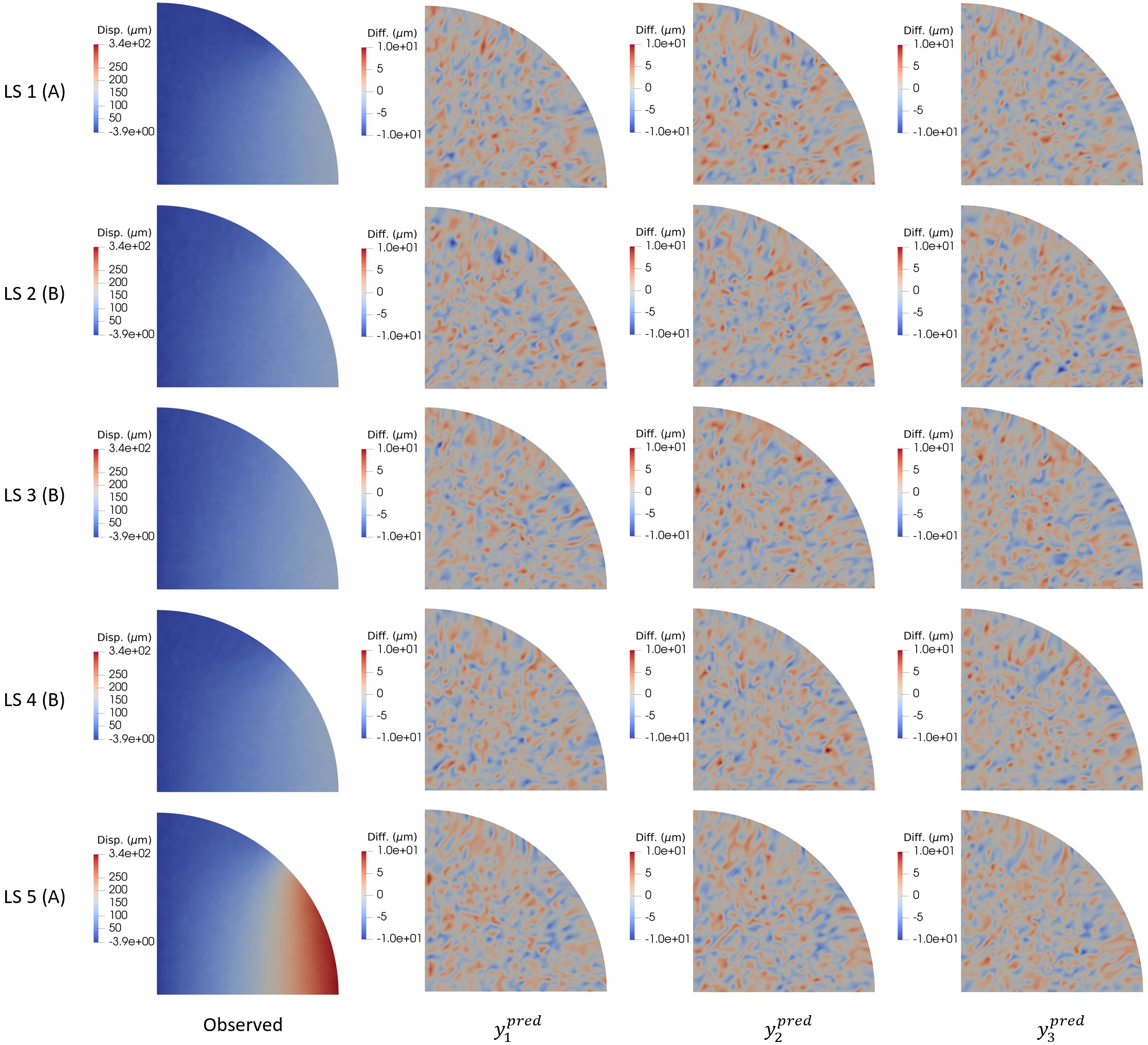}
    \caption{The error for three samples from the AAAAA posterior predictive distribution are shown for the $X$ directional component of the displacement field for the new ABBBA load path. The left column shows the observed data for each of the 5 load steps (LS) and the three right columns shows the error between three different predictive samples and the validation data.}
\label{fig: AAAAA ABBBA X field validation diff}
\end{figure}

\section{Log Predictive Density}\label{log_predictive_density.app}

An additional comparison can be made between the model calibration with an AAAAA load path vs an ABABA load path by comparing their predictive accuracy given a point estimate of the fitted model, $\hat{\pars}$---namely, the MAP parameter estimate which yields the best-fit log predictive density (BFLPD) (a.k.a. log-likelihood), $\log \pi(\data^{pred} \given \pars^{\text{MAP}})$ \cite{gelman1995bayesian}. This predictive score summarizes the fit of the model to the data and is proportional to the mean squared error in this case since a normal model and constant variance are assumed. Table~\ref{Tab: log predictive density} reports the log predictive density of the AAAAA- and ABABA-calibrated models for 16 new data sets that were not seen in calibration for the indicated load paths. For all but a few cases, the AAAAA-calibrated model yielded best-fit predictive densities that were greater than the ABABA-calibrated model. 

\begin{table}[ht!]
\centering
\resizebox{0.45\textwidth}{!}{\begin{tabular}{llll}
\hline
\multirow{2}{*}{\textbf{Prediction Path}} & \multirow{2}{*}{\textbf{Calibration Path}} & \textbf{BFLPD} \\ 
& & $\boldsymbol{(1\times 10^{3})}$ \\
\hlineB{4}
\multirow{2}{*}{AAAAA} & AAAAA & 113,959   \\
& ABABA & 114,072 \\ \hline
\multirow{2}{*}{AAAAB} & AAAAA  & 113,911 \\
& ABABA & 113,917 \\ \hline
\multirow{2}{*}{AAABA} & AAAAA & 114,304 \\
& ABABA & 114,287 \\ \hline
\multirow{2}{*}{AAABB} & AAAAA & 114,114 \\
& ABABA & 114,109 \\ \hline
\multirow{2}{*}{AABAA} & AAAAA &  114,206 \\
& ABABA & 114,206 \\ \hline
\multirow{2}{*}{AABAB} & AAAAA & 114,345 \\
& ABABA & 114,345 \\ \hline
\multirow{2}{*}{AABBA} & AAAAA & 114,491 \\
& ABABA & 114,489 \\ \hline
\multirow{2}{*}{AABBB} & AAAAA & 114,180 \\
& ABABA & 114,177 \\ \hline
\multirow{2}{*}{ABAAA} & AAAAA & 113,856 \\
& ABABA & 113,854 \\ \hline
\multirow{2}{*}{ABAAB} & AAAAA & 114,343 \\
& ABABA & 114,338 \\ \hline
\multirow{2}{*}{ABABA} & AAAAA & 114,510 \\
& ABABA & 114,270 \\ \hline
\multirow{2}{*}{ABABB} & AAAAA & 114,300 \\
& ABABA & 114,296 \\ \hline
\multirow{2}{*}{ABBAA} & AAAAA & 114,520 \\
& ABABA & 114,519\\ \hline
\multirow{2}{*}{ABBAB} & AAAAA & 114,183\\
& ABABA &  114,179 \\ \hline
\multirow{2}{*}{ABBBA} & AAAAA & 114,167 \\
& ABABA & 114,162  \\ \hline
\multirow{2}{*}{ABBBB} & AAAAA & 114,022 \\
& ABABA &  114,000 \\ \hline
\hline
\end{tabular}}
\caption{Best-fit log predictive density.}\label{Tab: log predictive density}
\end{table}

\newpage

\begin{center}
\textbf{\large Supplemental Material: Additional Figures and Tables}
\end{center}

\makeatletter
\setcounter{figure}{0}
\setcounter{equation}{0}
\setcounter{page}{1}
\setcounter{table}{0}
\setcounter{section}{0}
\renewcommand{\thetable}{S\arabic{table}}%
\renewcommand{\theequation}{S\arabic{equation}}
\renewcommand{\thefigure}{S\arabic{figure}}
\renewcommand{\bibnumfmt}[1]{[S#1]}
\renewcommand{\citenumfont}[1]{S#1}
\renewcommand{\thesection}{Supp \@arabic\c@section}

\section{Cruciform Specimen}\label{cruciform.supp}

\begin{figure}[h!]
    \centering
    \includegraphics[width=140mm]{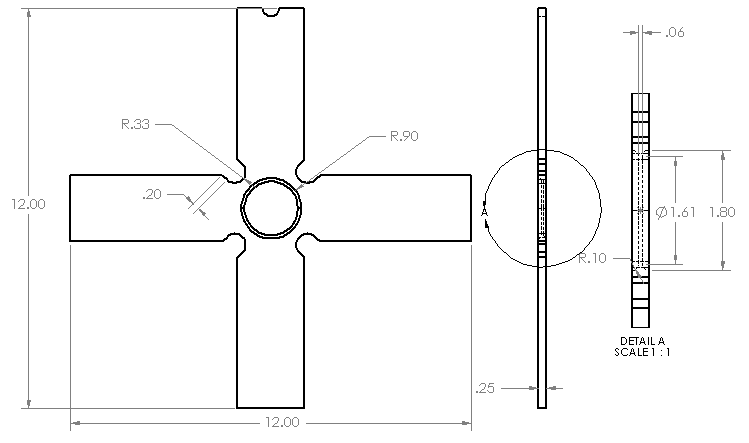}
    \caption{Abbreviated drawing of the cruciform specimen geometry.  Dimensions are in inches.}
    \label{fig:cruciform_geometry}
\end{figure}

The dimensions of the cruciform geometry simulated in this work are detailed in Fig.~\ref{fig:cruciform_geometry}.

\newpage

\section{Retained Principal Components of the Nodal Displacements}\label{pca_modes.supp}

\begin{figure}%
    \textbf{Node 1 Principal Components for X Directional Component}\par\medskip
    \centering
    \sidesubfloat[]{\includegraphics[width=45mm]{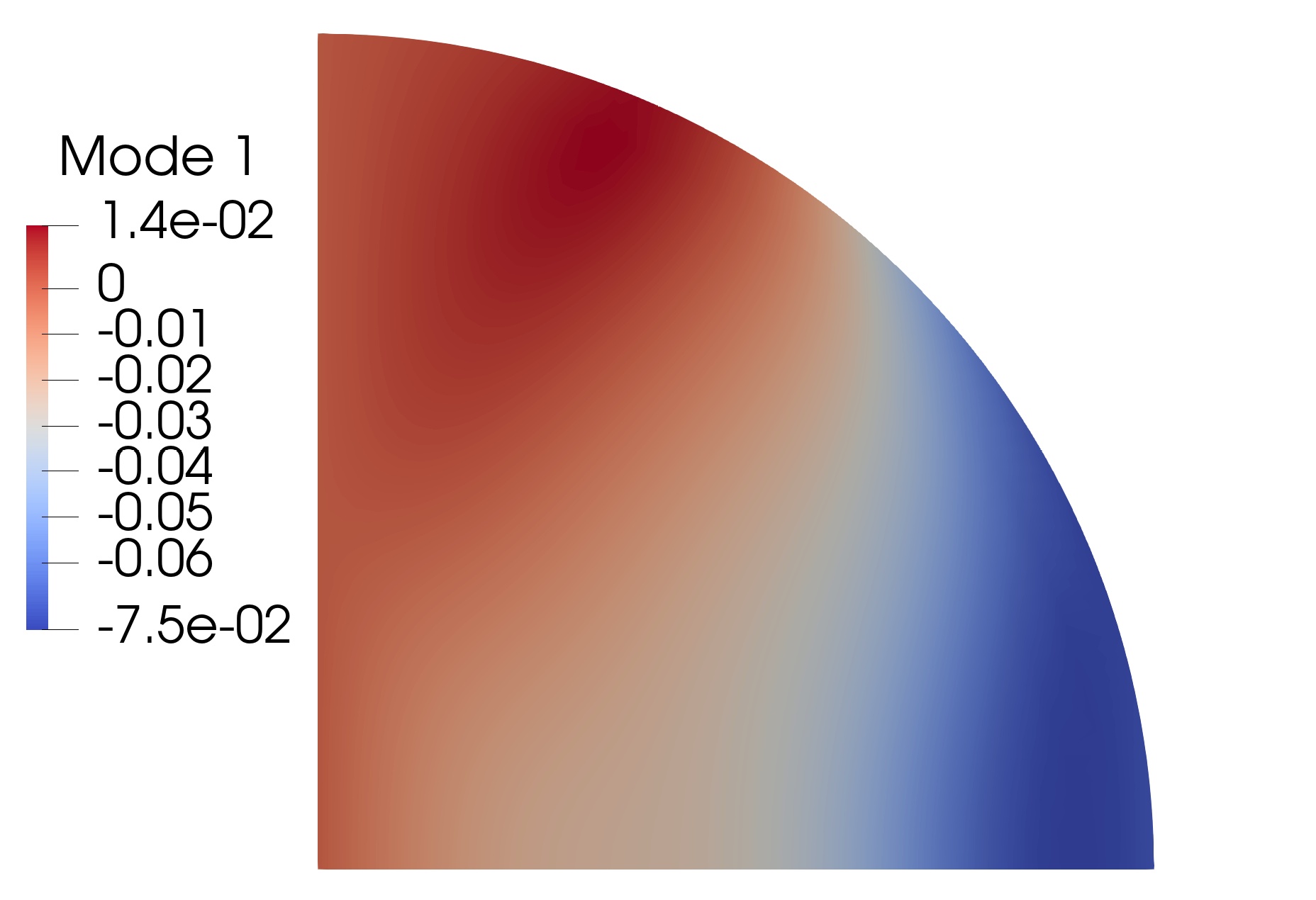}\label{fig: pca mode 1 x} }
    \sidesubfloat[]{\includegraphics[width=45mm]{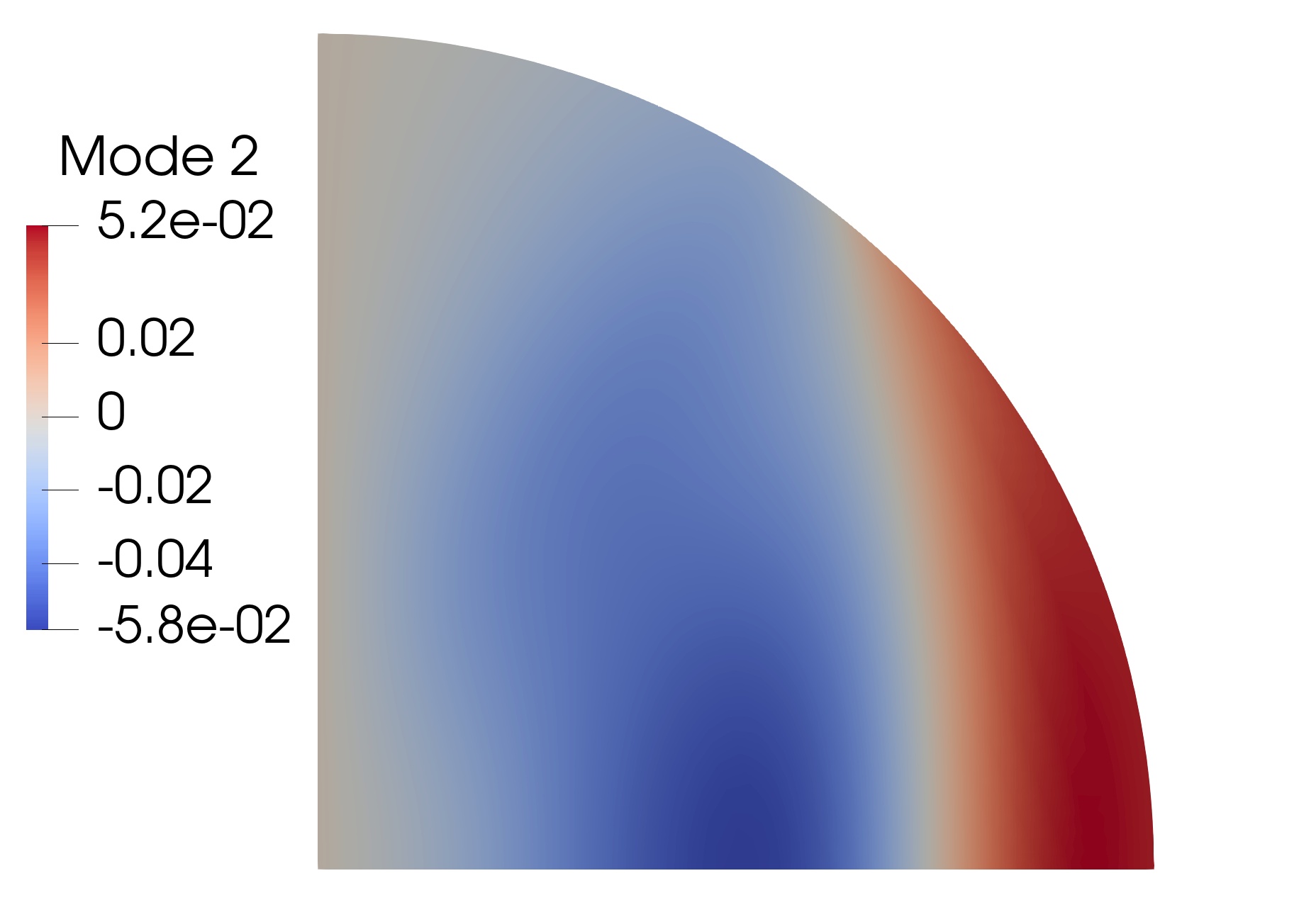}\label{fig: pca mode 2 x} }%
    \sidesubfloat[]{\includegraphics[width=45mm]{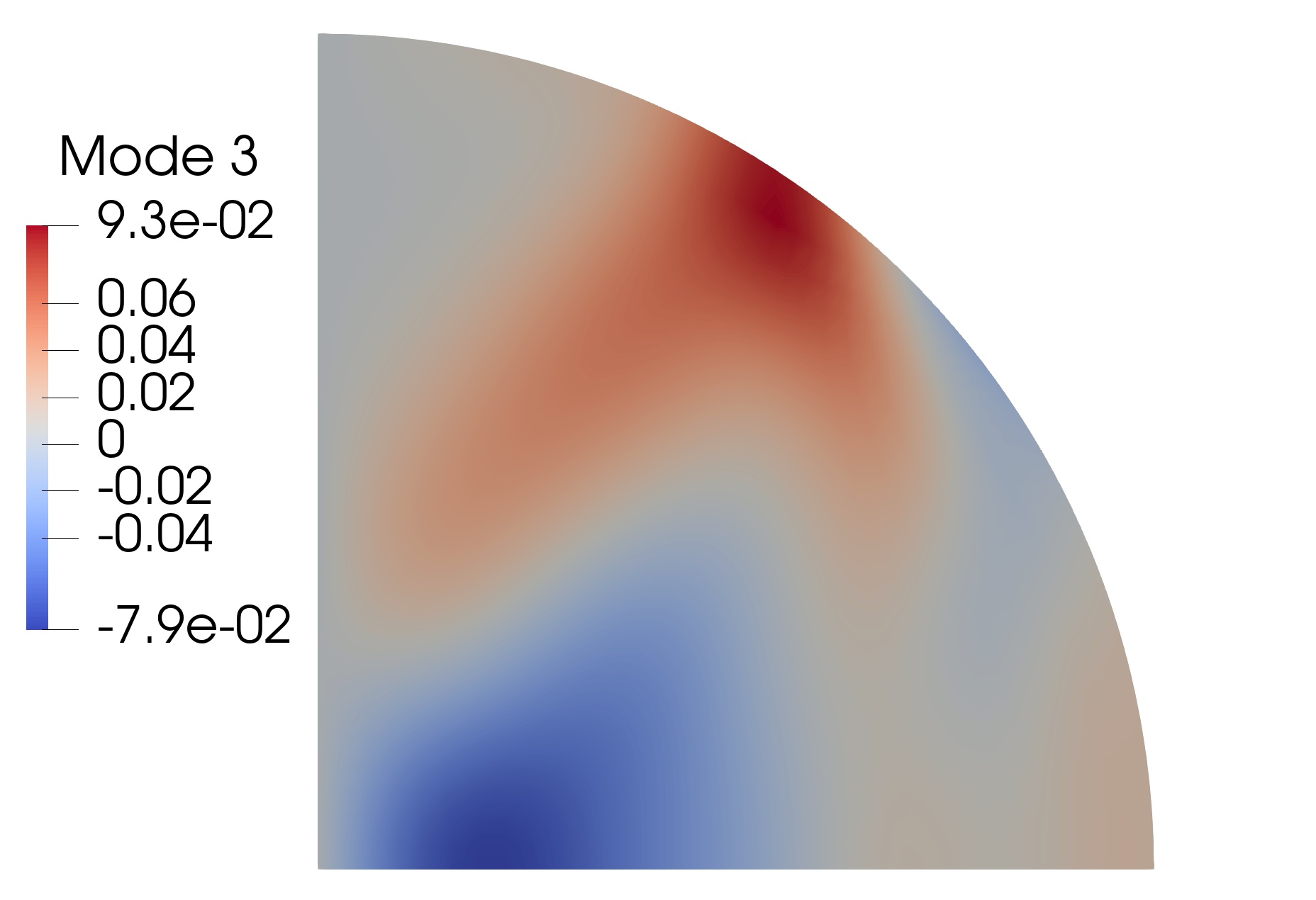}\label{fig: pca mode 3 x} } \\ %
    \sidesubfloat[]{\includegraphics[width=45mm]{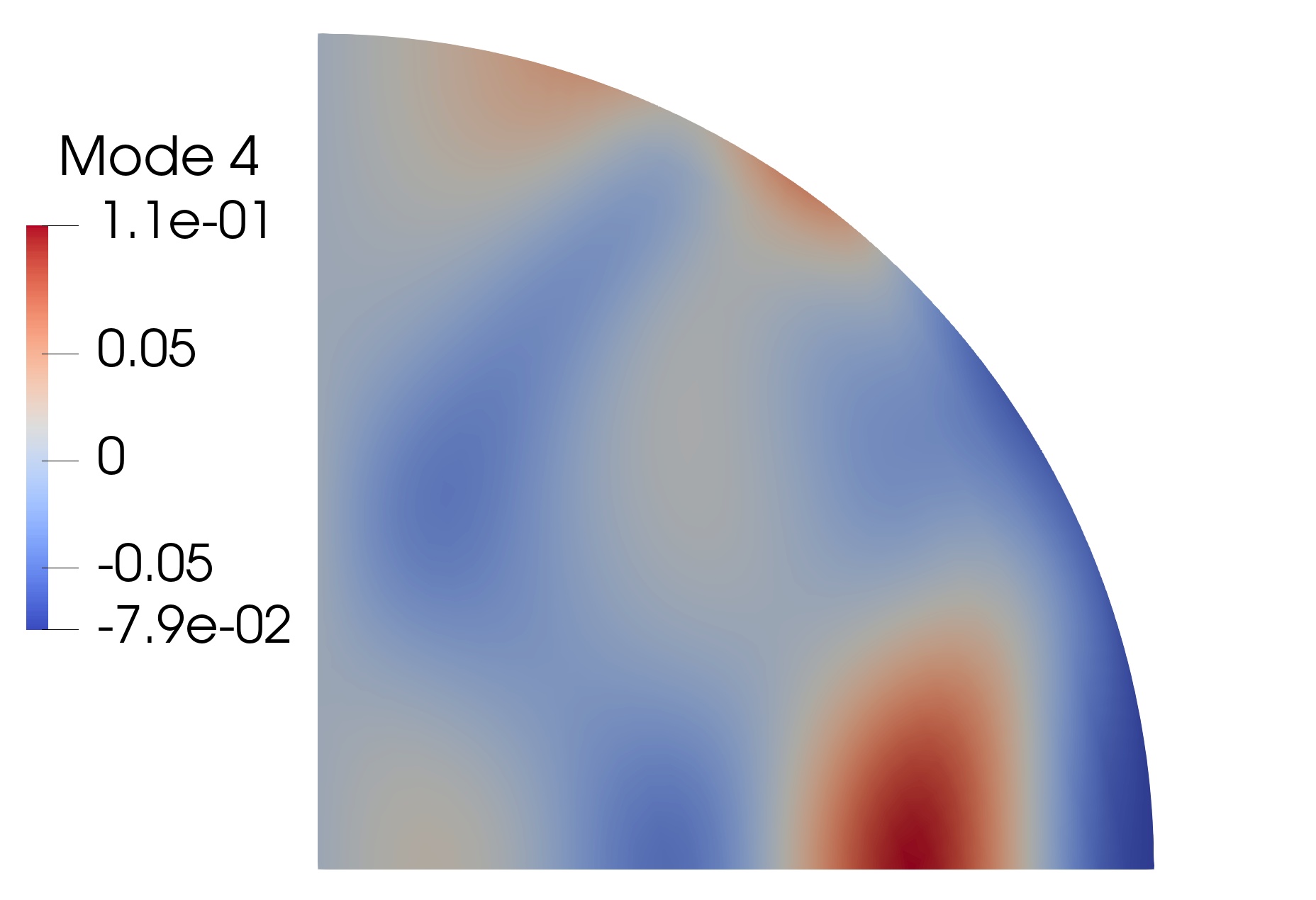}\label{fig: pca mode 4 x} }%
    \sidesubfloat[]{\includegraphics[width=45mm]{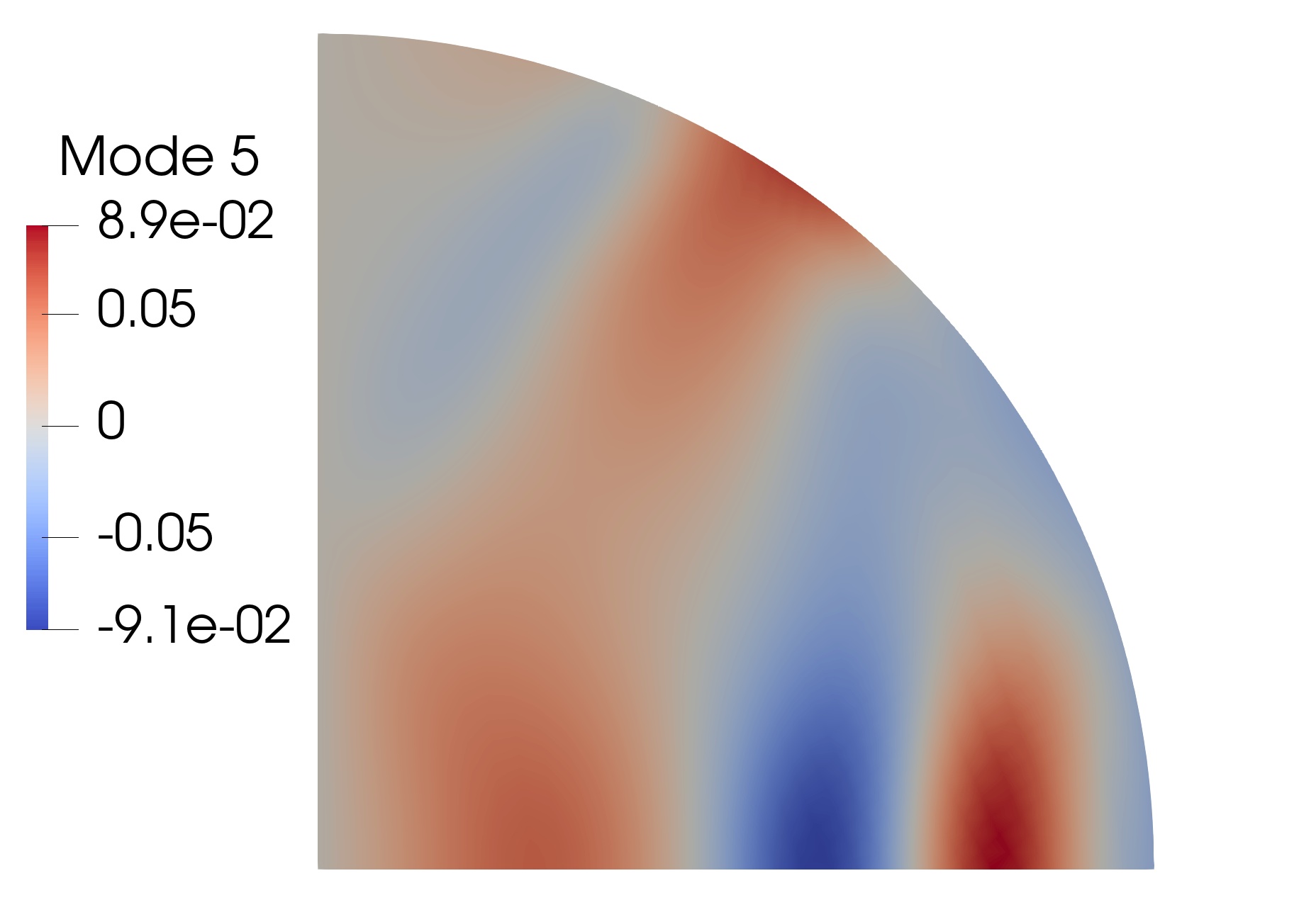}\label{fig: pca mode 5 x} }%
    \caption{The 5 retained principal components for the $X$ directional component of displacement after one load step along the X (horizontal) Axis (load step A) - i.e. Node 1 in the load path tree in Fig.~\ref{fig: loadpath_tree}}%
    \label{fig: x pca modes}%
\end{figure}

\begin{figure}%
    \centering
    \textbf{Node 1 Principal Components for Y Directional Component}\par\medskip
    \sidesubfloat[]{\includegraphics[width=45mm]{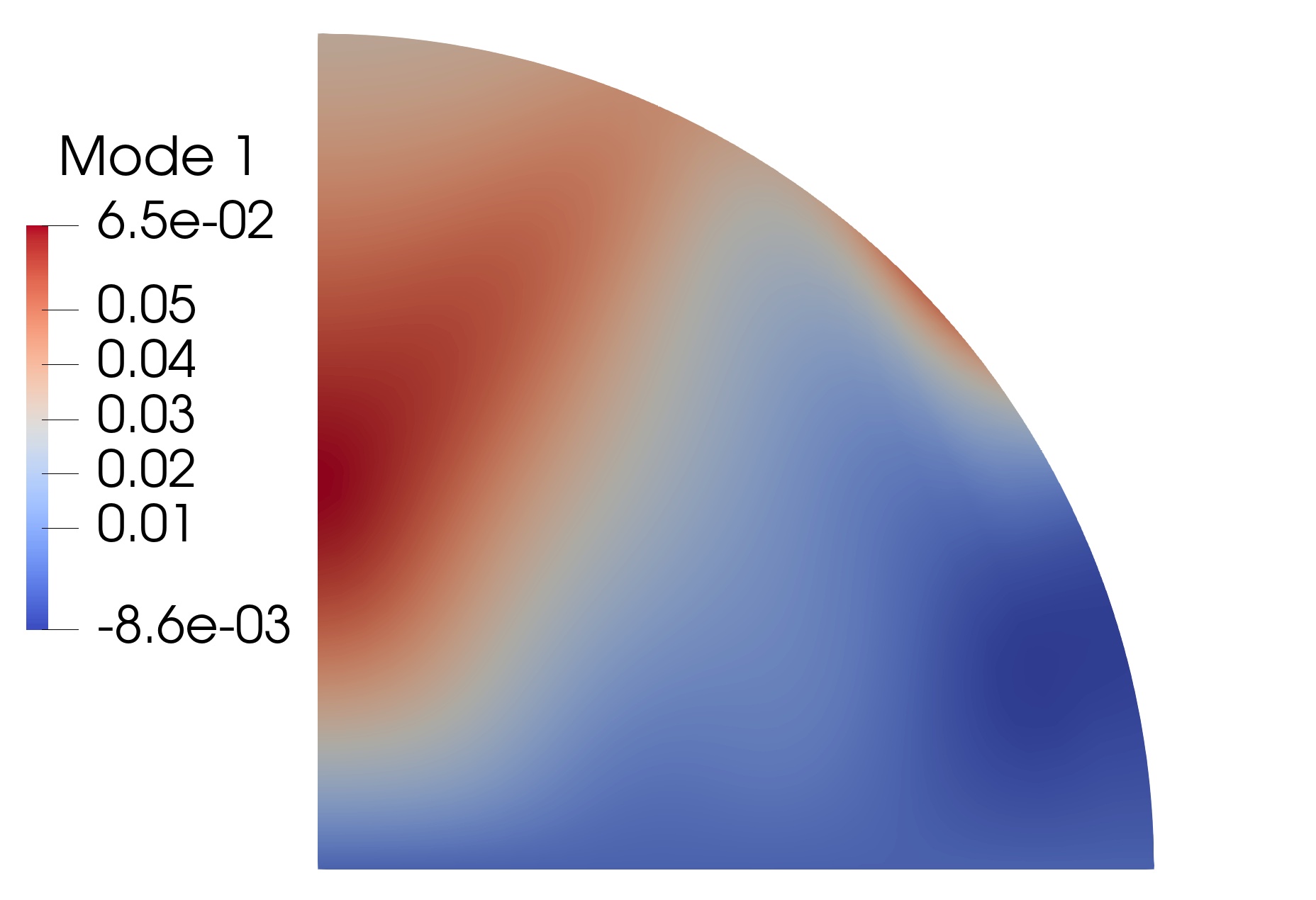}\label{fig: pca mode 1 y} }
    \sidesubfloat[]{\includegraphics[width=45mm]{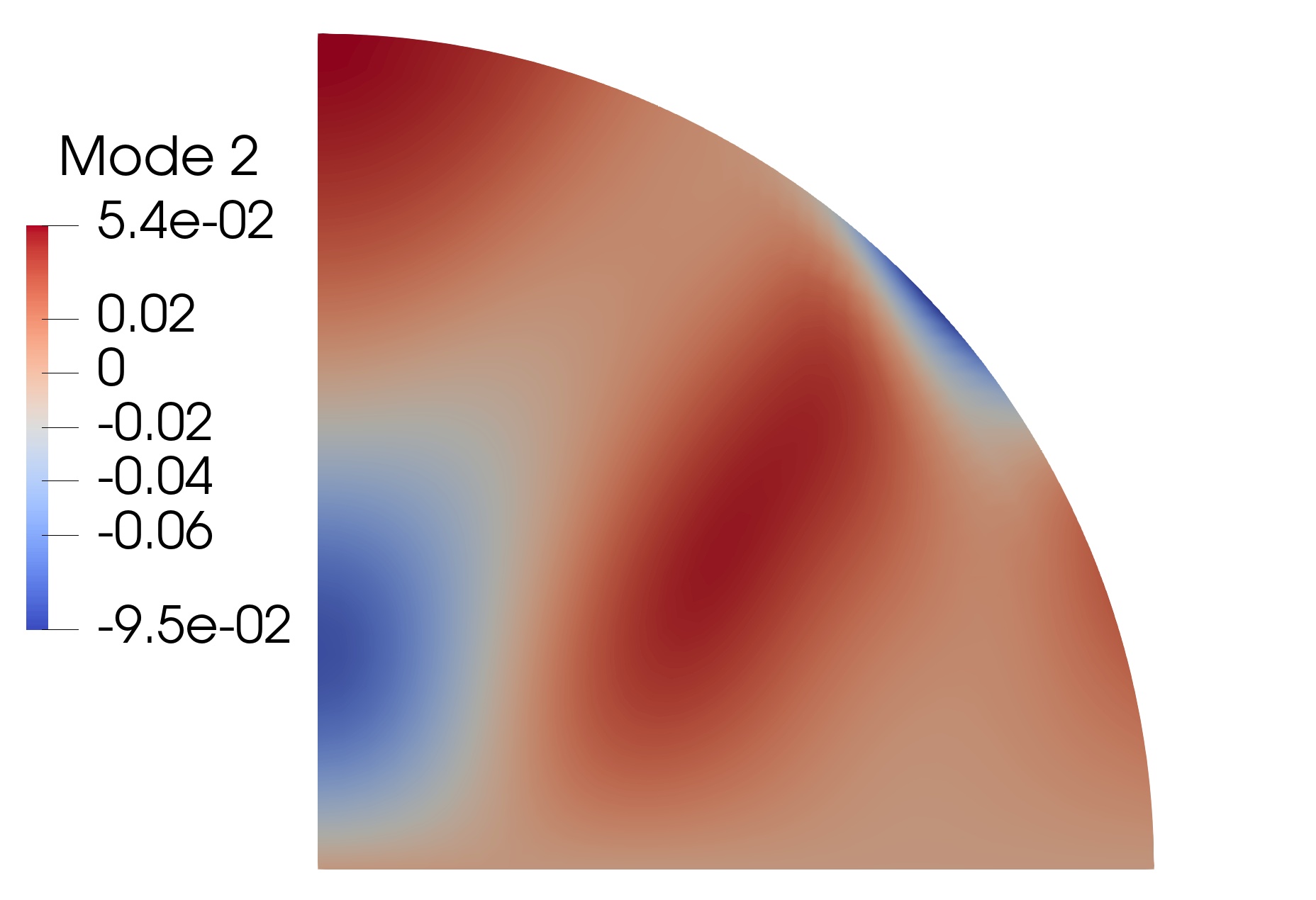}\label{fig: pca mode 2 y} }%
    \sidesubfloat[]{\includegraphics[width=45mm]{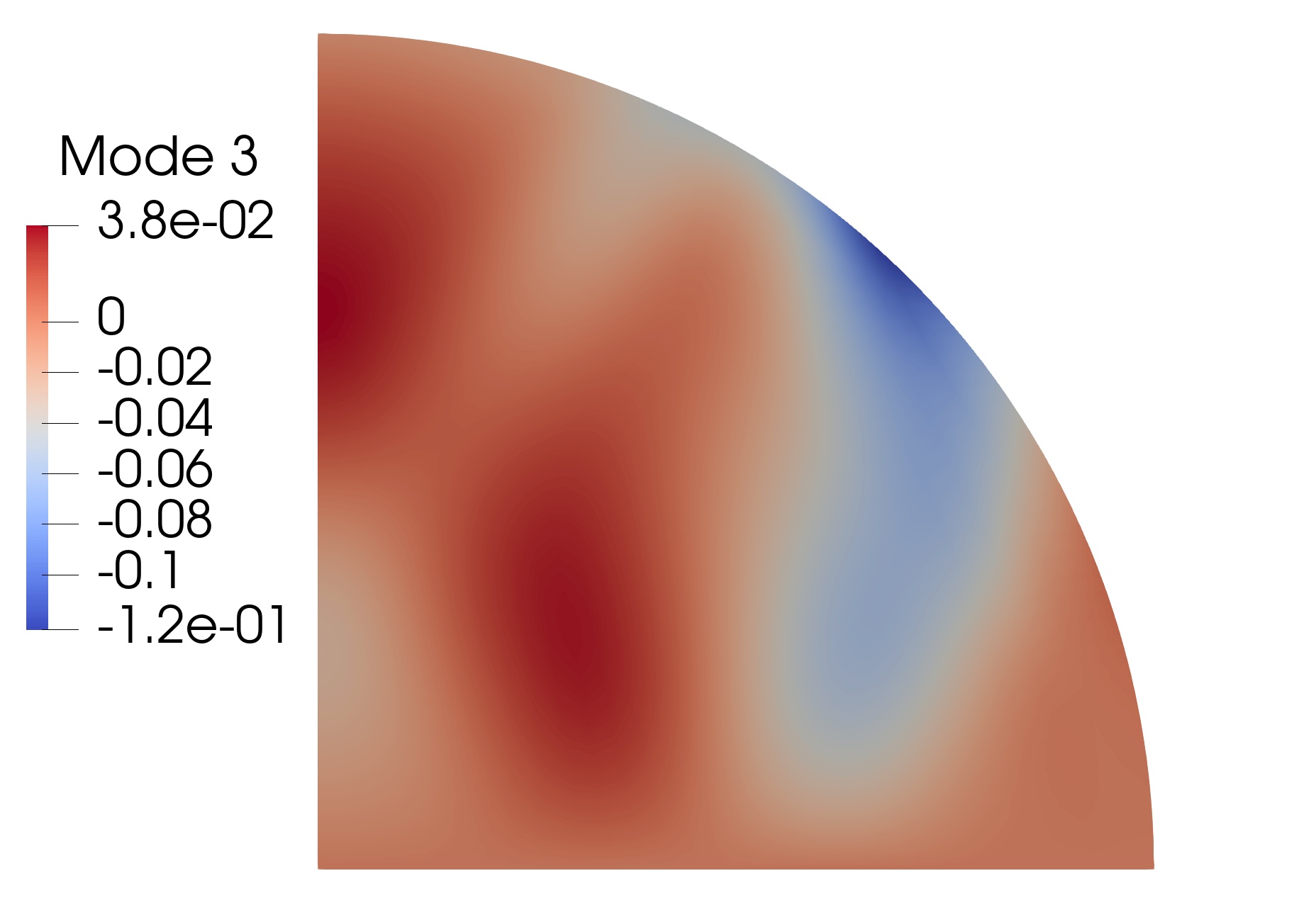}\label{fig: pca mode 3 y} } \\ %
    \sidesubfloat[]{\includegraphics[width=45mm]{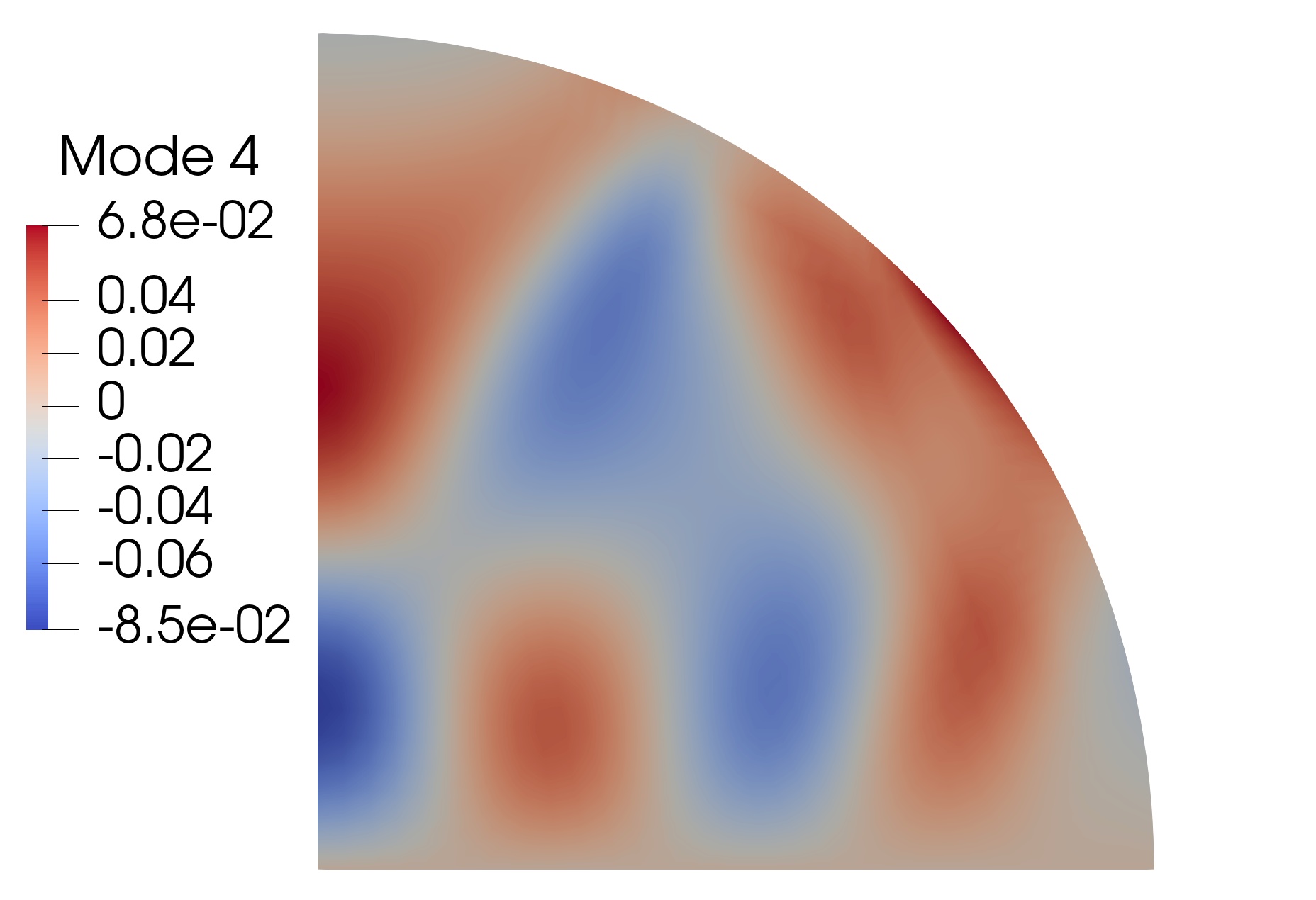}\label{fig: pca mode 4 y} }%
    \sidesubfloat[]{\includegraphics[width=45mm]{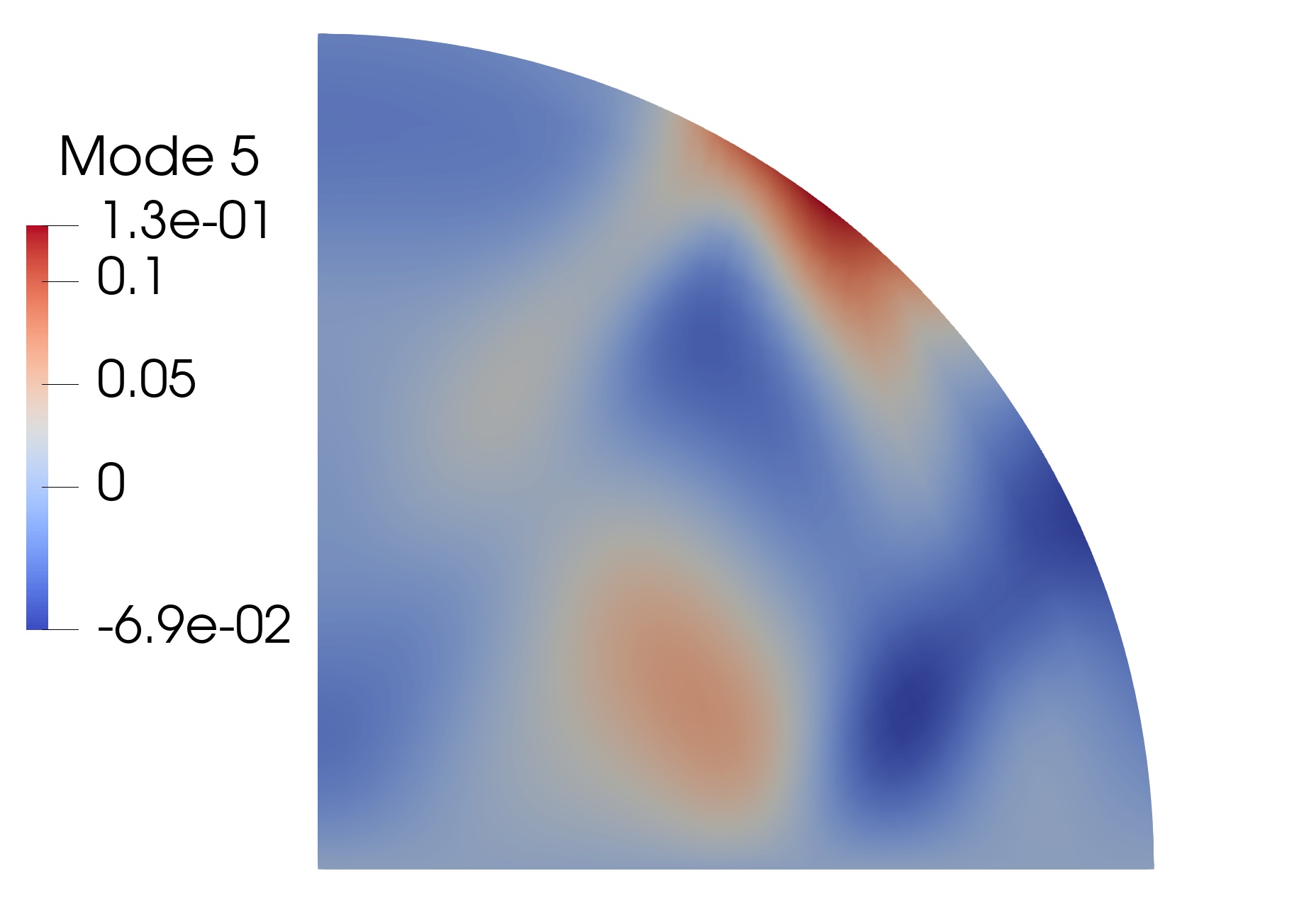}\label{fig: pca mode 5 y} }%
    \caption{The 5 retained principal components for the $Y$ directional component of displacement after one load step along the X (horizontal) Axis (load step A) - i.e. Node 1 in the load path tree in Fig.~\ref{fig: loadpath_tree}.}%
    \label{fig: y pca modes}%
\end{figure}

The retained principal components at Node 1 in the load path tree (Fig.~2 in the main body of the text) are shown for both directional components X and Y in Figs.~\ref{fig: x pca modes}~and~\ref{fig: y pca modes}, respectively.

\newpage

\section{Additional Figures and Tables from the Calibrations}\label{sec: additional.supp}

Additional figures and tables from the calibrations are contained here. These include the synthetic displacement data in \ref{cal_data.supp}, the posterior probability over the QoIs in \ref{sec: qoi_post_prob.supp} and the posterior predictive distribution over the QoIs in \ref{sec: ABBBA_val.supp}.

\subsection{Synthetic Nodal Displacement Data}\label{cal_data.supp}

\begin{figure}[h!]%
    \textbf{AAAAA Data X Directional Component}\par\medskip
    \centering
    \sidesubfloat[]{\includegraphics[width=50mm]{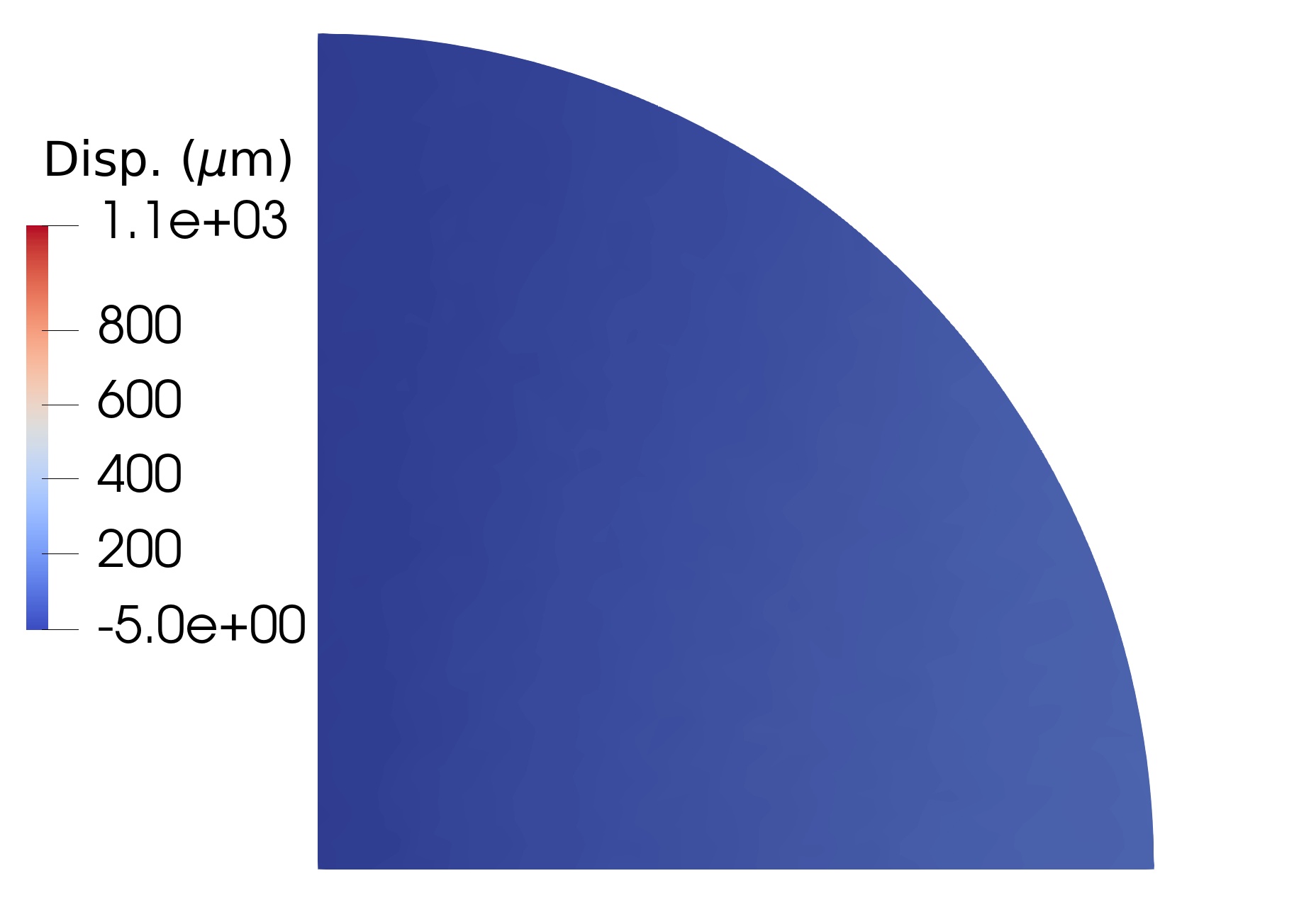}\label{fig: x data AAAAA step 0} }
    \sidesubfloat[]{\includegraphics[width=45mm]{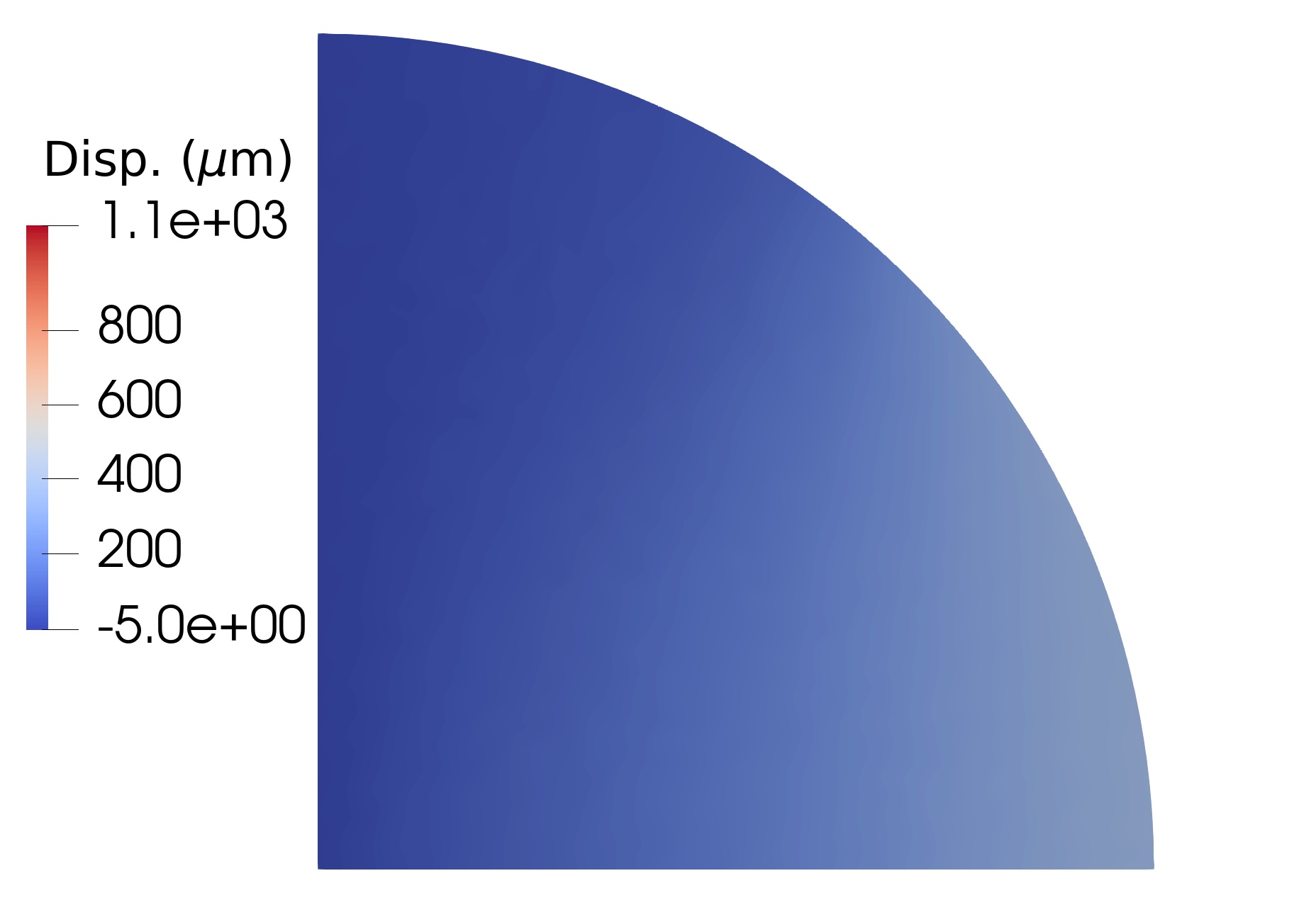}\label{fig: x data AAAAA step 1} }%
    \sidesubfloat[]{\includegraphics[width=45mm]{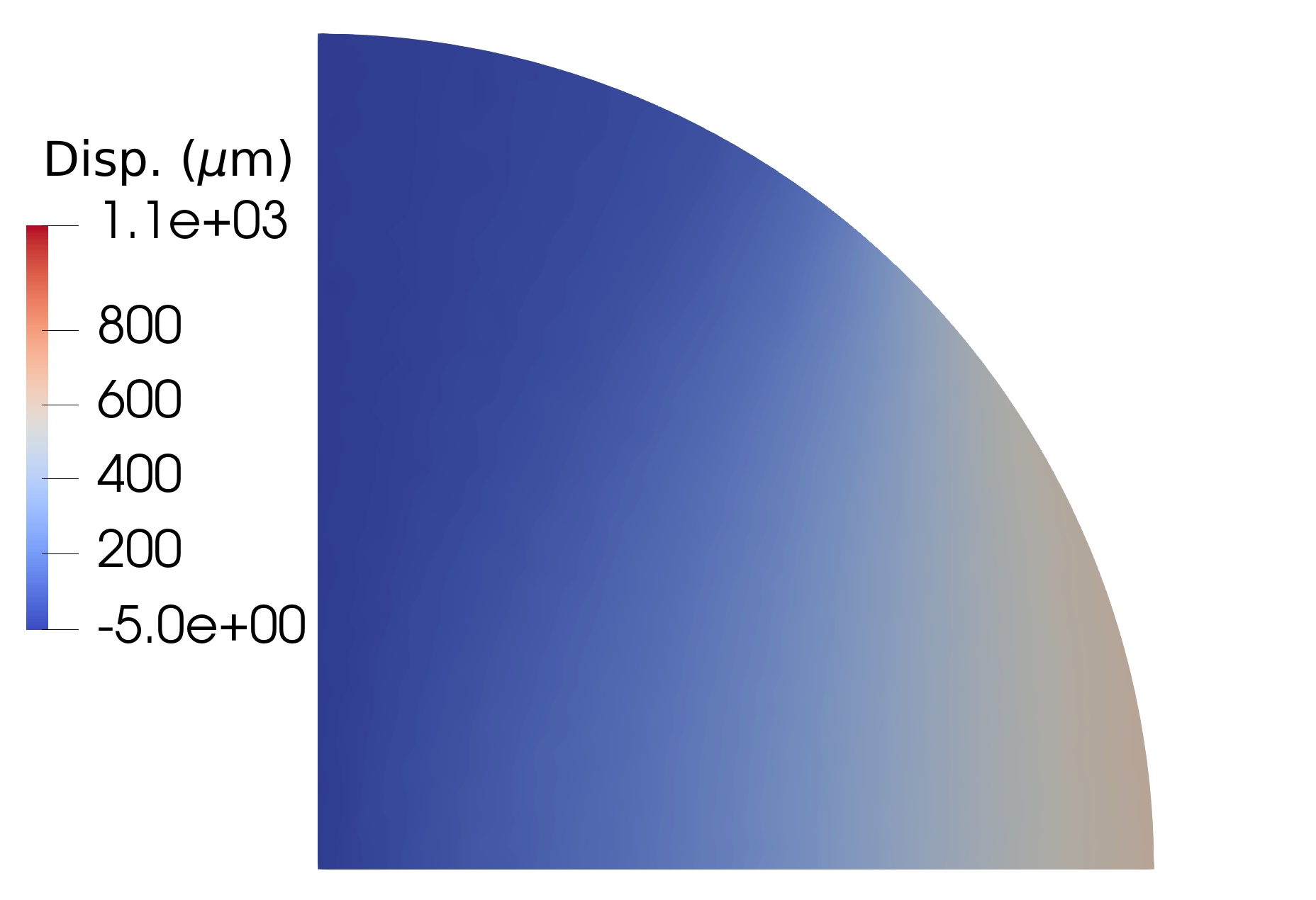}\label{fig: x data AAAAA step 2} } \\ %
    \sidesubfloat[]{\includegraphics[width=45mm]{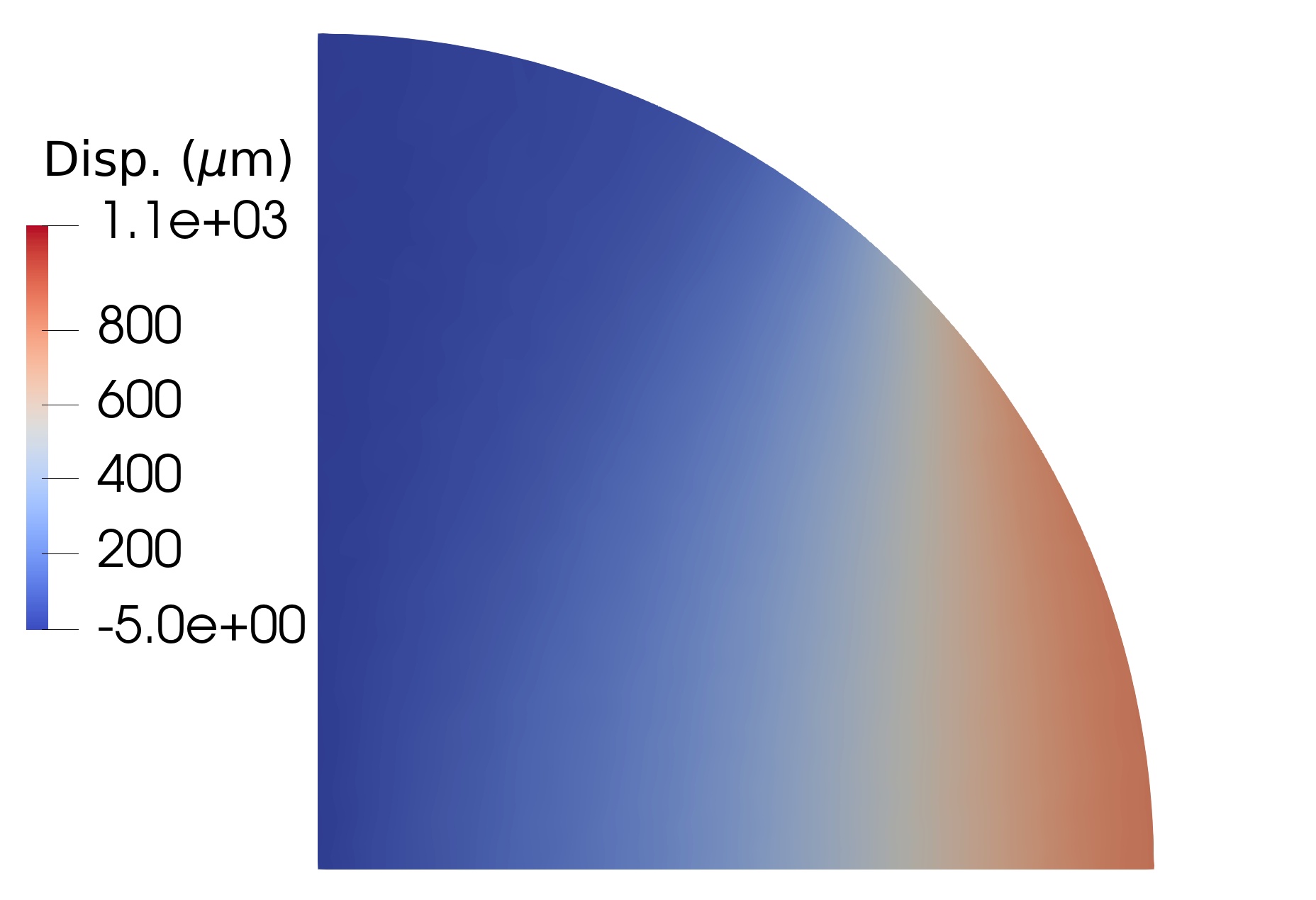}\label{fig: x data AAAAA step 3} }%
    \sidesubfloat[]{\includegraphics[width=45mm]{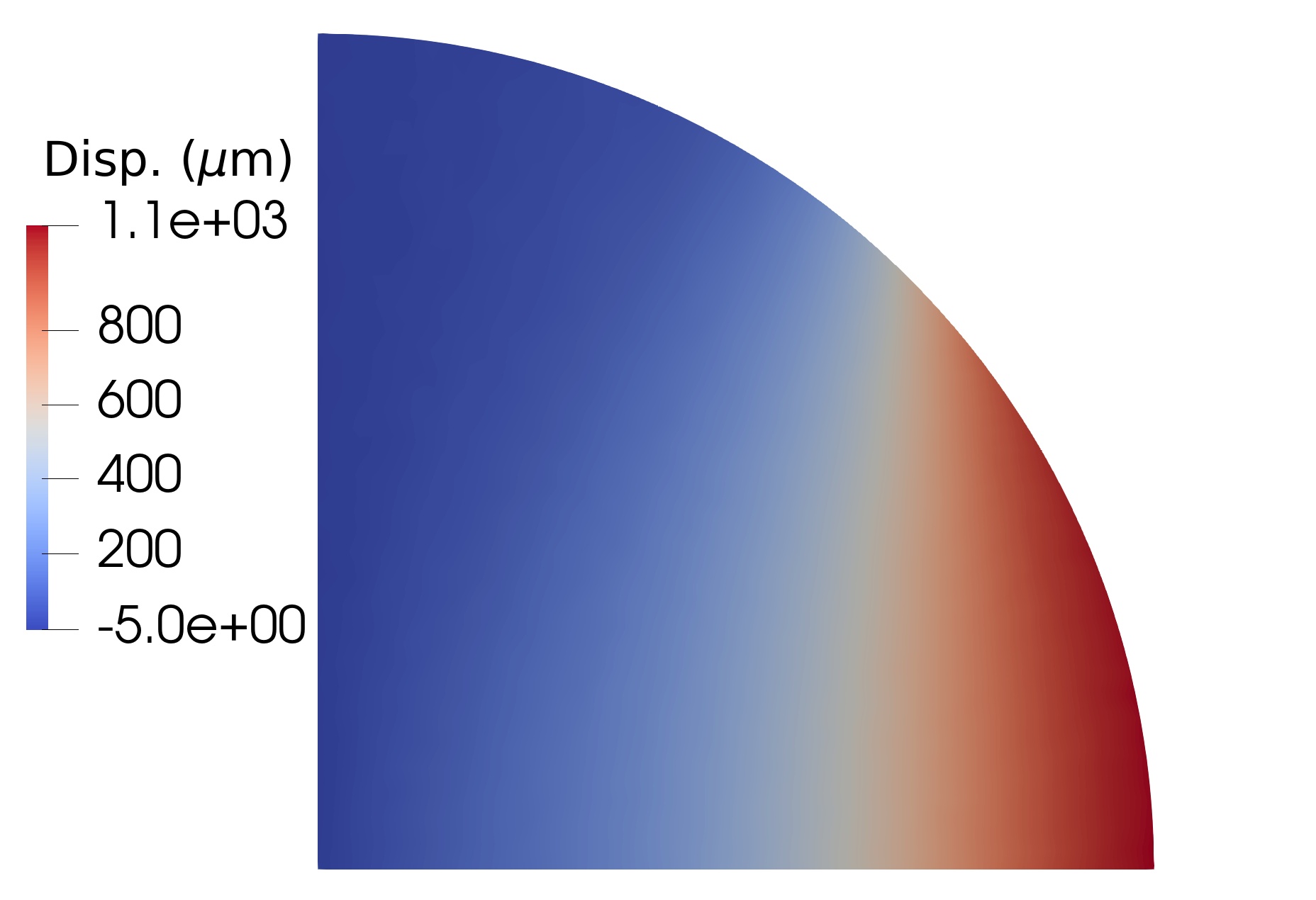}\label{fig: x data AAAAA step 4} }%
    \caption{Synthetic DIC data for the AAAAA load path. The X component of displacement is shown in microns for load steps 1-5 in plots (a)-(e), respectively.}%
    \label{fig: AAAAA data x}%
\end{figure}

\begin{figure}[h!]%
    \textbf{AAAAA Data Y Directional Component}\par\medskip
    \centering
    \sidesubfloat[]{\includegraphics[width=45mm]{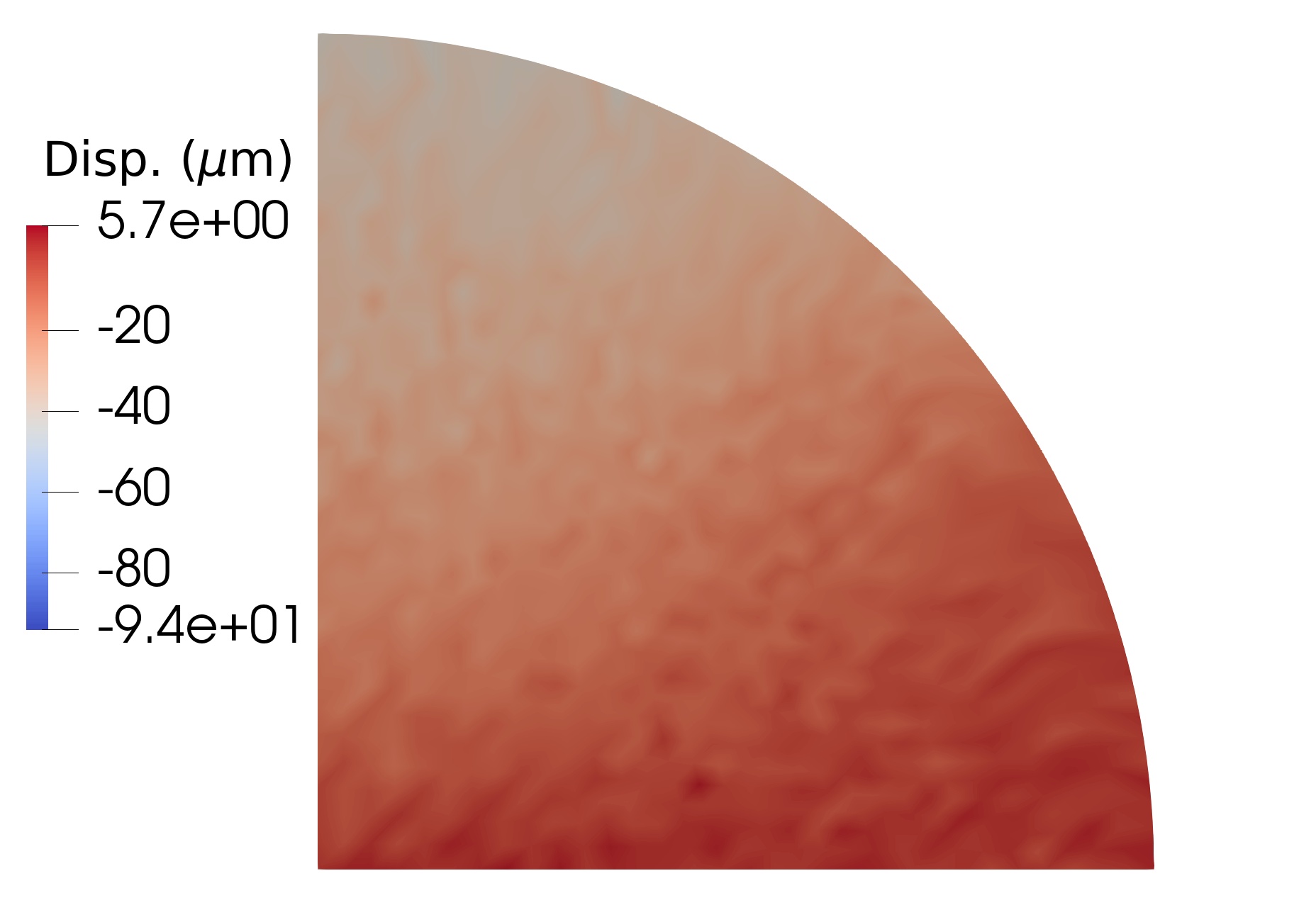}\label{fig: y data AAAAA step 0} }
    \sidesubfloat[]{\includegraphics[width=45mm]{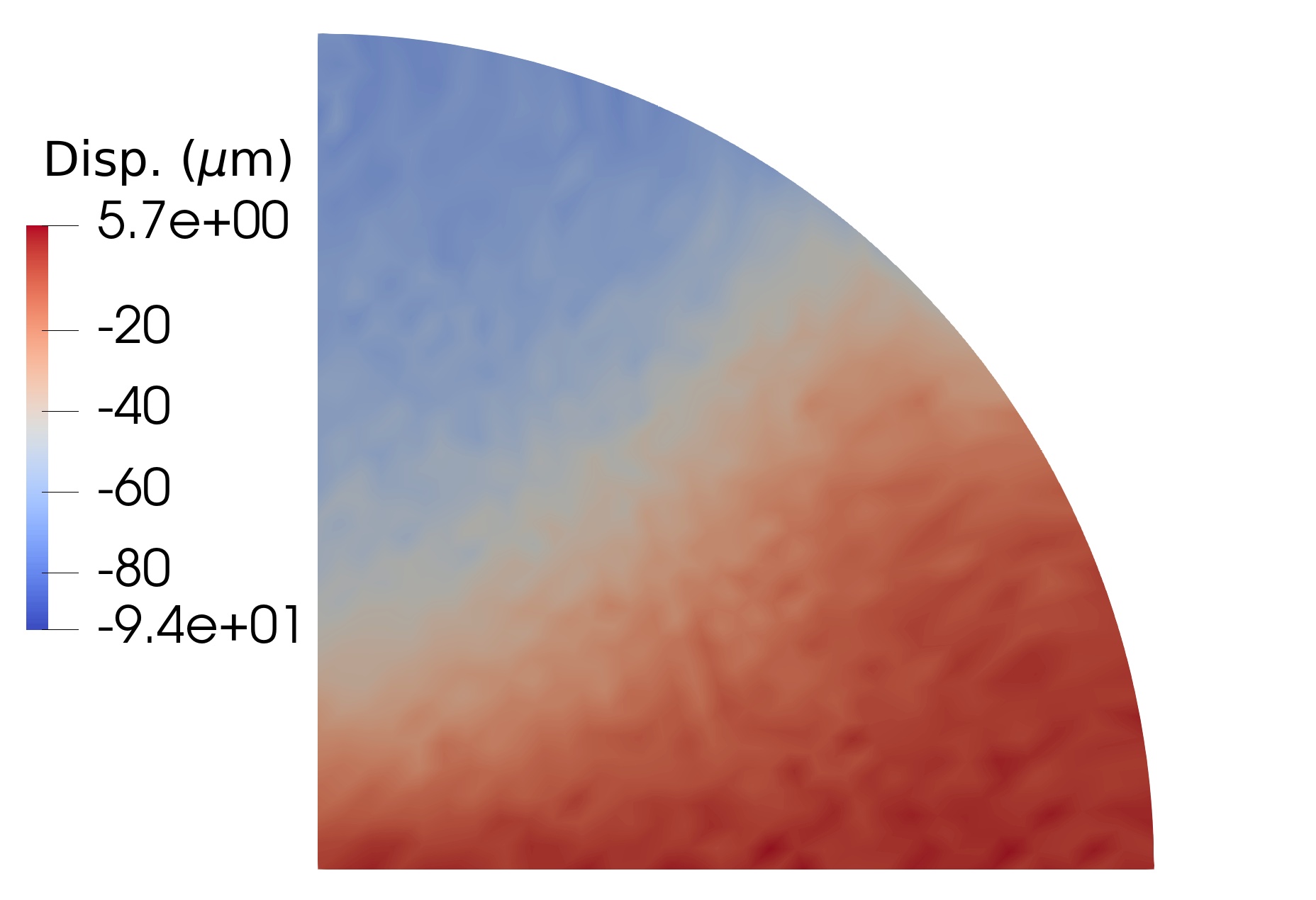}\label{fig: y data AAAAA step 1} }%
    \sidesubfloat[]{\includegraphics[width=45mm]{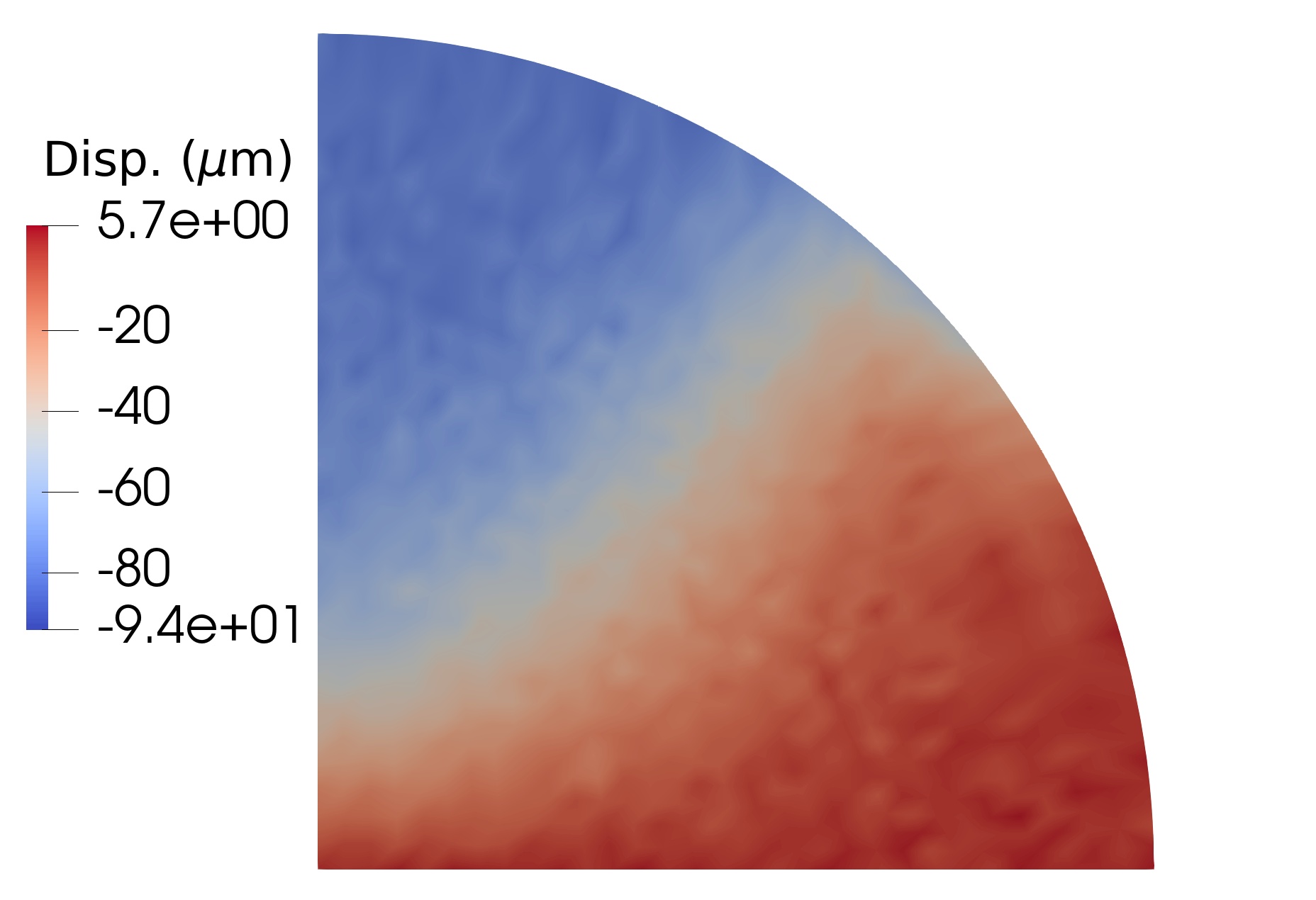}\label{fig: y data AAAAA step 2} } \\ %
    \sidesubfloat[]{\includegraphics[width=45mm]{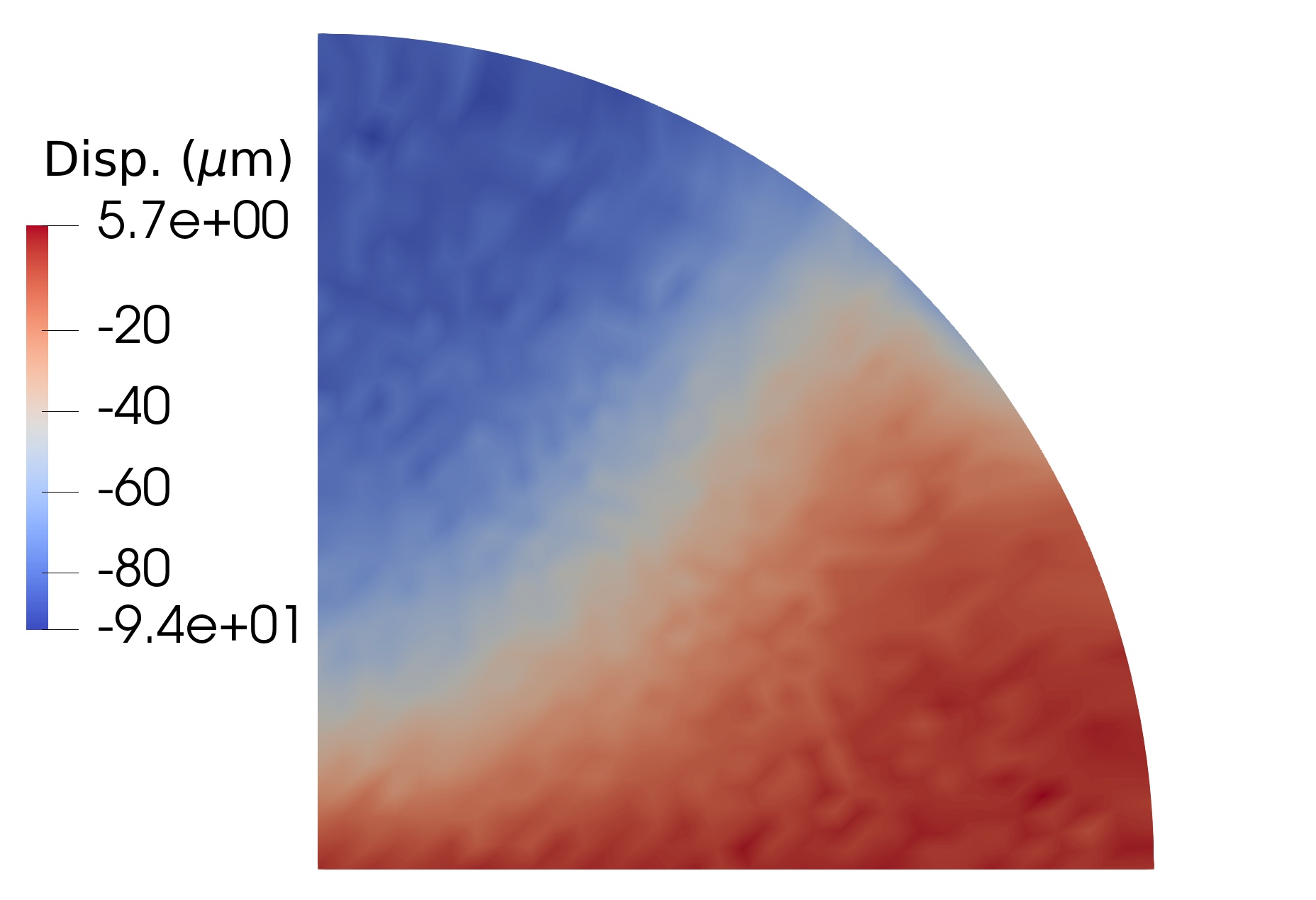}\label{fig: y data AAAAA step 3} }%
    \sidesubfloat[]{\includegraphics[width=45mm]{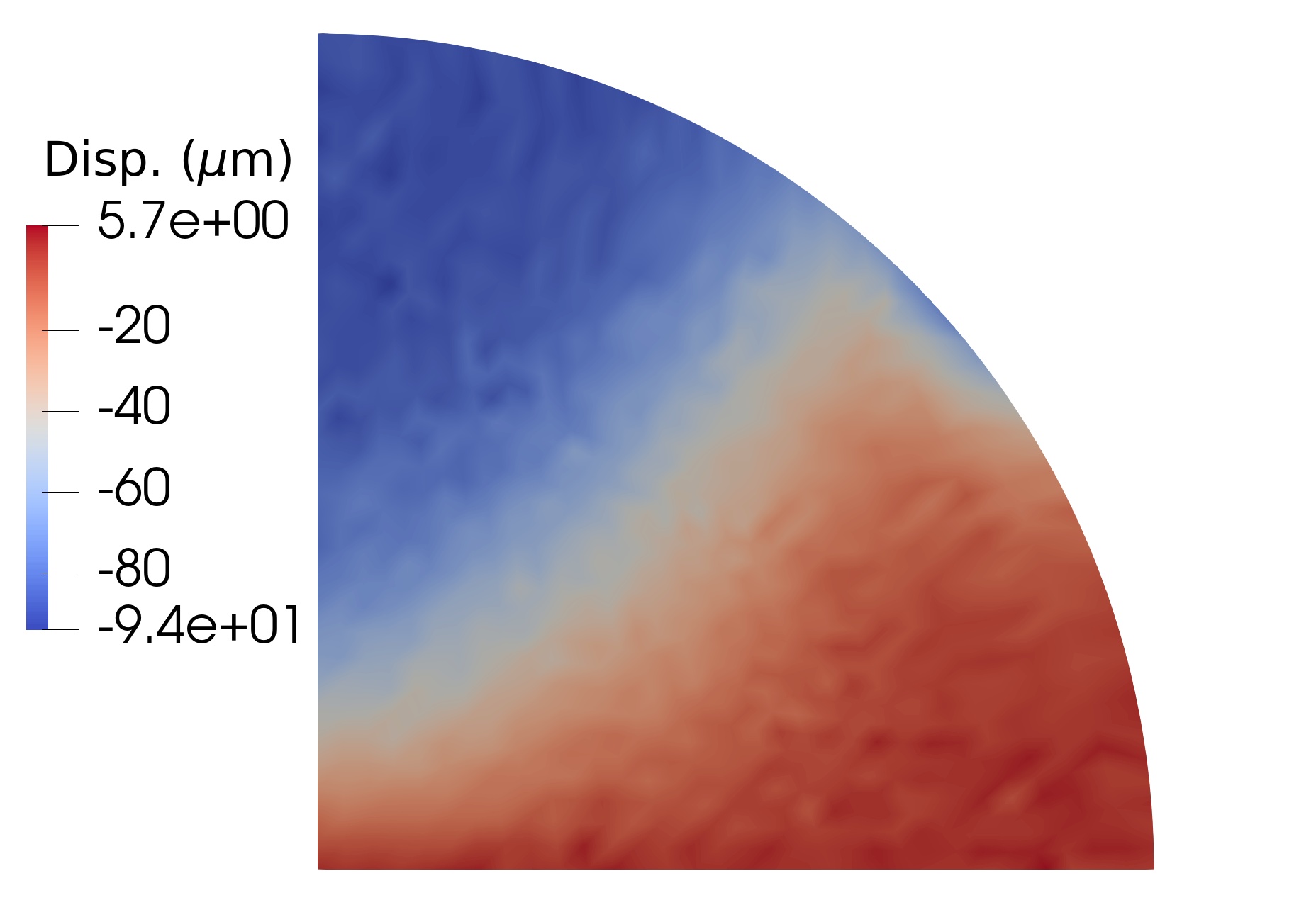}\label{fig: y data AAAAA step 4} }%
    \caption{Synthetic DIC data for the AAAAA load path. The Y component of displacement is shown in microns for load steps 1-5 in plots (a)-(e), respectively.}%
    \label{fig: AAAAA data y}%
\end{figure}

\begin{figure}[hb!]%
    \textbf{ABABA Data X Directional Component}\par\medskip
    \centering
    \sidesubfloat[]{\includegraphics[width=45mm]{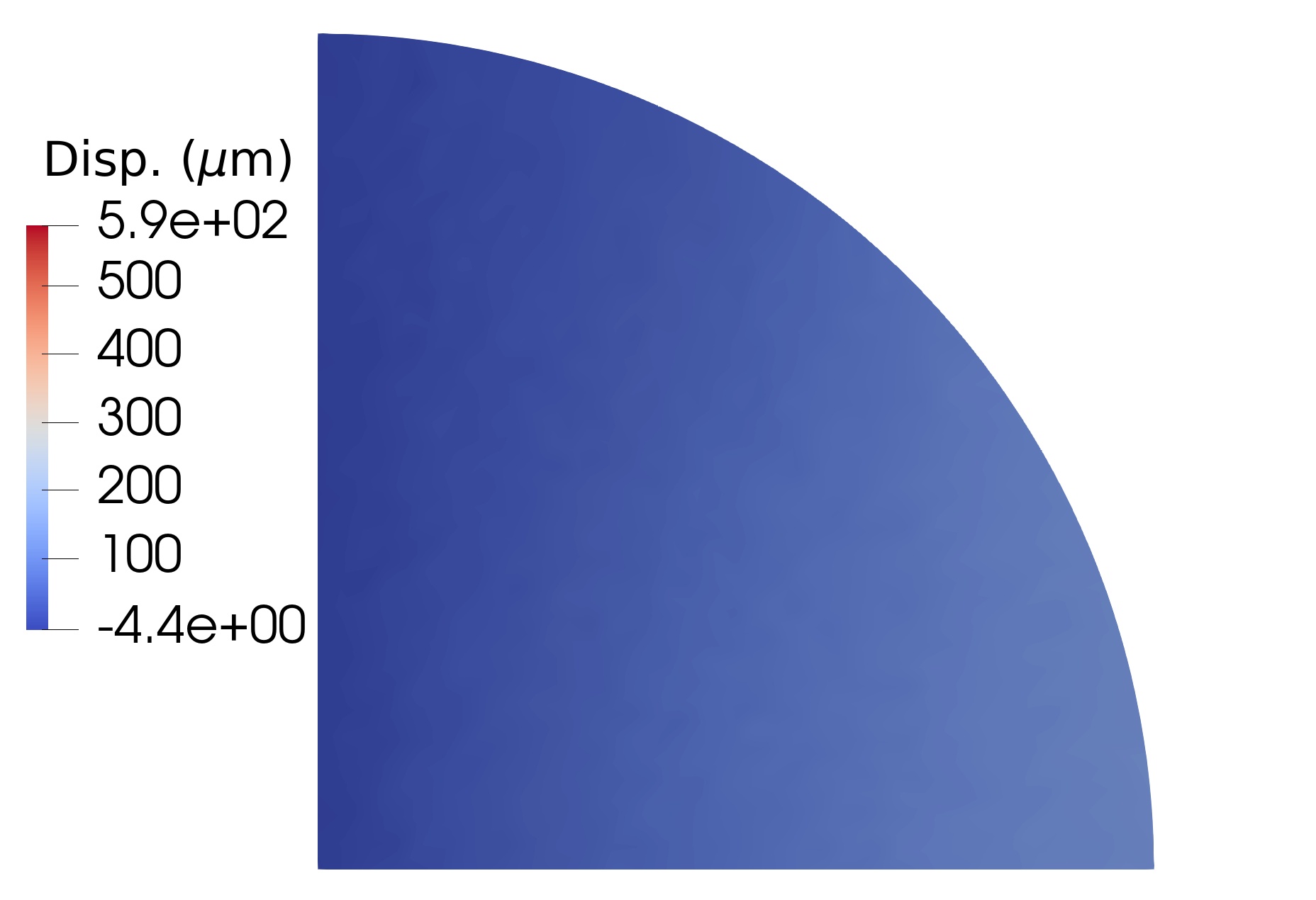}\label{fig: x data ABABA step 0} }
    \sidesubfloat[]{\includegraphics[width=45mm]{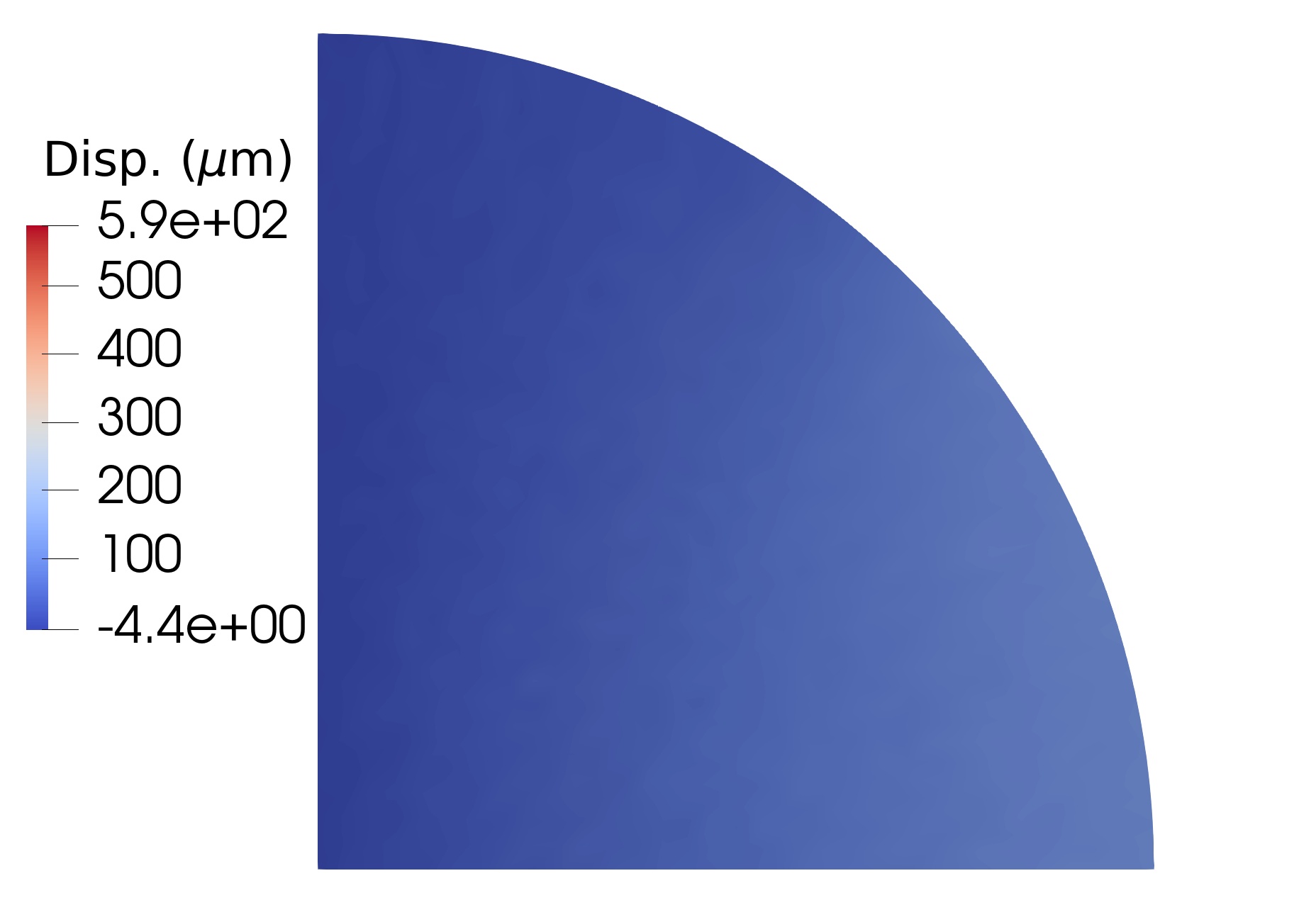}\label{fig: x data ABABA step 1} }%
    \sidesubfloat[]{\includegraphics[width=45mm]{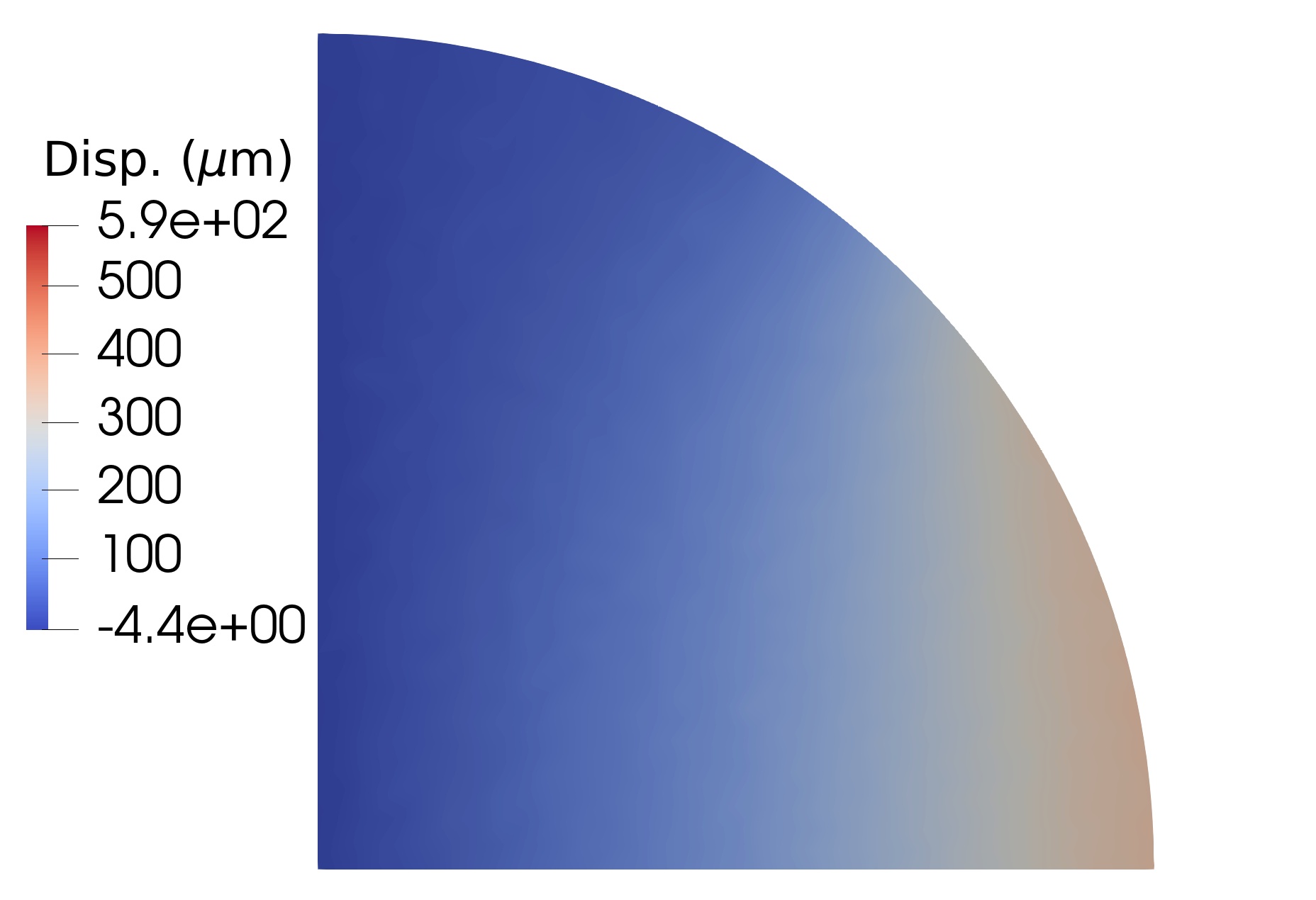}\label{fig: x data ABABA step 2} } \\ %
    \sidesubfloat[]{\includegraphics[width=45mm]{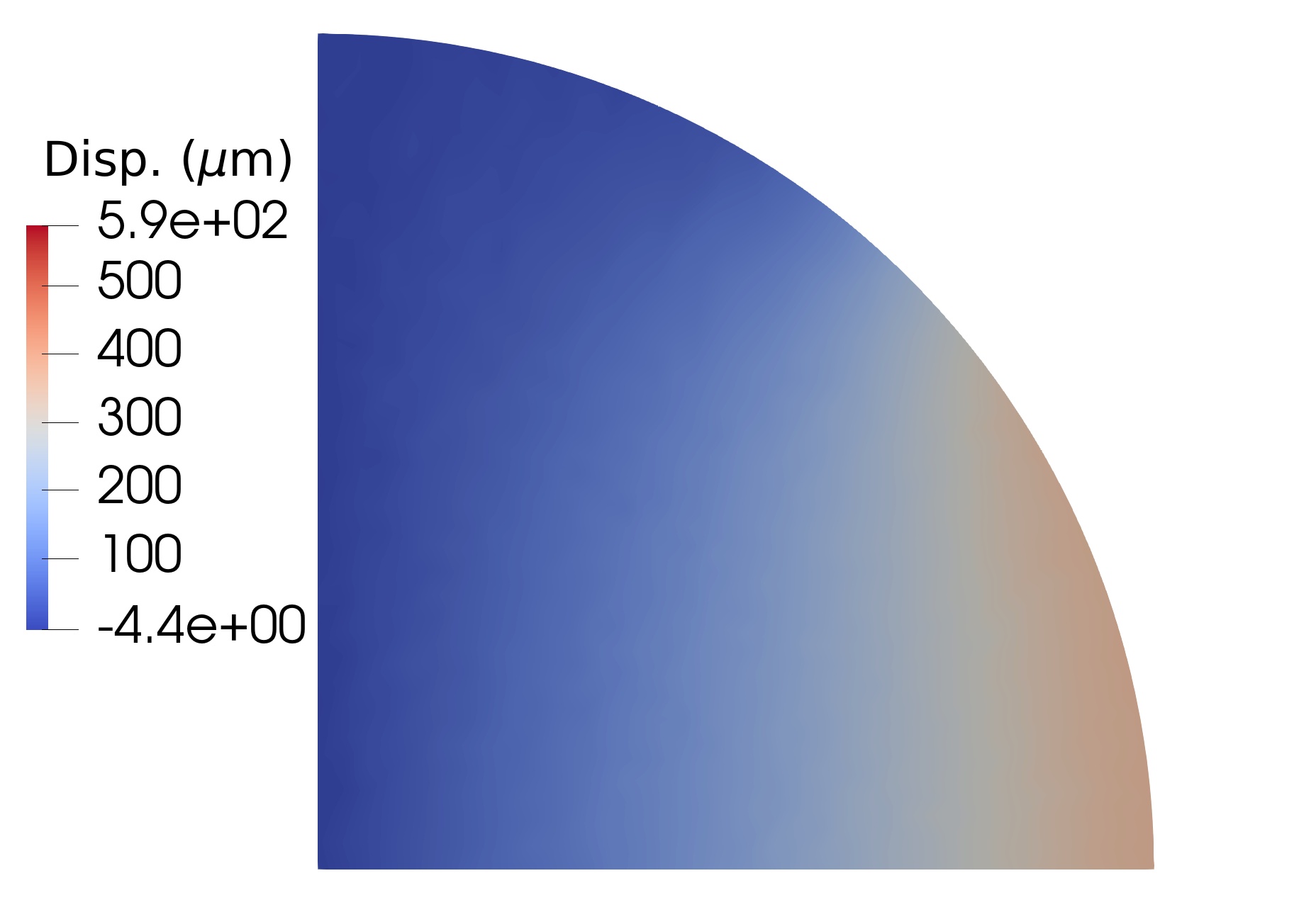}\label{fig: x data ABABA step 3} }%
    \sidesubfloat[]{\includegraphics[width=45mm]{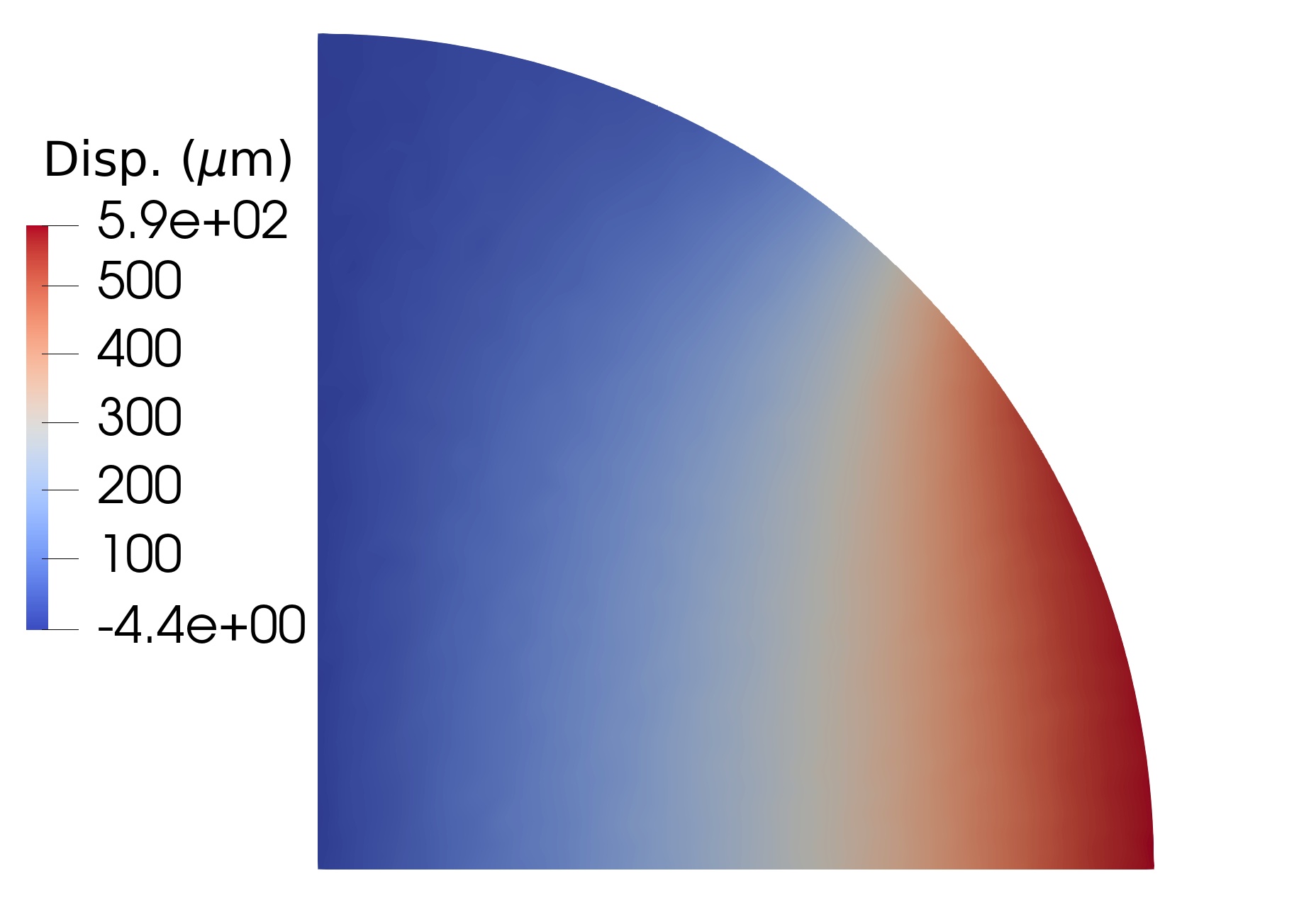}\label{fig: x data ABABA step 4} }%
    \caption{Synthetic DIC data for the ABABA load path. The X component of displacement is shown in microns for load steps 1-5 in plots (a)-(e), respectively.}%
    \label{fig: ABABA data x}%
\end{figure}

\begin{figure}%
    \textbf{ABABA Data Y Directional Component}\par\medskip
    \centering
    \sidesubfloat[]{\includegraphics[width=45mm]{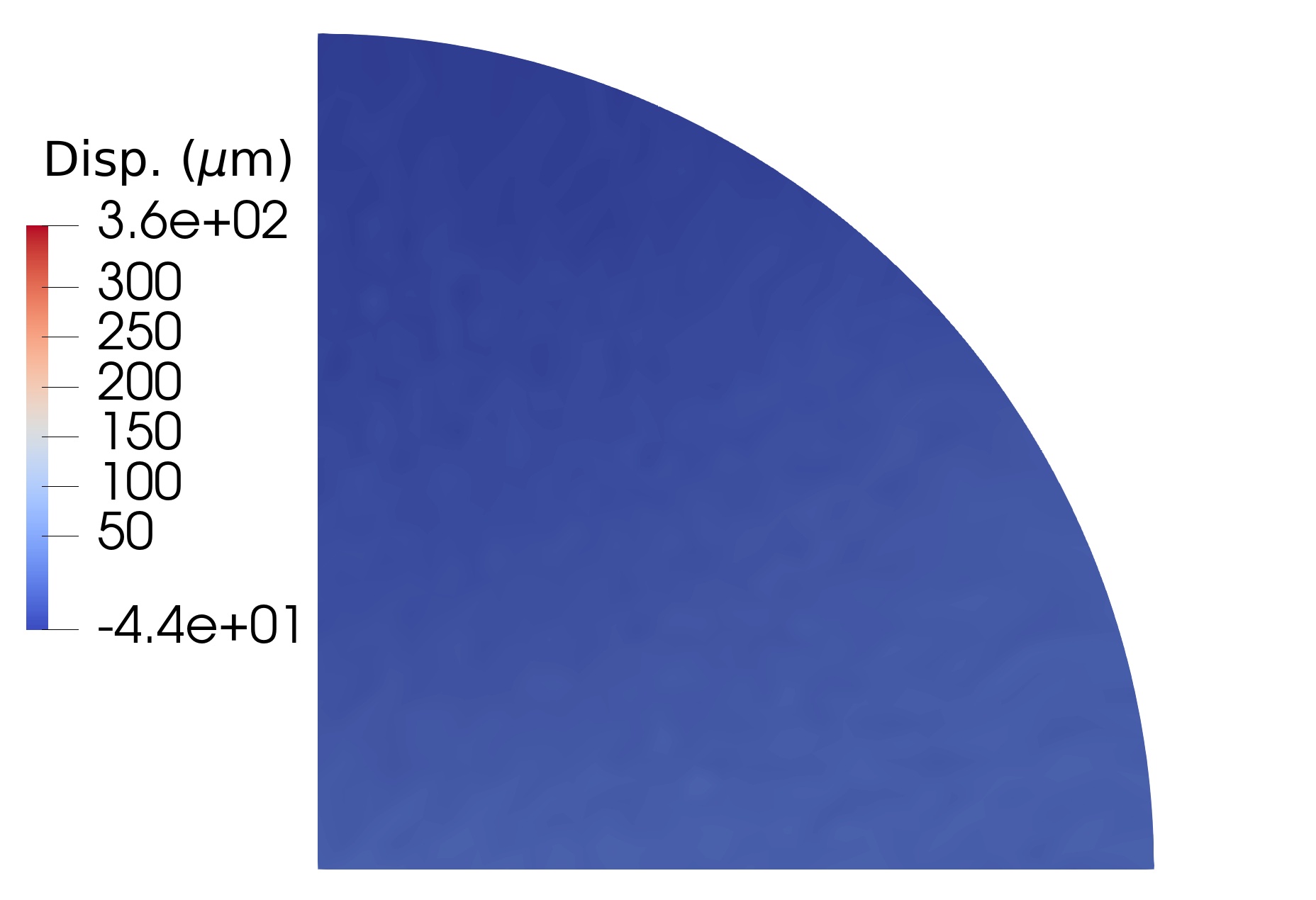}\label{fig: y data ABABA step 0} }
    \sidesubfloat[]{\includegraphics[width=45mm]{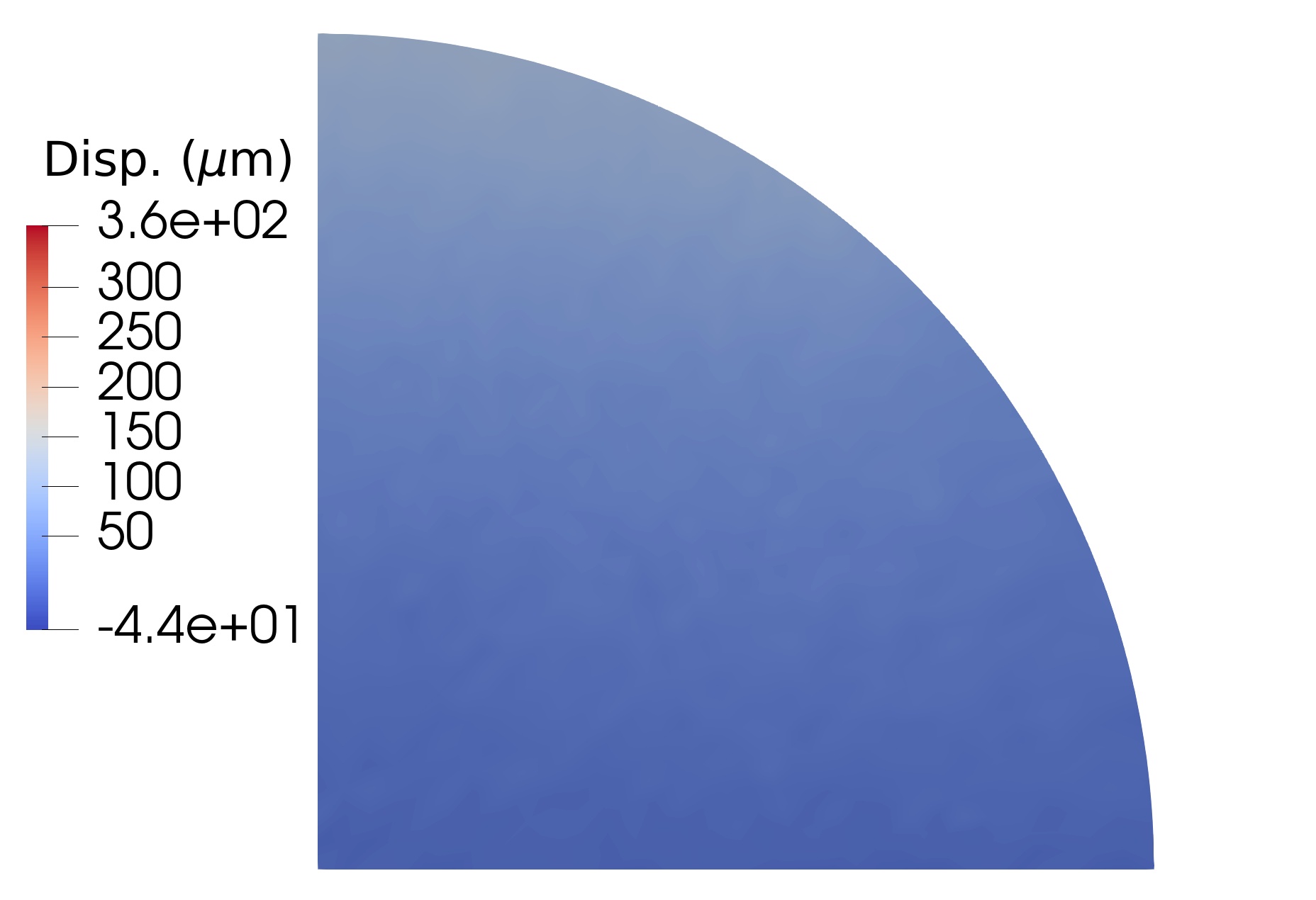}\label{fig: y data ABABA step 1} }%
    \sidesubfloat[]{\includegraphics[width=45mm]{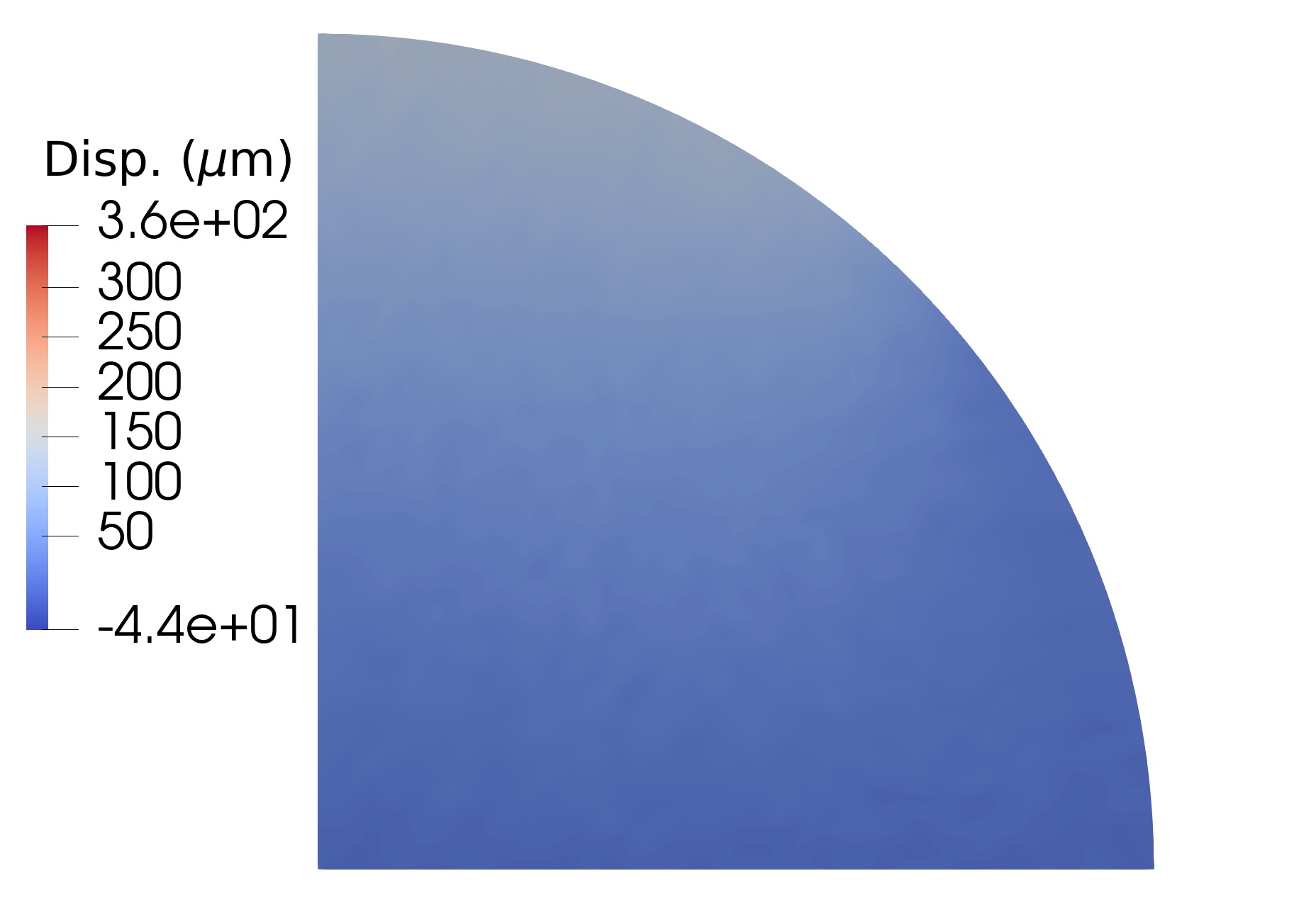}\label{fig: y data ABABA step 2} } \\ %
    \sidesubfloat[]{\includegraphics[width=45mm]{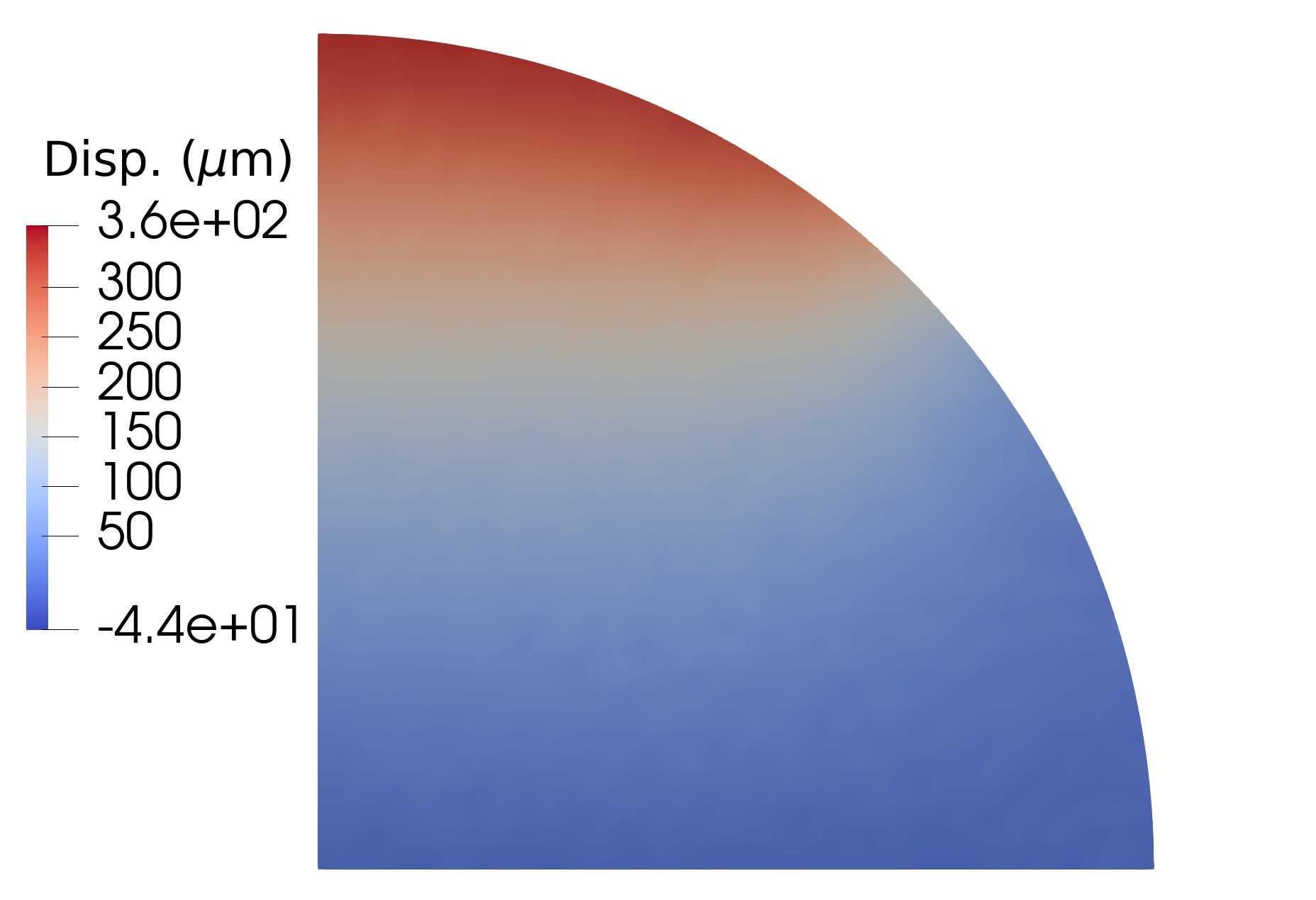}\label{fig: y data ABABA step 3} }%
    \sidesubfloat[]{\includegraphics[width=45mm]{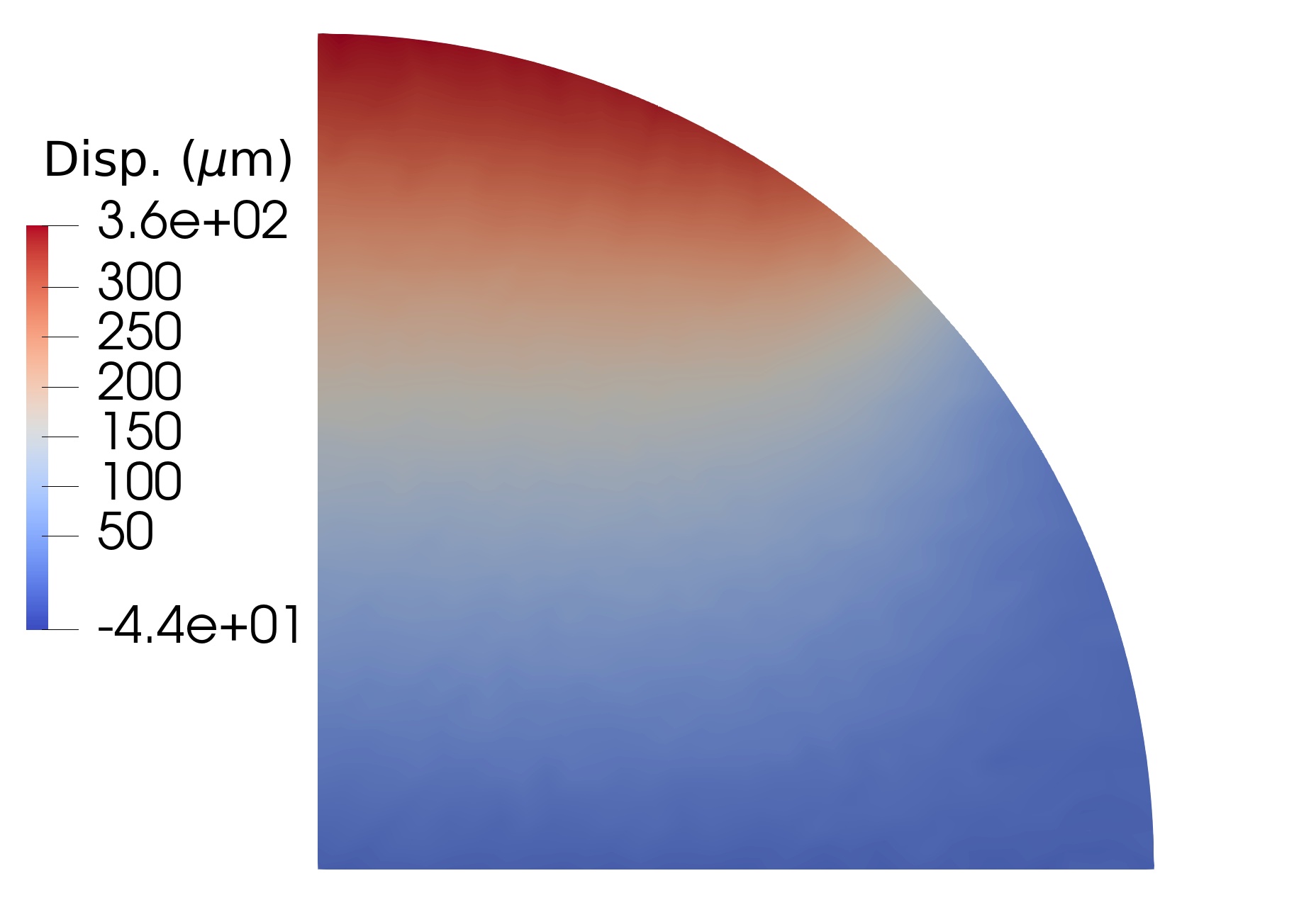}\label{fig: y data ABABA step 4} }%
    \caption{Synthetic DIC data for the ABABA load path. The Y component of displacement is shown in microns for load steps 1-5 in plots (a)-(e), respectively.}%
    \label{fig: ABABA data y}%
\end{figure}

\subsection{Posterior Probability Over the QoIs}\label{sec: qoi_post_prob.supp}
The posterior probability is visualized over the QoIs (load and nodal displacements) for both the AAAAA and ABABA calibrations in this section.

\begin{figure}[ht!]
\centering
\textbf{AAAAA Posterior Over Load}\par\medskip
\includegraphics[width=110mm]{load_hist_AAAAA.pdf}\\
\textbf{ABABA Posterior Over Load}\par\medskip
\includegraphics[width=110mm]{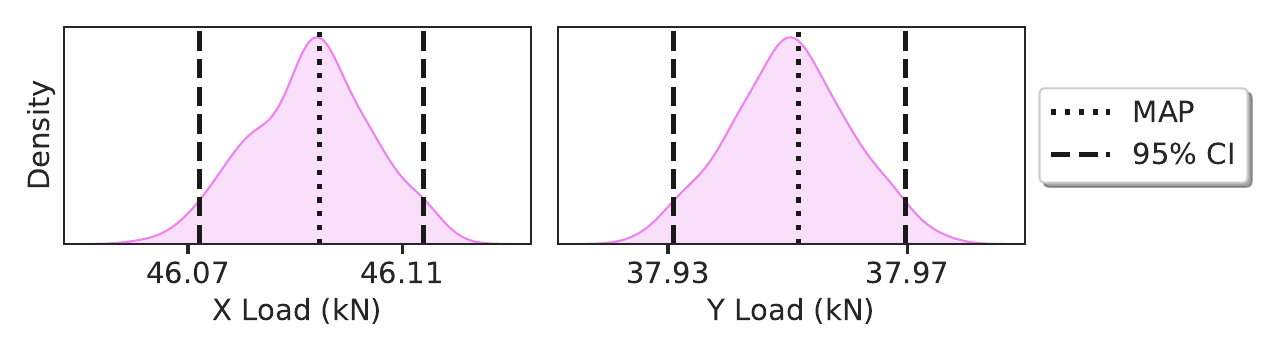}
\caption{Posterior probability over the load for the AAAAA (top row) and ABABA (bottom row) calibrations are shown for both the $X$ (left column) and $Y$ (right column) directional components. The load at the MAP parameter estimate is shown with a dotted black line, and the 95\% CI of the distribution over the load is shown with dashed black lines, indicating the interval in which the load values have a 95\% probability of residing within.}\label{fig: posterior probabilty over load}
\end{figure}

\begin{figure}%
    \centering
    \textbf{AAAAA Load Step 5 Data}\par\medskip
    \sidesubfloat[]{\includegraphics[width=55mm]{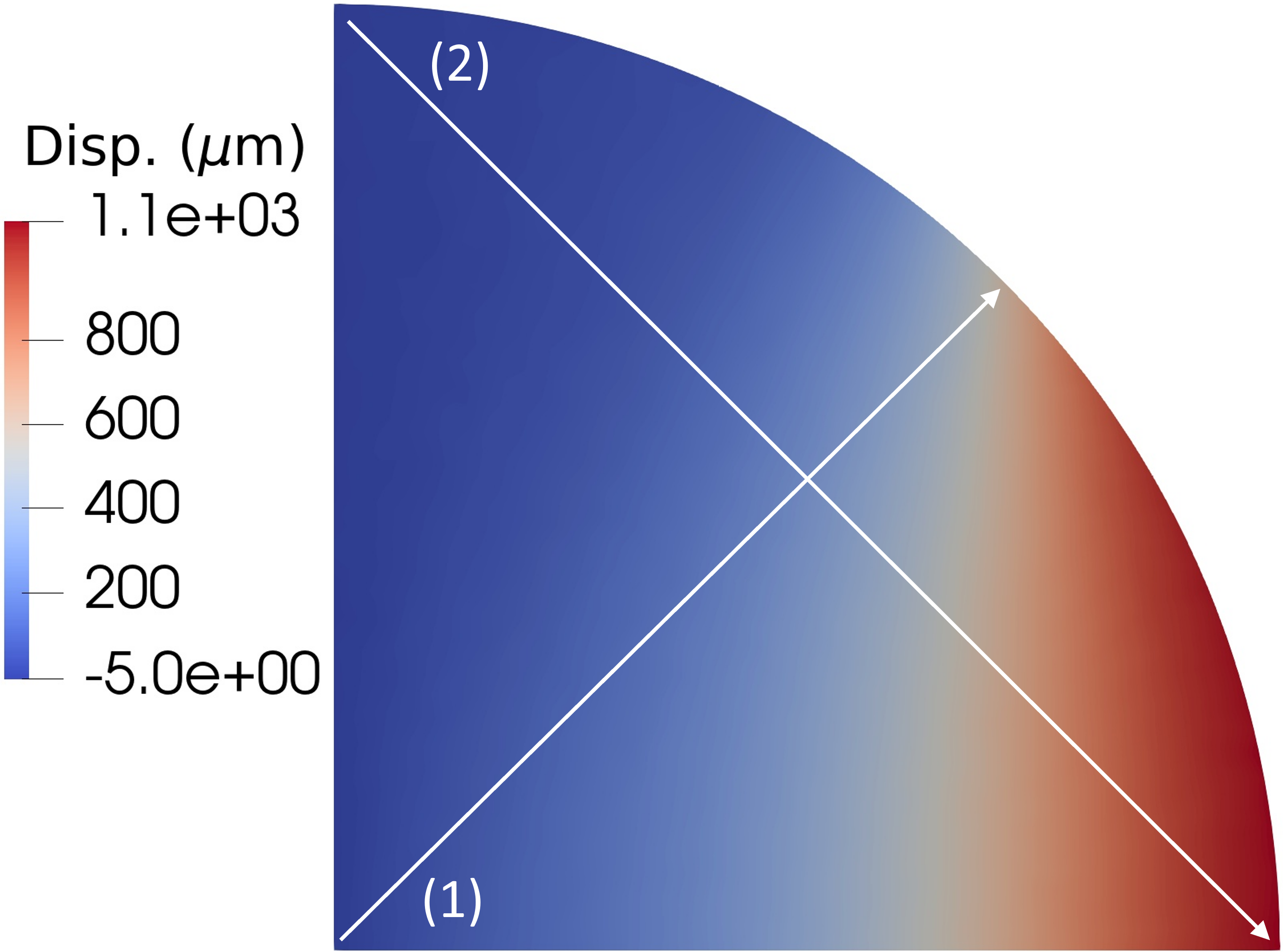}\label{fig: SUPP AAAAA ux cal data step 4}} %
    \sidesubfloat[]{\includegraphics[width=55mm]{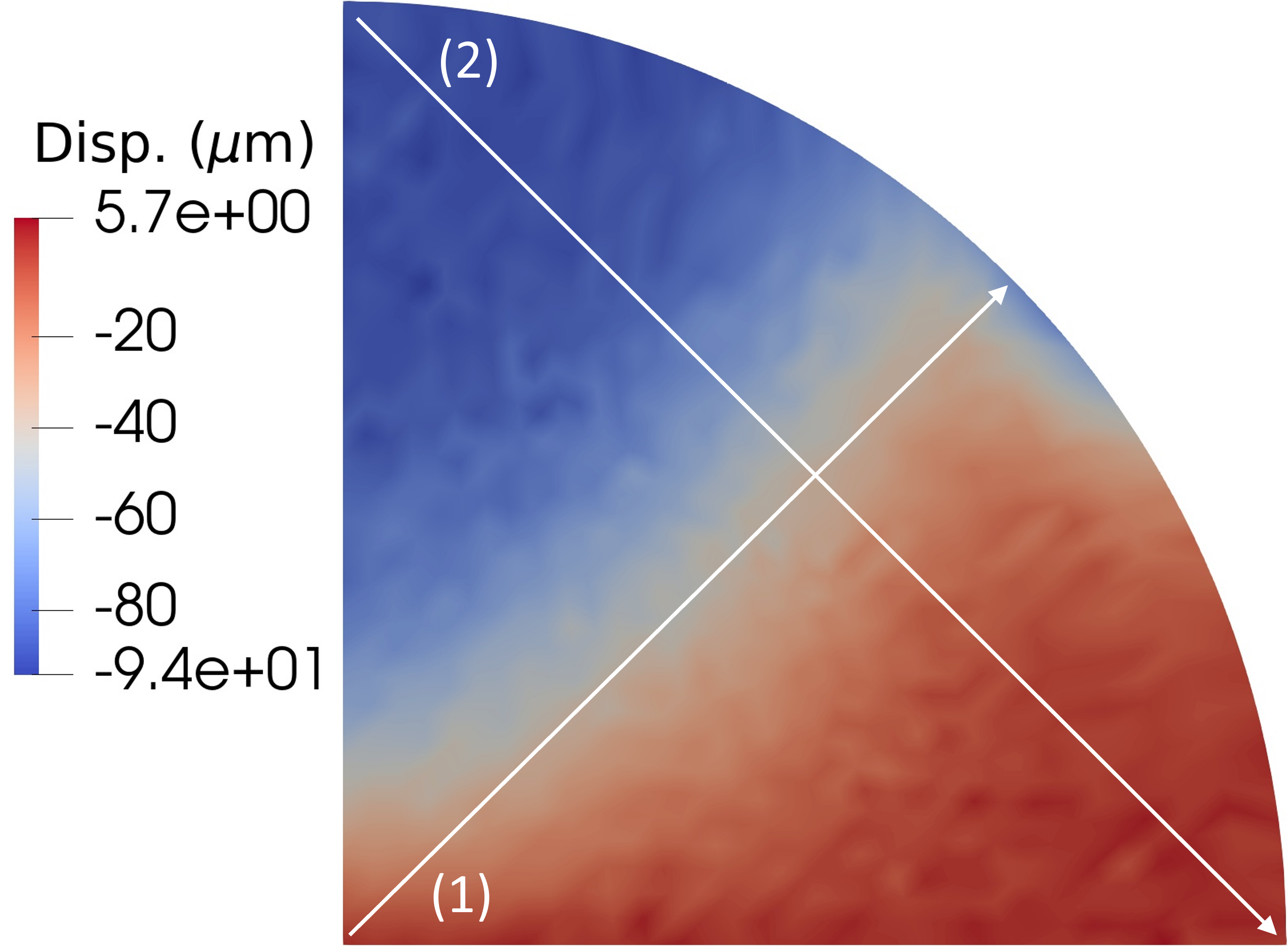}\label{fig: SUPP AAAAA uy cal data step 4}} 
    \caption{The observed displacement field at the final load step of load path AAAAA generated with $\pars^{true}$ and added noise for the (a) X directional component and (b) Y directional component reported in microns. Lines (1) and (2) indicate the space over which the posterior probability of the displacement field is plotted in Fig.~\ref{fig: SUPP posterior over displacement AAAAA}.}%
    \label{fig: SUPP AAAAA step 4 cal data with lines}%
\end{figure}

\begin{figure}%
    \centering
    \textbf{AAAAA Posterior Over Displacement: Lines 1 and 2}\par\medskip
    \sidesubfloat[]{\includegraphics[width=110mm]{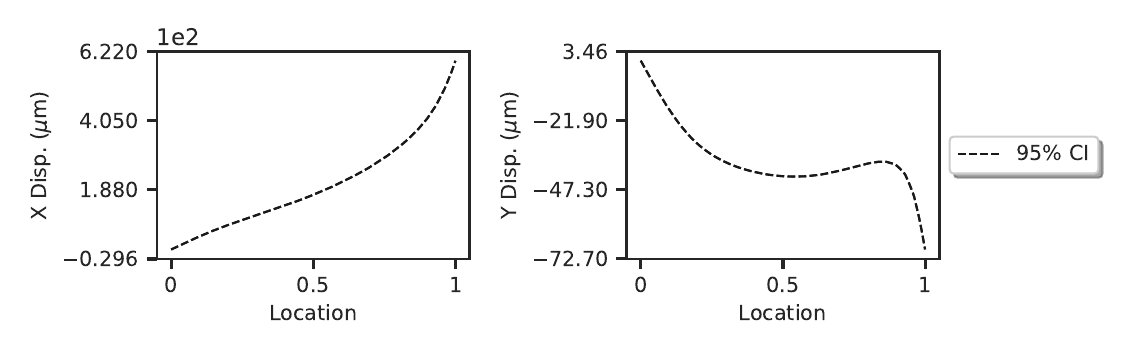}}\label{fig: SUPP AAAAA diag} \\ %
    \sidesubfloat[]{\includegraphics[width=110mm]{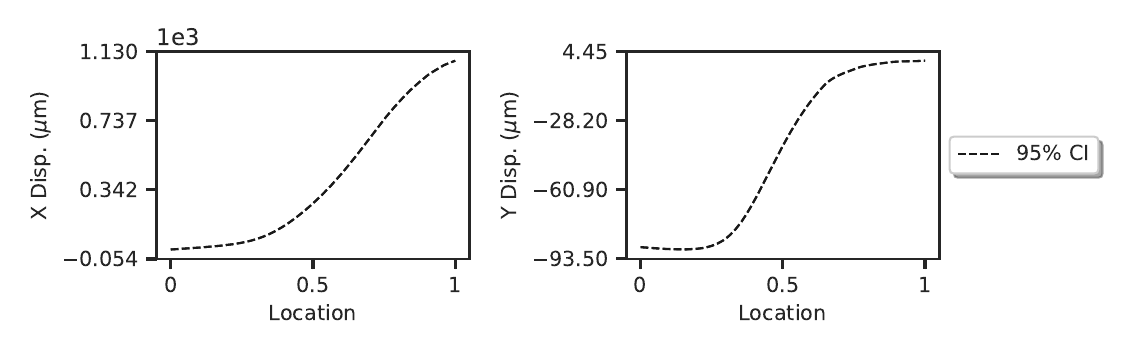}}\label{fig: SUPP AAAAA cross}%
    \caption{Posterior probability over the nodal displacements for the AAAAA calibration across the two lines indicated in Fig.~\ref{fig: SUPP AAAAA step 4 cal data with lines}. The $X$ (left column) and $Y$ (right column) directional components are shown along with Line 1 (top row) and Line 2 (bottom row). The 95\% CI of the distribution over the load is shown with dashed black lines. Given the relative scale of the displacements and the low uncertainty, the 95\% CI interval appears as a single line. The x-axis is the normalized location across each line scan.}%
    \label{fig: SUPP posterior over displacement AAAAA}%
\end{figure}

\begin{figure}%
    \centering
    \textbf{ABABA Load Step 5 Data}\par\medskip
    \sidesubfloat[]{\includegraphics[width=55mm]{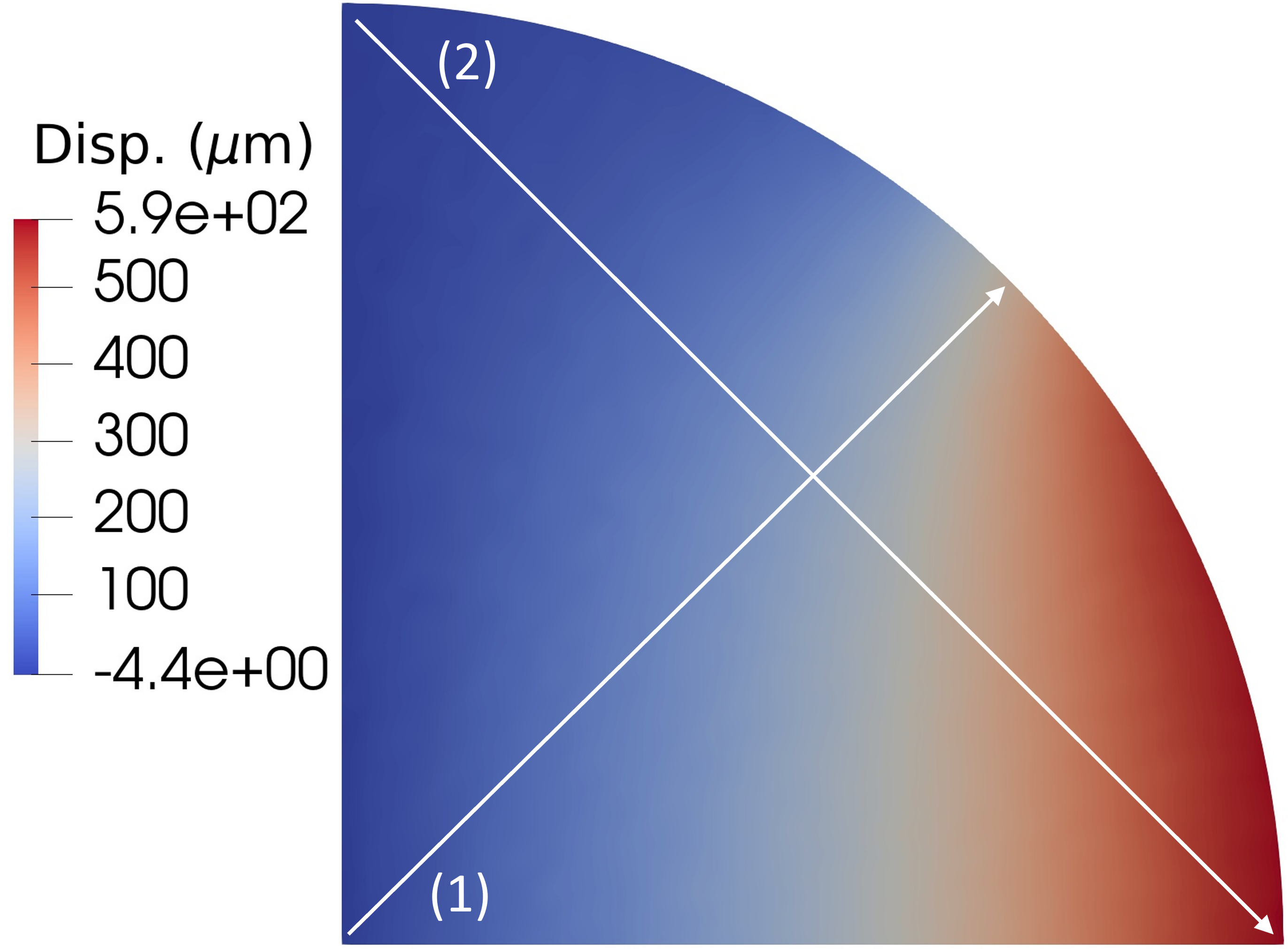}\label{fig: ABABA ux cal data step 4}} %
    \sidesubfloat[]{\includegraphics[width=55mm]{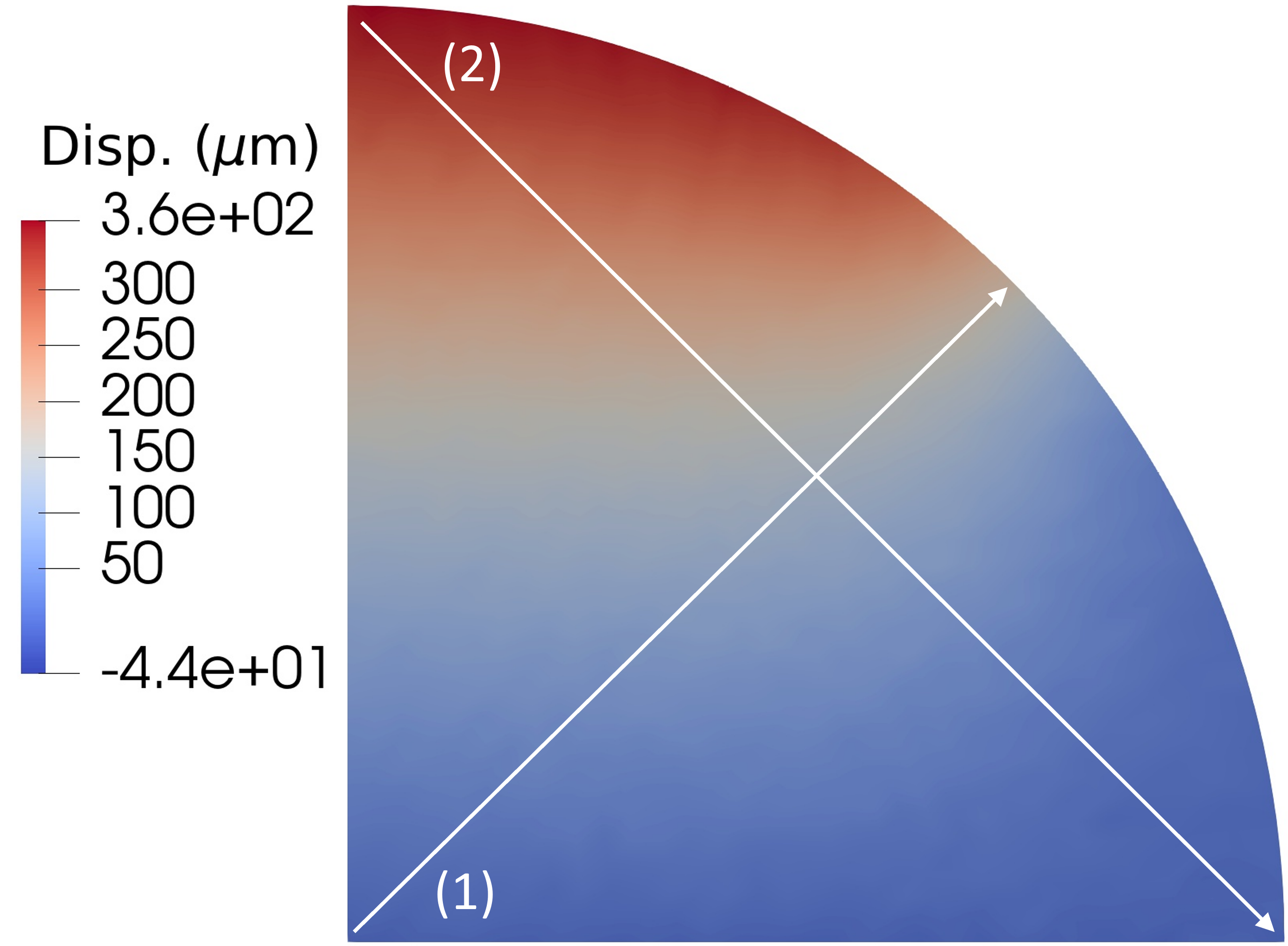}\label{fig: ABABA uy cal data step 4}} %
    \caption{The observed displacement field at the final load step of load path ABABA generated with $\pars^{true}$ and added noise for the (a) X directional component and (b) Y directional component reported in microns. Lines (1) and (2) indicate the space over which the posterior probability of the displacement field is plotted in Fig.~\ref{fig: posterior over displacement ABABA}.}%
    \label{fig: ABABA step 4 cal data with lines}%
\end{figure}

\begin{figure}%
    \centering
    \textbf{ABABA Posterior Over Displacement: Lines 1 and 2}\par\medskip
    \sidesubfloat[]{\includegraphics[width=110mm]{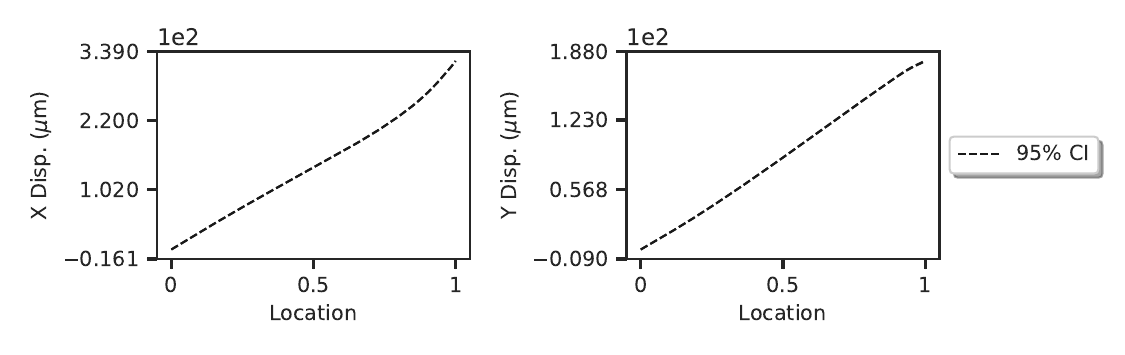}}\label{fig: ABABA diag} \\ %
    \sidesubfloat[]{\includegraphics[width=110mm]{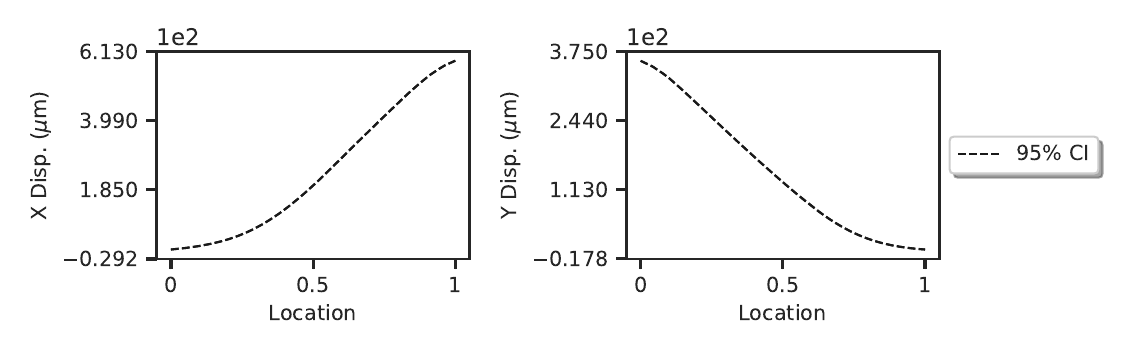}}\label{fig: ABABA cross}%
    \caption{Posterior probability over the nodal displacements for the ABABA calibration across the two lines indicated in Fig.~\ref{fig: ABABA step 4 cal data with lines}. The $X$ (left column) and $Y$ (right column) directional components are shown along with Line 1 (top row) and Line 2 (bottom row). The 95\% CI of the distribution over the load is shown with dashed black lines. Given the relative scale of the displacements and the low uncertainty, the 95\% CI interval appears as a single line. The x-axis is the normalized location across each line scan.}%
    \label{fig: posterior over displacement ABABA}%
\end{figure}

\subsection{Validation with ABBBA Load Path}\label{sec: ABBBA_val.supp}

Validation of the calibrated models was performed by making predictions about a new load path ABBBA. Samples from the posterior predictive distribution---Eqn. (15) in the main body of the text---are shown over the load and displacement field for both the AAAAA and ABABA-calibrated models. The left-most column of Figs.~\ref{fig: SUPP AAAAA ABBBA X field validation}, \ref{fig: SUPP AAAAA ABBBA Y field validation},  \ref{fig: ABABA ABBBA X field validation} and \ref{fig: ABABA ABBBA Y field validation} shows the observed displacement data for load path ABBBA and the next three columns show three different draws from the posterior predictive distribution. The distribution of the field over the line plots in Fig.~\ref{fig: SUPP ABBBA step 4 val data with lines} is shown in Figs.~\ref{fig: SUPP AAAAA prediction of ABBBA field diag line}, \ref{fig: SUPP AAAAA prediction of ABBBA field cross line}, \ref{fig: ABABA prediction of ABBBA field diag line} and \ref{fig: ABABA prediction of ABBBA field cross line} for both the AAAAA and ABABA calibration. The left plots show the 95\%~PI along with the observed data, and the right plots show the difference between the 95\%~PI and the observed data. The solid line represents the observations (centered at 0 difference) and the difference between the observations and the PIs are shown with dashed lines. Everywhere the observations fall within the 95\% PI coincides with locations where the upper bound difference is positive and the lower bound difference is negative; these regions are colored orange.

\begin{figure}[h!]
\centering
\textbf{Posterior Predictive Distribution Over ABBBA Load for the (Left) AAAAA--- and (Right) ABABA---Calibrated Model}\par\medskip
\includegraphics[width=140mm]{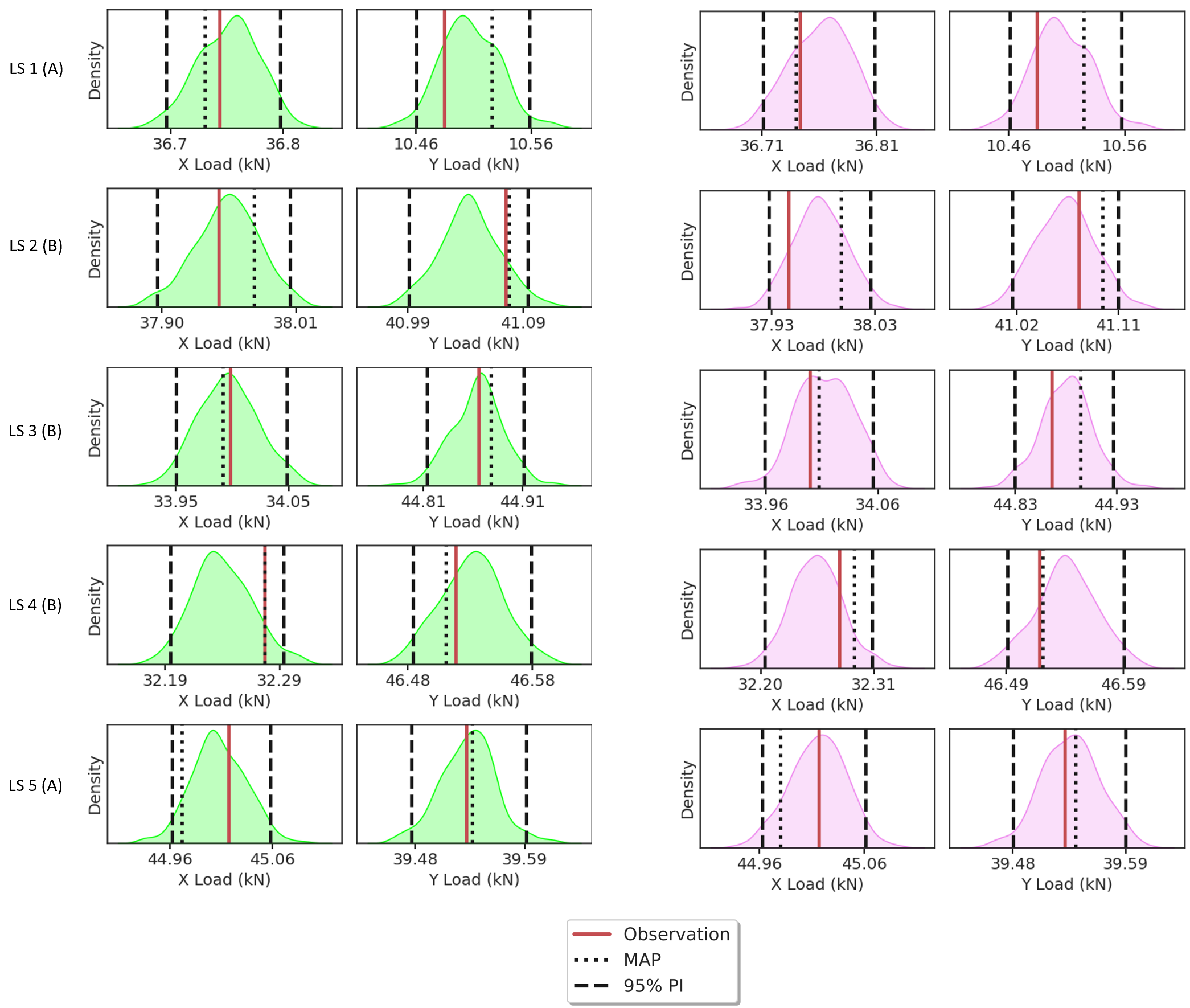}
\caption{The posterior predictive distribution of the (left) AAAAA--- and (right) ABABA---calibrated model is shown for the new ABBBA load path over the global load values for the $X$ and $Y$ directional components at each load step. The observed data is shown with a solid red line, the load at the MAP parameter estimate is shown with a dotted black line, and the 95\% highest density PI is shown with dashed black lines.}\label{fig: SUPP posterior predictive load for ABBBA}
\end{figure}

\begin{figure}[ht!]
\textbf{AAAAA Posterior Predictive Distribution Over ABBBA X Displacement}\par\medskip
\centering
\includegraphics[width=\linewidth]{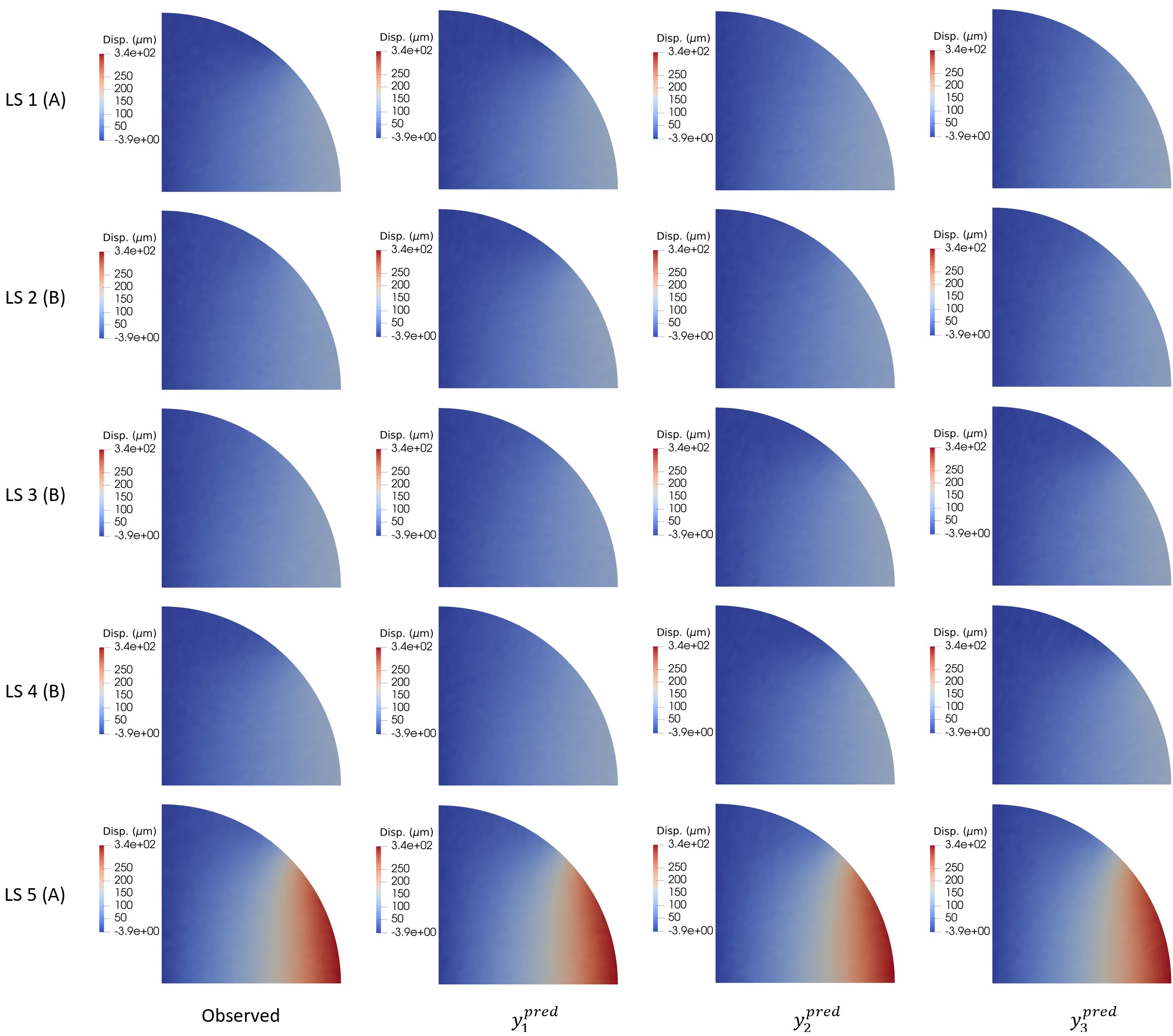}
\caption{Samples from the posterior predictive distribution of the AAAAA---calibrated model is shown for the $X$ directional component of the displacement field for a new ABBBA load path. The left column shows the observed data for each of the 5 load steps (LS) and the three right columns show three different predictive samples.}
\label{fig: SUPP AAAAA ABBBA X field validation}
\end{figure}

\begin{figure}[ht!]
\textbf{AAAAA Posterior Predictive Distribution Over ABBBA Y Displacement}\par\medskip
\centering
\includegraphics[width=\linewidth]{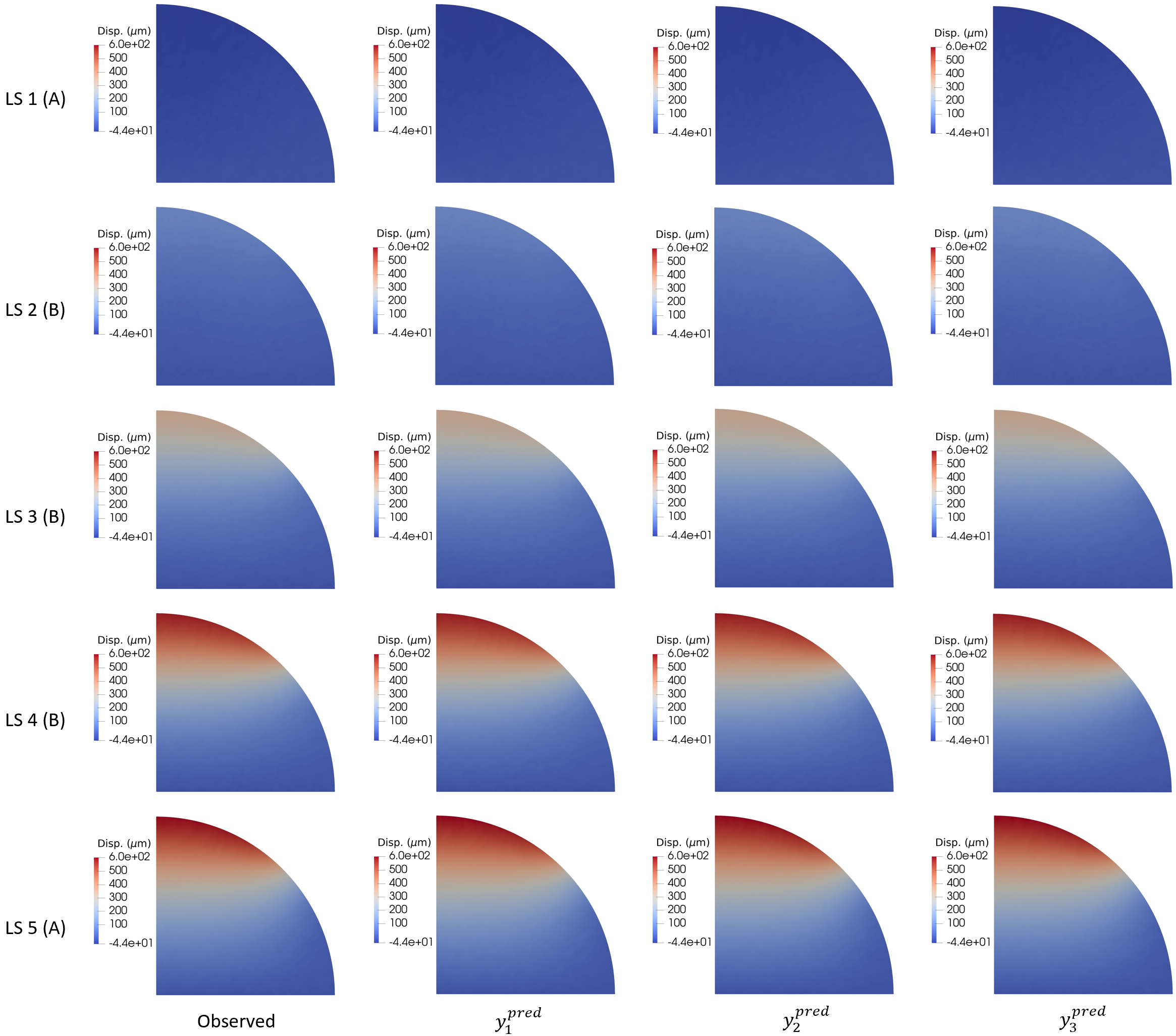}
\caption{Samples from the posterior predictive distribution of the AAAAA---calibrated model is shown for the $Y$ directional component of the displacement field for a new ABBBA load path. The left column shows the observed data for each of the 5 load steps (LS) and the three right columns show three different predictive samples.}
\label{fig: SUPP AAAAA ABBBA Y field validation}
\end{figure}

\begin{figure}[ht!]
\textbf{Error of AAAAA Posterior Predictive Distribution Over ABBBA X Displacement}\par\medskip
\centering
\includegraphics[width=\linewidth]{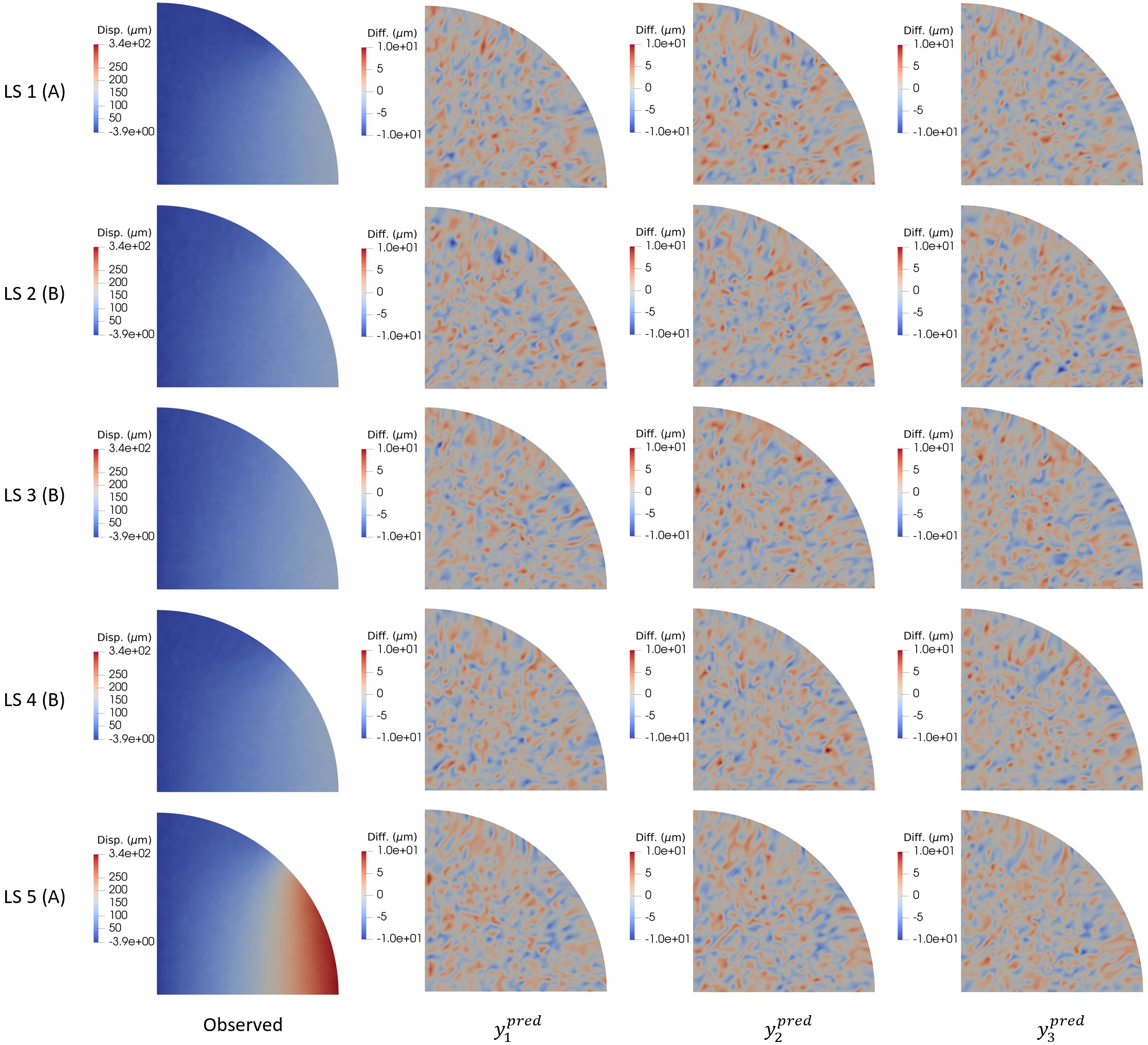}
\caption{The error for three samples from the AAAAA posterior predictive distribution are shown for the $X$ directional component of the displacement field for the new ABBBA load path. The left column shows the observed data for each of the 5 load steps (LS) and the three right columns shows the error between three different predictive samples and the validation data.}
\label{fig: SUPP AAAAA ABBBA X field validation diff}
\end{figure}

\begin{figure}[ht!]
\textbf{Error of AAAAA Posterior Predictive Distribution Over ABBBA Y Displacement}\par\medskip
\centering
\includegraphics[width=\linewidth]{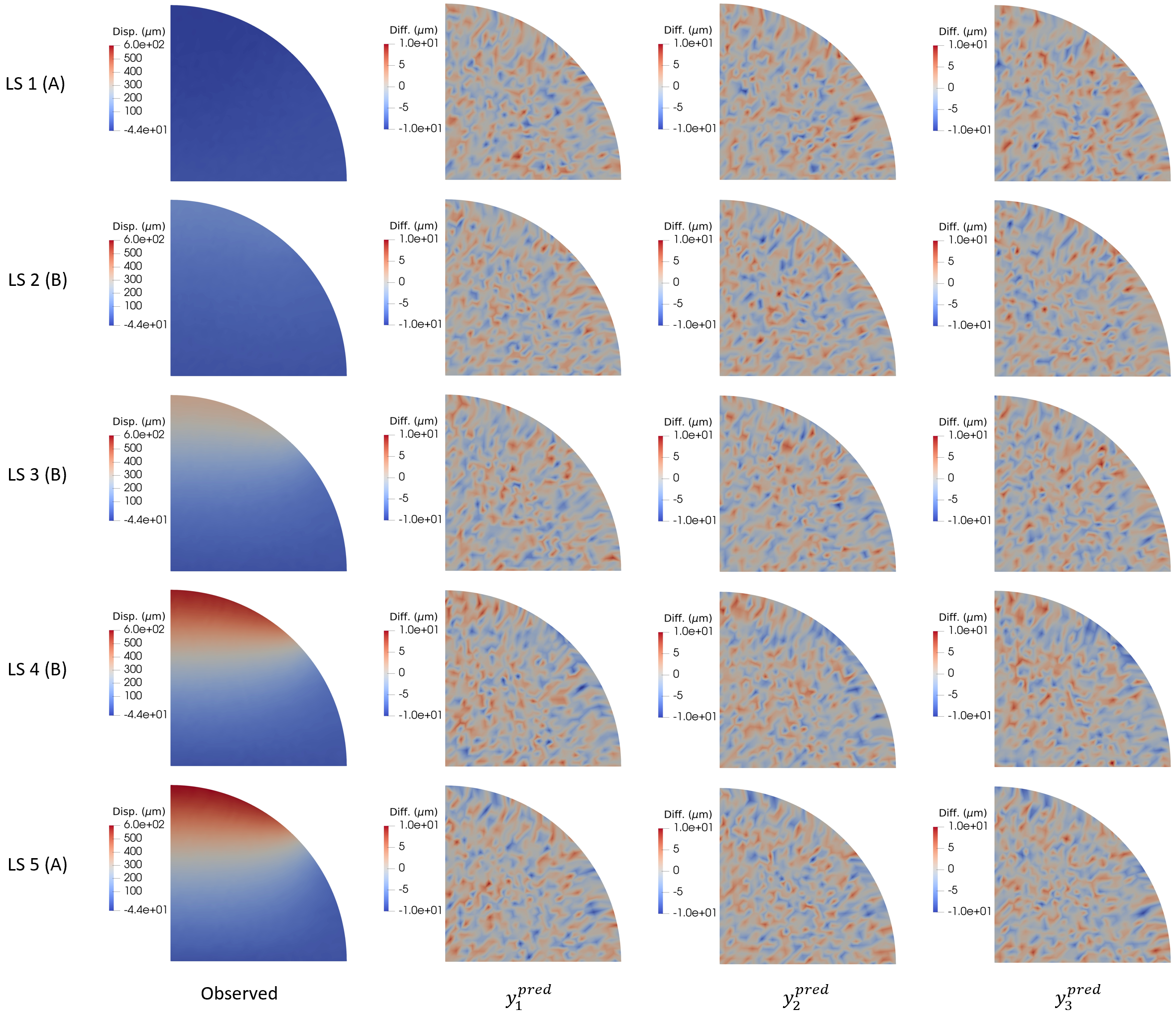}
\caption{Samples from the posterior predictive distribution of the AAAAA---calibrated model is shown for the $Y$ directional component of the displacement field for a new ABBBA load path. The left column shows the observed data for each of the 5 load steps (LS) and the three right columns show three different predictive samples.}
\label{fig: SUPP AAAAA ABBBA Y field validation diff}
\end{figure}

\begin{figure}%
    \centering
    \textbf{ABBBA Load Step 5 Data}\par\medskip
    \sidesubfloat[]{\includegraphics[width=55mm]{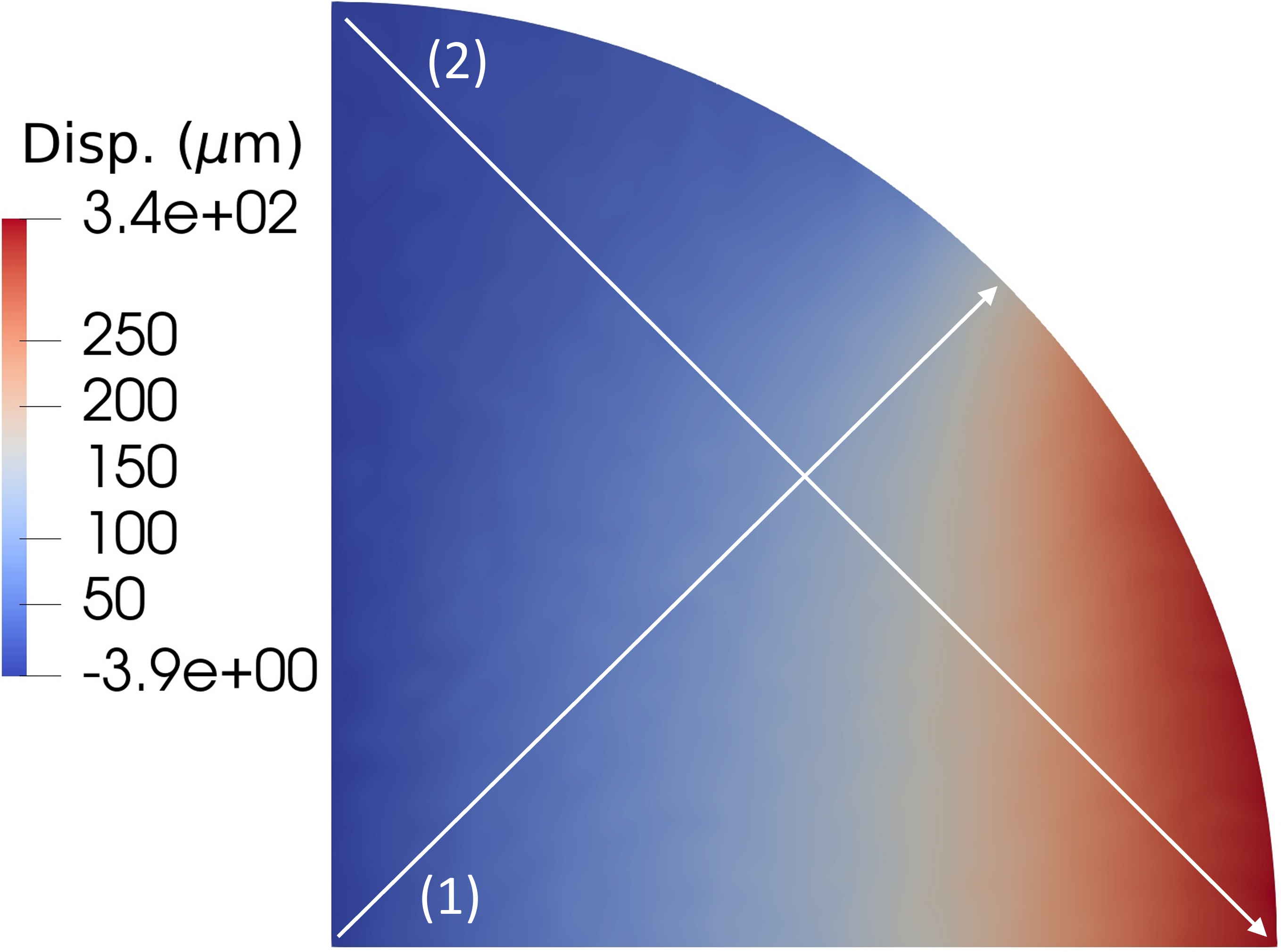}\label{fig: SUPP ABBBA ux cal data step 4}} %
    \sidesubfloat[]{\includegraphics[width=55mm]{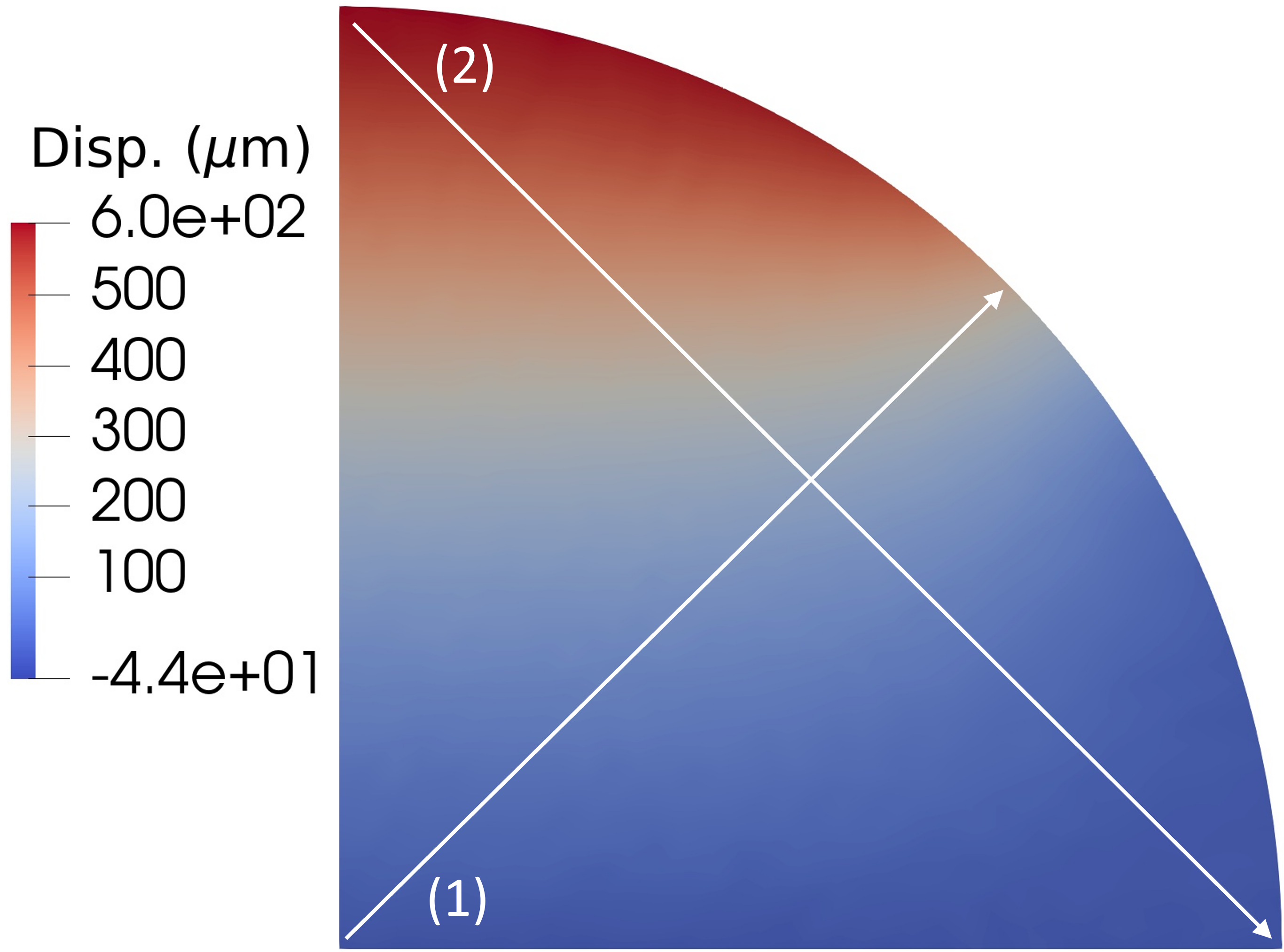}\label{fig: SUPP ABBBA uy cal data step 4}} 
    \caption{The observed displacement field at the final load step of load path ABBBA generated with $\pars^{true}$ and added noise for the (a) X directional component and (b) Y directional component reported in microns. Lines (1) and (2) indicate the space over which the posterior probability of the displacement field is plotted in the following figures.}%
    \label{fig: SUPP ABBBA step 4 val data with lines}%
\end{figure}

\begin{figure}%
\textbf{AAAAA Posterior Predictive Distribution Over ABBBA Displacements: Line 1}\par\medskip
    \centering
    \includegraphics[width=\linewidth]{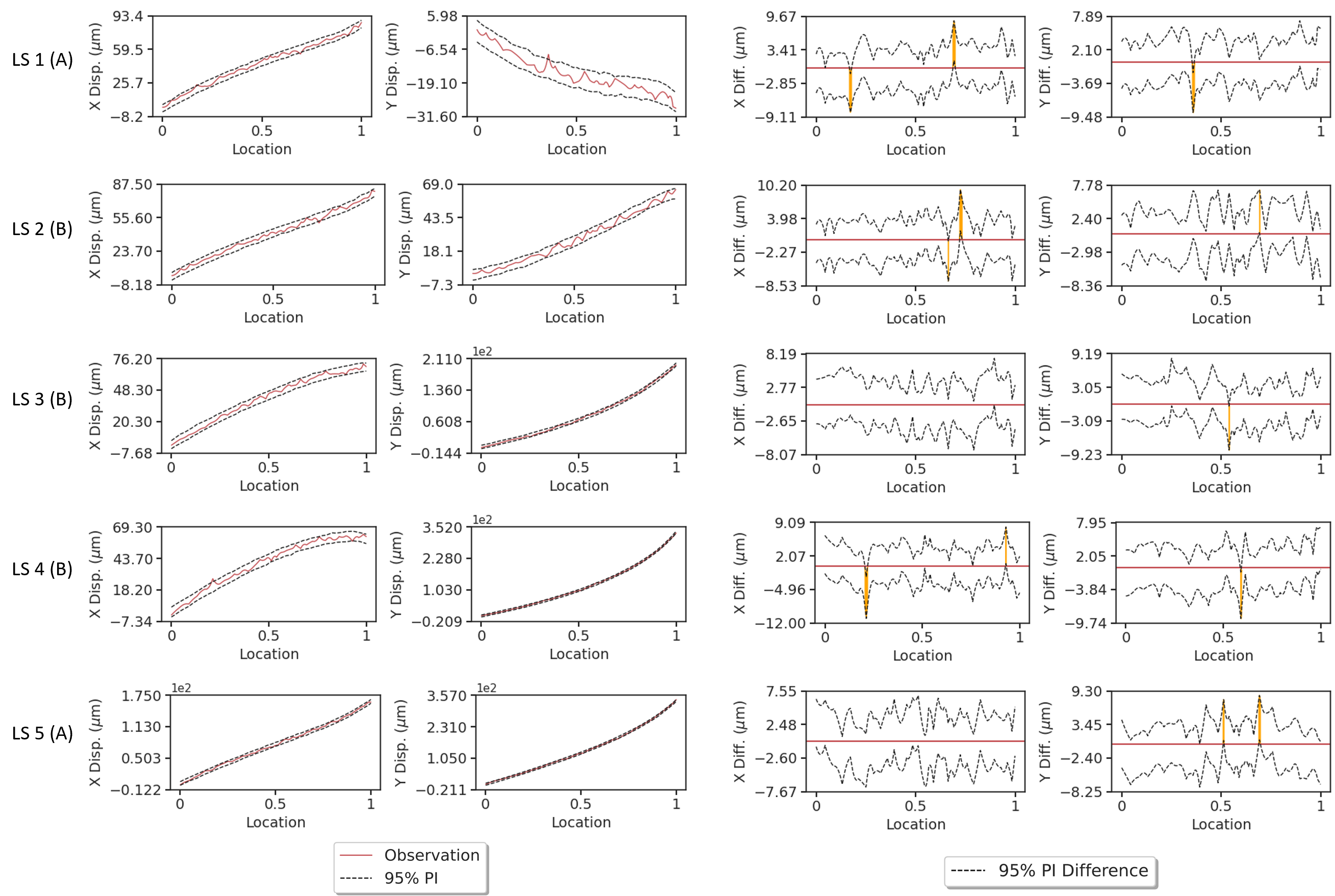}\label{fig: SUPP AAAAA post pred of ABBBA along line 1 (diag)} 
    \caption{The posterior predictive distribution of the AAAAA---calibrated model over Line 1 displacements (Fig.~\ref{fig: SUPP ABBBA step 4 val data with lines}) is shown for new load path ABBBA. The two left columns show the 95\%~PI of the distribution as well as the observations for the $X$ and $Y$ directional components for all 5 load steps (LS). The two right columns show the difference between the 95\%~PI and the observations. Regions where the observations fall outside the 95\%~PI are colored orange, and 0 is marked with a red line. The x-axis is the normalized location along Line 1.}%
    \label{fig: SUPP AAAAA prediction of ABBBA field diag line}%
\end{figure}

\begin{figure}%
\textbf{AAAAA Posterior Predictive Distribution Over ABBBA Displacements: Line 2}\par\medskip
    \centering
    \includegraphics[width=\linewidth]{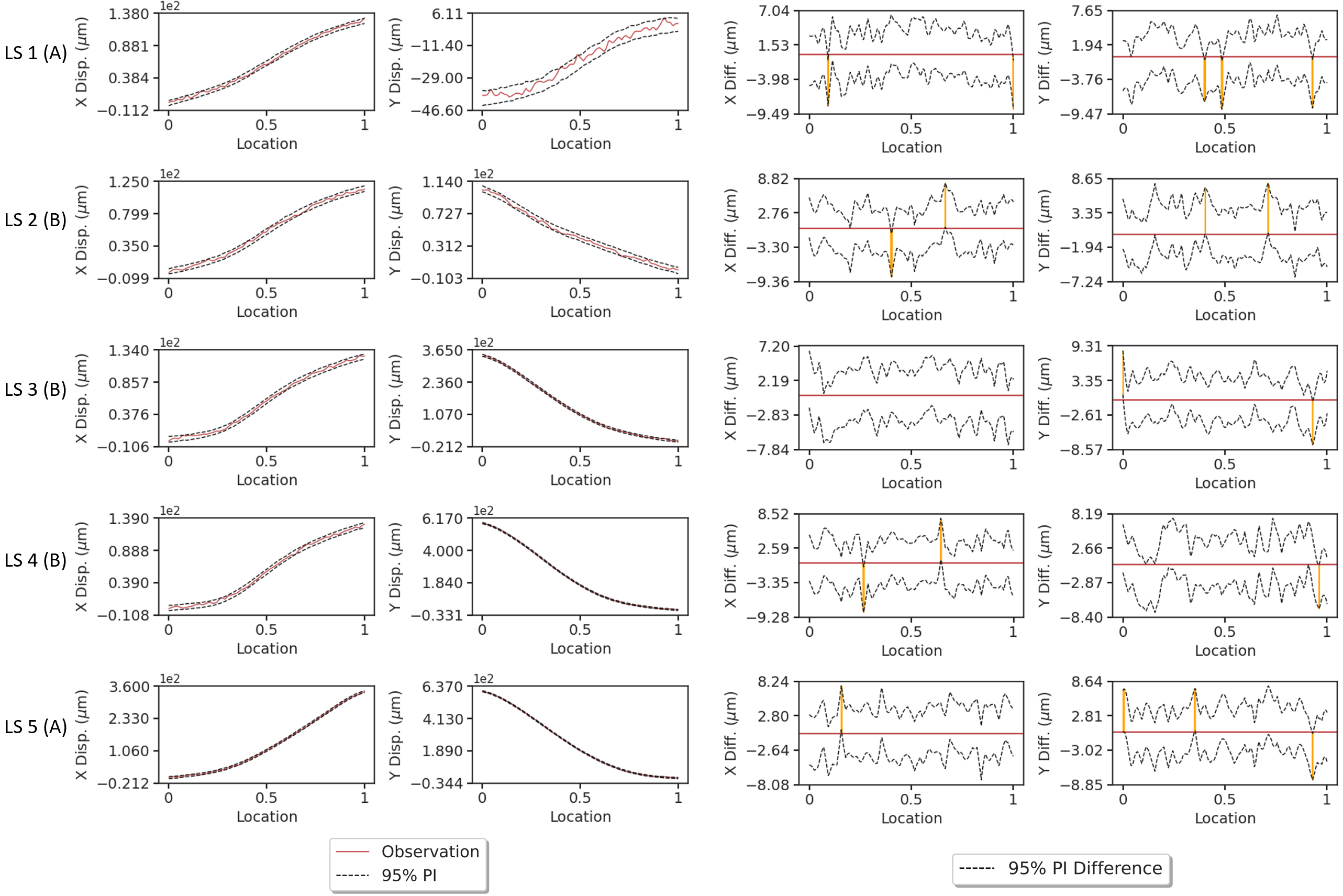}\label{fig: SUPP AAAAA post pred of ABBBA along line 2 (cross)} 
    \caption{The posterior predictive distribution of the AAAAA---calibrated model over Line 2 displacements (Fig.~\ref{fig: SUPP ABBBA step 4 val data with lines}) is shown for new load path ABBBA. The two left columns show the 95\%~PI of the distribution as well as the observations for the $X$ and $Y$ directional components for all 5 load steps (LS). The two right columns show the difference between the 95\%~PI and the observations. Regions where the observations fall outside the 95\%~PI are colored orange, and 0 is marked with a red line. The x-axis is the normalized location along Line 2.}%
    \label{fig: SUPP AAAAA prediction of ABBBA field cross line}%
\end{figure}

\begin{figure}[ht!]
\textbf{ABABA Posterior Predictive Distribution Over ABBBA X Displacement}\par\medskip
\centering
\includegraphics[width=\linewidth]{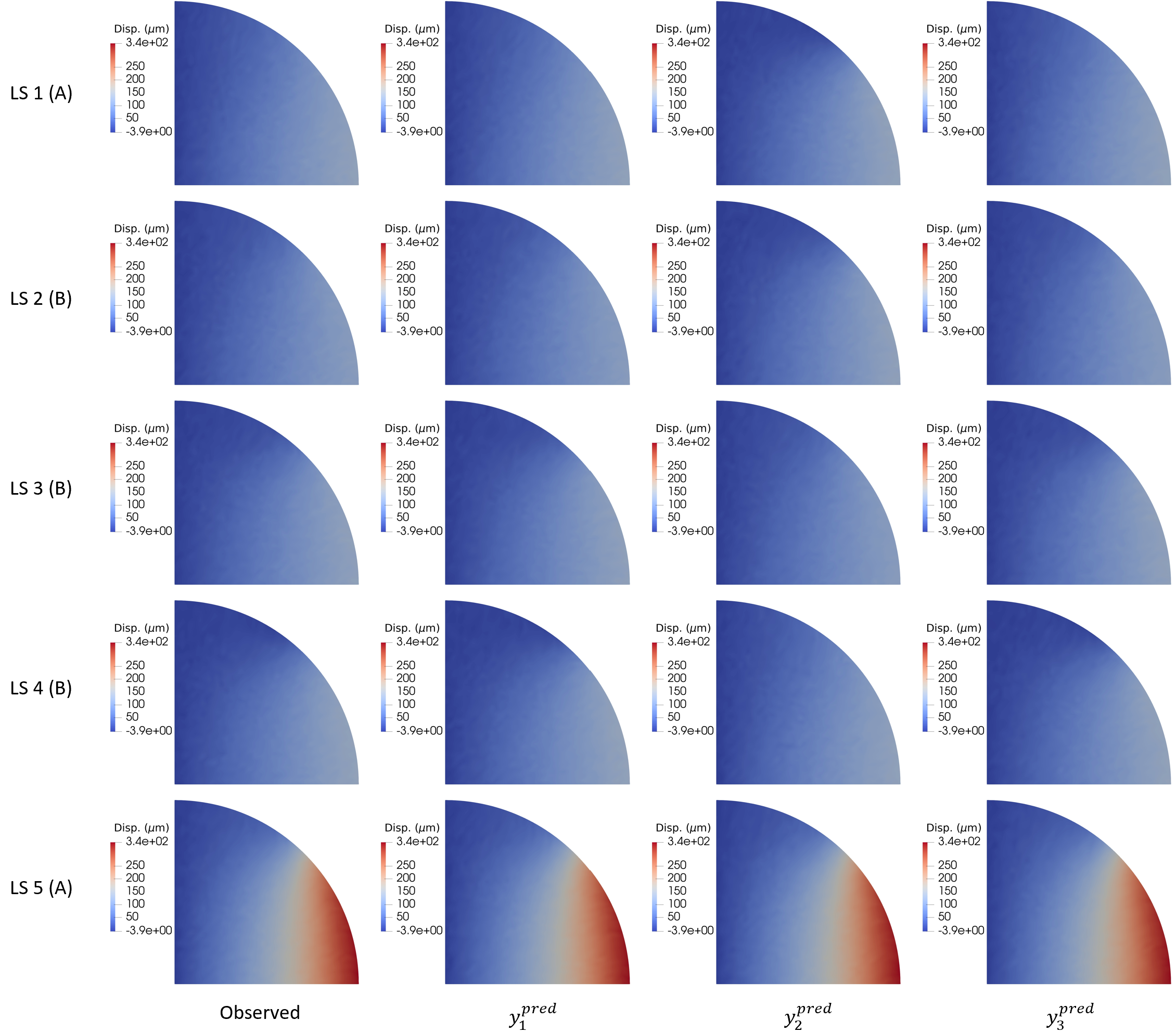}
\caption{Samples from the posterior predictive distribution of the ABABA---calibrated model is shown for the $X$ directional component of the displacement field for a new ABBBA load path. The left column shows the observed data for each of the 5 load steps (LS) and the three right columns show three different predictive samples.}
\label{fig: ABABA ABBBA X field validation}
\end{figure}

\begin{figure}[h!]
\textbf{ABABA Posterior Predictive Distribution Over ABBBA Y Displacement}\par\medskip
\centering
\includegraphics[width=\linewidth]{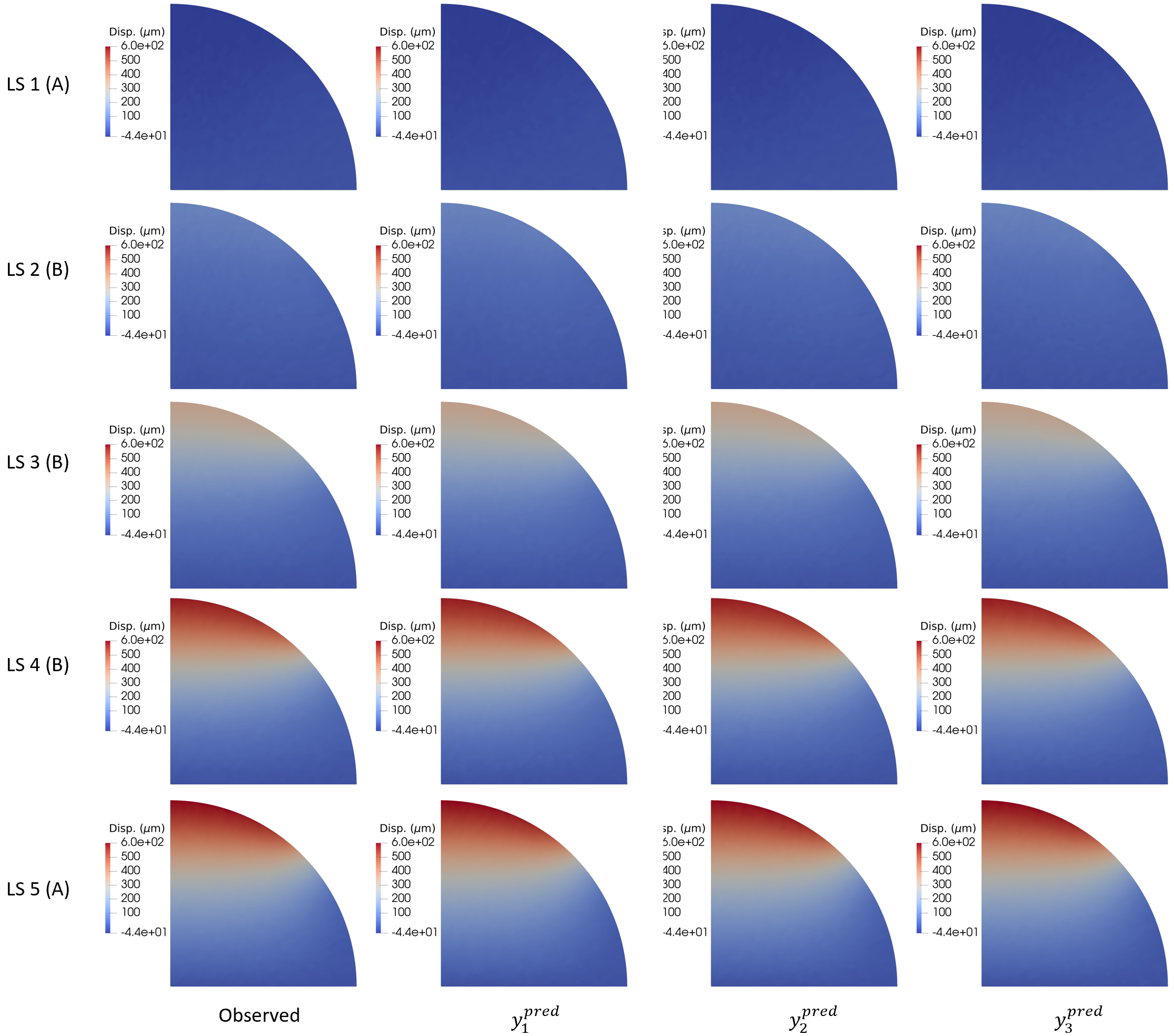}
\caption{Samples from the posterior predictive distribution of the ABABA---calibrated model is shown for the $Y$ directional component of the displacement field for a new ABBBA load path. The left column shows the observed data for each of the 5 load steps (LS) and the three right columns show three different predictive samples.}
\label{fig: ABABA ABBBA Y field validation}
\end{figure}

\begin{figure}[ht!]
\textbf{Error of ABABA Posterior Predictive Distribution Over ABBBA X Displacement}\par\medskip
\centering
\includegraphics[width=\linewidth]{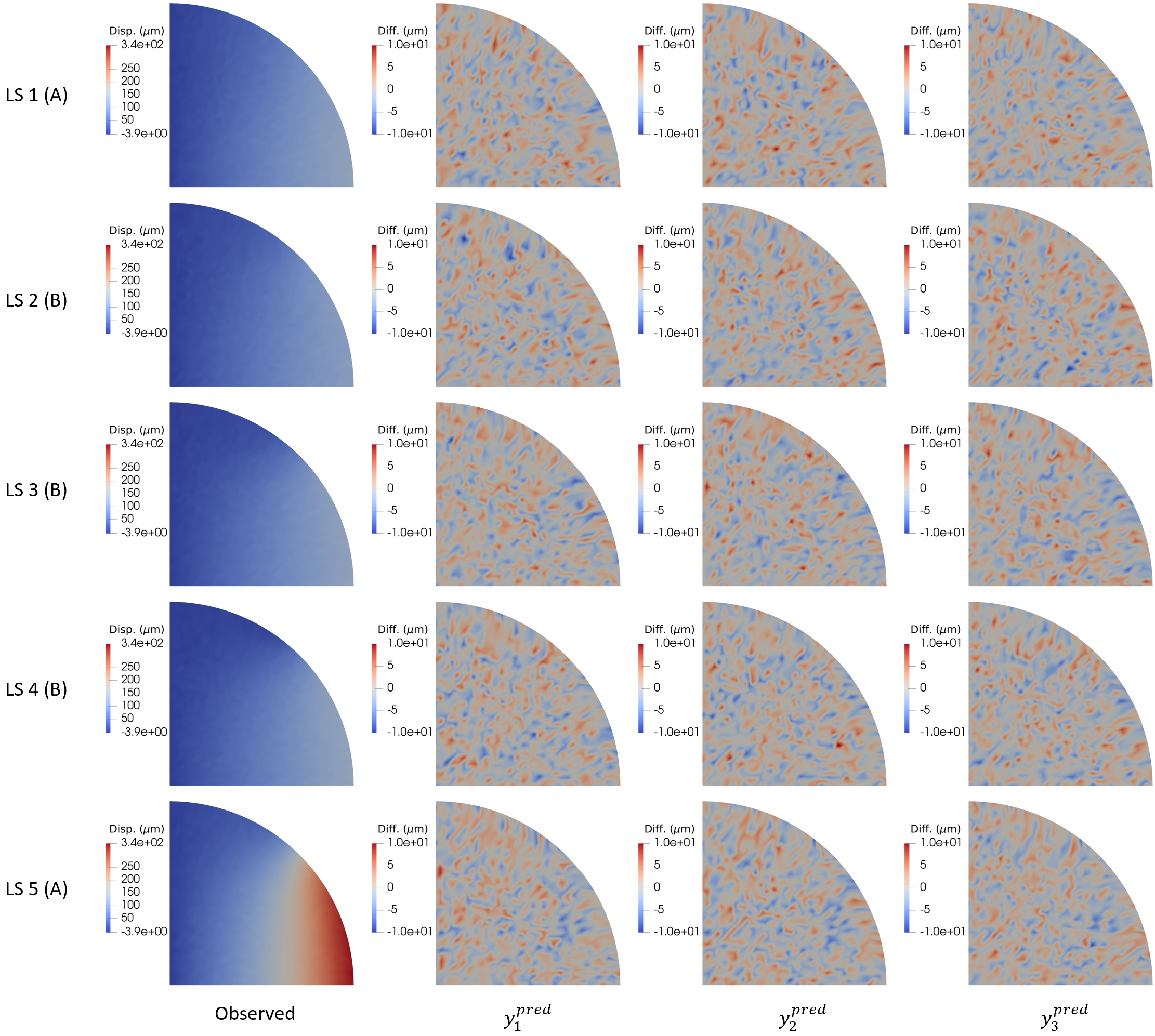}
\caption{The error for three samples from the ABABA posterior predictive distribution are shown for the $X$ directional component of the displacement field for the new ABBBA load path. The left column shows the observed data for each of the 5 load steps (LS) and the three right columns shows the error between three different predictive samples and the validation data.}
\label{fig: SUPP ABABA ABBBA X field validation diff}
\end{figure}

\begin{figure}[ht!]
\textbf{Error of ABABA Posterior Predictive Distribution Over ABBBA Y Displacement}\par\medskip
\centering
\includegraphics[width=\linewidth]{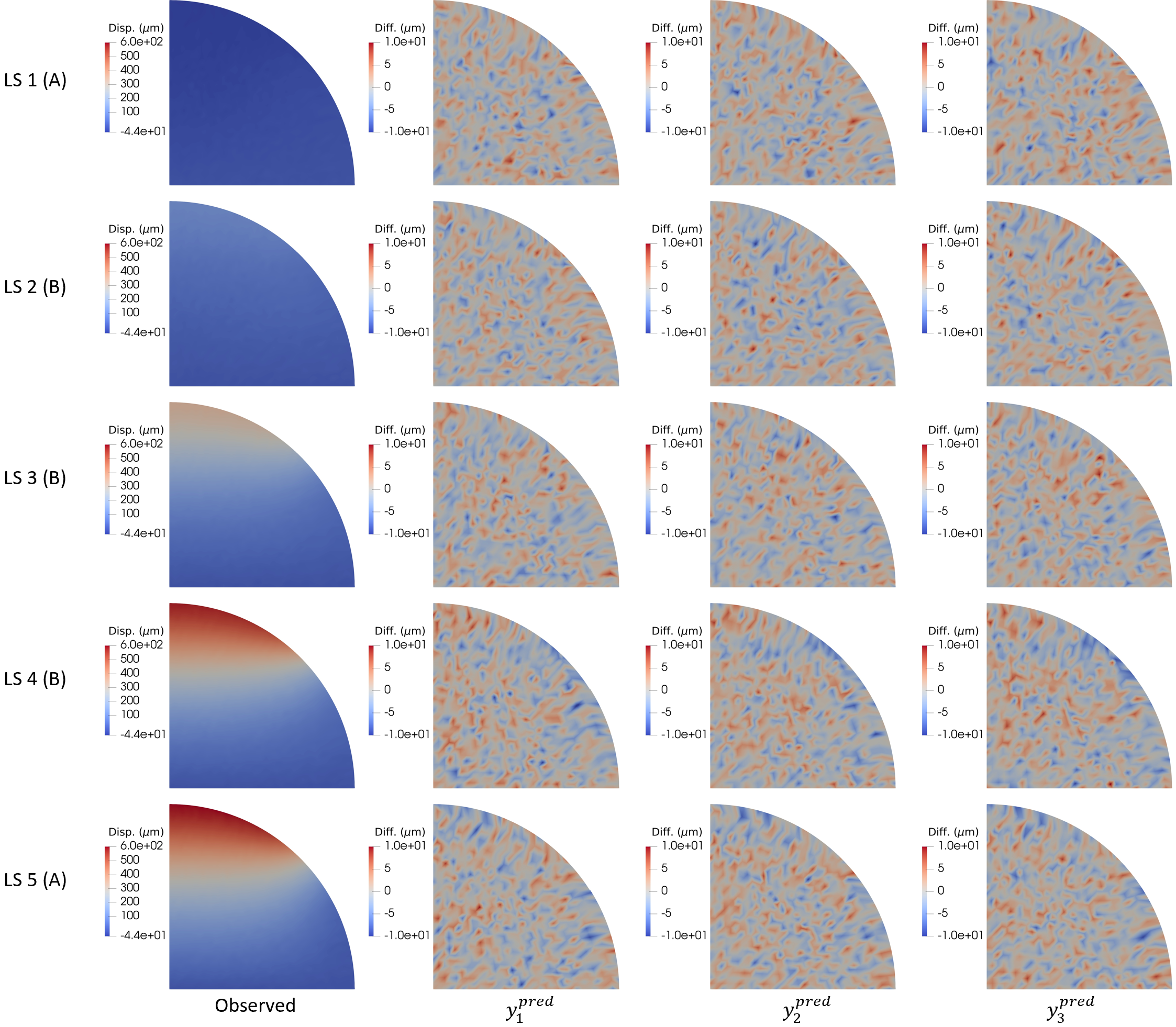}
\caption{The error for three samples from the ABABA posterior predictive distribution are shown for the $Y$ directional component of the displacement field for the new ABBBA load path. The left column shows the observed data for each of the 5 load steps (LS) and the three right columns shows the error between three different predictive samples and the validation data.}
\label{fig: SUPP ABABA ABBBA Y field validation diff}
\end{figure}

\begin{figure}[h!]%
    \textbf{ABABA Posterior Predictive Distribution Over ABBBA Displacements: Line 1}\par\medskip
    \centering
    \includegraphics[width=165mm]{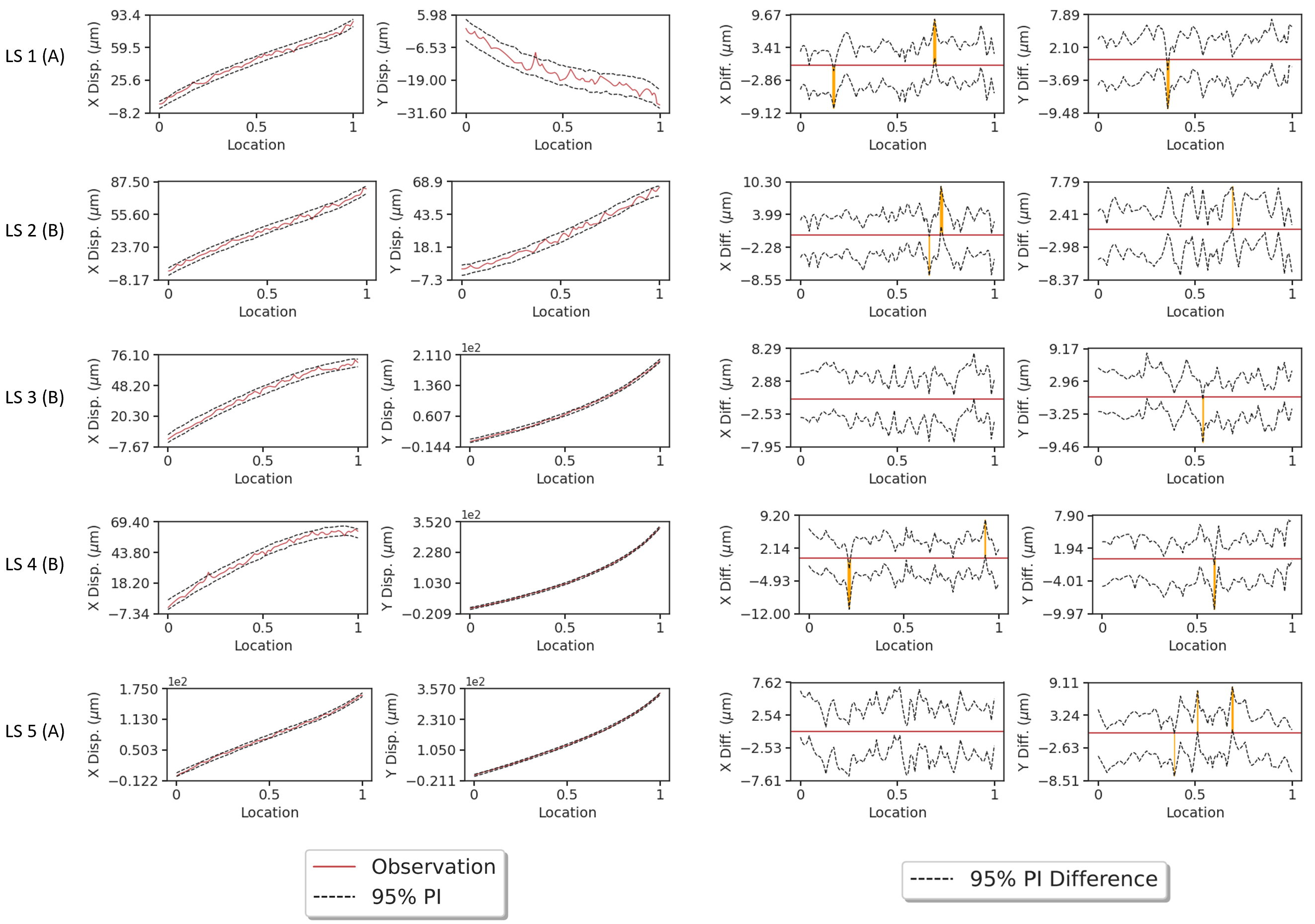}\label{fig: ABABA post pred of ABBBA along line 1 (diag)} 
    \caption{The posterior predictive distribution from the ABABA---calibrated model is displayed over the Line 1 displacements (Fig.~\ref{fig: SUPP ABBBA step 4 val data with lines}) for a new ABBBA load path. The two left columns show the 95\%~PI of the distribution as well as the observations for the $X$ and $Y$ directional components for all 5 load steps (LS). The two right columns show the difference between the 95\%~PI and the observations. Regions where the observations fall outside the 95\%~PI are colored orange, and 0 is marked with a red line. the x-axis is the normalized location along Line 1.}%
    \label{fig: ABABA prediction of ABBBA field diag line}%
\end{figure}

\begin{figure}[h!]%
    \textbf{ABABA Posterior Predictive Distribution Over ABBBA Displacements: Line 2}\par\medskip
    \centering
    \includegraphics[width=165mm]{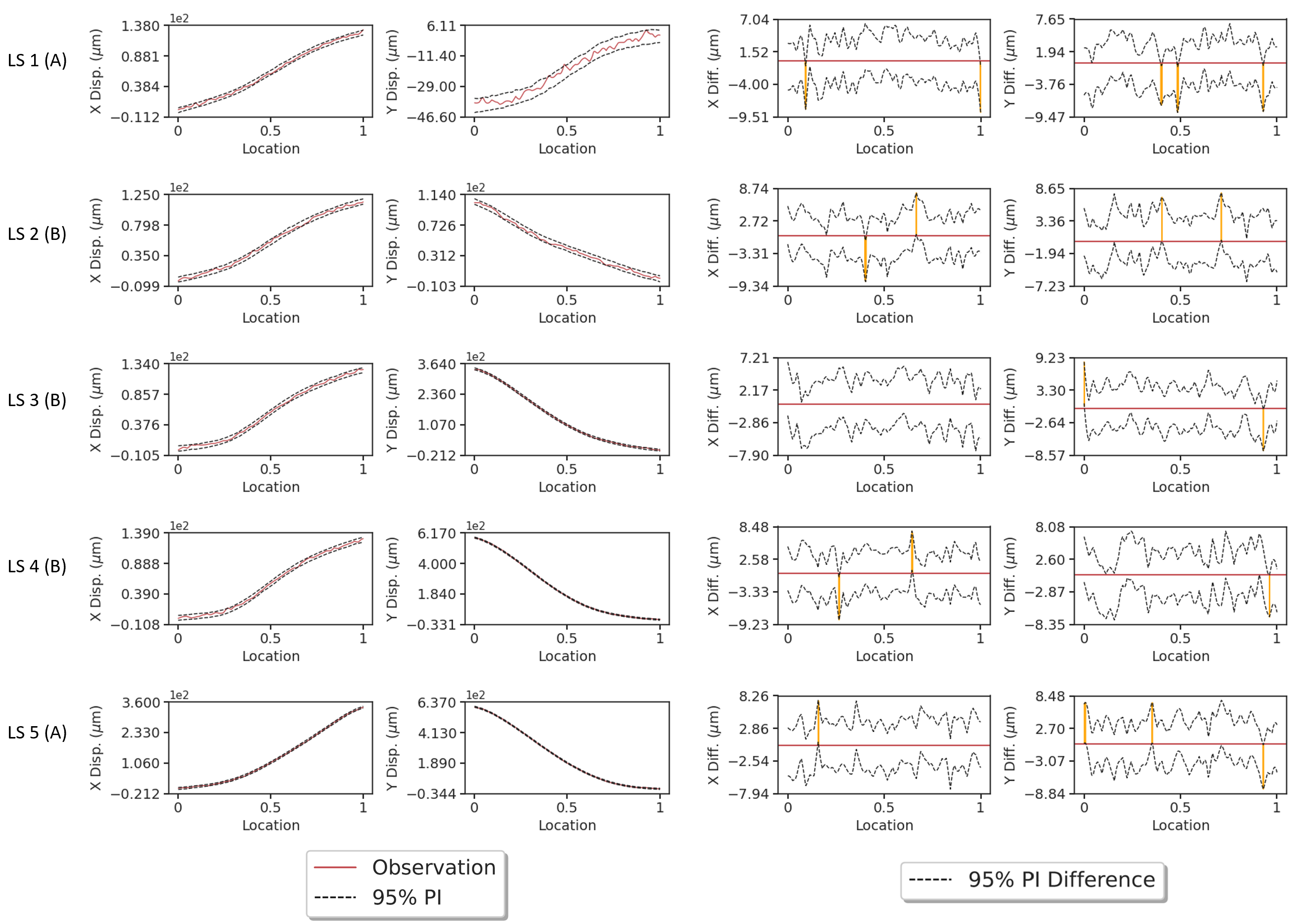}\label{fig: ABABA post pred of ABBBA along line 2 (diag)} 
    \caption{The posterior predictive distribution from the ABABA---calibrated model is displayed over the Line 2 displacements (Fig.~\ref{fig: SUPP ABBBA step 4 val data with lines}) for a new ABBBA load path. The two left columns show the 95\%~PI of the distribution as well as the observations for the $X$ and $Y$ directional components for all 5 load steps (LS). The two right columns show the difference between the 95\%~PI and the observations. Regions where the observations fall outside the 95\%~PI are colored orange, and 0 is marked with a red line. the x-axis is the normalized location along Line 2.}%
    \label{fig: ABABA prediction of ABBBA field cross line}%
\end{figure}

\end{document}